\documentclass[12pt]{article}
\usepackage{amsmath,amssymb,epsfig,amsfonts}
\usepackage{graphicx,subfigure}
\usepackage[usenames, dvipsnames]{color}
\usepackage[backref]{hyperref}
\usepackage[nosort]{cite}
\usepackage{lscape}
\usepackage{rotating}

\usepackage{extarrows}

\usepackage{verbatim} 

\makeatletter

\newcommand{\be}{\begin{equation}}
\newcommand{\ee}{\end{equation}}
\newcommand{\ba}{\begin{aligned}}
\newcommand{\ea}{\end{aligned}}

\newenvironment{mathbox}
{\par\smallskip\centering\begin{lrbox}{0}%
\begin{minipage}[c]{0.3\textwidth}}
{\end{minipage}\end{lrbox}%
\framebox[0.25\textwidth]{\usebox{0}}%
\par\medskip
\ignorespacesafterend}

\newcommand{\bea}{\begin{eqnarray}}
\newcommand{\eea}{\end{eqnarray}}

\newcommand{\hyperplane}{H} 
\newcommand{\rootsystem}{\Phi} 
\newcommand{\weight}{L} 
\newcommand{\coweight}{e} 

\def\sign{\mathop{\mathrm{sign}}\nolimits}

\def\unit{{1\kern-.65ex {\rm l}}}
\def\1{{1\kern-.65ex {\rm l}}}

\newcount\hour \newcount\minute
\hour=\time \divide \hour by 60
\minute=\time
\def\now{%
\ifnum \hour<13
  \ifnum \hour=0 \advance \hour by 12 \number\hour:\else \number\hour:\fi%
     \ifnum \minute<10 0\fi%
     \number\minute%
\ A.M.%
\else \advance \hour by -12 \number\hour:%
  \ifnum \minute<10 0\fi%
  \number\minute%
  \ P.M.%
\fi%
}

\makeatother

\begin{document}

\baselineskip=18pt  
\numberwithin{equation}{section}  
\allowdisplaybreaks  


%
%


\thispagestyle{empty}

\vspace*{-2cm} 
\begin{flushright}
\begin{center} 
{\tt IFT-UAM/CSIC-14-005}\qquad\qquad
{\tt KCL-MTH-14-01} \qquad \qquad 
{\tt UCSB Math 2014-07} \\
\end{center}
\end{flushright}

\vspace*{0.8cm} 
\begin{center}
{\Huge  Box Graphs and Singular Fibers}\\

 \vspace*{1.5cm}
{Hirotaka Hayashi$\,^1$, Craig Lawrie$\,^2$, David R. Morrison$\,^3$, and Sakura Sch\"afer-Nameki$\,^2$}\\

 \vspace*{1.0cm} 
$^1$  {\it Instituto de Fisica Teorica UAM/CSIS, \\
Cantoblanco, 28049 Madrid, Spain }\\
{\tt h.hayashi csic.es
}\\
\vspace*{0.5cm}
$^2$ {\it Department of Mathematics, King's College, London \\
 The Strand, London WC2R 2LS, England }\\
 {\tt {gmail:$\,$ craig.lawrie1729, sakura.schafer.nameki}}\\
\vspace*{0.5cm}
$^3$ {\it Departments of Mathematics and Physics, \\
University of California, Santa Barbara, CA 93106, USA}\\
{\tt drm physics.ucsb.edu}

\vspace*{0.8cm}
\end{center}
\vspace*{.1cm}

\noindent
We determine the higher codimension fibers of elliptically fibered Calabi-Yau fourfolds with section
 by studying the three-dimensional $\mathcal{N}=2$ supersymmetric  gauge theory with matter which describes
 the low energy effective theory of M-theory compactified on the associated
 Weierstrass model, a singular model of the fourfold.  Each phase of the
 Coulomb branch of this theory corresponds to a particular resolution of the
 Weierstrass model, and we show that these have a concise description in terms
 of decorated box graphs based on the representation graph of the matter
 multiplets, or alternatively by a class of convex paths on said graph.
 Transitions between phases have a simple interpretation as ``flopping" of the
 path, and in the geometry correspond  to actual flop transitions. This
 description of the phases enables us to enumerate and determine the
 entire network between them, with various matter representations for all
 reductive Lie groups. Furthermore, we observe that each network of phases
 carries the structure of a (quasi-)minuscule representation of a specific Lie algebra. 
Interpreted from a geometric point of view, this analysis determines the generators of the cone of effective curves as well as the network of flop transitions between crepant resolutions of singular elliptic Calabi-Yau fourfolds. From the box graphs we determine all fiber types in codimensions two and three, and we find new, non-Kodaira, fiber types for $E_6$, $E_7$ and $E_8$.


\tableofcontents

\newpage

\section{Introduction and Overview}

The Kodaira-N\'eron classification of fibers in nonsingular elliptic surfaces associates to each singular fiber a decorated affine Dynkin diagram corresponding to a simple Lie algebra $\mathfrak{g}$, where the decoration indicates the multiplicities of 
the irreducible fiber components \cite{Kodaira, Neron}. When the nonsingular elliptic surface is the resolution of a (singular) Weierstrass model, the Dynkin diagram can be associated with the singularity.  For higher-dimensional elliptically fibered geometries the analysis of fibers in codimension one is very similar but a natural question arises: is the Kodaira-N\'eron classification  still applicable to fibers in  higher codimension?
In this paper we answer this question for elliptically fibered Calabi-Yau varieties and show that the fibers in codimensions two and three have a classification in terms of decorated representation graphs, so-called {\it decorated box graphs}, associated to a representation ${\bf R}$ of the Lie algebra $\mathfrak{g}$. 
These box graphs contain the information about the higher-codimension fiber type, which in general goes beyond the Kodaira-N\'eron classification. In particular they specify the extremal rays of the cone of effective curves of the resolved geometry, and thereby the network of possible flop transitions among different resolutions.

The correspondence between decorated box graphs and singular fibrations is inspired by M-theory/F-theory duality, which implies a characterization of crepant resolutions of an elliptically fibered Calabi-Yau variety in terms of the Coulomb phases of a three-dimensional  $\mathcal{N}=2$ supersymmetric gauge theory,  which describes the low energy effective theory of M-theory compactified on the fourfold 
\cite{Cadavid:1995bk,Ferrara:1996hh,WitMF,Ferrara:1996wv,Becker:1996gj,SVW,Morrison:1996xf,Witten:1996md,fiveDgauge, Diaconescu:1998ua,Gukov:1999ya}. 
If the Calabi-Yau variety has an elliptic fibration with a section, then we can also take the F-theory limit by shrinking the size of the elliptic fiber \cite{Vafa:1996xn,Morrison:1996na,Morrison:1996pp,Bershadsky:1997zs}. Compactification on the resolved Calabi-Yau fourfold  with fiber type $G$ in codimension one realizes the Coulomb branch with gauge group broken to $U(1)^r$, where $r$ is the rank of $G$;  inclusion of matter introduces a substructure in the Coulomb branch \cite{deBoer:1997kr, Aharony:1997bx}. A crepant resolution of the Calabi-Yau variety then corresponds to a Coulomb phase of the three-dimensional theory.
The study of this correspondence was initiated in \cite{Morrison:1996xf, fiveDgauge} in the case of Calabi-Yau threefolds, and further pursued in the case of Calabi-Yau fourfolds in \cite{Diaconescu:1998ua, Grimm:2011fx,  Cvetic:2012xn, Hayashi:2013lra}. More concretely, any crepant resolution will resolve the codimension-one Kodaira fibers and the corresponding exceptional curves can be labeled by the simple roots of $\mathfrak{g}$, intersecting according to the (affine) Dynkin diagram of $\mathfrak{g}$. 
Along codimension-two loci, some of these curves become reducible, corresponding to roots splitting into weights of ${\bf R}$, as observed in \cite{gorenstein-weyl, MS}, and in codimension three these further split into each other in a way compatible with the Yukawa couplings.

The main idea of this paper is to use the correspondence between crepant resolutions of elliptic Calabi-Yau varieties and the Coulomb phases of the gauge theory to find a purely representation theoretic description of the fibers in codimensions two and three, as well as their network of flop transitions. To this effect, we first analyze the structure of the Coulomb phases of a three-dimensional  $\mathcal{N}=2$ supersymmetric gauge theory obtained by compactification of M-theory on a Calabi-Yau fourfold (or likewise the five-dimensional analog for Calabi-Yau threefolds) and prove the correspondence between Coulomb phases and decorated box graphs. More precisely, consider a gauge theory with gauge group $G$ specified by the Kodaira fiber type in codimension one of the elliptic fibration. In addition consider matter in a representation ${\bf R}$ of the gauge group $G$. This is modeled in the Calabi-Yau by codimension-two loci in the base of the elliptic fibration, in particular, the Kodaira fiber type in codimension one  can degenerate further in higher codimension \cite{gorenstein-weyl,Katz:1996xe}. The type of degeneration depends on the representation, but also on the precise embedding of the cone of effective curves in codimension one into that in codimension two. In terms of the Coulomb phases of the three-dimensional gauge theory this corresponds to choosing a cone inside the Weyl chamber of the gauge group $G$. We show that  this choice is characterized in terms of a coloring of the representation graph of ${\bf R}$, which we refer to as decorated box graph.

Returning to the geometry, we then show that the decorated box graphs fully characterize the fibers in higher codimension and can be used to (re)construct the fibers: the box graphs contain the information about the extremal generators (or rays) of the cone of effective curves,  as well as their intersection data, and thereby the analog of the Kodaira-N\'eron intersection graph for the fiber. The box graphs furthermore contain the information about flop transitions, which map topologically distinct small resolutions into each other. 
Schematically the correspondence we use is as follows

\be
\begin{array}{ccccc}
&&
\begin{mathbox}
\begin{center}
Decorated \\
Box Graphs
\end{center}
\end{mathbox} &&\cr
&\nearrow\swarrow  &&\searrow\nwarrow& \cr
\begin{mathbox}
\begin{center}
Coulomb Phases of \\
$d=3,\  \mathcal{N}=2 $\\
supersymmetric \\
gauge theories
\end{center}
\end{mathbox}
&&
\xleftrightarrow[{}]{\quad \text{M-theory compactification}\quad } 
&& \begin{mathbox}
\begin{center}
Crepant resolutions \\
 of elliptically \\
 fibered
  Calabi-Yau fourfolds\\
 \end{center}
\end{mathbox}
\end{array}
\ee


With this correspondence in place we find the following implications for the fibers in higher codimension. The fiber types generically are not of Kodaira type, and the description in terms of box graphs gives a full classification of the types of fibers that can arise. In particular for the case of rank one enhancements we show that the different fiber types are obtained by deleting nodes in the Kodaira fiber. 
This is in accord with known examples of crepant resolutions of elliptic Calabi-Yau varieties, where it has been known that the fibers in higher codimension need not belong to Kodaira's list \cite{Morrison:2011mb,Esole:2011sm, MS, Lawrie:2012gg} (for earlier examples illustrating a related issue, see \cite{birgeoRDP}).  Detailed studies of such resolutions for Calabi-Yau fourfolds, mostly focusing on an $SU(5)$ gauge group, i.e., $I_5$ Kodaira fiber in codimension one, appeared in \cite{Esole:2011sm, MS, Lawrie:2012gg} using algebraic methods, and in \cite{Blumenhagen:2009yv, Grimm:2009yu, Grimm:2010ez, Krause:2011xj, Grimm:2011fx} using toric resolutions.

The network of small resolutions connected by flop transitions follows directly from the decorated box diagrams, by flops of the extremal generators of the cone of curves.  The intriguing correspondence that we find from the identification with box graphs is that in many cases, the network of flops is given in terms of representation graphs of so-called (quasi-)minuscule representations of a Lie algebra $\widetilde{\mathfrak{g}} \supset \mathfrak{g}$, where $\mathfrak{g}$ is the Lie algebra associated to the gauge group $G$. 

The box graphs allow in addition the analysis of multiple matter representations, which we exemplify for $G=SU(5)$ with matter in the fundamental and anti-symmetric representations. The complete network of small resolutions  for $SU(5)$ was determined in \cite{Hayashi:2013lra}, where it was observed that neither toric nor standard algebraic resolutions are sufficient to map out all topologically inequivalent resolutions, and some of these can be reached only by flop transitions along fiber components, that exist only in codimension two (above ``matter loci"). The present method using the decorated representation graph gives a systematic characterization of these networks, and in particular reproduces the flop network for $SU(5)$.

There are various equivalent descriptions of the decorated box graphs or the Coulomb phases of the three-dimensional gauge theory. To be more precise, the general situation we consider is a Lie subalgebra $\mathfrak{g}\subset \widetilde{\mathfrak{g}}$ with trivial center, and commutant $\mathfrak{g}_{\perp}$
of $\mathfrak{g}$ in $\widetilde{\mathfrak{g}}$.  The adjoint representation
of $\widetilde{\mathfrak{g}}$ decomposes as
\be\ba\label{gtildeg}
\widetilde{\mathfrak{g}} &\quad\rightarrow \quad \mathfrak{g} \oplus \mathfrak{g}_{\perp}\cr
\hbox{Adj} (\widetilde{\mathfrak{g}} ) &\quad\rightarrow \quad  \hbox{Adj} ({\mathfrak{g}} ) 
\oplus \hbox{Adj} (\mathfrak{g}_{
\perp}) \oplus {\bf R} \oplus  \overline{{\bf R}} \,, 
\ea\ee
with ${\bf R}\oplus \overline{{\bf R}}$ the analogue of a bifundamental
representation.  In the case that $\mathfrak{g}_{\perp}=\mathfrak{u}(1)$, for instance, 
we show the equivalence between each of the following points
\begin{itemize}
\item Coulomb phases of $d=3$ $\mathcal{N}=2$ $\mathfrak{g}\oplus \mathfrak{u}(1)$ gauge theory with matter in ${\bf R}\oplus \overline{{\bf R}}$
\item Elements of the Weyl group quotient $W_{\widetilde{\mathfrak{g}}}/W_{\mathfrak{g}}$ (with so-called Bruhat ordering)
\item Codimension-two fibers of elliptic Calabi-Yau varieties, with codimension-one Kodaira fiber type 
corresponding to $\mathfrak{g}$ with an additional section realizing $\mathfrak{u}(1)$
\item Decorated box graphs constructed from the 
representation graph of ${\bf R}$ with decoration by signs (colorings),  obeying so-called {\it flow rules}
\item (Anti-)Dyck paths\footnote{{\em Dyck paths} are staircase paths on a (representation) graph, 
which are not allowed to cross the diagonal \cite{enumerative-combinatorics-2}.} on the representation graph: 
paths on the representation graph (which for the case of simple gauge group, such as $\mathfrak{su}(N)$, 
have to cross the diagonal, thus the {\it Anti}-Dyck). 
\end{itemize}
From this list of equivalent descriptions, the decorated box graphs or anti-Dyck paths are particularly  simple and elegant ways to describe the phases, and allow the determination of the entire network of phases by simple, combinatorial rules. Transitions between phases are described in terms of sign changes or ``flops" of corners of the paths. We prove and exemplify this method in a  large class of matter representations for $\mathfrak{su}(N)$, $\mathfrak{so}(N)$, $\mathfrak{sp}(N)$, and the exceptional Lie algebras. 
Furthermore we draw a connection between the network of flop transitions and (quasi-)minuscule representations of $\widetilde{\mathfrak{g}}$. 

Geometrically the situation described above is less generic as it requires an additional rational section. We therefore also discuss the decorated box graphs and corresponding paths in the case when the gauge algebra is just $\mathfrak{g}$, in which case further restrictions have to be imposed on the allowed colorings. For instance for $\mathfrak{g}= \mathfrak{su}(n)$, the tracelessness condition implies that the box graphs have to satisfy an additional diagonal condition, which in terms of paths on the representation graph corresponds to restricting to anti-Dyck paths. 

The fibers in higher codimension are of particular interest when $\mathfrak{g}_{\perp} = \mathfrak{su}(2)$, which allows  additional monodromy, for instance, $\mathfrak{su}(6)\oplus \mathfrak{su}(2)\subset \mathfrak{e}_6$. In this case additional monodromy is possible, which is characterized by phases which are invariant under the action of the  Weyl group of $\mathfrak{g}_{\perp}$.  Reconstructing the fibers from the  decorated box graphs in codimensions two and three, it follows that  for non-trivial monodromy these are not of Kodaira type. In general, they have fewer components than a Kodaira type fiber and we refer to these fibers as being {\em monodromy-reduced}.  For example for  $\widetilde{\mathfrak{g}}=\mathfrak{e}_6$ both    in codimension two (from $\mathfrak{su}(6)$) or codimension three (from $\mathfrak{su}(5)$) we will determine all the fiber types and show that there are a few new non-Kodaira fibers that can occur, which go beyond the ones obtained by Esole-Yau in \cite{Esole:2011sm}. The possible fibers are summarized in figures \ref{fig:SU6Lambda3Fibs} and \ref{fig:SU5AFCodim3Fibs}. In fact we show that all monodromy-reduced fibers are obtained by deleting one of the non-affine nodes of the Kodaira fiber. For $E_7$ and $E_8$ this is shown in appendix \ref{app:EPhases}. 

Finally, we should point out a few nice conclusions and a somewhat curious observation regarding the networks of flop transitions. First, we should highlight the fact that in many cases, the network of phases or, equivalently, small resolutions  form a so-called {\it minuscule representation of } $\widetilde{\mathfrak{g}}$, where the representation structure is exactly given by the flop transitions. I.e. not only is the structure of the fibers determined in terms of representation data, but also the network of flop transitions has a representation structure under the higher rank Lie algebra  $\widetilde{\mathfrak{g}}$. We shall discuss this correspondence in detail in section  \ref{sec:MiniRep}. Whenever the commutant $\mathfrak{g}_{\perp}=\mathfrak{su}(2)$, we further show that the flop diagram forms the (non-affine) Dynkin diagram of the Lie algebra $\widetilde{\mathfrak{g}}$. 
The case of somewhat more peculiar nature is when $\mathfrak{g}_{\perp}=\mathfrak{u}(1)$. The phases of the gauge theory with gauge algebra $\mathfrak{g}\oplus \mathfrak{u}(1)$ form a minuscule representation of $\widetilde{\mathfrak{g}}$. However, the phases of the theory without the additional abelian factor  seem to form pairs of Dynkin diagrams of the Lie algebra $\widetilde{\mathfrak{g}}$, glued together as for instance shown in figure \ref{fig:U5A} for $\mathfrak{su}(5)$ with ${\bf 10}$ matter, figure \ref{fig:SU5AFPhaseDiag} for $\mathfrak{su}(5)$ with ${\bf 10}$ and ${\bf 5}$ matter\footnote{In this case there are three ways to cut the flop diagram resulting in pairs of $E_6$, $D_6$ and $A_6$ Dynkin diagrams, respectively.}, and $\mathfrak{e}_6$ with ${\bf 27}$ matter in figure 
\ref{fig:E6Flops27}. These are certainly curious observations that require further investigation.

In the mathematics literature, considerations of Weyl group actions as flops in the context of the Minimal Model program have appeared in Matsuki \cite{MatsukiWeyl}. The main difference with the present work is in that we do not restrict our attention to normal crossing singularities and address global issues of the resolution.  Furthermore, our main object of study is the structure of fibers in higher codimension.


The paper is organized as follows. We begin with a lightning review of the Coulomb branch of $d=3$, $\mathcal{N}=2$ supersymmetric gauge theories with matter. The subsequent parts of sections \ref{Sec:WGQ} determine the equivalence of the characterization of these phases in terms of Weyl group quotient, Bruhat order, and box graphs. Furthermore we show that in many cases the networks of flop transitions correspond to the representation graphs of certain (quasi-)minuscule representations.  In section \ref{sec:SUnPhases} we introduce the correspondence to anti-Dyck paths for the discussion of phases of $\mathfrak{su}(n)$ gauge theories with matter\footnote{After this work appeared, the phases and geometric resolutions for $\mathfrak{su}(n)$ for $n=2, 3, 4$ were discussed in \cite{Esole:2014bka}.} in the fundamental, in the anti-symmetric and in  both representations. In particular, we show how this offers confirmation of the results obtained from geometry in  \cite{Hayashi:2013lra} in  the case of $SU(5)$ with ${\bf 5}$ and ${\bf 10}$ matter. Important properties such as extremal generators, flops and codimension-three behavior are discussed in section \ref{sec:StructureofPhases}. In section \ref{sec:Count} we count the phases for $\mathfrak{su}(n)$ with various matter representations. Other groups and the case of monodromy are discussed in section \ref{sec:OtherPhases}. 

Finally in section 
\ref{Sec:PhasesGeom} we draw the relation to the geometry and discuss the detailed map between phases and resolutions, in particular determining explicitly the effective curves, extremal rays, and flops in the geometry from the decorated box graphs or anti-Dyck paths. We determine from the box graphs the fibers in codimensions two and three in section \ref{sec:EllipticFibsCodim} and exemplify this by determining all the $E_6$ fiber types that arise in $SU(6)$ in codimension two, and $SU(5)$ in codimension three, including the flop transitions among them. Likewise we determine the monodromy-reduced $E_7$ fibers from $SO(12)$. 
Global issues related to the existence of flops into phases of $U(n)$ versus $SU(n)$ gauge theories and the relation to the existence of additional rational sections are explained and exemplified in section \ref{sec:GlobalFib}. 
Details of the representation theory of Lie groups and our conventions are summarized in appendix \ref{app:Group}. Two useful tables with the set of effective curves for the phases of the $U(5)$ gauge theory with fundamental  and with anti-symmetric representation are given in appendix \ref{app:U5}, and the phases of $SO(2n)$ with fundamental matter are discussed in appendix \ref{app:SO}. Finally, in appendix \ref{app:EPhases} the phases of $E$-type gauge group are discussed and the fiber types of $E_8$ are determined from $E_7$ in codimension one with monodromy.


\section{Phases from Weyl group quotients and Box Graphs}
\label{Sec:WGQ}

In this section we start by briefly reviewing the classical Coulomb phase of $d=3$ $\mathcal{N}=2$ gauge theories with matter. We then give three equivalent descriptions of it in terms of either Weyl group quotient, Bruhat order, or decorated box graphs. Furthermore, we show that in many cases, the phases form a so-called quasi-minuscule representation.  
These provide the framework for all subsequent sections discussing the phase structure of these gauge theories. 

\subsection{Phases of $d=3$ $\mathcal{N}=2$ gauge theories}
\label{sec:phasegauge}

Let us first review the classical phase structure of three-dimensional $\mathcal{N}=2$ supersymmetric gauge 
theories \cite{deBoer:1997kr, Aharony:1997bx}. We consider vector multiplets $V$ whose components are in the adjoint representation of a gauge group $G$. The scalar components of $V$ are a three-dimensional vector potential $A$ 
and a real scalar $\phi$. In addition, we have $N_f$ chiral multiplets $Q_f$ whose components are in a representation ${\bf R}_f$ of the gauge group $G$. We assume that there are no classical real mass terms nor classical complex mass terms for the chiral multiplet. Since we also do not introduce a classical Chern-Simons term, we consider an appropriate set of chiral multiplets which does not break the parity anomaly.  

When the adjoint scalar $\phi$ gets a vacuum expectation value (vev) in the Cartan subalgebra of $G$, the gauge group $G$ breaks into $U(1)^{r}$ where $r = \text{rank}(G)$. Then, the vev of $\phi$ takes value 
in a Weyl chamber $\mathcal{C}^* = \mathbb{R}^{r}/W_G$, where $W_G$ is the Weyl group of $G$. 

The presence of a chiral multiplet $Q_f$ adds an additional structure to the Coulomb branch. The vev of the adjoint scalar gives rise to a real mass term for the chiral multiplet. 
However, the mass becomes zero along a real codimension-one subspace inside the Weyl chamber, characterized by
\be\label{boundary}
\langle \phi, \lambda \rangle = 0  \,,
\ee
where the massless chiral multiplet  transforms in the representation ${\bf R}_f$ of $G$ with weight $\lambda$. Hence, the Weyl chamber is further divided by the real codimension-one walls \eqref{boundary} for all the weights. A phase of the three-dimensional gauge theory  corresponds to one of these subwedges of the Weyl chamber. 

In the bulk of the Coulomb branch, the real scalar $\phi$ may be complexified by using a scalar $\gamma$ which is dual to a photon coming from $U(1)^{r}$. The scalar $\gamma$ is subject to a shift symmetry, and the charge quantization restricts it to be compact. Hence, $\gamma$ lives on an $r$-dimensional torus\footnote{Due to this construction, the chiral multiplet obtained by dualizing the vector potential into the scalar $\gamma$ always has a $U(1)_J$ symmetry which shifts $\gamma$.}. The classical Coulomb branch then becomes the total space of the $r$-dimensional torus fibration over a subwedge of the Weyl chamber. However, the radius of the torus vanishes along \eqref{boundary} due to quantum corrections \cite{deBoer:1997kr, Aharony:1997bx}. Therefore, \eqref{boundary} becomes a complex codimension-one wall. The structure of the quantum Coulomb branch may be further altered depending on $N_f$. 

Since we will relate the phase structure to a resolution of a singular geometry, the only information that is relevant for this purpose is the classical moduli space parametrized by the vev of $\phi$. Hence, we focus on determining the subwedges of the Weyl chamber where the boundary is given by \eqref{boundary} for all the weights in the representation ${\bf R}_f$.

\subsection{Phases from Weyl group quotients}\label{Sec:WeylG}

In the following we give various representation theoretic characterizations of the Coulomb phases. The first correspondence we explicate is between phases and the Weyl group quotient $W_{\widetilde{\mathfrak{g}}}/W_{\mathfrak{g}}$ with $\mathfrak{g}$ the Lie algebra of the gauge group $G$, and $\widetilde{\mathfrak{g}}$ as in (\ref{gtildeg}). 

Let $\mathfrak{g}$ be a (simple) Lie algebra, and $\mathfrak{h}$ its Cartan
subalgebra. We set out our notation and conventions as well as some useful properties of Lie algebras and representations in appendix \ref{app:Group}. 
Denote by $\mathfrak{h}^*$ the dual of the Cartan subalgebra, which
can be identified with the 
root space of $\mathfrak{g}$. Furthermore, let $\rootsystem$ be the set of
roots of $\mathfrak{g}$, and an ordering of the roots is determined by a
linear functional $\mu$ on the root space, which determines
$\rootsystem =  \rootsystem^+_\mu \cup \rootsystem^-_\mu$, where
$\rootsystem^+_\mu = \{\alpha\in \rootsystem;\ \mu(\alpha) >0\}$. The elements in
$\rootsystem^+_\mu$ are called positive roots and linear combinations of the
elements in $\rootsystem^+_\mu$ with non-negative coefficients forms a simplicial
cone $\mathcal{C}_\mu$. The generators of the cone are called
simple roots.

A Weyl chamber is determined by the ordering $\mu$
\be
\mathcal{C}^*_{\mu} =\left\{ \phi\in \mathfrak{h}\,,\quad  \langle \phi ,\alpha \rangle >0 \,, \quad \hbox{for all} \quad \alpha\in\rootsystem^+_\mu \right\}\,.
\label{weyl}
\ee
Note that $\mathcal{C}^*$ is a subset of $\mathfrak{h}$, which
is identified with the {\em coroot space}\/ in the conventions of
appendix \ref{app:Group}.  

The Weyl group acts simply transitively, with trivial stabilizers, on the set of orderings $\rootsystem^+$
and on the set of Weyl chambers. The number of Weyl chambers is thus
equal to the order of the Weyl group. 

Let $\lambda_I$, $I=1, \cdots, r$ be the weight vectors of a given
representation ${\bf R}$ of dimension $r$. Then a \textit{phase} is defined
as a non-empty subwedge in a Weyl chamber of $\mathfrak{g}$ such that the
inner product with any weight of the representation has a definite fixed
sign
\be
\sign_\phi:\qquad
\ba
 {\bf R} \quad & \rightarrow\quad  \mathbb{Z}_2 \cr
 {\lambda_I} \quad &\mapsto\quad  \sign(\langle\phi, \lambda_I\rangle)  \,.
 \label{phase}
\ea\ee
A phase is then labelled by a fixed vector ${\epsilon_1\cdots \epsilon_r}$ of
signs $\epsilon_I =\pm1$
\be
\Phi_{\epsilon_1\cdots \epsilon_r}^\mu=  \left\{
\phi\in \mathcal{C}^*_\mu: \quad  \sign(\langle\phi, \lambda_I\rangle)  = \epsilon_I \,,\quad I= 1,\cdots, r\,
\right\} \,.
\ee
This clearly depends on the choice of the Weyl chamber, and an arbitrary
choice of the signs is not allowed. We will fix the Weyl chamber for the
phases to be that given by the ordering with respect to the Weyl vector $\mu=\rho$.

To state our claim,
consider a simple Lie algebra $\widetilde{\mathfrak{g}}$, of one rank higher\footnote{In some instances we will also consider decompositions with $\mathfrak{su}(2)$ or higher rank enhancements. }
than $\mathfrak{g}$, with
\be
\widetilde{\mathfrak{g}} \quad \supset \quad {\mathfrak{g}} \oplus \mathfrak{u}(1) \,,
\ee
 whose adjoint has a decomposition as a representation of $\mathfrak{g} \oplus \mathfrak{u}(1)$
\be\label{gtadj}
\hbox{adj} (\widetilde{\mathfrak{g}} ) \quad \longrightarrow \quad \hbox{adj}(\mathfrak{g}) \oplus \hbox{adj}(\mathfrak{u}(1)) \oplus {\bf R}_+ \oplus \overline{{\bf R}}_- \,.
\ee
Let $\widetilde{\rootsystem}$ be the roots of $\widetilde{\mathfrak{g}}$. The isomorphism
(\ref{gtadj}) gives an embedding of the roots  $\rootsystem$ of $\mathfrak{g}$
and the weights $\lambda_I$ of ${\bf R}$ into $\widetilde{\rootsystem}$.

Each ordering $\widetilde{\mu}$ of the roots $\widetilde{\rootsystem}$ gives an ordering
on $\hbox{adj}(\mathfrak{g})$ and $({\bf R} \oplus {\bf \overline{R}})$ from
the decomposition (\ref{gtadj}). An ordering on $({\bf R} \oplus {\bf
\overline{R}})$ is equivalent to a choice of signs on ${\bf R}$,
consistent with the ordering on $\hbox{adj}(\mathfrak{g})$. The phases
are defined with respect to one particular Weyl chamber, which we choose
above to be that coming from the ordering $\mu=\rho$. Let us fix the
functional $\widetilde{\mu}$ such that it reduces to $\rho$ when considered
on $\hbox{adj}(\mathfrak{g})$. Then there is a one-to-one map between the phases for fixed ordering $\mu = \rho$ and orderings (linear functionals) on $\widetilde\Phi$, i.e.
\begin{equation}
    \Phi_{\epsilon_1\cdots\epsilon_r}^\rho  \quad\leftrightarrow \quad  \widetilde{\mu} \,.
\end{equation}

The Weyl group acts transitively on the set of orderings, $W_\mathfrak{g}$
acts thus on the orderings of $\hbox{adj}(\mathfrak{g})$, and fixing that
ordering to $\rho$ involves quotienting $W_{\widetilde{\mathfrak{g}}}$ by this
action. In summary, we find that the number of distinct phases of the $\mathfrak{g} \oplus \mathfrak{u}(1)$ to be given by the order of the Weyl group quotient 
\begin{equation}
    \#\Phi_{\epsilon_1\cdots\epsilon_r}^\rho =
    \frac{|W_{\widetilde{\mathfrak{g}}}|}{|W_\mathfrak{g}|} \,,
\end{equation}
where $|W_\mathfrak{g}|$ denotes the order of the Weyl group of
$\mathfrak{g}$.
By the simple transitivity of the action of the Weyl group on the orderings
and Weyl chambers we have the following one-to-one maps
\begin{equation}
    \label{eq:OneToOneF}
    \Phi_{\epsilon_1\cdots\epsilon_r}^\rho \quad\leftrightarrow \quad 
    \widetilde{\rootsystem}^+_{\widetilde{\mu}}  \quad\leftrightarrow \quad 
    \widetilde{\mathcal{C}}^*_{\widetilde{\mu}}  \quad\leftrightarrow \quad 
    [w_{\widetilde{\mu}}] \,,
\end{equation}
where $[w_{\widetilde{\mu}}]$ represents an element of the quotiented Weyl group $W_{\widetilde{\mathfrak{g}}}/W_{\mathfrak{g}}$.

Considering the (Cartan-Weyl) ordering with respect to $\widetilde{\rho}$,  the Weyl vector of $\widetilde{\mathfrak{g}}$, one finds that
\begin{equation}
    \widetilde{\mathcal{C}}^*_{\widetilde{\rho}}  \quad\leftrightarrow \quad  \Phi^\rho_{++\cdots+} \,.
\end{equation}
Since the simple roots are the extremal rays of the simplicial cone, 
the Weyl reflections with
respect to them map $\widetilde{\mathcal{C}}^*_{\widetilde{\rho}}$ to adjoining Weyl chambers.

The Weyl group acts simply transitively on sets of simple roots, so one can
start with $\widetilde{\rho}$ and perform Weyl reflections by simple roots that
preserve $\hbox{adj}(\mathfrak{g})$ to generate the phases that share a real
codimension-one wall, and repeat to generate all the phases. Using the same
procedure one associates to each phase an element of the quotiented Weyl
group, the combination of Weyl reflections taking $\Phi^\rho_{++\cdots+}$ to
that phase, as expected from (\ref{eq:OneToOneF}).

To show how this works explicitly, we provide two examples in appendix \ref{app:U5} for $\mathfrak{u}(5)$  with matter in the fundamental ${\bf 5}$ and antisymmetric representation ${\bf 10}$.


\subsection{Network of Phases and Bruhat order}

The Coulomb phases or, equivalently, the Weyl group quotients have a natural ordering, known as the {\it Bruhat order}. The entirety of the phases with this ordering will correspond, in terms of the geometry, to the network of flop transitions and so characterizes them in terms of representation-theoretic data. 
We shall now provide a short summary of the Bruhat order. 

The element of the Weyl group which corresponds to the phase
$\Phi_{\epsilon_1 \cdots \epsilon_r}^{\rho}$ of the theory with
$\mathfrak{g} \oplus \mathfrak{u}(1)$ has a nice mathematical
characterization. Let $S$ be the set of  Weyl reflections with respect to
the simple roots of $\widetilde{\mathfrak{g}}$, and $W$ be the whole Weyl group
of $\widetilde{\mathfrak{g}}$. From $S$, we take a subset $J \subset S$ which
are the Weyl reflections with respect to the roots corresponding to the
simple roots of $\mathfrak{g}$. Then, let $W_J$ be the subgroup of $W$
generated by the elements in $J$, which is in fact a parabolic subgroup.
Furthermore, define 
\be
W^J := \{w \in W \; | \quad l(sw) > l(w) \quad \text{for all $s \in J$}\},
\ee
where $l(w)$ is the length of $w$. Decomposing an element $w \in W$ by
in terms of the  generators $s_i$ as $w = s_1 \cdots s_r$, the length $l(w)$ is the smallest
such $r$, and the corresponding decompositions with the smallest $r$ is called reduced. The length
of the identity is defined to be zero.

In fact, the elements which correspond to the phase $\Phi_{\epsilon_1 \cdots
\epsilon_r}^{\rho}$ are precisely characterized by the elements in $W^J$. In
order to see that, we use the following two claims. First, if $w_{\alpha} \in S$,
then $w_{\alpha}(\widetilde{\rootsystem}^+\setminus \{\alpha\}) =
\widetilde{\rootsystem}^+\setminus \{\alpha\}$, and $\alpha$ becomes the negative root
$-\alpha$ after the Weyl reflection $w_{\alpha}$. Second,  if one
fixes a positive root system, then the number of positive roots sent to
negative roots by $w$ is equal to $l(w)$. From the algorithm to associate an
element $w$ of the Weyl group $W$ to the phases $\Phi_{\epsilon_1 \cdots
\epsilon_r}^{\rho}$, the roots correspond to the simple roots of
$\mathfrak{g}$ are still positive with respect to the positivity of
$\widetilde{\rootsystem}^{+}_{\widetilde{\rho}}$ after the Weyl action $w$ to the
positive root system $\widetilde{\rootsystem}^{+}_{\widetilde{\rho}}$. Therefore, any
element $w$ corresponding to the phases $\Phi_{\epsilon_1 \cdots
\epsilon_r}^{\rho}$ satisfies $l(sw) = l(w)+ 1$ for all the elements $s \in
J$, which means that $w$ exactly satisfies the definition of $W^J$. We still
need to prove that $w$ indeed exhausts all the elements of $W^J$. Note that,
if there is a reduced expression of $w \in W^J$, then $sw$ should not be
inside $W^J$ for all the elements $s \in W_J$. Therefore, the maximum number
of $|W^J|$ is $\frac{|W|}{|W_J|}$. Since the algorithm gives
$\frac{|W|}{|W_J|}$ number of $w$, it exhausts all the elements in $W^J$.

Given this setup, we can now define the {\it (left) Bruhat order}, in which we are interested in\footnote{See for example \cite{CombinatoricsOfCoxeter} for some more details on Bruhat order and related matters. }. For $u, w \in W$,
Bruhat order $u \leq w$ means that there exist $w_i \in W$ such that
\be
u = w_0 \rightarrow w_1 \rightarrow \cdots \rightarrow w_{k-1} \rightarrow w_k = w.
\ee
The arrow $w_i \rightarrow w_{i+1}$ means that $w_{i+1} = tw_{i}$ and $l(w)
< l(w_{i+1})$ where $t$ is a reflection element of $T$ defined as
\be
T := \{ wsw^{-1}\; : \; w \in W, \quad s \in S \}.
\ee
Then, from the construction of $w$ corresponding to the phase
$\Phi_{\epsilon_1 \cdots \epsilon_r}^{\rho}$, it obeys the Bruhat order.
Namely, starting from the identity which corresponds to the phase $\Phi_{+
\cdots +}^{\rho}$, the length of the element $w$ increases by one when one
performs a Weyl reflection with respect to a root corresponding to the
weights of ${\bf R}$.


\begin{figure}
\centering
\includegraphics[width=4cm]{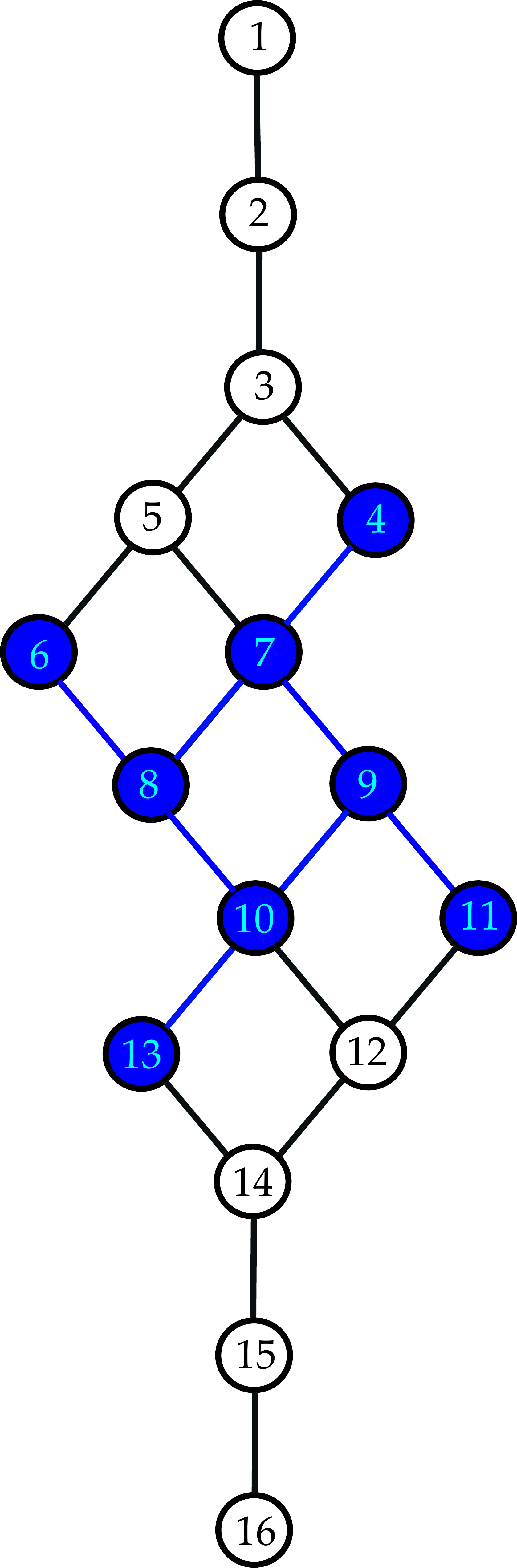}
\caption{Example for Bruhat ordering of the phases or equivalently Weyl group quotients, for the $\mathfrak{u}(5)$ theory with antisymmetric ${\bf 10}$ representation, where each edge represents a Weyl reflection. The nodes $4, 6,7, 8, 9, 10, 11, 13$ correspond to the phases of $SU(5)$ with ${\bf 10}$ matter, shown in blue. Note that as explained in section \ref{sec:MiniRep}, the phase diagram for the $U(5)$ theory corresponds exactly to the representation graph of the {\bf 16} representation of $SO(10)$. }
\label{fig:U5A} 
\end{figure}


Based on the Bruhat order for  $W^J$ we define a diagram. The nodes correspond to
the elements in $W^J$. The lines between the nodes means that the elements
are ordered by Bruhat order and the difference between their lengths is one.
From the properties of the Bruhat order and the quotient $W^J$, this diagram
matches with the phase diagram of $\Phi_{\epsilon_1 \cdots
\epsilon_r}^{\rho}$ of the theory with $\mathfrak{g} \oplus
\mathfrak{u}(1)$. Figure \ref{fig:U5A} exemplifies this for $U(5)$ with antisymmetric  ${\bf 10}$ representation. 
Furthermore the reader will find the graphs for various Weyl group quotients throughout the paper in figures \ref{fig:SU6E6Flops} and \ref{fig:E6Flops27}.



\subsection{Box Graphs and Flow Rules}
\label{Sec:FlowRules}

The most elegant and compact description of the phases is in terms of what we refer to as {\it decorated box graphs}. 
The box graphs are based on the representation graph and contain all the relevant information about the phases, or equivalently the geometry.

Let us consider the algebra $\mathfrak{su}(n)\oplus\mathfrak{u}(1)$.
\noindent The positive roots can be written in terms of $L_i$, $i=1, \cdots, n$ as explained in the appendix \ref{app:Group},
\begin{equation}
    \rootsystem^+ = \{ L_i - L_j \, | i = 1,\cdots,\, n ; \,  j > i\} \,.
\end{equation}
The weights of the fundamental representation of dimension $n$ are
\begin{equation}
    V = \{L_i \, | i = 1,\cdots,n \} \,.
\end{equation}
and the weights of the anti-symmetric representation of dimension $ n(n-1)/2$ are
\begin{equation}
    \Lambda^2V = \{L_i + L_j \, | i = 1,\cdots,\, n; \,  j > i\} \,.
\end{equation}
These correspond to roots and weights of $\mathfrak{su}(n)$ subject to the condition
\begin{equation}\label{TLNC}
    \sum_{i=1}^nL_i = 0 \,,
\end{equation}
which we often refer to as the tracelessness condition. 
If this is not satisfied, the generator $\sum_i L_i$ corresponds to an additional $\mathfrak{u}(1)$ generator. 


\begin{figure}
\centering
\includegraphics[width=6cm]{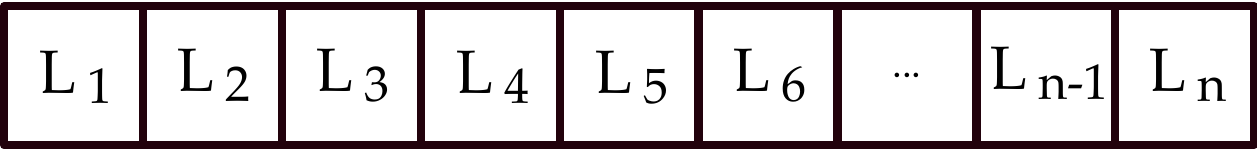}
\caption{The representation ${\bf n}$ for $\mathfrak{su}(n)$. 
Each box represents a weight, $L_i$ and the walls separating the boxes represent the action of the negative (positive) simple roots on the weights.}
\label{fig:SU8F} 
\end{figure}

\begin{figure}
\centering
\includegraphics[width=7cm]{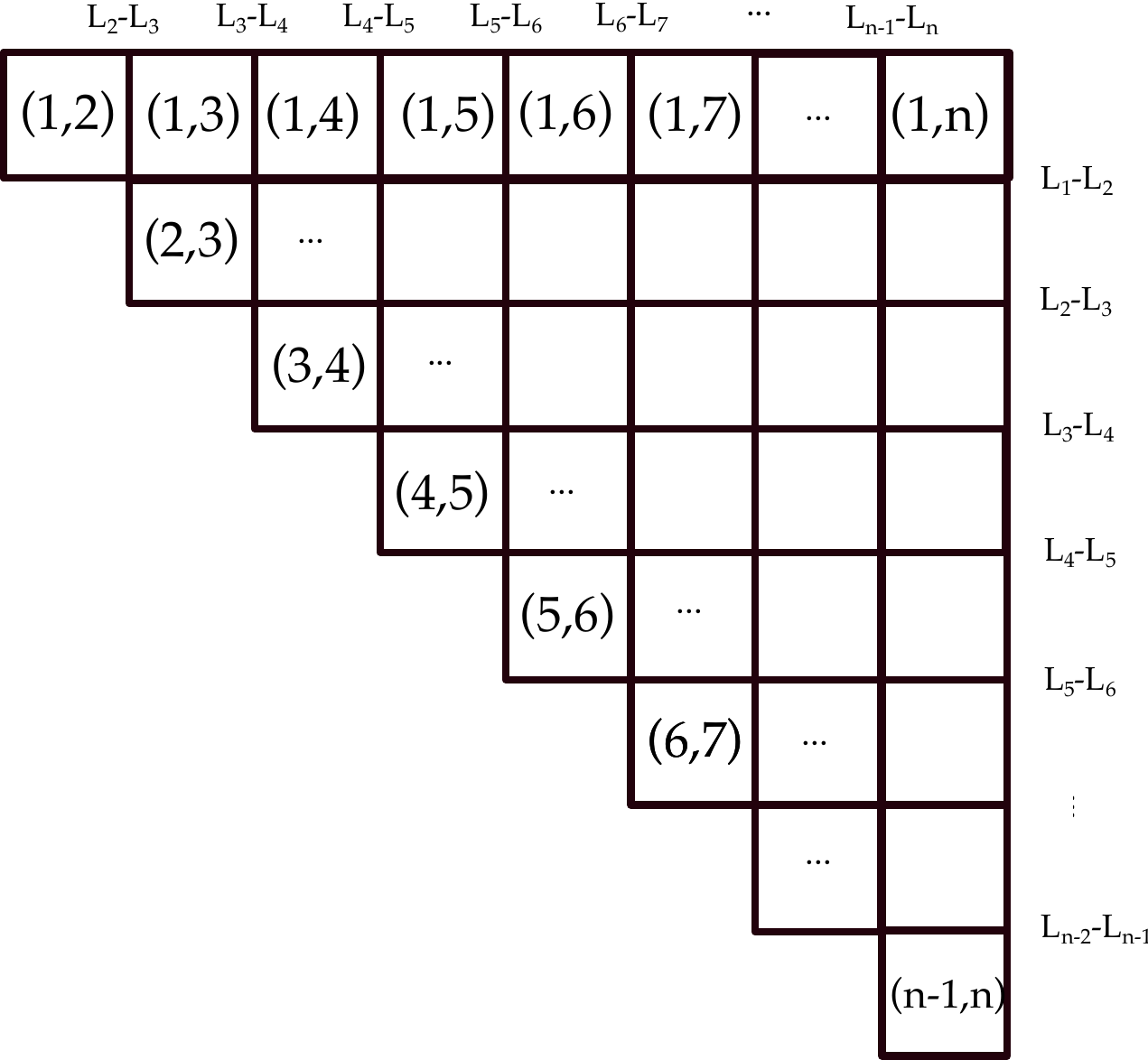} 

\caption{The representation $\Lambda^2 {\bf n}$ for $\mathfrak{su}(n)$. 
Each box represents a weight, $L_i+ L_j$ of the representations labeled by $(i,j)$, the walls separating the boxes represent the action of the negative (positive) simple roots on the weights.}
\label{fig:SU8A} 
\end{figure}


One can present the fundamental (resp. antisymmetric) representation by the
box graph given in figure \ref{fig:SU8F} (resp. \ref{fig:SU8A}). Each box
can be decorated by a sign (or equivalently by a coloring). 
We can then ask which such decorated box graphs correspond to a phase,
$\Phi^\rho_{\epsilon_1\cdots\epsilon_r}$, where the
$\epsilon_i$ are the signs decorating the box corresponding to the $i$th weight of the representation.

We show the existence of the following \textit{flow rules} governing the
placement of signs which are a necessary condition for any decorated box graph to correspond 
to a consistent phase (or a non-empty subwedge of the Weyl chamber). The flow rules are 
\begin{equation}\label{FlowsRulesF1}
    \begin{array}{lc}
        \text{Fundamental: } \qquad &
        \begin{array}{cc}
            \boxed{+} \quad \leftarrow& \boxed{+} \cr
                \cr
            \boxed{-} \quad \rightarrow& \boxed{-} \cr
        \end{array} \cr\cr\cr
        \text{Antisymmetric: } \qquad &
        \begin{array}{cc}
            \boxed{+} \quad \leftarrow& \boxed{+} \cr
            \cr
                 & \uparrow \cr
                     &            \cr
                         & \boxed{+}
        \end{array} \qquad
        \begin{array}{cc}
            \boxed{-} \quad \rightarrow& \boxed{-} \cr
            \cr
                 & \downarrow \cr
                     &            \cr
                         & \boxed{-}
        \end{array} \cr\cr
\end{array}
\end{equation}
The arrows indicate that if the sign is specified at the nock (the
end of the arrow opposite the arrowhead) then the sign flows
through the diagram in the direction of the arrow.

An alternative description to the representation graph decorated with signs that follow the flow rules is to consider the path that separates the $+$ and $-$ sign boxes. These paths will play a particularly important role for the case of $\mathfrak{su}(n)$, and will allow a  simple description of flop transitions and the counting of the phases.

These rules are proved separately for each representation. Consider the
fundamental representation, and assume that the flow rules given above are
violated. Then one has $L_i > 0$ and $-L_j > 0$ with $j < i$. By taking
positive linear combinations we get $L_i + (-L_j) + (L_j - L_i) = 0$. Thus the
subwedge of the Weyl chamber with respect to this sign assignment is empty.

Again for the antisymmetric representation assume that the flow rules are
violated. We shall consider here only the vertical arrows, a similar argument
holds for the horizontal arrows. The violation tells us that $(L_i + L_k) > 0$
and $-(L_j + L_k) > 0$ with $j < i$. Using positive linear combinations one
generates $(L_i + L_k) + -(L_j + L_k) + (L_j - L_i) = 0$. The subwedge of the
Weyl chamber $\mathcal{C}^{\ast}_{\mu}$ with respect to this sign assignment is again empty.

Combinatorics allows us to count the decorated box graphs obeying the flows
rules, in terms of monotonous staircase paths in the representation graph, starting at the point $S$ and ending at one of the green nodes along the diagonal in figure \ref{fig:ExamplePhase}. 
For the $i$th node we count the number of paths in an $i\times (n-1-i)$ rectangular grid, which is given by ${n \choose i}$. Thus the total number of paths is $\sum_{i=0}^{n-1} {n-1 \choose i} = 2^{n-1}$, which equals 
 $|W_{{\mathfrak{so}(2n)}}|/|W_{\mathfrak{su}(n)}|$. 
More generally the number of phases agrees with the cardinality of the Weyl group quotients of $\widetilde{\mathfrak{g}}$ and $\mathfrak{g}$, as described in section \ref{Sec:WeylG}. 

This, combined with the above argument, shows the
sufficiency and necessity of the flow rules in the determination of the
phases.


\begin{table}
\begin{center}
\begin{tabular}{|c||c|c|c|c|c|}
\hline
$\widetilde{\mathfrak{g}}$ &$\mathfrak{su}(n)$ & $\mathfrak{so}(2n)$ & $\mathfrak{e}_6$ & $\mathfrak{e}_7$ & $\mathfrak{e}_8$ \cr
\hline\hline
Minuscule $\varpi_{\mathfrak{g}}$ &  
$\varpi_1,  \varpi_{n-1}$& $\varpi_1$, $\varpi_{n-1}$, $\varpi_n$ & $\varpi_1$, $\varpi_5$ & $\varpi_6$ & $-$ \cr
\hline
Dim of $V_{\varpi_{\mathfrak{g}}}$  & 
$n,  \ n $ & $2n$ , $2^{n-1}$, $2^{n-1}$  &$27$, $27$ &  $56$& $-$ \\
\hline
$\mathfrak{g}$ & 
$\mathfrak{su}(n{-}1),  \mathfrak{su}(n{-}1)$& $\mathfrak{so}(2n-2)$, $\mathfrak{su}(n)$, $\mathfrak{su}(n)$ & $\mathfrak{so}(10)$, $\mathfrak{so}(10)$ &  $\mathfrak{e}_6$ & $-$\\
\hline
\end{tabular}
\end{center}
\caption{Simply-laced Lie algebras $\widetilde{g}$ and their minuscule 
representations $V_{\varpi_\mathfrak{g}}$, relevant for the rank one
embeddings. We first list the highest weight
$\varpi_{\mathfrak{g}}$ of the minuscule representation of $\widetilde{g}$ 
(in terms of the fundamental weights $\varpi_i$) and then the dimension 
of the corresponding representation $V_{\varpi_{\mathfrak{g}}}$. In the 
last row we list the Lie algebras $\mathfrak{g}$ which have a rank one 
embedding into $\widetilde{\mathfrak{g}}$ such that the Weyl group quotient 
$W_{\widetilde{\mathfrak{g}}}/W_{\mathfrak{g}}$ is the corresponding 
minuscule representation.  The labeling of the highest weights
is done in a standard way, as discussed in Appendix~\ref{app:Group}.
For $\mathfrak{su}(n)$, every 
fundamental weight is a minuscule weight, but we only 
include the ones relevant for rank one embeddings. 
\label{tb:minuscule}}
\end{table}

\subsection{Minuscule Representations and Weyl group quotients}
\label{sec:MiniRep}

We have seen that the Weyl group quotient identifies the phase $\widetilde{\rootsystem}^+_{\widetilde{\mu}}$ with the corresponding Weyl chamber $\widetilde{\mathcal{C}}^*_{\widetilde{\mu}}$. Furthermore, a transition to adjacent phases corresponds to a Weyl reflection with respect to simple roots that preserves the fundamental Weyl chamber of $\mathfrak{g}$.  In fact, this point of view reveals an intriguing relation between the network of phases and a representation graph of $\widetilde{\mathfrak{g}}$. For example, let us consider the phases of the $U(5)$ gauge theory with matter in the antisymmetric representation ${\bf 10}$. The phase network is depicted in figure~\ref{fig:U5A}, and corresponds to the Weyl group quotient $W_{\mathfrak{so}(10)}/W_{\mathfrak{su}(5)}$, as explained in the previous section. 
We  observe that this phase diagram  is in fact identical to the representation graph of the ${\bf 16}$ spinor representation of $\mathfrak{so}(10)$. This is not a coincidence but the relation will hold for so-called minuscule representations of simply laced Lie algebras. 

Let us consider embeddings satisfying $\text{rank}(\widetilde{\mathfrak{g}}) = \text{rank}(\mathfrak{g}) + 1$ and  summarize the  correspondences that we find in this case. A {\it minuscule representation} of a Lie algebra is defined as an  irreducible representation with the property that the Weyl group acts transitively on all weights occurring in the representation \cite{gal-78}. For all the simply-laced Lie algebras these are listed in table \ref{tb:minuscule}, where the $\varpi_i$ are the fundamental weights 
\be
\langle \alpha_i^{\vee}, \omega_j \rangle = \delta_{ij} \,.
\ee
Furthermore, a {\it quasi-minuscule representation} is one such that the Weyl group acts transitively on the nonzero weights (see for instance \cite{QuasiMinirepRef}).  In fact there is a unique quasi-minuscule representation for the simply-laced Lie algebras, which has as highest weight the unique dominant root. The zero weights of this representation are one-to-one with the simple roots. In particular for all the ADE type Lie algebras, the quasi-minuscule representations are given by the adjoint representations.  

For simply-laced Lie algebras the minuscule representations listed in table \ref{tb:minuscule} can be obtained in terms of the Weyl group quotient $W_{\widetilde{\mathfrak{g}}}/W_{\mathfrak{g}}$ for $\mathfrak{g}$ as given in the last row in table \ref{tb:minuscule}, and the phase structure is precisely reproduced by the $\widetilde{\mathfrak{g}}$ representation structure on these minuscule representations. 
Let {\bf R} be the representation appearing in the decomposition of the adjoint 
\be
\widetilde{\mathfrak{g}} \rightarrow \mathfrak{g} \oplus\mathfrak{u}(1)\,.
\ee
We show the following equivalences between phases, Weyl group quotients and the minuscule representations 
\be\label{MiniCorresp}
\begin{tabular}{c}
\hbox{Phases of $\mathfrak{g} \oplus \mathfrak{u}(1)$ with matter in the representation ${\bf R}$} 
\cr 
\cr
$\updownarrow $
\cr
\cr
$W_{\tilde{\mathfrak{g}}}/W_{\mathfrak{g}}$ 
\cr 
\cr
$\updownarrow $
\cr
\cr
Minuscule representation $V_{\omega_\mathfrak{g}}$ of $\tilde{\mathfrak{g}}$
\end{tabular}
\ee
Example diagrams of this type are shown for triplets $(\widetilde{\mathfrak{g}}, \mathfrak{g}, V_{\omega_{\mathfrak{g}}})$ for $(\mathfrak{so}(10), \mathfrak{su}(5), {\bf 16})$ in figure \ref{fig:U5A}, $(\mathfrak{e}_7, \mathfrak{e}_6, {\bf 56})$ in figure  
\ref{fig:E6Flops27}. 

A similar correspondence holds for the quasi-minuscule representations, which for the ADE Lie algebras (including $\mathfrak{e}_8$) are simply the adjoint representations. In this case the decompositions are of the type 
\be\label{GtGSU2}
\widetilde{\mathfrak{g}} \quad \rightarrow\quad  \mathfrak{g} \oplus \mathfrak{su}(2) \,.
\ee
The quasi-minuscule representations arise in terms of Weyl group quotients, with the subtlety that the zero-weights are not realized in the quotient. The simple Lie algebras for which this occurs are
\be\label{QuasiMiniE}
\ba
\mathfrak{e}_6 & \quad \rightarrow  \quad\mathfrak{su}(6)\oplus\mathfrak{su}(2)\cr
\mathfrak{e}_7 & \quad \rightarrow   \quad\mathfrak{so}(12)\oplus\mathfrak{su}(2) \cr
\mathfrak{e}_8  & \quad \rightarrow  \quad\mathfrak{e}_7\oplus\mathfrak{su}(2)\,.
\ea
\ee
The Weyl group $\mathbb{Z}_2$ of the non-abelian rank one commutant of $\mathfrak{g}$ can act on the representation, and we show that the invariant phases are precisely given in terms of the simple roots of the quasi-minuscule representation. Furthermore, their phase diagram is exactly the Dynkin diagram of the simple Lie algebra $\widetilde{\mathfrak{g}}$. In summary we show for representation ${\bf R}$ appearing in the decomposition (\ref{GtGSU2})
\be\label{MiniCorresp2}
\begin{tabular}{c}
\hbox{Phases of $\mathfrak{g} \oplus \mathfrak{su}(2)$} \cr
\hbox{with matter in the representation ${\bf R}$}  \cr 
\cr
$\updownarrow $
\cr
\cr
$W_{\tilde{\mathfrak{g}}}/W_{\mathfrak{g}}$ 
\cr 
\cr
$\updownarrow $
\cr
\cr
Quasi-minuscule  (adjoint) representation \cr  
except zero weights of  $\tilde{\mathfrak{g}}$
\end{tabular}
\qquad 
\begin{tabular}{c}
\hbox{$\mathbb{Z}_2$-invariant phases of $\mathfrak{g} \oplus \mathfrak{su}(2)$} \cr
\hbox{with matter in the representation ${\bf R}$}  \cr 
\cr
$\updownarrow $
\cr
\cr
Dynkin diagram of $\widetilde{\mathfrak{g}}$ 
\cr 
\cr
$\updownarrow $
\cr
\cr
Simple roots of the quasi-minuscule \cr
representation (adjoint) of  $\tilde{\mathfrak{g}}$
\end{tabular}
\ee
We analyze the case of $\mathfrak{e}_6$, $\mathfrak{e}_7$ and $\mathfrak{e}_8$ in (\ref{QuasiMiniE}) in detail in section \ref{Sec:Su6L3} and appendix \ref{sec:E756}. The phase diagrams for these theories are in figures  \ref{fig:SU6E6Flops} and \ref{fig:E7Phases56}, and the subdiagram from the invariant phases, given by the Dynkin diagrams is  discussed in figures \ref{fig:SU6Lambda3Fibs}, \ref{fig:SO12E7} and \ref{fig:E8MonoFibs}, respectively. 

We now prove these correspondences. 
In order to see the relation (\ref{MiniCorresp}), let us first show an equality about  Dynkin labels $l_i$ of a  weight $\omega$ of a representation {\bf R}
\be
l_i = \langle \alpha_{i}^0 , \omega \rangle= \langle \alpha_i, \omega^0 \rangle. 
\label{Dynkinlabel}
\ee
Since we focus on simply-laced Lie algebras, we will not distinguish a coroot from a root. The $\alpha^0_i$s are the canonical simple roots of $\mathfrak{g}$, and hence the first equality is equivalent to the definition of the Dynkin label. On the other side of the equality,  $\omega^0$ is the highest weight of the representation {\bf R}, and the  $\alpha_i$ are a set of simple roots after performing an appropriate number of Weyl reflections on the set of the canonical simple roots. Since we will associate a subtraction of a canonical simple root from a weight in the construction of the representation {\bf R} with a Weyl reflection, the number is  related to the number of the canonical simple roots we subtract from the highest weight $\omega^0$ to get the weight $\omega$ (up to some subtlety which arises when the Dynkin label is greater than one, which we will discuss later). The proof of the second equality in \eqref{Dynkinlabel} can be done by induction. When $\omega = \omega^0$, then the second equality trivially holds. So, let us assume that it holds for some weight $\omega$. If $l_j > 0$, then a descendant weight $\omega^{\prime}$ can be obtained by $\omega^{\prime} = \omega - l_j\alpha_{j}^0$. Correspondingly, we consider a new set of simple roots $\alpha^{\prime}_i$ by performing a Weyl reflection with respect to $\alpha_j$ on the set of the simple roots $\alpha_i$
\be
\alpha^{\prime}_{i} = \alpha_i - \langle \alpha_{j}, \alpha_i \rangle \alpha_j.
\label{Weylreflection}
\ee
We then need to show, assuming \eqref{Dynkinlabel}, that 
\be
\langle \alpha_{i}^0 , \omega^{\prime} \rangle= \langle \alpha_i^{\prime}, \omega^0 \rangle \,.
\label{induction}
\ee
 The left-hand side of \eqref{induction} becomes 
\bea
\langle \alpha^0_i, \omega^{\prime}\rangle &=& \langle \alpha^0_i, \omega - l_j\alpha^0_j\rangle\nonumber\\
&=&\langle \alpha_i, \omega^0 \rangle - C_{ij} l_j, \label{induction1}
\eea
where $C_{ij} =  \langle \alpha_i^0, \alpha_j^0 \rangle$ is the Cartan matrix. On the other hand, by using the new set of the simple roots \eqref{Weylreflection}, the right-hand side of \eqref{induction} becomes
\bea
\langle \alpha^{\prime}_i, \omega^0 \rangle &=& \langle \alpha_i, \omega^0 \rangle -  \langle \alpha_{j}, \alpha_i \rangle \langle \alpha_j, \omega^0 \rangle\nonumber\\
&=&\langle \alpha_i, \omega^0 \rangle - C_{ji} l_j, \label{induction2}
\eea
where we used $\langle \alpha_j, \alpha_i \rangle = \langle \alpha_j^0, \alpha_i^0 \rangle$. Eq.~\eqref{induction1}--\eqref{induction2} implies that \eqref{induction} holds, which completes the proof of \eqref{Dynkinlabel}.

From the relation \eqref{Dynkinlabel}, we can associate a Weyl reflection with respect to a simple root that preserves the fundamental Weyl chamber of $\mathfrak{g}$ on the Weyl group quotient side to a subtraction of a canonical simple root on the representation side if the Dynkin label satisfies $l_i \leq 1$. In order to achieve such a correspondence, we first need to associate $l_i^0 = 0$ to $\alpha_i^0$ which is a canonical simple root of $\mathfrak{g}$, and also associate $l_i^0 = 1$ to $\widetilde{\alpha}^0$ which we define as a simple root of $\widetilde{\mathfrak{g}}$ but not a simple root of $\mathfrak{g}$. Here $l^0$ denotes the Dynkin label of the highest weight. Once we determine the correspondence between $l^0 = [0, \cdots, 0, 1]$ and $\{\alpha^0_1, \cdots, \alpha^0_{n-1} \widetilde{\alpha}^0\}$, then the same correspondence holds in all the subsequent steps due to the relation \eqref{Dynkinlabel}. At some step corresponding to a weight $\omega$, we have a set of simple roots $\{\alpha_1, \cdots, \alpha_n\}$, which can be obtained by performing Weyl reflections on the canonical simple roots $\{\alpha_1^0, \cdots, \alpha^0_{n-1}, \widetilde{\alpha}\}$. If a simple root $\alpha_i \in \{\alpha_1, \cdots, \alpha_n\}$ is a simple root of $\mathfrak{g}$, then $l_i = \langle \alpha^0_i, \omega \rangle = \langle \alpha_i, \omega^0 \rangle = 0$. If a simple root $\alpha_j \in \{\alpha_1, \cdots, \alpha_n\}$ is a root of $\widetilde{\mathfrak{g}}$ but not a root of $\mathfrak{g}$, then  
\be
l_i = \langle \alpha_j^0, \omega \rangle = \langle \alpha_j, \omega^0 \rangle 
= \langle \widetilde{\alpha}^0 + \sum_k a_k \alpha_k^0, \omega^0 \rangle = 1 \,.
\ee
For the second last equality, we assume that we have only one kind of representation from the decomposition as in  \eqref{gtadj}. This is in fact true for the cases we consider. From this construction, we can determine which representation appears, namely its highest weight by considering an embedding of the canonical simple roots of $\mathfrak{g}$ inside the canonical simple roots of $\widetilde{\mathfrak{g}}$.

In this correspondence, it is important to assume that the Dynkin label is less than two. From the proof of the relation \eqref{Dynkinlabel}, a Weyl reflection with respect to $\alpha_i$ corresponds to subtracting $l_i$  times the canonical simple root $\alpha^0_i$ in one go. Although the $l_i$ weights obtained by the subtraction of the canonical simple root $\alpha^0_i$ one by one appears as the weights of the representation {\bf R}, the Weyl reflection cannot see the intermediate states since it corresponds to the subtraction of $l_i \times \alpha^0_i$ at one time. Therefore, the dimension of a representation does not match with $|W_{\widetilde{\mathfrak{g}}}|/|W_{\mathfrak{g}}|$ when the representation has a weight whose Dynkin label is greater than one. The condition of $l_i \leq 1$ can be translated into $\langle \alpha_i, \omega^0 \rangle \leq 1$ for the roots $\alpha_i$ of $\widetilde{\mathfrak{g}}$ due to the relation \eqref{Dynkinlabel}. If a fundamental weight $\omega$ satisfies $2\frac{\langle \beta, \omega \rangle}{\langle \beta, \beta \rangle} \leq 1$ for all the positive roots $\beta$ of $\widetilde{\mathfrak{g}}$, then $\omega$ is called minuscule, which is equivalent to the definition we gave earlier \cite{gal-78}. The list of the minuscule fundamental weights of simply-laced Lie algebras, which appear from rank one embeddings is depicted in table \ref{tb:minuscule}. Therefore, for a representation specified by a highest weight listed in table \ref{tb:minuscule}, the representation graph is identical to the network of phases from the corresponding Weyl group quotient.

The relation may be extended to a quasi-minuscule representation whose nonzero weights correspond to elements of the Weyl group quotient, as summarized in (\ref{MiniCorresp2}). The quasi-minuscule representations of simply laced Lie algebras, which come from a rank one embedding are the adjoint representations of $\mathfrak{e}_6, \mathfrak{e}_7$ and $\mathfrak{e}_8$ arising from the embedding $\mathfrak{su}(6)\oplus\mathfrak{su}(2) \subset \mathfrak{e}_6,\mathfrak{so}(12)\oplus\mathfrak{su}(2) \subset \mathfrak{e}_7, \mathfrak{e}_7\oplus\mathfrak{su}(2) \subset \mathfrak{e}_8$,  respectively. In the adjoint representation, the weights corresponding to the canonical simple roots have  Dynkin label $2$, and the intermediate state arising from the subtraction of one canonical simple root from the weights is a zero weight. Therefore, the dimension of the relevant quasi-minuscule representation of the simply-laced Lie algebras $\mathfrak{e}_6, \mathfrak{e}_7, \mathfrak{e}_8$ satisfies a relation
\be
\text{dim}\;{\bf \text{adj}} - \text{rank}(\widetilde{\mathfrak{g}}) = \frac{|W_{\widetilde{\mathfrak{g}}}|}{|W_{\mathfrak{g}}|} \,.
\ee
In the cases of the quasi-minuscule representations, we can consider the phases which are invariant under the Weyl group action of $\mathfrak{su}(2)$ in the decompositions. Those $\mathbb{Z}_2$ invariant phases also have a characterization in terms of the weights in the representation graph. For such $\mathbb{Z}_2$ invariant phases, a root of $\mathfrak{su}(2)$ appears as a generator of the Weyl chamber $\widetilde{\mathcal{C}}_{\widetilde{\mu}}$. Since the highest weight of the adjoint representation of $\mathfrak{e}_6, \mathfrak{e}_7, \mathfrak{e}_8$ corresponds to the  simple root of $\mathfrak{su}(2)$, such Weyl chambers correspond to the weights, whose Dynkin label has a component of $\pm 2$, namely the canonical simple roots of the Lie algebras or the negative of them. Whether the phase corresponds to the canonical simple root or its negative  is related to the sign of the simple root of $\mathfrak{su}(2)$ in the embedding. Its sign is not relevant for the phase of the gauge theory with the Lie algebra $\mathfrak{g}$, hence the number of the $\mathbb{Z}_2$ invariant phases is 
\be
\# \text{ distinct }\mathbb{Z}_2\text{ invariant phases} = \text{rank}\;\widetilde{\mathfrak{g}}.
\ee

In fact, we can also determine the network of these phases. Note that if the simple root of $\mathfrak{su}(2)$ appears as a generator of the Weyl chamber of $\widetilde{\mathcal{C}}_{\widetilde{\mu}}$, this means that we have two roots which are related by the $\mathbb{Z}_2$ action as generators of the Weyl chamber at the previous step. Suppose we have a weight whose Dynkin label is $[\cdots, 2, \cdots, -1, \cdots]$. If we perform a Weyl reflection with respect to a root corresponding to the Dynkin label $-1$, then the weight becomes $[\cdots, 1, \cdots, 1, \cdots]$. The roots corresponding to the two $1$'s in the Dynkin label are related by the $\mathbb{Z}_2$ action of $\mathfrak{su}(2)$. Performing a Weyl reflection with respect to the root corresponding to the first $1$ in the Dynkin label,  the weight becomes $[\cdots, -1, \cdots, 2, \cdots]$. These two Weyl reflections relate the two adjacent phases of the $\mathbb{Z}_2$ invariant phases. Therefore, the network of phases is the same as the intersection graph of the  canonical simple roots of $\widetilde{\mathfrak{g}}$, i.e. the network of  $\mathbb{Z}_2$ invariant phases from the decomposition of the adjoint representation of $\mathfrak{e}_6, \ \mathfrak{e}_7,\  \mathfrak{e}_8$ are nothing but the Dynkin diagrams of the Lie algebra $\mathfrak{e}_6,\  \mathfrak{e}_7,\  \mathfrak{e}_8$,  respectively, thus proving the right hand side of (\ref{MiniCorresp2}).


\section{Phases of the $SU(n)$ Theory with Matter}
\label{sec:SUnPhases}

The phases of the $SU(n)$ theories with matter are described in terms of one of the three equivalent characterizations that we have given so far (Weyl group quotient, Bruhat order and  decorated box graph). The flow rules are determined to characterize the $\mathfrak{su}(n) \oplus \mathfrak{u}(1)$ phases. There will be additional constraints on the phases once we impose in addition the tracelessness condition\footnote{Our conventions are those of \cite{FultonHarris}.}
\be\label{TracelessNess}
\sum_{i=1}^n L_i =0 \,,
\ee 
which reduces the gauge algebra to  $\mathfrak{su}(n)$. 
We first explain how this is implemented and then discuss all phases for $\mathfrak{su}(n)$ with fundamental, anti-symmetric, as well as the combined representations. A similar but much simpler discussion for $SO(2n)$ can be found in appendix \ref{app:SO}, whereas some of the exceptional cases are covered in appendix \ref{app:EPhases}. 

\subsection{Reduction to $SU(n)$ Phases}

The additional constraint compared to the $\mathfrak{u}(n)$ theory is (\ref{TracelessNess})
We now show how to impose this tracelessness condition on the phases. 

From the point of view of the Weyl chamber description, the
 reduction of the extra $\mathfrak{u}(1)$ can be done by restricting the Weyl chamber (\ref{phase}) to a hypersurface $\Sigma$ in the space of $\phi$. If the subcone of the Weyl chamber (\ref{phase}) still has a non-empty region after the restriction on $\Sigma$, then it corresponds to the phase of the theory with $\mathfrak{g}$, without the extra $\mathfrak{u}(1)$. In the dual weight space language, the condition can be understood as whether a vector perpendicular to the hypersurface $\Sigma$ is inside a cone or not. More precisely, if the subcone shares a  non-empty region with the hypersurface $\Sigma$, then the vector $v$, which is perpendicular to $\Sigma$, should neither be inside $\widetilde{\mathcal{C}}_{\widetilde{\mu}}$ nor $\widetilde{\mathcal{C}}_{\widetilde{\mu}}^-$ (which is the span of the negative roots). Suppose the vector $v$ is inside 
 $\widetilde{\mathcal{C}}_{\widetilde{\mu}}$, then the hypersurface is defined as $\langle\phi ,v\rangle = 0$. On the other hand, the subcone of the Weyl chamber is defined as a region in $\mathfrak{h}$ such that $\langle \phi , \alpha\rangle > 0 $ for all the vectors $\alpha$ in $\widetilde{\mathcal{C}}_{\widetilde{\mu}}$. Therefore, the hypersurface does not intersect with $\widetilde{\mathcal{C}}^*_{\widetilde{\mu}}$. Moreover, the hypersurface does not intersect with $\widetilde{\mathcal{C}}^*_{\widetilde{\mu}}$ only if $v$ is inside $\widetilde{\mathcal{C}}_{\widetilde{\mu}}$ or $\widetilde{\mathcal{C}}_{\widetilde{\mu}}^-$. Hence $v$ should be outside both $\widetilde{\mathcal{C}}_{\widetilde{\mu}}$ and $\widetilde{\mathcal{C}}_{\widetilde{\mu}}^-$ for the phase of the theory with $\mathfrak{g}$ after the reduction of the extra $\mathfrak{u}(1)$.

In the following sections we will give a description of the phases of the $\mathfrak{su}(n)$ theory in terms of the flow rules and provide a combinatorial enumeration of them for the fundamental, anti-symmetric and combined representation case.


\subsection{Fundamental Representation}\label{Sec:SUNF}

For the fundamental representation, we label the weights by $L_i$, $i=1 \cdots n$, and we have shown that the $\mathfrak{u}(n)$ theory has phases given by the signs 
\be\label{UnF}
(+  \cdots + )\,,\ ( +  \cdots   +-) \,,\quad  \cdots \quad \,, \  (+ - \cdots -)\,,\  (-\cdots -) \,.
\ee
Recall that a sign $\epsilon_i=+$ means that $L_i>0$ is positive, and $\epsilon_i =- $ means that $L_i <0$. 
There are precisely $n+1$ phases for the $\mathfrak{u}(n)$ theory, which can be counted by the Weyl group quotient
\be
\frac{|W_{\mathfrak{su}(n+1)}|}{|W_{\mathfrak{su}(n)}|} = n+1 \,.
\ee
These are shown for $\mathfrak{u}(5)$ in table \ref{tab:SU5FunTab} in appendix  \ref{app:U5}.


\begin{figure}
\centering
\includegraphics[width=5cm]{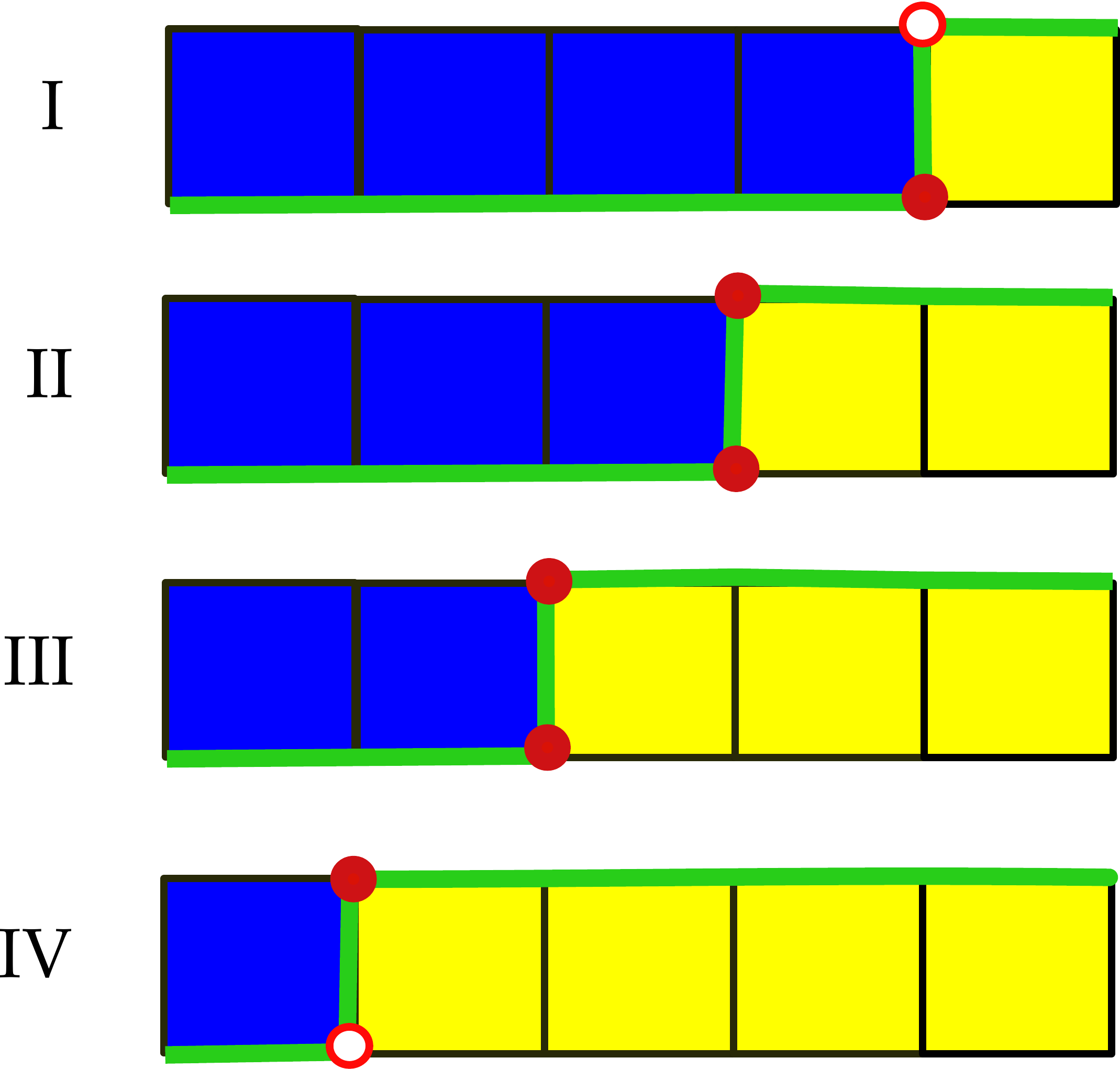}
\caption{Phases of the $\mathfrak{su}(5)$ theory with fundamental representation ${\bf 5}$ matter. The blue/yellow boxes correspond to decorating with $\pm$.  The green lines and red dots will play a role later on in understanding the flops between these phases: red dots correspond to extremal points that can be flopped, whereas white dots correspond to flops that would map out of the $\mathfrak{su}(5)$ phases to the $\mathfrak{u}(5)$ phases. }
\label{fig:SU5FPhases}
\end{figure}


However, it is clear that due to (\ref{TracelessNess}), the two phases $ (+ \cdots +)$ and $(- \cdots -)$ mean that 
$\sum_{i=1}^n L_i >0$ and $<0$ respectively, and therefore do not respect the tracelessness condition. 
The $\mathfrak{su}(5)$ phases are shown in figure  \ref{fig:SU5FPhases}.

All remaining phases are consistent $\mathfrak{su}(n)$ phases: for this it is enough to show that positive linear combinations of the elements in the cone do not give rise to $ L_1 +\cdots + L_n >0 $ or $<0$. Indeed, any phase that is not $ (+ \cdots +)$ or $(- \cdots -)$ will have at least one element $L_i<0$.
First recall that the flow rules are

\be\label{FlowsRulesF}
\includegraphics[width=2.5cm]{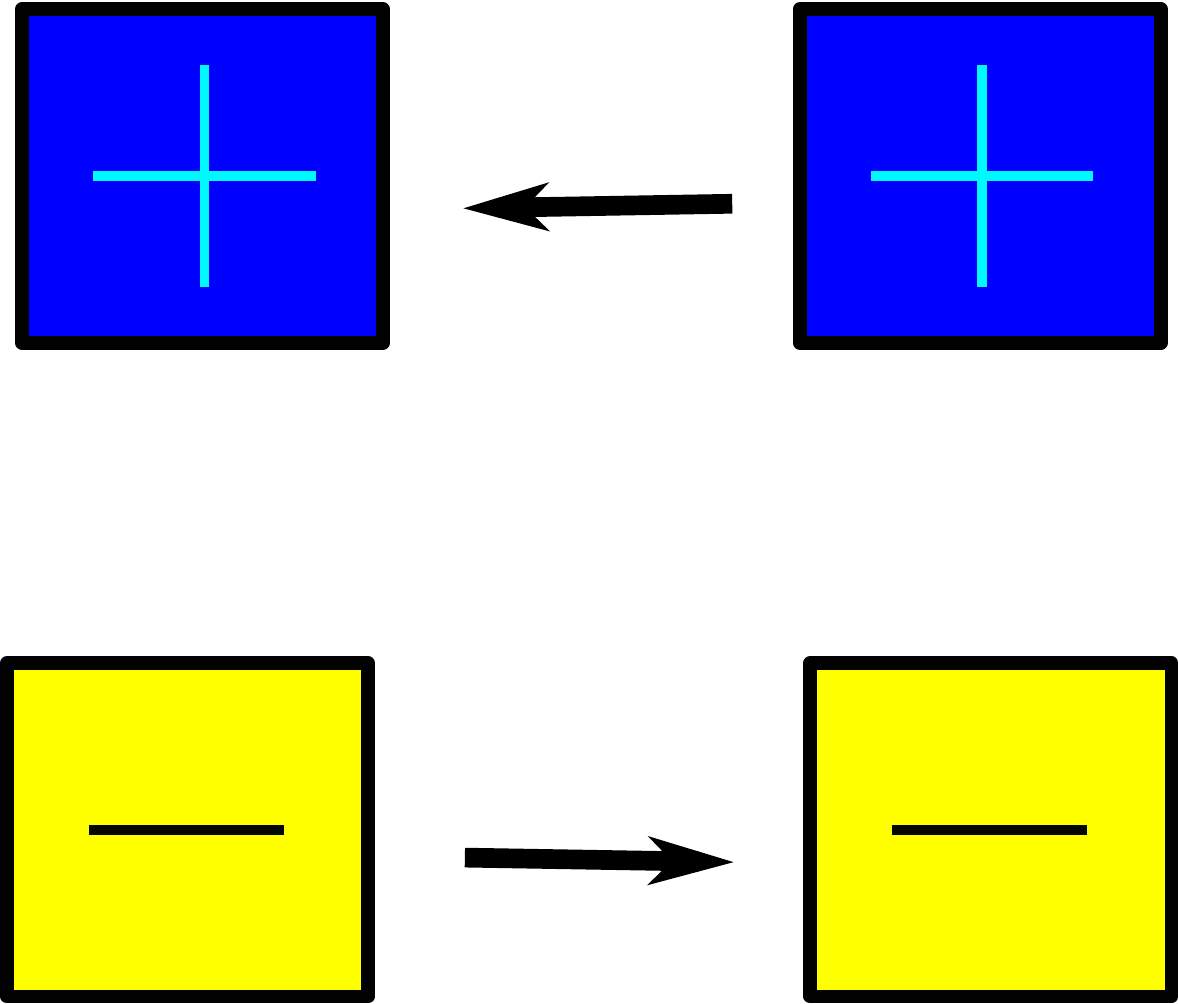}
\ee

Let $i$ be the  smallest entry with $L_i<0$, 
it follows that $L_{j}<0$ for all $j>i$. Then $-\sum_{l=i}^n L_l >0$. However, it is not possible to linear combine using only positive roots and $-L_l$ the terms $-(L_1 +\cdots + L_{i-1})$. Likewise, one can get $\sum_{l=1}^{{i-1}} L_l>0$, however, then there is no combination that gives rise to $L_i +\cdots +L_n>0$. Thus, a phase in (\ref{UnF}) that is not  $(+ \cdots +)$ or $(- \cdots -)$ will correspond to an $\mathfrak{su}(n)$ phase. 

The number of phases for the $\mathfrak{su}(n)$ theory with fundamental representation is therefore 
\be
SU(n)\,, \hbox{ with  } {\bf n}:\qquad  \# \hbox{Phases} =  n-1 \,.
\ee


\subsection{Antisymmetric Representation}
\label{subsec:AR}

Next we consider the phases of a $\mathfrak{u}(n)$ gauge theory with the antisymmetric representation $\Lambda^2 {\bf n}$. As we showed in the last section these are characterized in terms of the Weyl group quotient
\be
{W_{{\mathfrak{so}(2n)}} \over W_{\mathfrak{su}(n)} } \qquad \longleftrightarrow \qquad  \left\{ \Phi^{\mathfrak{u}(n)}_{\epsilon_1 \cdots \epsilon_{d}} \right\} \,,
\ee
where $d={n(n-1)/2}$ is the dimension of the representation. The number of such phases is 
\be
\frac{|W_{{\mathfrak{so}(2n)}|}}{|W_{\mathfrak{su}(n)}|} = 2^{n-1} \,.
\ee
The weights of the antisymmetric representation are labeled by $L_i + L_j $ with $i < j$ and $i, j=1\cdots n$. 
In figure \ref{fig:SU8A} we depict the weights of the representation, ordered in terms of the $i$th row corresponds to $L_i + L_j$, $j=i+1 \cdots n$. 
The weights are arranged such that each separating line corresponds to the action of a negative simple root $L_{i+1} - L_{i}$.

\begin{figure}
\begin{tabular}{lr}
\includegraphics[width=6cm]{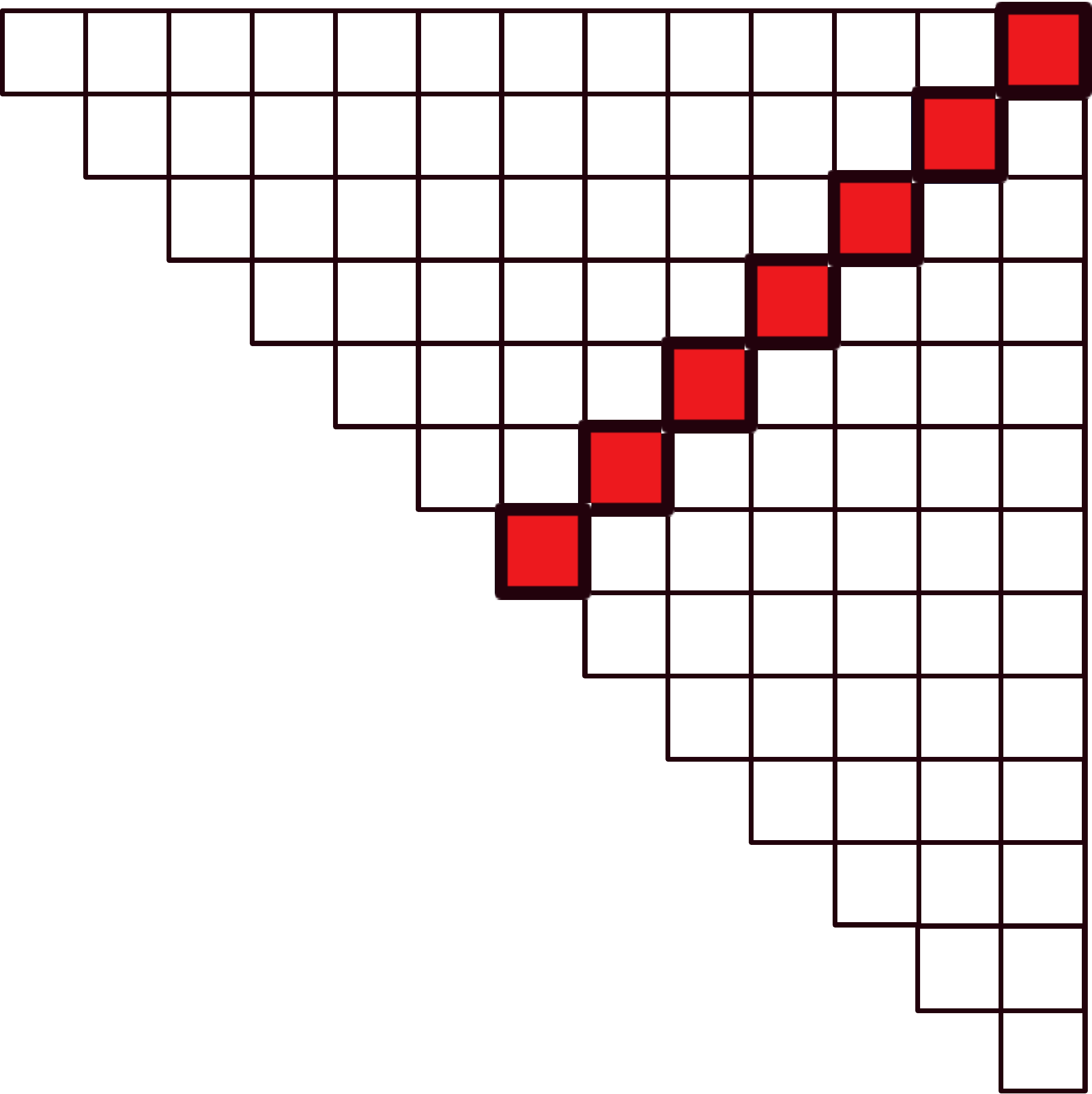} & $\qquad \qquad $
\includegraphics[width=6cm]{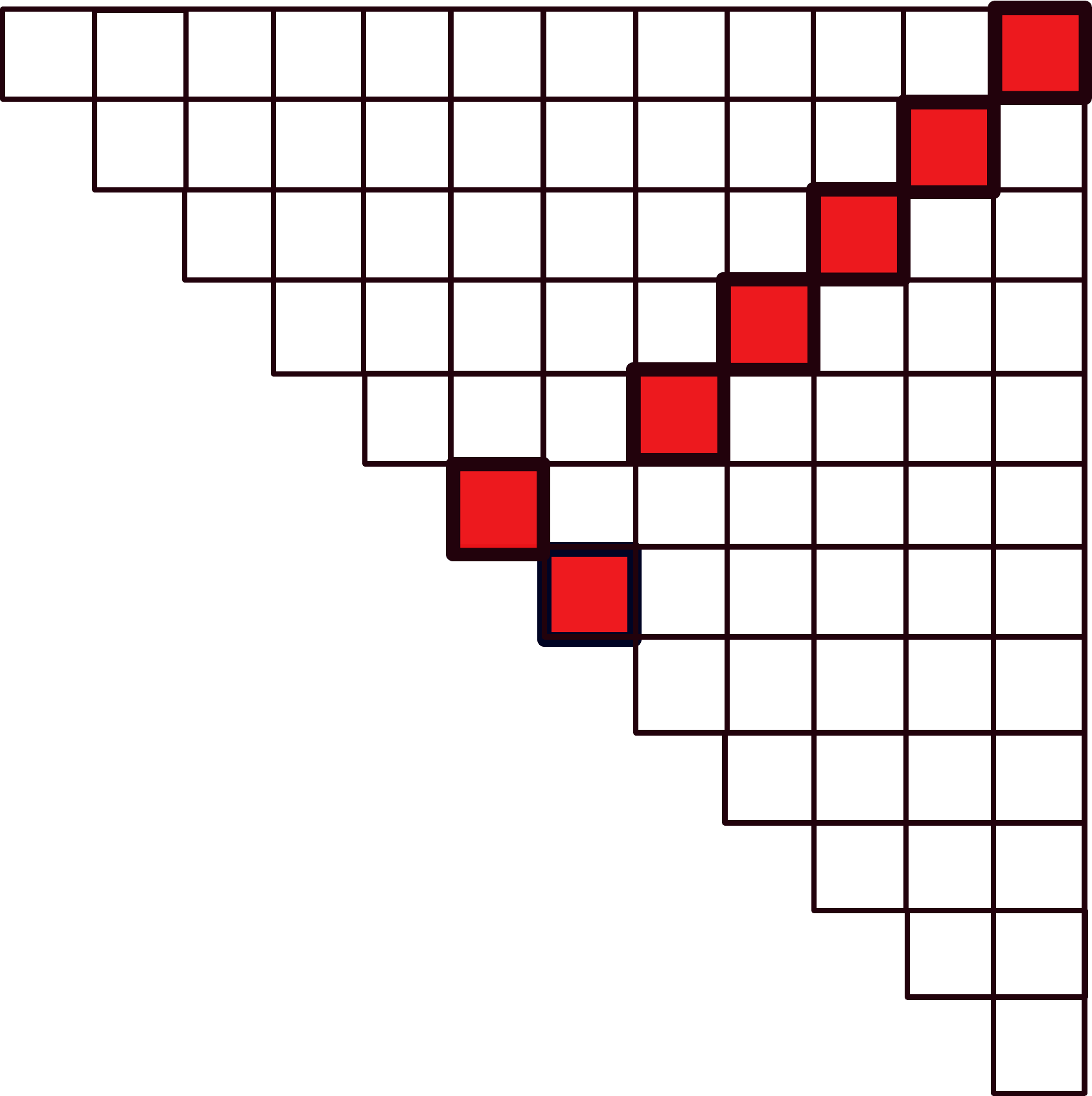} 
\end{tabular}
\caption{Weight diagrams for $n$ even (LHS) and odd (RHS). The weights marked in red are the ones appearing in the sign constraints (\ref{SUevenSignCond}) and (\ref{SUoddSignCond}), which determine the $\mathfrak{su}(n)$ phases. The examples drawn here are $\mathfrak{su}(14)$ and $\mathfrak{su}(13)$. The nodes represent the weights as explained in figure \ref{fig:SU8A}.} 
\label{fig:Diags}
\end{figure}

Each phase corresponds to a sign assignment for this weight diagram, and we label the signs as 
\be
\sign (L_i + L_j) \equiv \epsilon_{ij} \,.
\ee
For $n=2k$ even, the phases of the  $\mathfrak{su}(n)$ theory are characterized by the subset of phases of the $\mathfrak{u}(n)$ theory, which satisfy in addition
\be\label{SUevenSignCond}
SU(2k) :\qquad  
\ba
&  \mathcal{E}_{2k} \not= (+\, , \cdots , \, +) \cr
& \mathcal{E}_{2k} \not= (-\,, \cdots\, , -) 
\ea
\ee
where 
\be
\mathcal{E}_{2k}\equiv 
(\epsilon_{1\, 2k}, \epsilon_{2 \, 2k-1}, \cdots , \epsilon_{k\, k+1 })  \,.
\ee
Likewise, for $n= 2k+1$ odd, the condition is 
\be\label{SUoddSignCond}
SU(2k+1) :\qquad 
\ba
&\mathcal{E}_{2k+1} \not= (+\, , \cdots , \, +) \cr
&\mathcal{E}_{2k+1} \not=  (-\,, \cdots\, , -) 
\ea
\ee
with the diagonal defined as 
\be
\mathcal{E}_{2k+1}\equiv
(\epsilon_{1\, 2k+1}, \epsilon_{2 \, 2k}, \cdots , \epsilon_{k-1\, k+3}, \, \epsilon_{k\, k+1},\, \epsilon_{k+1, k+2})  \,.
\ee
These conditions are depicted in terms of red boxes in figure \ref{fig:Diags}. 
Note  that the signs ``flow" as explained in (\ref{FlowsRulesF1}), i.e. a consistent phase sign assignment will always respect the following sign implications (flow rules),  
\be\label{Flows1}
\includegraphics[width=6cm]{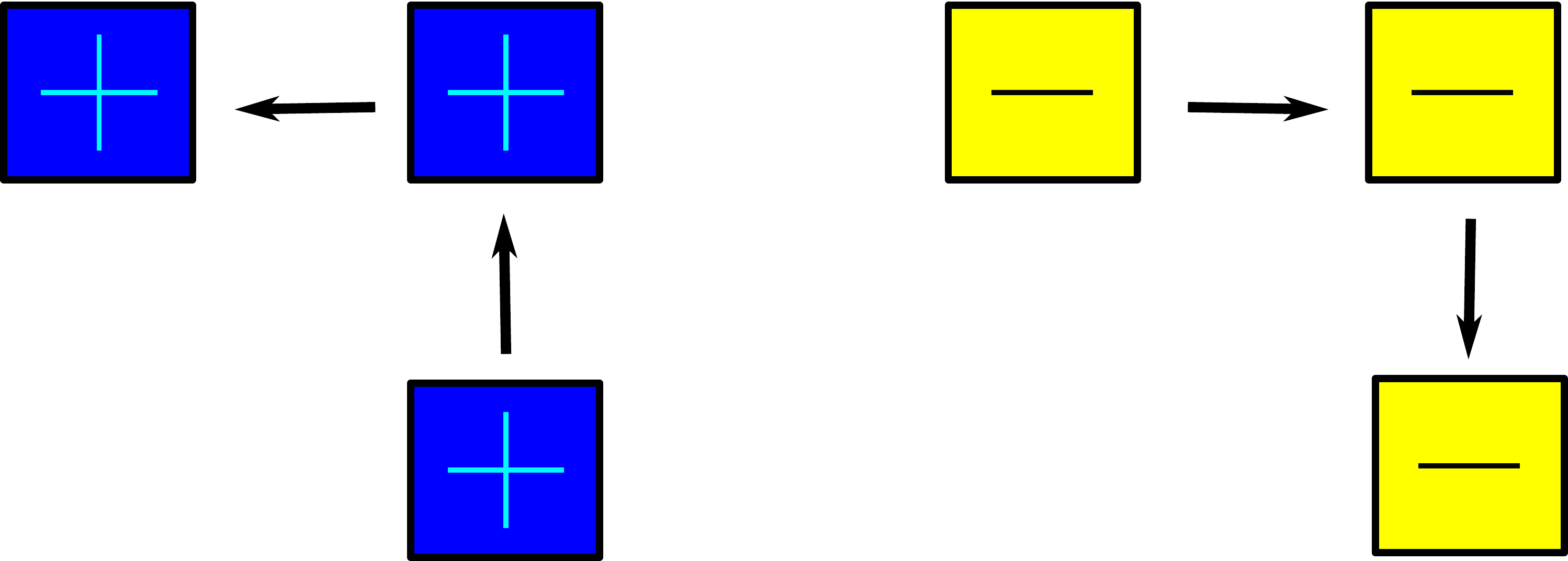}
\ee

\begin{figure}
\centering
\includegraphics[width=5cm]{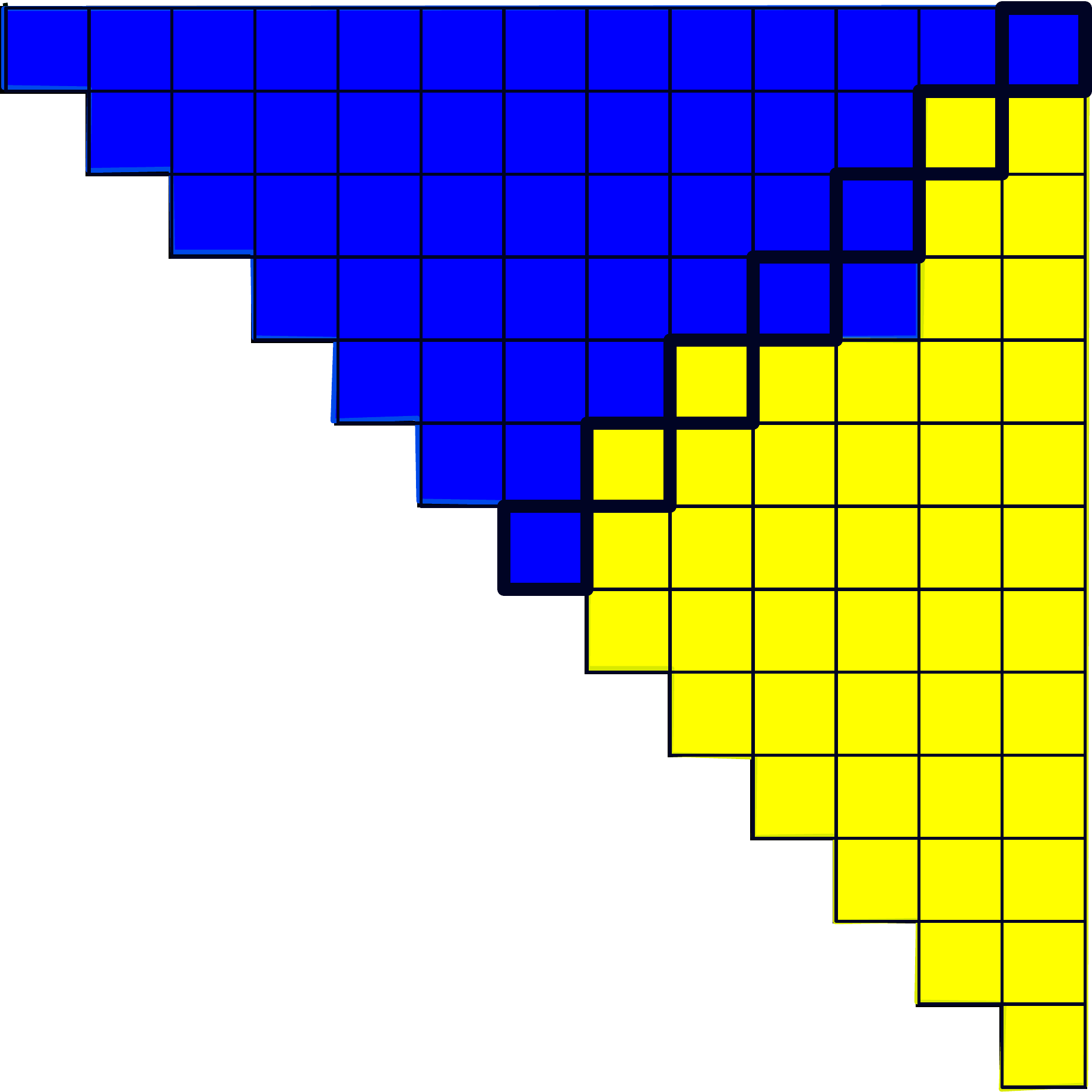}
\qquad\qquad 
\includegraphics[width=5.1cm]{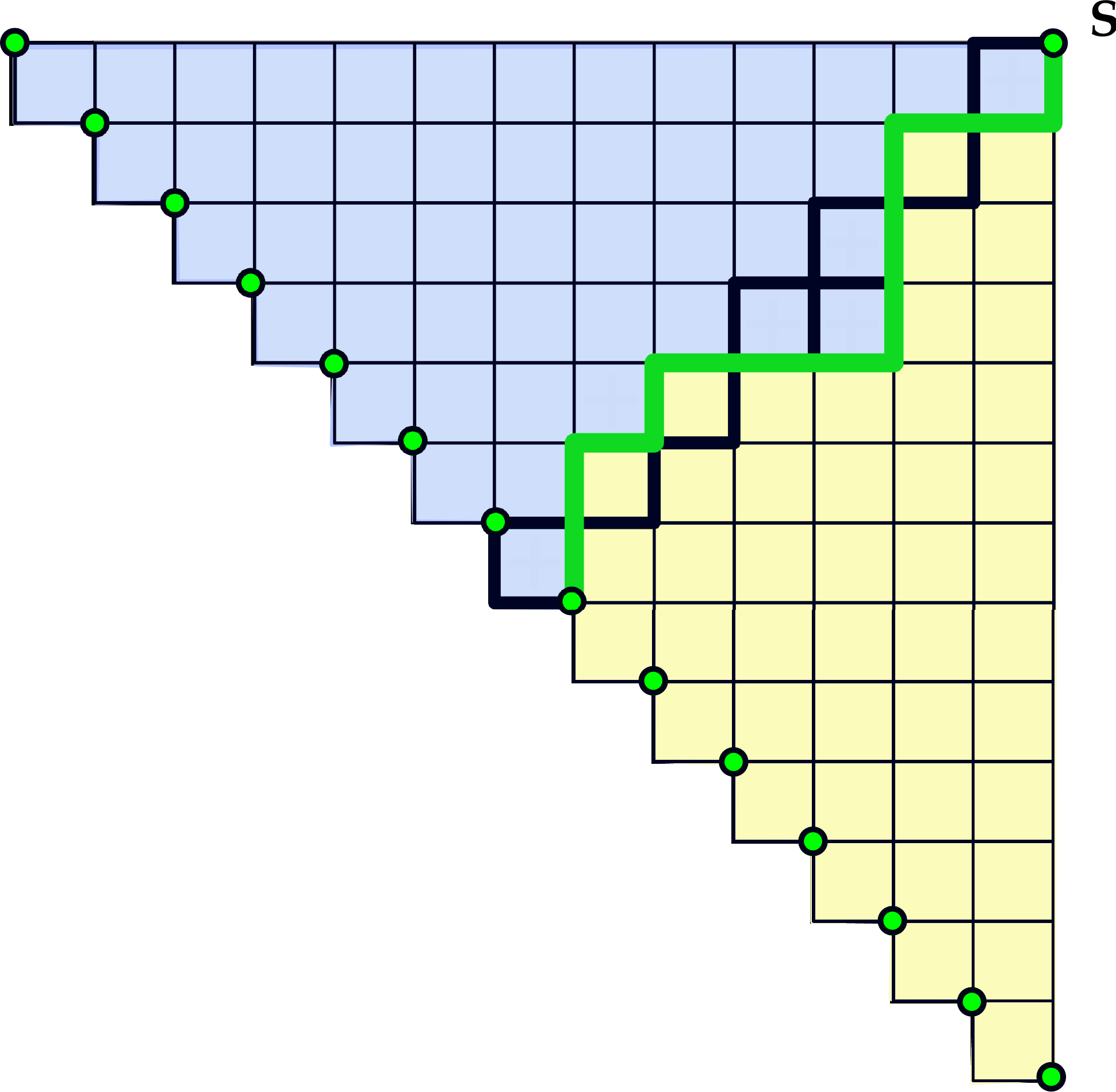}
\caption{Example phase diagram, where blue are $+$ and yellow $-$. This is a consistent phase, as the entries $\mathcal{E}_{n}$, indicated by the bold-face boxes, are not all of the same sign. On the right the same phase is characterized in terms of a convex path (green) starting at $S$ and ending at one of the green nodes and crossing the NE to SW diagonal at least once, i.e., an anti-Dyck path. }
\label{fig:ExamplePhase}
\end{figure}

\begin{figure}
\centering
\includegraphics[width=13cm]{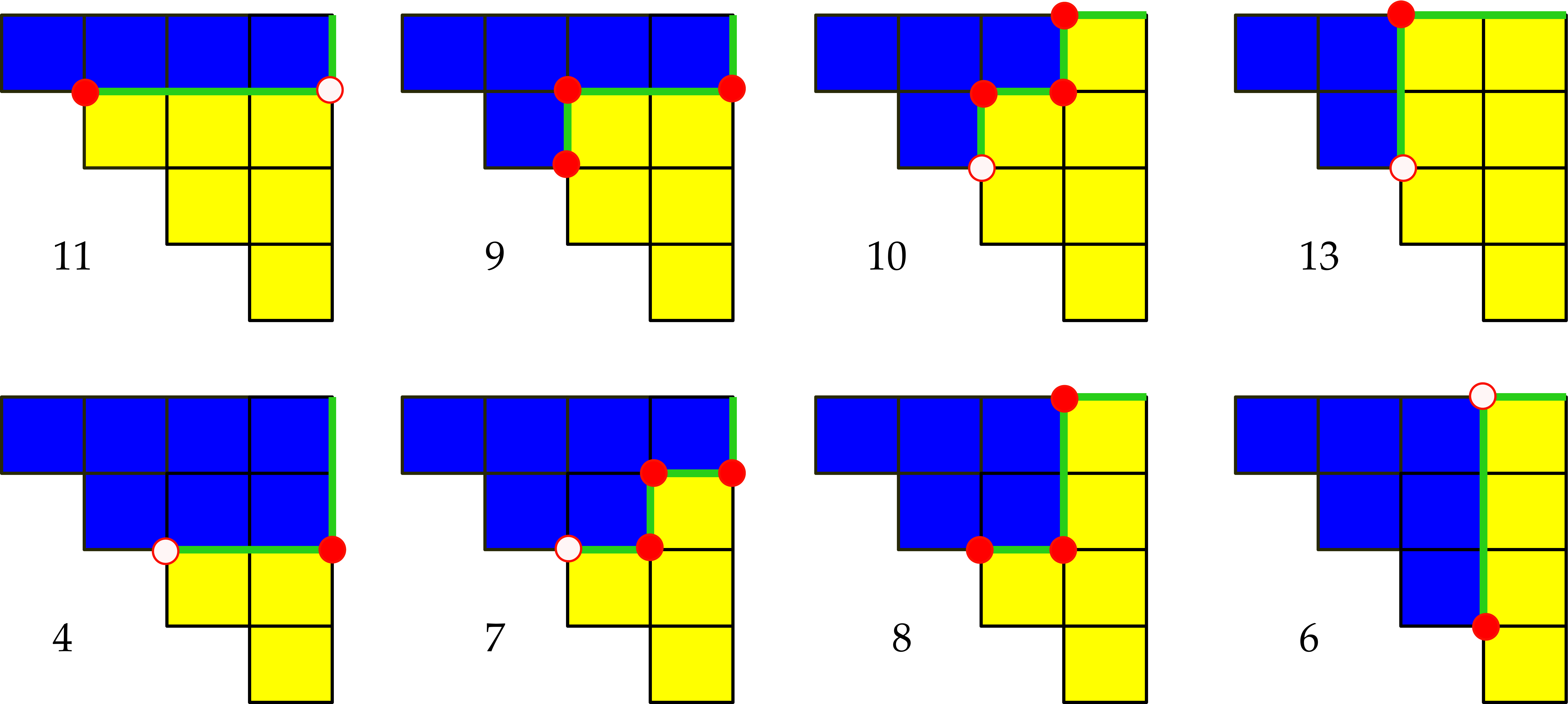}
\caption{Phases of $SU(5)$ with anti-symmetric representation, with $\pm$ corresponding to blue/yellow boxes, 
including the corresponding anti-Dyck paths (green lines) and extremal points (red nodes), which will be discussed in the section on flops. 
All phases satisfy  the anti-Dyck property with respect to the diagonal defined by $\mathcal{E}_5$ in (\ref{SUoddSignCond}). The numbering is as in figure \ref{fig:U5A}.}
\label{fig:SU5Flops}
\end{figure}

A phase for $\mathfrak{su}(n)$ with the antisymmetric representation is characterized by a representation diagram as in figure \ref{fig:ExamplePhase}, i.e., a sign assignment which is consistent with the flow rules (\ref{Flows1}) and respects the sign conditions (\ref{SUevenSignCond}, \ref{SUoddSignCond}). An entirely equivalent way to characterize this setup is in terms of
\be
\hbox{Phases of $\mathfrak{su}(n)$ with $\Lambda^2{\bf n}$} \quad\stackrel{1:1}{\longleftrightarrow} \quad \hbox{Anti-Dyck Paths}\,,
\ee
where we define 
an {\it anti-Dyck path} as  a monotonous path in the representation graph, starting at the top NE corner (denoted by $S$ in figure \ref{fig:ExamplePhase}), and ending at one of the points along the NW to SE diagonal (shown in green in the figure) and crossing the diagonal defined by $\mathcal{E}_n$ at least once.  An example anti-Dyck path is shown on the right of figure \ref{fig:ExamplePhase}.
For $\mathfrak{su}(5)$ all phases satisfying the flow rules and the diagonal condition are shown in figure \ref{fig:SU5Flops}. 


To prove that these conditions are necessary and sufficient, consider first $n=2k$ even. First we show necessity of the sign condition, (\ref{SUevenSignCond}), i.e., if it is violated, then this implies a  $\mathfrak{u}(n)$ phase, that is not an $\mathfrak{su}(n)$ phase. This can be easily seen noting that the sum 
$\sum_{ij \in \mathcal{E}} L_i + L_j = (L_1 + L_n) + (L_2 + L_{n-1}) + \cdots + (L_{\frac{n}{2}} + L_{\frac{n}{2}+1}) > 0$ or $< 0$ due to \eqref{SUevenSignCond}. 

On the other hand, to show that it is a sufficient condition, we show that if the sign condition holds, then the phase is an $SU(2k)$  phase. For this it is enough to show that positive linear combinations of the weights in the cone (i.e., the weights with the sign as prescribed for this specific cone) do not give rise to $L_1 +\cdots+ L_n>0$ or $<0$\footnote{This requirement follows by noting that in $\mathfrak{su}(n)$ the tracelessness condition  implies $L_1 +\cdots L_n =0$, and this element would not be in the cone, however, in a $\mathfrak{u}(n)$ phase, that is not an $\mathfrak{su}(n)$ phase, this element would have a definite sign. }. 
As the sign condition (\ref{SUevenSignCond}) holds by assumption, at least one
of the entries $\epsilon_{ij}$ in $\mathcal{E}_{2k}$ is negative (and at least
one {$\epsilon_{lm}$} is positive). Denote this by $L_i + L_j <0$. By the
``flow'' rules this implies that $L_k + L_m <0$ for all $k\geq i$ and $m \geq j$. In the representation graph, these are all the weights below and to the right of $L_i + L_j$.  
However, then it is not possible to linear combine $L_1 +\cdots+ L_n>0$ as any linear combination of positive weights will require that some $L_{m}$ appears at least twice for some $m= 1,\cdots  i-1$ {or there is no positive root of the form $L_l + L_m$ where $l = 1, ...m-1$, for fixed $m$ such that $j \leq m \leq n$}. 
Similarly we can argue for the case $L_1 +\cdots+ L_n<0$, since at least one entry $\epsilon_{lm}$ in $\mathcal{E}_{2k}$ is positive.

For $n=2k+1$, a similar argument applies. First note that the sign of $L_{k} + L_{k+2}$ is determined once the signs for the entries  $\mathcal{E}_{2k+1}$ are fixed (and is the same as theirs) . 
Without loss of generality consider the case when $\mathcal{E}_{2k+1} = (+ \cdots +)$. 
Then it is clear that $L_1 + L_{2k+1} = - \sum_{i=2}^{2k} L_i >0$ and adding the some of the remaining entries in $\mathcal{E}_{2k+1}$ {as well as $L_k + L_{k+2}$ which are all positive,  it follows that $-L_{k+1} >0$. Note that $L_{k+1} + L_{k+2}>0$, and thus $L_{k+2}>0$. However, this implies that adding this to the simple root $L_{k+1} - L_{k+2}>0$, that we can linear combine $0$ as a positive linear combination, and thus the sign choice 
$\mathcal{E}_{2k+1} = (+ \cdots +)$ does not give rise to a cone, and thus to an $\mathfrak{su}(n)$ phase. Similarly, reversing the signs, it is straight forward to show that $\mathcal{E}_{2k+1} = (- \cdots -)$ also does not give rise to $\mathfrak{su}(n)$ phases. 
The argument that this is also a sufficient condition is identical to the $n$ even case.


\begin{figure}
\centering
\includegraphics[width=4cm]{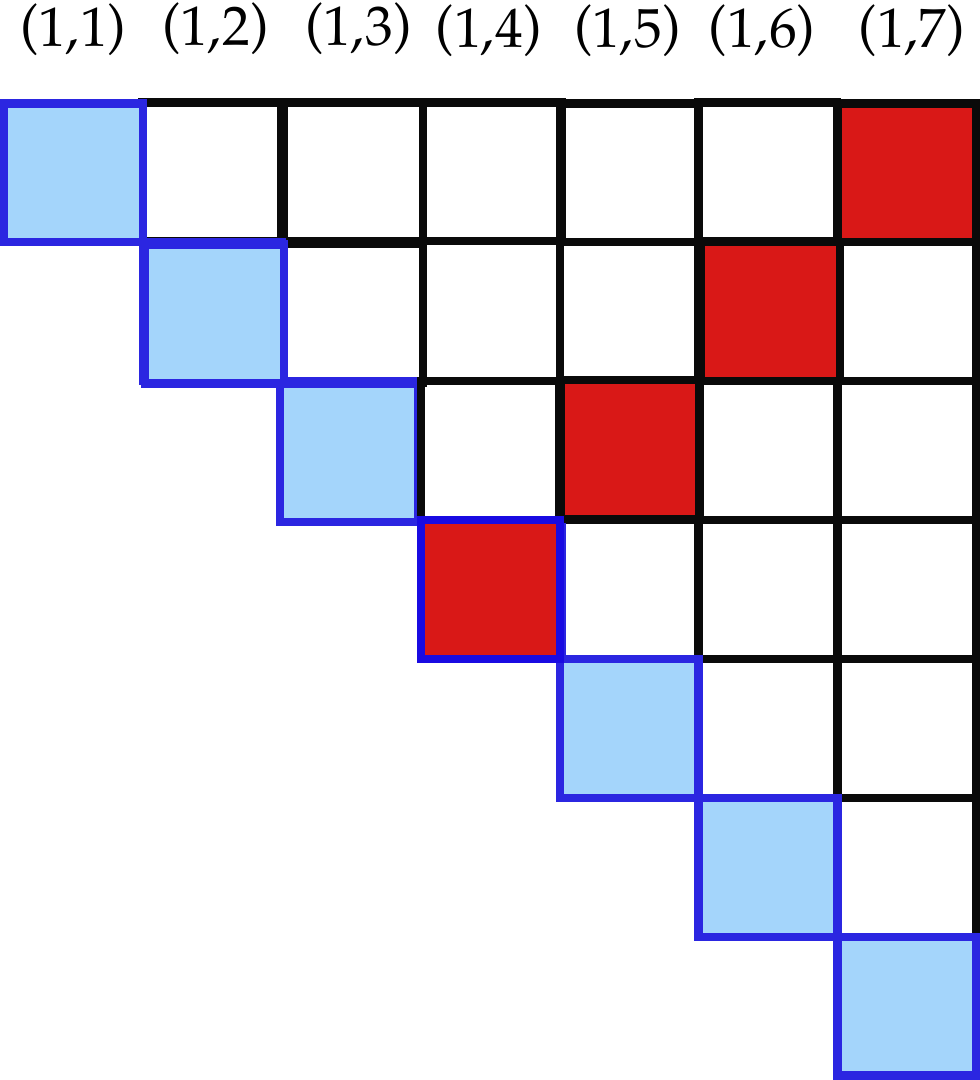}
\caption{Setup for phases for antisymmetric and fundamental representation. The black boxes denote the antisymmetric representation for $SU(2k+1)$, the blue boxes depict how the fundamental is attached to this diagram. The resulting diagram is a consistent phase of $SU(2k+1)$ as long as the sign constraint holds for the diagram viewed as an $SU(2k+2)$ diagram. The red boxes correspond precisely to $\mathcal{E}_{2k+2}$. }
\label{fig:SUAFBox}
\end{figure}



\subsection{Antisymmetric and Fundamental Representations}
\label{subsec:AFPhases}

The phases for the gauge theory with chiral multiplets in both antisymmetric and fundamental representation can be characterized by decorated box graphs as well\footnote{This is setup is of particular interest for recent developments in constructing realistic F-theory compactifications based in $SU(5)$ grand unified theories.}.  One procedure to do this is as follows: 
First we consider the representation diagrams for the antisymmetric  representation for $\mathfrak{su}(2k+1)$. 
To this diagram we attach the weights of the fundamental embedded as $L_i + L_i$, $i=1, \cdots, n$ along the diagonal, as depicted  in figure \ref{fig:SUAFBox}{\footnote{This is nothing but the weight diagram of the symmetric representation of $\mathfrak{su}(2k+1)$. It is clear from the gauge theory analysis that the phase structure of an $\mathfrak{su}(n)$ gauge theory with the antisymmetric representation and the fundamental representation is the same as the phase structure of an $\mathfrak{su}(n)$ gauge theory with the symmetric representation.}.}
It is clear, first of all, that unless the phases of the antisymmetric and the fundamental are separately consistent phases, the resulting combined phase will not be a consistent $\mathfrak{su}(n)$ phase. 
However, not all combinations are consistent.

The consistency condition is that the combined diagram is consistent with 
\begin{itemize}
\item[(i)] Flow rules of signs in (\ref{Flows1})
\item[(ii)] The resulting diagram, interpreted as an $\mathfrak{su}(2k+2)$ antisymmetric representation satisfies the sign constraints (\ref{SUevenSignCond}), i.e., the diagonal is not all $+$ or all $-$ signs. 
\end{itemize}
To exemplify the method, we show  the phases for $\mathfrak{su}(5)$ with fundamental and anti-symmetric representation in figure \ref{fig:SU5FlopsAF}, where we also discuss the flops among these phases. 

To prove that these are consistent $\mathfrak{su}(2k+1)$ phases, we need to again show sufficiency of these conditions. We show that if the sign condition (ii) holds then the phase is an $SU(2k+1)$ phase, i.e., the element $L_1+ \cdots+ L_{2k+1}>0$ or $<0$ is not satisfied, and is thereby not in the cone. However, this we have shown to be true for $\mathfrak{su}(n)$ with $n$ even in section  \ref{subsec:AR}.


\begin{figure}
\centering
\includegraphics[width=13cm]{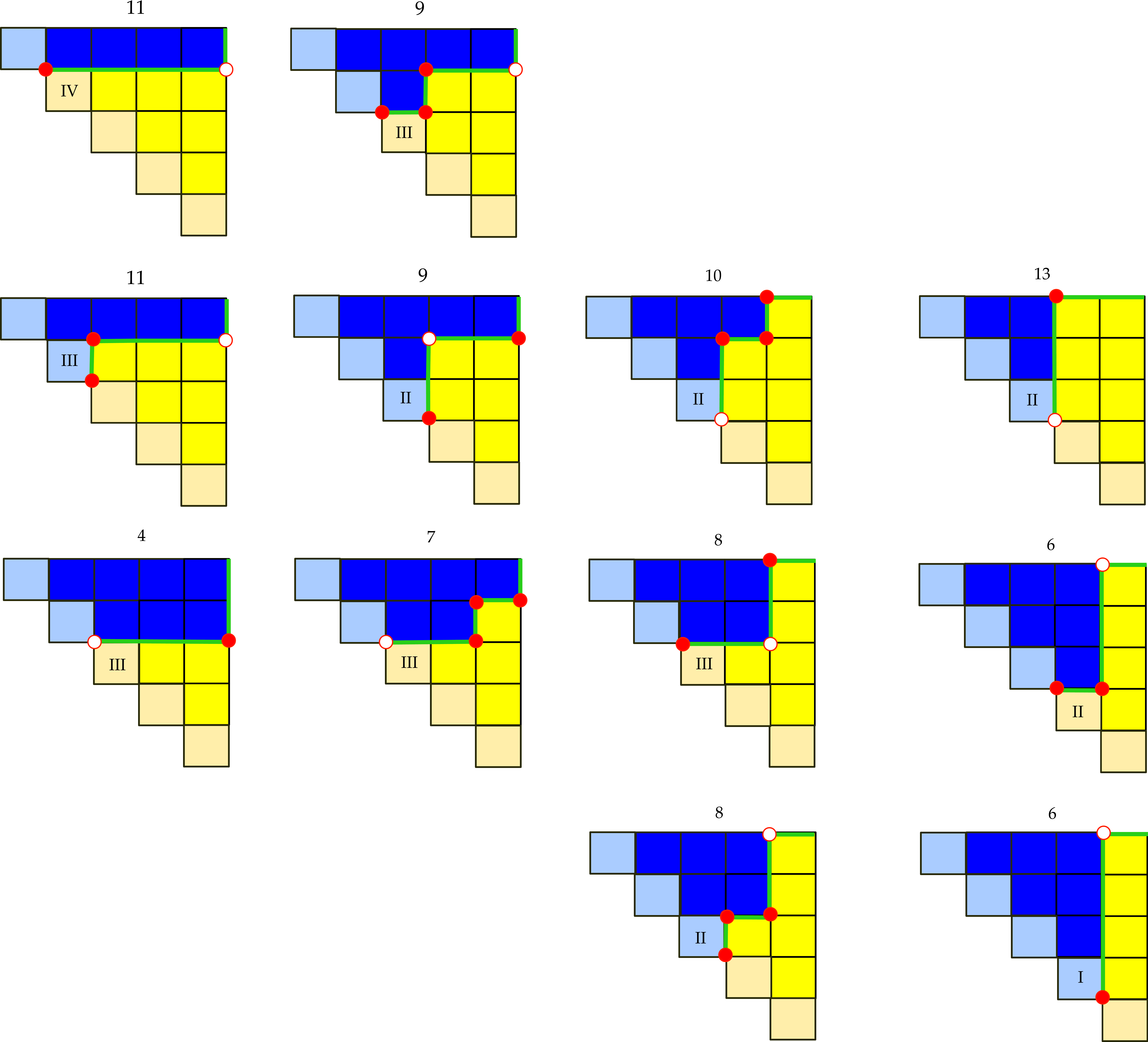}
\caption{Phases of $SU(5)$ with anti-symmetric (dark blue and yellow boxes) as in figure \ref{fig:SU5Flops} and fundamental representation (light blue and yellow boxes) as in figure \ref{fig:SU5FPhases}, including the corresponding anti-Dyck paths (green lines) and extremal points (red nodes). Note that the flops need to retain the anti-Dyck property with respect to the diagonal defined by $\mathcal{E}_6$ in (\ref{SUevenSignCond}), which would be violated if the white nodes would be flopped. The numbering is as in figures   \ref{fig:SU5FPhases}  and  \ref{fig:SU5Flops}, and following the flops along the red nodes of the anti-Dyck path, we reproduce the phase diagram \ref{fig:SU5AFPhaseDiag}. }
\label{fig:SU5FlopsAF}
\end{figure}



\section{Structure of Phases from Decorated Box Graphs}
\label{sec:StructureofPhases}

Decorated box graphs are an extremely efficient way to  characterize  phases of $d=3$ $\mathcal{N}=2$ gauge theories with matter and thereby the geometry of singular elliptic fibrations.  They contain however much more information than a simple book-keeping device. As we will see, the box graphs contain all the relevant information about the network of phases, transitions (flops) between the phases, and codimension-three loci,  each of which will have a geometric counterpart.

\subsection{Extremal Generators}
\label{subsec:Extremal}

Recall from section \ref{subsec:AR} that 
a phase for $\mathfrak{su}(n)$ with the antisymmetric representation is characterized by a decorated box graph, i.e., the representation graph with a  sign assignment which is consistent with the flow rules (\ref{Flows1}) and respects the sign conditions (\ref{SUevenSignCond}, \ref{SUoddSignCond}). Alternatively we can characterize them by 
anti-Dyck paths, which are  monotonous path in the representation graph, ending at the top NE corner (denoted by $S$ in figure \ref{fig:ExamplePhase}), and crossing the diagonals $\mathcal{E}_n$}.

From the diagram we can read off the extremal generators of the cones. They are either weights, or simple roots determined as follows:

\begin{itemize}
\item Weights that can be sign changed while retaining the anti-Dyck property of the path (we will refer to the corner along which the sign changes happens as an extremal point).  These are indicated by the red dots in the phase diagrams. 

\item A simple root is part of the extremal set, if adding it to any weight does not cross the anti-Dyck path. In fact, any other simple root, which crosses the anti-Dyck path is reducible, and can be written in terms of the two weights that are on either side of the anti-Dyck path. For instance in phase 9, figure \ref{fig:SU5Flops} the simple root $\alpha_3$ crosses the anti-Dyck path, $-(L_2+L_3) \rightarrow L_2+L_4$ and is therefore not an extremal generator, but is obtained as the  linear combination of these two weights, which are in the extremal set. 

\end{itemize}

Note that in figure \ref{fig:SU5Flops} the red dots indicate the extremal points, however the white dots are extremal points only of the $\mathfrak{u}(5)$ phase, not of the $\mathfrak{su}(5)$, i.e., sign changes that would violate the diagonal condition/anti-Dyck property of the monotonous path. 
Note that except for phases 8 and 9 in figure \ref{fig:SU5Flops} all phases have one white node, which means that the number of generators of the cone is reduced compared to $U(5)$, in fact in each of these cases there are four generators. 

In figure \ref{fig:SU5FlopsAF} all the phases of $SU(5)$ with fundamental and anti-symmetric representation are shown. In this case all phases contain white nodes, i.e., the extremal set has one less element than for the corresponding $U(5)$ phase, which is something that was already observed in \cite{Hayashi:2013lra}, where each of these phases was shown to have four generators.


\subsection{Flops between Phases and Extremal Points}

\label{sec:flopgauge}


\begin{figure}
\centering
\includegraphics[width=7cm]{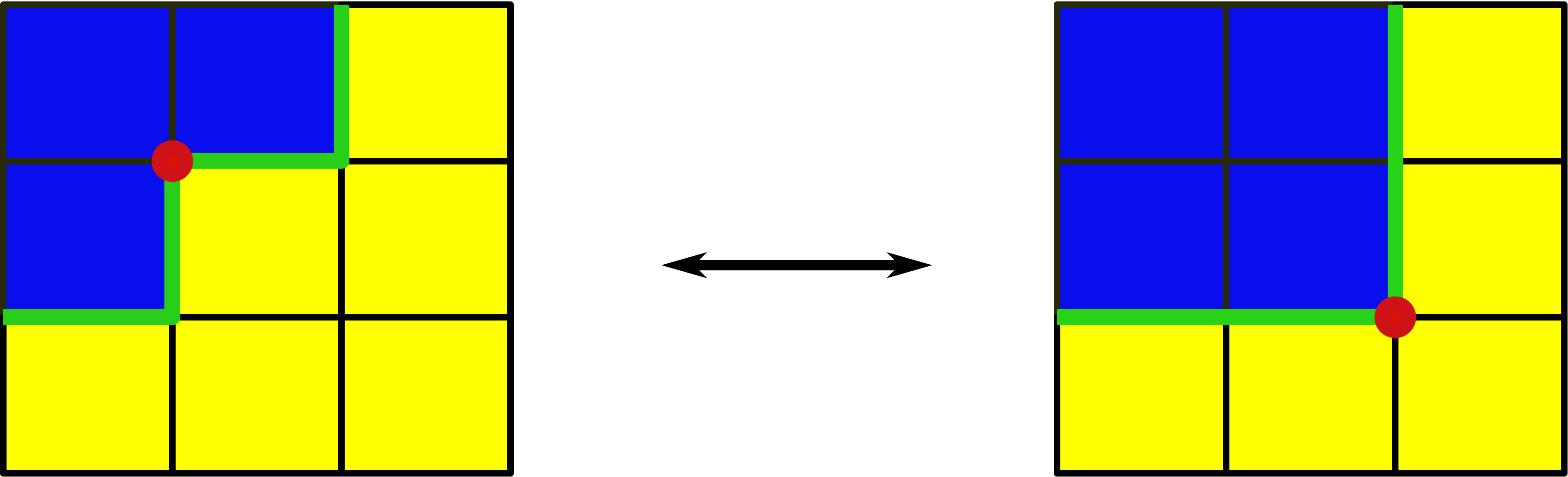}
\caption{A flop is the transition between these two decorated box graphs (which are details of a representation graph) and correspond to single-box sign changes (blue and yellow boxes), or flopping the corner along the anti-Dyck paths (green). The red dots are the extremal points on the path. }
\label{fig:FlopPic}
\end{figure}
 

 We define a {\it flop} (or flop transition) as 
 \begin{itemize}
\item Decorated box graphs:  a single-box sign change, that maps the representation graph to another representation graph, i.e., it retains compatibility with the flow rules  (\ref{Flows1}) and the sign conditions (\ref{SUevenSignCond}, \ref{SUoddSignCond}).
 \item Anti-Dyck path: a flop of a corner of the path which maps an anti-Dyck path on the representation graph into another one. We will refer to the corners associated to such flops by  {\it extremal points}.  
 \end{itemize}
The two descriptions are equivalent, and  depicted in figure \ref{fig:FlopPic}. The red node is the extremal point on the anti-Dyck path. This corner gets flopped, crossing over the box in the representation graph which changes sign under this flop. The resulting new corner of the anti-Dyck path carries an extremal point, which indicates the reverse flop transition. 

From the anti-Dyck path we can read off the extremal rays for each $\widetilde{\mathcal{C}}_{\widetilde{\mu}}$ corresponding to a phase. Each extremal point is associated to a weight in the decorated box graph. Each of these generate an extremal ray.  
In addition, the simple roots, which can be added to weights in the box graph, without changing their sign, i.e., adding or subtracting them does not cross the anti-Dyck path,  are also contained within $\widetilde{\mathcal{C}}_{\widetilde{\mu}}$. In fact, the simple roots, which do not cross the anti-Dyck path, are the other extremal rays, whereas the simple roots which cross the anti-Dyck path do not correspond to the extremal rays but are inside the cone. For example, with this rule we can reproduce, from the diagrams in figure \ref{fig:SU5Flops}, the table \ref{SU5ATable} included in appendix \ref{app:U5}, which was obtained from the Weyl group action.  


\begin{figure}
\centering
\includegraphics[width=8cm]{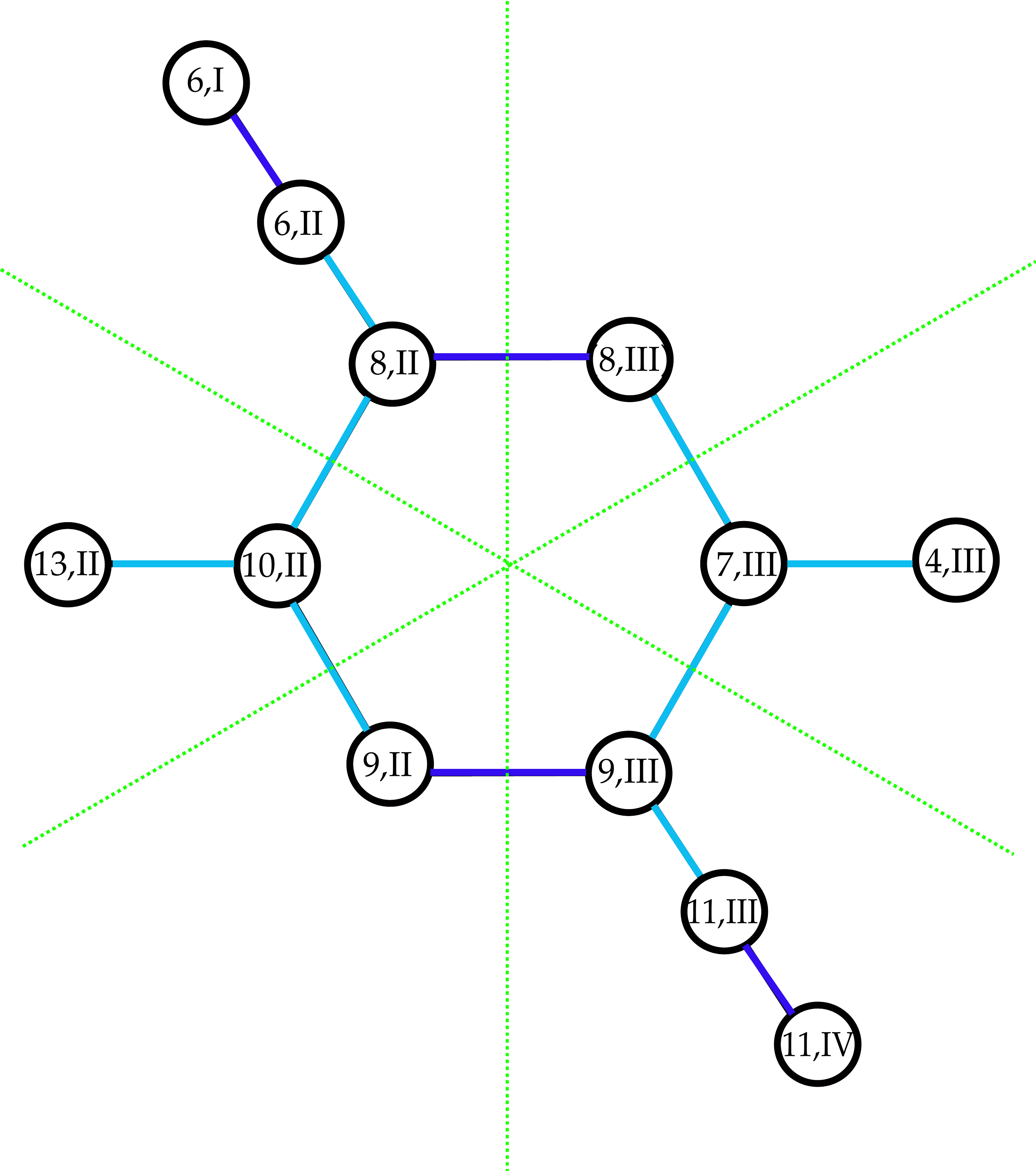}
\caption{The phase diagram for $SU(5)$ with fundamental and anti-symmetric matter, as derived from flops among the decorated box graphs/anti-Dyck paths in figure \ref{fig:SU5FlopsAF}. Identification along the dark blue lines yields the phase diagram of the anti-symmetric representation, identification of nodes along the light blue lines yields the phase diagram, which is a line, for the fundamental representation.  The curious observation which is drawn in terms of the green dotted lines is, that this diagram can be cut up into pairs of Dynkin diagrams of $E_6$, $D_6$ and $A_6$, respectively. The case of the $E_6$ flop diagram (corresponding to cutting the graph by the NE to SW diagonal) is discussed later in terms of the $E_6$ codimension-three fibers in section \ref{subsec:E6Codim3}. }
\label{fig:SU5AFPhaseDiag}
\end{figure}
 

Given that a flop will have to retain monotony of the paths, it is clear that extremal points will only appear along corners of the path. For $SU(5)$ this results in the paths and extremal points given in figure \ref{fig:SU5Flops}. Not all corners correspond to extremal points, as the resulting flop would violate the sign conditions (\ref{SUevenSignCond}, \ref{SUoddSignCond}), and thus yield a non-$\mathfrak{su}(n)$ phase. 
For $\mathfrak{u}(n)$ all corners can be flopped.

Thus a flop is characterized by either a sign change of a single box which is consistent with the flow rules  (\ref{Flows1})  and conditions (\ref{SUevenSignCond}, \ref{SUoddSignCond}) or  in terms of anti-Dyck paths, they correspond to flopping a corner, which contains an extremal point, i.e., such that the path remains an anti-Dyck path.

To exemplify this consider $SU(5)$ with fundamental and anti-symmetric representation, where the phases were obtained in \cite{Hayashi:2013lra}. As explained in section \ref{subsec:AFPhases}, the phases for the combined representation case are constructed out of the consistent phases for each representation, glued together to obey the flow rules and the diagonal condition. The resulting phases are shown in figure \ref{fig:SU5FlopsAF}: The dark blue and yellow boxes are the phases of the $SU(5)$ with anti-symmetric representation as in figure \ref{fig:SU5Flops}. Attached to it are the phase diagrams for the fundamental (light blue and cream-colored boxes), which are consistent with the flow rules and diagonal condition. In some cases, the diagonal condition allows both choices of signs such as in the case (11; III) and (11;IV). However for (4, III) the sign cannot be changed as it would result in the violation of the diagonal condition. 
In total there are 12 phases, and the flops are indicated by red dots in figure \ref{fig:SU5FlopsAF}. The phase diagram that follows from this is exactly the one obtained by direct computation in \cite{Hayashi:2013lra}, shown here in figure \ref{fig:SU5AFPhaseDiag}.


\subsection{Compatibility and Reducibility} \label{subsec:reducibility}

In section \ref{subsec:AFPhases} we have seen that for $\mathfrak{su}(n)$ we can study the phases of the theory with combined anti-symmetric and fundamental representations. This can be thought of as embedding  $\mathfrak{g}$ into $\widehat{\mathfrak{g}}$ in such a way
that there are several different intermediate subalgebras
$\mathfrak{g}\subset \widetilde{\mathfrak{g}}_j \subset \widehat{\mathfrak{g}}$,
with $\operatorname{rank}(\mathfrak{g})<\operatorname{rank}(\widetilde{\mathfrak{g}}_j)<\operatorname{rank}(\widehat{\mathfrak{g}})$.
Each $\mathfrak{g}\subset \widetilde{\mathfrak{g}}_j$ embedding will exhibit the 
codimension-two phenomenon that we have been studying, with phases associated to a matter representation. However, combining the phase information for two or more extended algebras typically generates further restrictions. In fact, some extremal 
generators of the phases of the intermediate enhancements to $\widetilde{\mathfrak{g}}_j$ cease to be extremal in the phase of the combined representation, and therefore become reducible.  
As we will see later, it also modifies the resolution of the singular fiber, and in fact will correspond to the splitting of codimension-two fibers that is observed along codimension-three singular loci. 

\begin{figure}
\centering
\includegraphics[width=9cm]{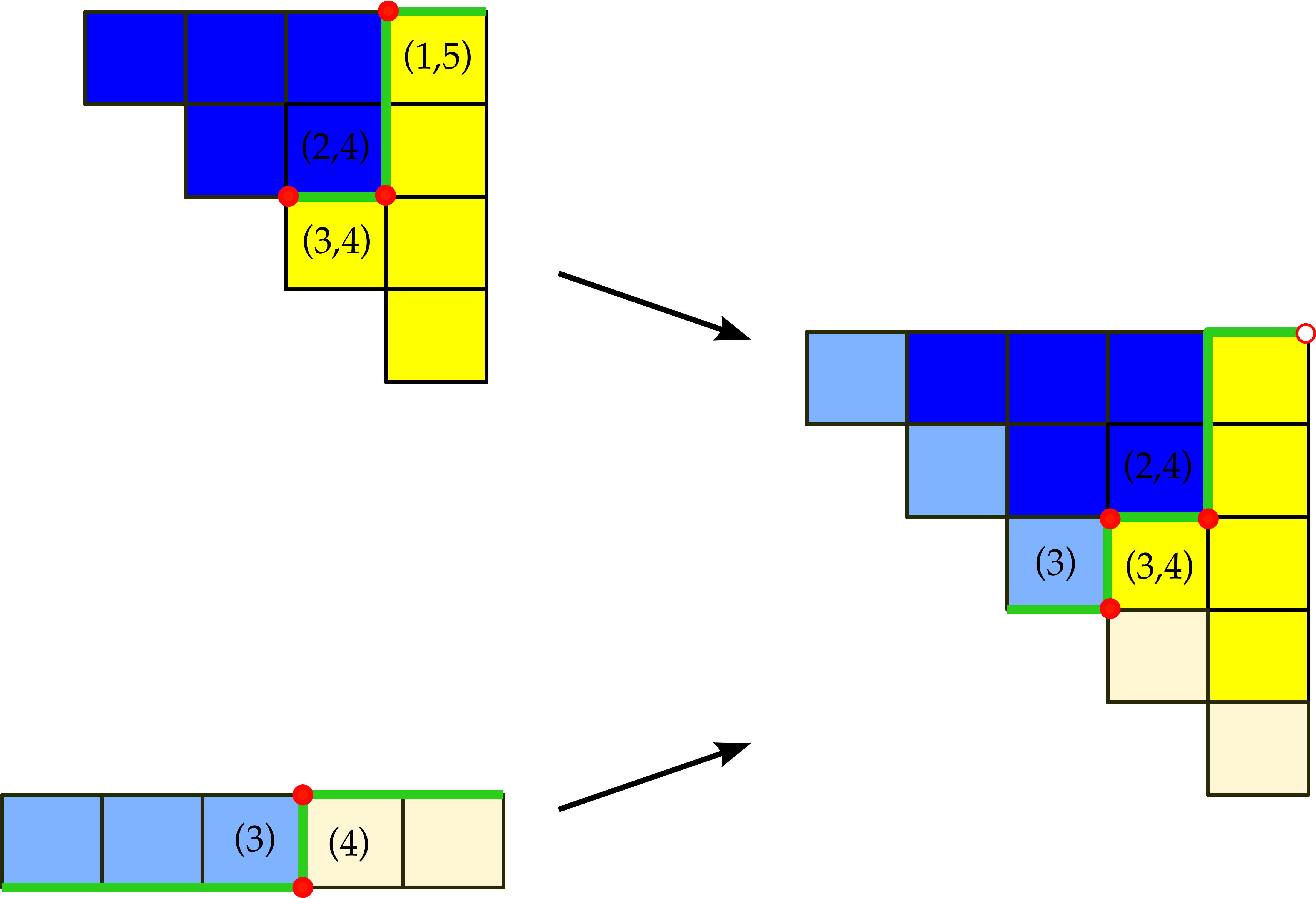}
\caption{$SU(5)$ with fundamental and anti-symmetric representation, by attaching a $(+++--)$ fundamental phase (which is phase II) to phase $8$ for the anti-symmetric representation. In each diagram the red nodes corresponds to the extremal points, and we labeled the corresponding extremal weights by $(i,j) = L_i + L_j$ and $(i) = L_i$. 
By joining the diagrams, to obtain the phase of the anti-symmetric and fundamental theory, $L_4$ and $L_1+L_5$ cease to be extremal and split. 
  }
\label{fig:SU5AFSplits}
\end{figure}

For example, consider $\mathfrak{su}(5)$ with both ${\bf 5}$ and ${\bf 10}$ matter, starting with phase 8 for $SU(5)$ with  ${\bf 10}$ and augmenting it with the ${\bf 5}$ in the phase II, which has signs $(+++--)$. 
In figure \ref{fig:SU5AFSplits} both matter phases are shown including the extremal points, corresponding to
$-(L_1+ L_5)$, $L_2+ L_4$ and $-(L_3+L_4)$ for ${\bf 10}$ matter, and $L_3$ and $-L_4$ for {\bf 5}. 

Joining the two diagrams to give the phase of the theory with both types of matter, the anti-Dyck paths simply join, however,  the  flops of $-L_4$ and $-(L_{1}+ L_5)$ are now disallowed, as they either correspond to flops that violate the flow rules or the diagonal condition. In fact, these weights become ``reducible" and can be expanded in terms of the extremal generators as follows
\be
\ba
-(L_1+ L_5)  	&\quad \rightarrow \quad  - (L_2 +L_4) - L_3 \cr
-L_4 		&\quad \rightarrow \quad  - (L_3+ L_4) +L_3 \,.
\ea
\ee
Let us remark in view of later discussions of the geometry that despite not using any information about  codimension-three singularities, these are exactly the splittings of the matter along the $E_6$ and $SO(12)$ codimension-three singular loci, which realize in the dual four-dimensional gauge theory obtained from an F-theory compatification the ${\bf 10} \times {\bf 10} \times {\bf 5}$ and ${\bf \bar{5}} \times {\bf {\bar{5}}}\times{\bf 10}$ couplings of matter. We will connect this to the fiber geometry in codimension three in sections \ref{subsec:FibCodim3} and \ref{subsec:E6Codim3}. In particular there we will see that the box graphs contain all the information about the possible codimension-three fiber types, and we will uncover several new non-Kodaira fiber types from them.


\section{Counting Phases of the $SU(n)$ Theory with Matter}
\label{sec:Count}

The box graphs and anti-Dyck paths also provide nice combinatorial way to 
count the phases of the $\mathfrak{su}(n)$ theories with matter. In fact, it 
turns out, it is easier to count the phases which violate the anti-Dyck 
property, and thus correspond to Dyck paths, and take the complement of 
these in the phases of the $\mathfrak{u}(n)$ theory, that was 
determined from the Weyl group quotient. 


\subsection{$SU(n)$ with antisymmetric representation}

We can determine the number of $\mathfrak{su}(n)$ phases with antisymmetric 
representation by counting the complementary phases which violate the sign 
conditions (\ref{SUevenSignCond}, \ref{SUoddSignCond}). Note that the number 
of phases with $\mathcal{E}_n = (+\cdots +)$ is the same as for 
$\mathcal{E}_n =(- \cdots -)$. By following the flow rules
(\ref{Flows1}) the number of phases for 
$\mathcal{E}_n = (+\cdots +)$, for instance, can be determined by a simple 
combinatorial argument. We consider the cases where $n$ is even and odd
separately. 


The total number of $\mathfrak{su}(n)$ phases for $n$ even is given by
\begin{equation}\label{SUevenCount}
    SU(n= 2k)\,, \hbox{ with  } \Lambda^2{\bf n}:\qquad  \# 
        \hbox{Phases} =  2^{2k-1} - 2 {2k-1 \choose k-1 } \,.
\end{equation}
To prove this, we consider induction in $n$ from $n$ to $n+2$. The induction 
starting point is easily shown to be correct as there is no phase for $SU(2)$.


\begin{figure}
    \centering
    \includegraphics[width=10cm]{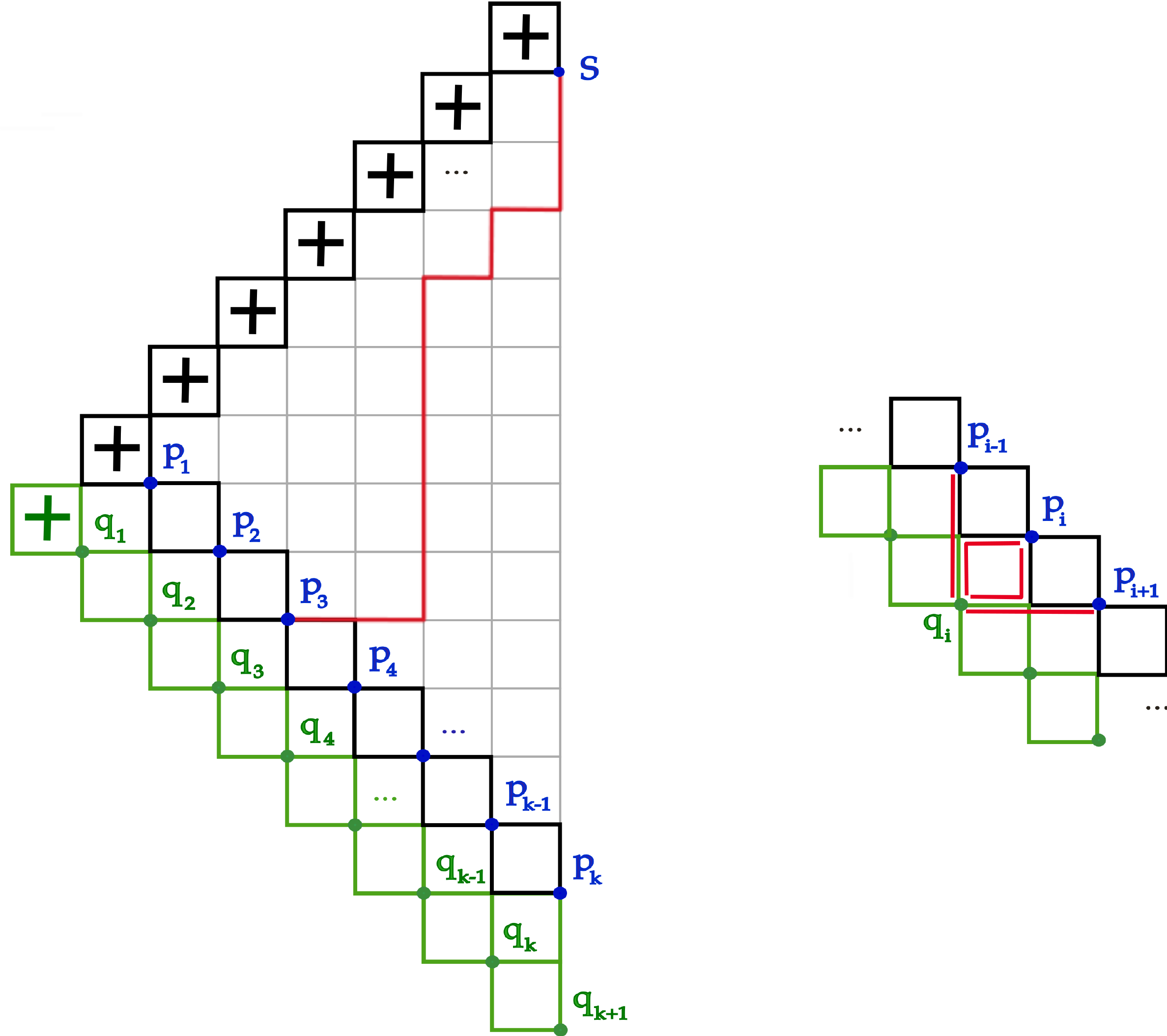}
    \caption{Induction step for $\mathfrak{su}(n)$ with $n=2k$ even. Each box in this 
    graph represents one of the weights in the lower diagonal of the weight 
    diagram of the anti-symmetric representation.}
    \label{fig:IndEven3}
\end{figure}



\begin{figure}
    \centering
    \includegraphics[width=8cm]{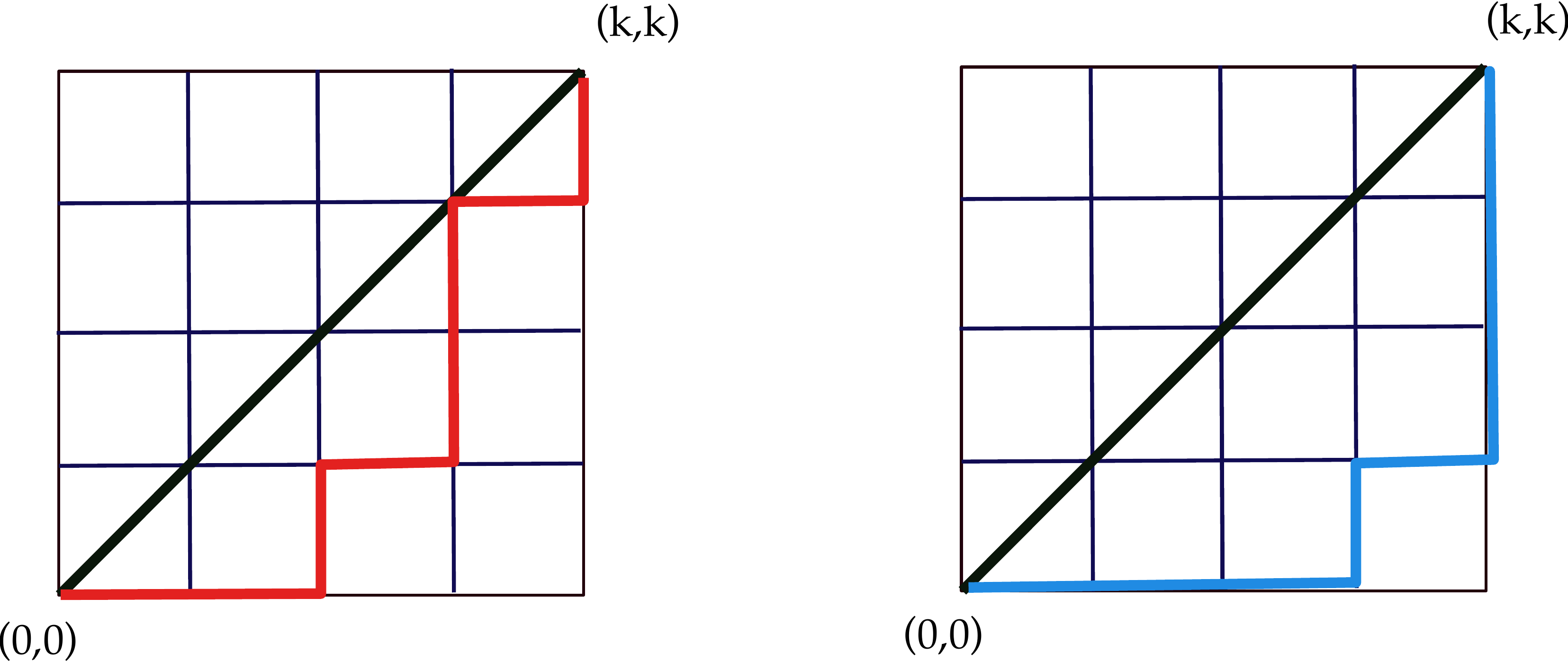}
    \caption{Example of two Dyck paths for $k= 4$. These are staircase paths, 
        starting at $(0,0)$ and ending at $(k,k)$ which do not cross, 
        but can touch, the diagonal.}
    \label{fig:Dyck}
\end{figure}

We know that $2^{n-1}$ is the total number of $\mathfrak{u}(n)$ phases as this is the
order of the quotiented Weyl group from section \ref{Sec:WeylG}. We count 
the complement by counting the number of phases, which do not respect the 
sign condition in (\ref{SUevenSignCond}). Without loss of generality 
consider the case with the diagonal $\mathcal{E}_n$ being all $+$ signs. 
For the induction step, we proceed as follows: each phase is characterized 
by a path, which separates the $+$ from the $-$ weights in the lower half of 
the triangle, as is depicted in figure \ref{fig:IndEven3}. These are paths 
that start at the point $s$ and end at some $p_i$. Let $a_{n,k}$ be the 
number of paths from $s$ to $p_k$ for $\mathfrak{su}(n)$, $n$ even. 
Then for all but $a_{n+2, 1}$ and $a_{n+2, n}$, we observe
\begin{equation}
    a_{n+2, k} = a_{n, k-1} + 2 a_{n, k} + a_{n, k+1} \,,
\end{equation}
which is easily seen by connecting a path to $p_i$ with a path to $q_i$ in
figure \ref{fig:IndEven3}. The two outlier cases are 
\begin{equation}
    a_{n+2, 1}  = 2 a_{n, 1} + a_{n, 2} \,,\qquad 
    a_{n+2, n}  =  2 a_{n, n}  + a_{n, n-1 } \,,\qquad
    a_{n+2,n+2} = a_{n,n} = 1  \,.
\end{equation}
In particular, the total number of paths 
\begin{equation}
    \# \hbox{paths from $s$ to any of the $p_k$}= b_n 
            = \sum_{k=1}^{n} a_{n,k}  \,,
\end{equation}
satisfies the recursion
\begin{equation}\label{bnan}
    b_{n+2} = 4b_n - a_{n,1} \,.
\end{equation}
This is seen by noting that every path ending in a point $p_k$ induces 4 
paths that end at one of the points $q_i$, except for the first one, $p_1$, 
which only induces $3$ paths. Thus we need to figure out the number of paths 
that go from $s$ to $p_1$. Happily this is related to the problem of counting 
so-called Dyck paths, from $(0,0)$ to $(n/2,n/2)$, which are staircase paths 
that do not cross the diagonal, but are allowed to touch it. Two examples of Dyck paths 
are shown in figure \ref{fig:Dyck}. These are counted by the Catalan numbers
\cite{fibonacci-lattice}
\begin{equation}\label{Cat}
    \# \hbox{Dyck paths from $(0,0)$ to $(k, k)$} = C_{k} 
        =  {(2k)! \over k! \left(k+1\right)!} \,.
\end{equation}
We can now prove that 
\begin{equation}
    b_n = {n-1 \choose {n\over 2} -1 } \,.
\end{equation}
The induction starting point is $b_2 =1$.  The induction step is
\begin{equation}
    4 b_{n} - a_{n,1} =  4 {n-1 \choose {n\over 2} -1}  
        -  {n! \over \left({n\over 2}\right)! \left({n\over 2}+1\right)!}
        = {(n+1)!\over \left({n\over 2}\right)! \left({n\over 2}+1\right)!} 
        = b_{n+2} \,.
\end{equation}
Applying the same argument for the case when the diagonal is all $-$, and 
subtracting these from the number of total $\mathfrak{u}(n)$ phases yields 
(\ref{SUevenCount}). 


\begin{figure}
    \centering
    \includegraphics[width=5cm]{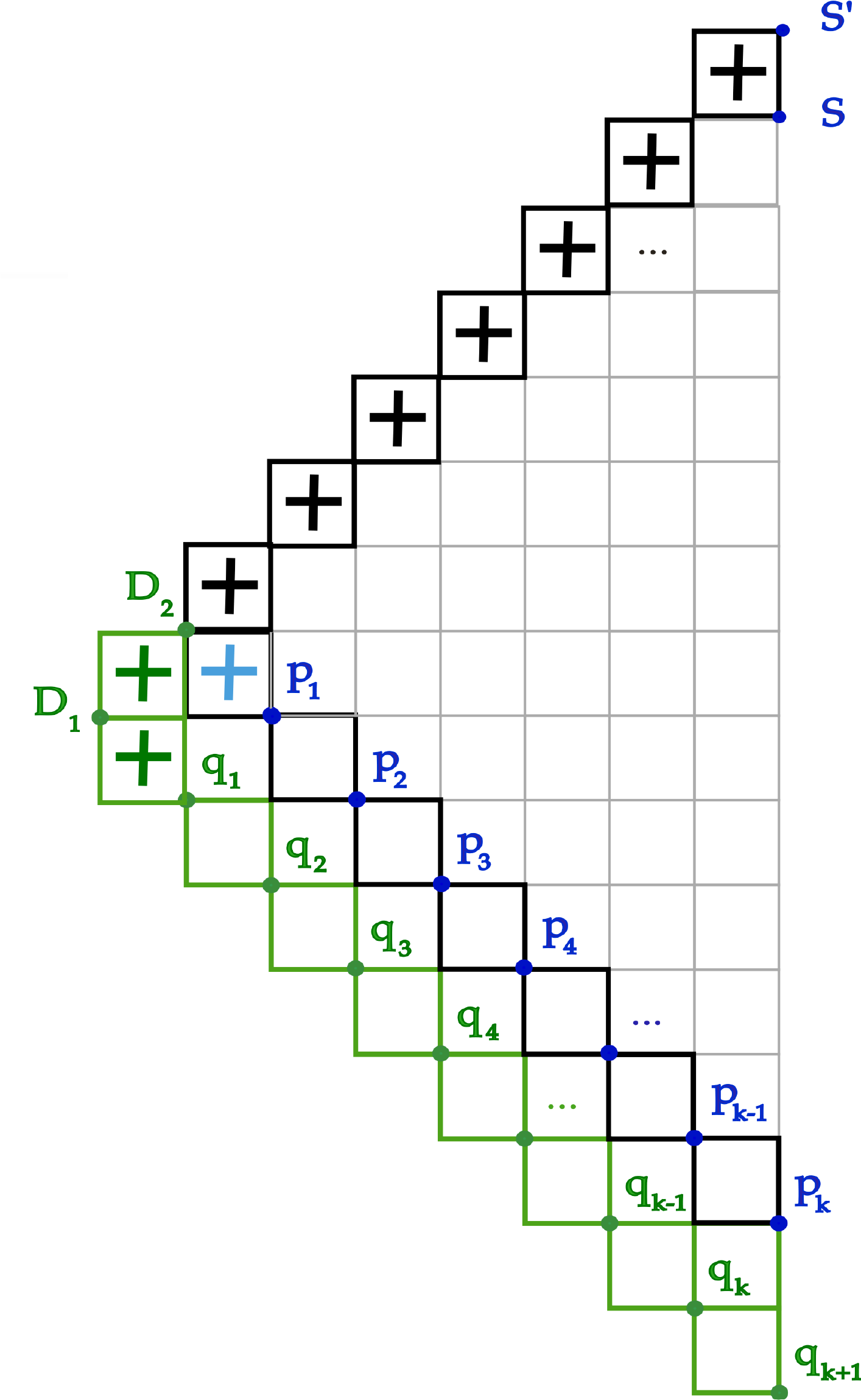}
    \caption{Induction for $n=2k+1$. The condition that 
        $\mathcal{E}_{2k+1} =(  + \cdots +)$ is given by the black $+$ and the 
        blue $+$. In the induction step, indicated by the additional green 
        boxes, the blue $+$ is not required  by the condition 
        $\mathcal{E}_{2k+3} =  (+ \cdots +)$.}
    \label{fig:SUoddInduction}
\end{figure}

For $n$ odd, the number of phases is
\begin{equation}
    \label{SUoddCount}
    SU(n= 2k+1)\,, \hbox{ with  } \Lambda^2{\bf n}:\qquad  
        \# \hbox{Phases} =  2^{2k} - 2 {2k \choose k -1} 
    \,.
\end{equation}
To prove this, again consider the non-$\mathfrak{su}(n)$ phases, which violate 
(\ref{SUoddSignCond}). First consider again the case with $\mathcal{E}_{2k+1}$ 
all $+$. The sign assignment is given in figure \ref{fig:SUoddInduction}. 
Again we count the paths from $s$ to $p_i$, $i=1, \cdots, n$. In this case,
however, there is a subtlety: starting with $n$ and passing to $n+2$, we obtain 
the extension of the diagram as shown in figure \ref{fig:SUoddInduction}. 
Most paths in the diagram for $\mathfrak{su}(n)$ will again give rise to paths for 
$n+2$, however, the blue $+$ sign, does not have to be $+$ in the case of 
$SU(n+2)$. Thus, we need to count the paths, which go to the point $p_1$ 
twice, as both sign choices are allowed in the $SU(n+2)$ diagram.

The recursion relation for the 
\begin{equation}
    \# \hbox{paths from $s$ to any of the $p_k$} =  b_n \,,
\end{equation}
is, again using the Catalan numbers (\ref{Cat}),
\begin{equation}
    b_{n+2} = 4 b_n - \left(C\left({n+1\over 2}\right) - 
        C\left({n-1\over 2}\right) \right)  + C\left({n-1\over 2}\right) \,.
\end{equation}
The last term is precisely the contribution that counts the number of 
paths that account for the sign choice one has in $SU(n+2)$ given by the 
blue $+$ in figure \ref{fig:SUoddInduction}. The terms that are subtracted 
correspond to the contribution $a_{n,1}$, which, as in (\ref{bnan}), has to be 
subtracted. Note that $a_{n,1}$ can be computed by observing that it is 
precisely the Dyck paths between $D_1-s'$ and $D_2 -s'$, which are given in 
terms of the two Catalan numbers
\begin{equation}
    a_{n,1} =  \left(C\left({n+1\over 2}\right) - 
        C\left({n-1\over 2}\right) \right)  \,.
\end{equation}
Again, applying the same type of argument to count the number of phases with 
all $-$ signs in the constraint (\ref{SUoddSignCond}), which yields the same 
number, and subtracting these from the total number of $\mathfrak{u}(n)$ phases results 
in (\ref{SUoddCount}).

The total number of phases for the $\mathfrak{u}(n)$ and $\mathfrak{su}(n)$ theories for some small 
values of $N$ are collected in table \ref{table:PhaseCounting}.


\begin{table}
    \begin{equation*}
        \begin{array}{c|c|c}
            \hbox{$n$} & \hbox{Phases of $\mathfrak{u}(n)$ with $\Lambda^2{\bf n }$} & 
                \hbox{Phases of $\mathfrak{su}(n)$ with $\Lambda^2{\bf n }$} \cr\hline
            5    &   16  &    8\cr
            6     &  32    &  12\cr
            7      & 64      &34\cr
            8      & 128    & 58\cr
            9      & 256   &  144\cr
            10     & 512   &  260\cr
            11      &1024  &  604\cr
            12     & 2048  &  1124\cr
            13     & 4096   & 2512\cr
            14     & 8192   & 4760\cr
            15     & 16384  & 10378\cr
        \end{array}
    \end{equation*}
    \caption{The number of phases for small values of $n$ for $\mathfrak{u}(n)$ and $\mathfrak{su}(n)$.}
    \label{table:PhaseCounting}
\end{table}


\subsection{$SU(n)$ with antisymmetric and fundamental representation}

For the phases of the $\mathfrak{su}(n)$ with fundamental and antisymmetric, we claim 
the following counting: for $n = 2k+1$ odd, the number of phases is 
\begin{equation}
    \begin{aligned}
        \label{CountSUAF}
        SU(n= 2k+1)\,, \hbox{ with  } \Lambda^2{\bf n} \hbox{ and }  {\bf n} : 
        \qquad \# \hbox{Phases }&=   2^{k+1}- 2 {2k+1 \choose k }\cr &= 
        \# \hbox{Phases } SU(2k+2) \,, \hbox{ with }\Lambda^{2} {\bf (2k+2)}\,.
    \end{aligned}
\end{equation}

From the combined diagrams in section \ref{subsec:AFPhases}, figure
\ref{fig:SUAFBox}, we obtain the counting for phases with both representations
\begin{equation}\label{AFCount}
    \begin{aligned}
        &SU(n= 2k+1)\,, \hbox{ with  } \Lambda^2{\bf (2k+1)} \hbox{ and }  {\bf (2k+1)} : \qquad \cr
        &\qquad \qquad \# \hbox{Phases }=  2 \left(\# \hbox{Phases } SU(2k+1)  \hbox{ with } \Lambda^2 ({\bf 2k+1})\right) - 2 C(k)\,,
    \end{aligned}
\end{equation}
where $C(k)$ is again the Catalan number, and this expression agrees 
straightforwardly with (\ref{CountSUAF}). 

\begin{figure}
\centering
\includegraphics[width=3.5cm]{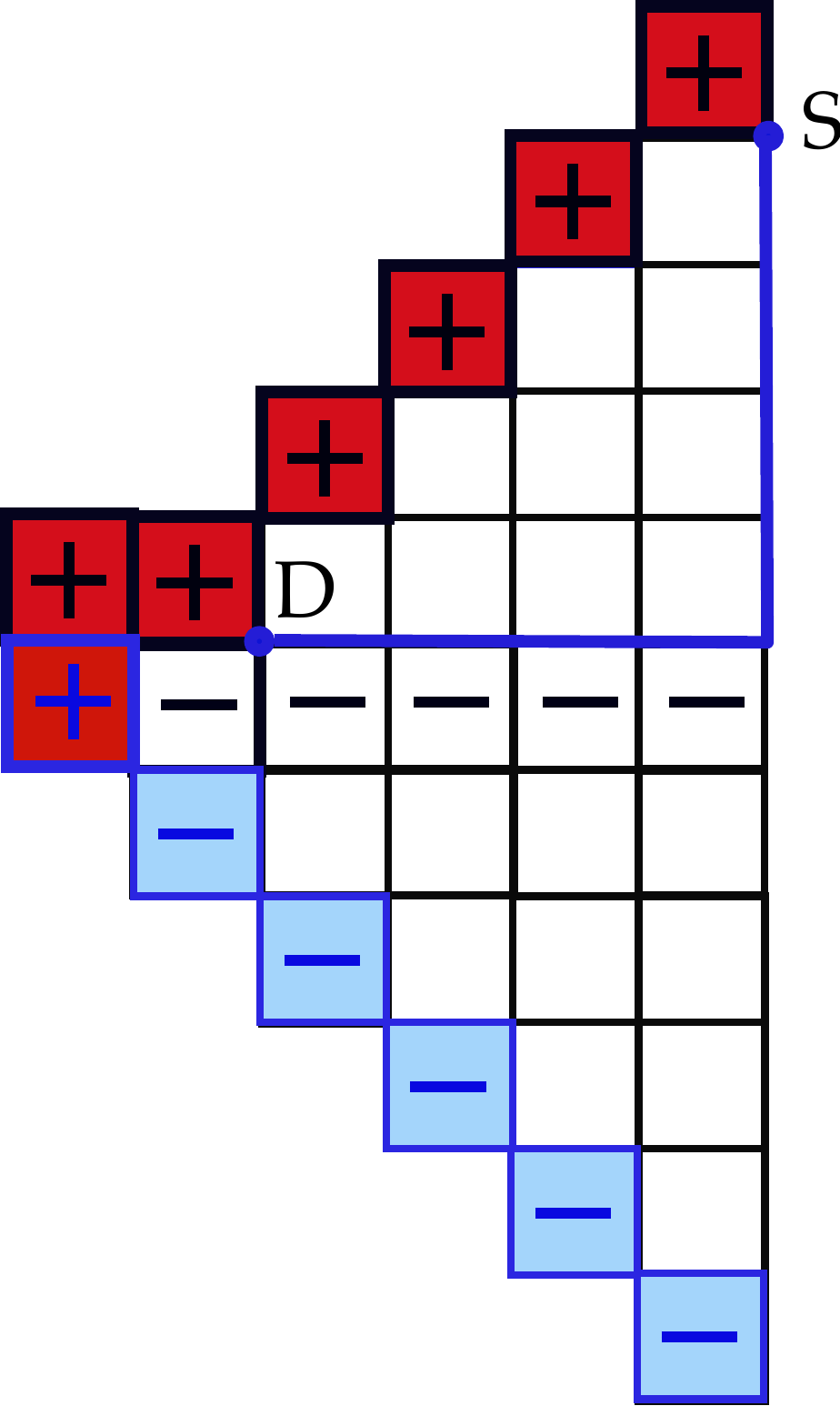}
\caption{Counting of phases for the combined antisymmetric and fundamental representation. The relevant box graph is shown in figure \ref{fig:SUAFBox}. Here we depict the counting of the phases, which are consistent phases of $SU(2k+1)$ with anti-symmetric representation and with a fundamental representation, however, the resulting combined diagram has $\mathcal{E}_{2k+2}= (+ \cdots +)$ and is therefore not a consistent $SU(2k+1)$ phase. The number of such phases is counted by the Dyck paths from $S$ to $D$. }
\label{fig:SUAFCount}
\end{figure}

To prove this counting formula, note that the NW to SE diagonal $L_i+ L_{i+1}$ 
for each consistent sign assignment for the antisymmetric representation has 
a sign change over from $+$ to $-$. The only consistent way to extend this 
with the fundamental representation is if the sign change over is matched. 
There is generically the choice of two signs for the fundamental given in 
terms of $L_i + L_i$ to attach to the $L_i+ L_{i+1}$ diagonal. This 
explains the first term in (\ref{AFCount}). However, this still counts phases, 
which violate the sign condition (\ref{SUevenSignCond}). The number of these 
is easily seen to be equal (via the, by now standard, map to Dyck paths) to the
paths given on the RHS in figure \ref{fig:SUAFCount}, which are precisely the 
Catalan numbers $C(k)$. Likewise we need to subtract the cases where
$\mathcal{E}_{2k+1} = (- - - \cdots - - + -), \epsilon_{k, k+2} = -$ and the 
fundamental representation is attached with $\epsilon_{k+1, k+1} = -$. The 
number of those cases can be also counted by $C(k)$, which explains the 
multiplicity $2$ in front of $C(k)$ in \eqref{AFCount}.

For $SU(2k)$ with fundamental and antisymmetric representation the 
construction of the consistent signs from the phases is the same as in the 
odd case. However, this time due to the flow rules, any consistent diagram 
for $SU(2k)$ with anti-symmetric representation satisfying the sign 
constraint (\ref{SUevenSignCond}), joined with the fundamental representation, 
gives rise to a consistent $SU(2k+1)$ diagram that satisfies the sign 
condition (\ref{SUoddSignCond}). So we arrive at
\begin{equation}
    \begin{aligned}
        &SU(n= 2k)\,, \hbox{ with  } \Lambda^2{\bf (2k)} \hbox{ and }  {\bf (2k)} : \qquad \cr
        &\qquad \qquad \# \hbox{Phases }=  2 \left(\# \hbox{Phases } SU(2k)  \hbox{ with } \Lambda^2 ({\bf 2k})\right) \,.
    \end{aligned}
\end{equation}


\newpage

\section{Phases with non-trivial Monodromy}
\label{sec:OtherPhases}

Monodromies in singular elliptic fibrations occur when there is an additional discrete group, usually some outer automorphism, that is acting on the fiber components. Likewise for phases there exists a similar notion, which occurs when the commutant $\mathfrak{g}_{\perp}$ of $\mathfrak{g}$ in $\widetilde{\mathfrak{g}}$ is non-abelian. So far we considered the case when the commutant is $\mathfrak{u}(1)$ only. This leads to different phase structures, depending on whether the Weyl group of $\mathfrak{g}_{\perp}$ acts trivially or not. We discuss this in the case of $\mathfrak{su}(6) \oplus \mathfrak{su}(2) \subset \mathfrak{e}_6$. Another instance of monodromy occurs when there is an outer automorphism which acts to reduce the gauge group, for example, the outer automorphism reducing $\mathfrak{su}(2n)$ to $\mathfrak{sp}(n)$. 
\subsection{Monodromy}

So far the commutant of the gauge group inside the higher rank group was
assumed to be $U(1)$, as in (\ref{gtildeg}). 
If the commutant is a non-abelian Lie group, e.g. $SU(2)$ for $SU(6)$ inside
$E_6$, then there is an additional group acting on the phases, which we will
refer to as {\it monodromy} from the action of the non-trivial Weyl group of the commutant.  In general, the decomposition of the Lie algebras is 
\begin{equation}
    \widetilde{\mathfrak{g}} \supset \mathfrak{g} \oplus \mathfrak{g}_{\perp} \,,
\end{equation}
where $\mathfrak{g}_{\perp}$ is a rank one non-abelian Lie algebra\footnote{Note it can be higher rank, however we will mainly consider the case of $\mathfrak{su}(2)$. }. The existence of a non-abelian commutant $\mathfrak{g}_{\perp}$ 
results in two differences compared to the no monodromy cases discussed so far. One is that we have 
a Weyl group associated with the root system of $\mathfrak{g}_{\perp}$. 
Hence we need to take into account the action of the Weyl group when we 
consider the phase structure of the theory of the gauge algebra 
$\mathfrak{g}$. In fact, the phases of such a theory will need to be 
invariant under the action of the Weyl group. The other is that we have 
singlets under $\mathfrak{g}$, which are roots of $\mathfrak{g}_{\perp}$. The presence of the singlets means that they are neutral massless chiral multiplets even in the bulk of the Coulomb branch. We cannot assign a definite sign for the singlets. This can be remedied if a $\mathfrak{u}(1) \subset \mathfrak{g}_{\perp}$ remains unbroken. The $\mathfrak{u}(1)$ symmetry gives a charge to the singlets, and they have a definite sign in the bulk of the Coulomb branch.
These two differences have clear interpretations on the geometry side, which we will 
see later in section \ref{sec:GlobalFib}.  An  example for this is discussed in the next subsection. 

Another instance of monodromy occurs when the gauge group arises from a
quotient of a simply-laced Lie group, for which the phases can be obtained as
invariant phases under the quotienting. Again the presence of zero weights
prevents the existence of a Coulomb branch, as there are additional massless
modes. An example of this is $Sp(n)$ obtained as a $\mathbb{Z}_2$ quotient of
$SU(2n)$, which we will discuss later in this section. 


\subsection{$SU(6)$ with the $\Lambda^3{\bf 6}$ representation}
    \label{Sec:Su6L3}

\begin{figure}
    \centering
    \includegraphics[width=8cm]{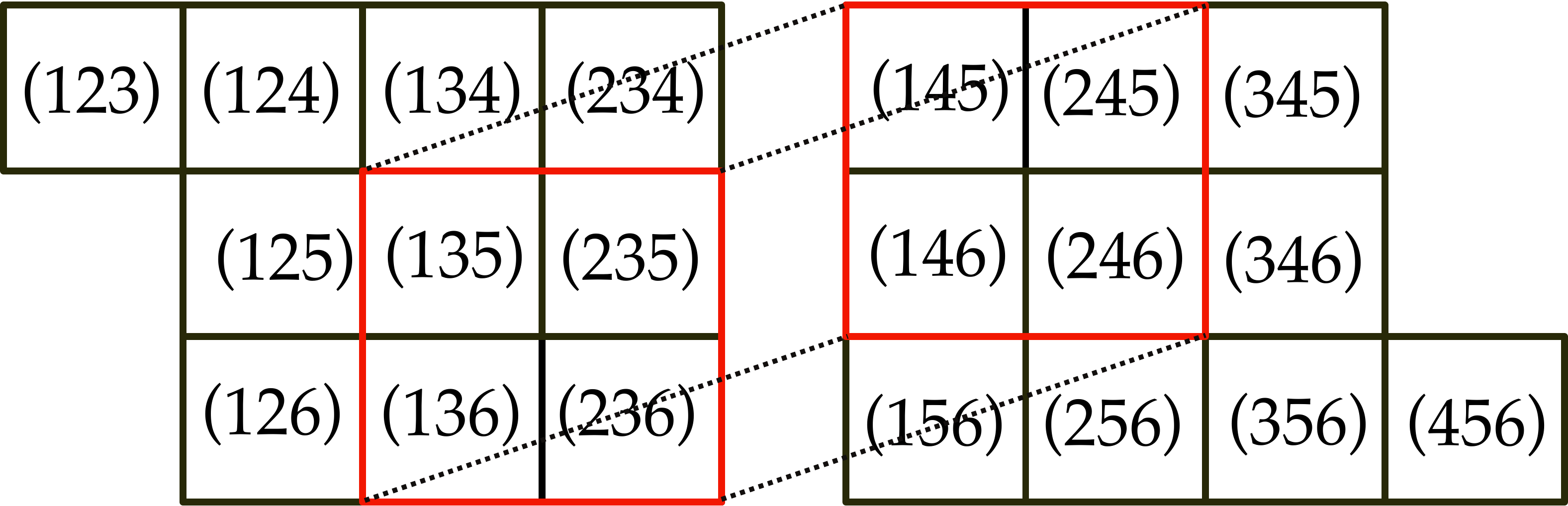}\qquad 
    \includegraphics[width=6cm]{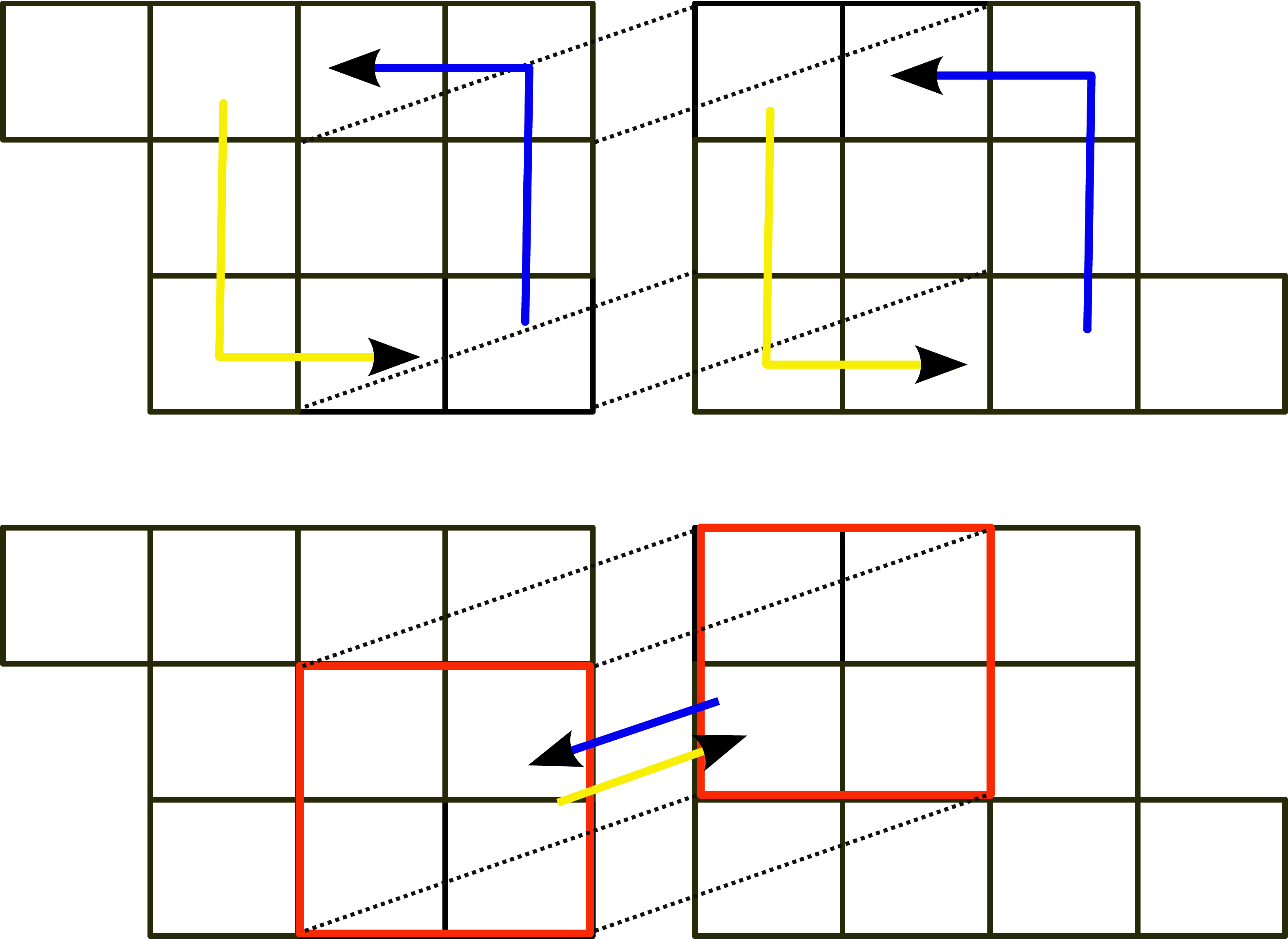}
    \caption{Representation graph for  $\Lambda^3{\bf 6}$ of $SU(6)$. The 
    weights in each box are $(ijk)= L_{i}+ L_j+ L_k$, and they are aligned 
    so that the action of the simple roots corresponds to the edges of the 
    diagram. The dotted lines indicate the action of the simple root $L_4-L_3$, and 
    the representation graph is really three-dimensional, with the two red boxes 
    being on top of each other. The right hand side shows the flow rules, 
    with $+$ corresponding to blue and $-$ to yellow.}
    \label{fig:SU6Lambda3FlowRules}
\end{figure}

We consider an exceptional example of an $SU(6)$ gauge theory with matter 
fields in the $\Lambda^3{\bf 6}={\bf 20}$ representation. This theory arises 
from the embedding of $SU(6)$ into $E_6$. The decomposition of the adjoint 
representation of $E_6$ is as follows,
\begin{equation}
    \begin{aligned}
        \mathfrak{e}_6 \quad & \rightarrow \quad \mathfrak{su}(2) 
            \oplus \mathfrak{su}(6) \cr
        {\bf 78} \quad & \rightarrow \quad ({\bf 3}, {\bf 1}) \oplus 
            ({\bf 1}, {\bf 35}) \oplus ({\bf 2}, {\bf 20}) \,. 
        \label{decompE6}
    \end{aligned}
\end{equation}
The weights of the representation ${\bf 20}$ can be written as 
$L_{i}+ L_j+ L_k$ with $i < j < k$. The representation graph is shown in 
figure \ref{fig:SU6Lambda3FlowRules}. The phases are governed again by flow 
rules, which are also shown in figure \ref{fig:SU6Lambda3FlowRules}. 

In this example, $\mathfrak{g}$ and $\mathfrak{g}_{\perp}$ are $\mathfrak{su}(6)$ 
and $\mathfrak{su}(2)$ respectively. 
One can easily understand this decomposition from the roots of $\mathfrak{e}_6$. One 
useful way to construct the roots of $\mathfrak{e}_6$ is to make use of two-cycles in 
the del Pezzo  surface $dP_6$\footnote{For further details on this construction we refer the reader to the appendix \ref{app:Group}.}. Let $L_0, L_1, \cdots, L_6$ be the bases of 
a seven-dimensional vector space and we introduce a bilinear form 
$\text{diag}(-1, 1, \cdots, 1)$
\footnote{The bilinear form here is in fact the negative of the standard 
    bilinear form on $H_2(dP_6, \mathbb{R})$. Then the inner product of 
    two-cycles in $H_2(dP_6, \mathbb{R})$ gives the same sign as the one 
    from the pairing $\langle \cdot, \cdot \rangle$ introduced in the root 
    space.}
. Then the root space of $\mathfrak{e}_6$ can be identified with the orthogonal 
complement of $k = -3L_0 + L_1 + \cdots + L_6$. We can choose six 
independent bases in the orthogonal complement as
\begin{equation}
    L_i - L_{i+1} \quad (i=1,\cdots, 5), \qquad L_0 - (L_1 + L_2 + L_3) \,,
\end{equation}
which are the canonical simple roots of $\mathfrak{e}_6$. The roots of $\mathfrak{e}_6$ are then
\begin{equation}
    L_i - L_j, \qquad \pm\left(L_0 - (L_i+L_j + L_k)\right), 
        \qquad\pm(2L_0 - (L_1 + \cdots + L_6)) \,,
\end{equation}
where $1 \leq i \neq j \neq k \leq 6$. 

The embedding of $\mathfrak{su}(6)$ into $\mathfrak{e}_6$ may be understood by identifying 
$L_i - L_{i+1}, (i = 1, \cdots, 5)$ with the simple roots of $\mathfrak{su}(6)$. Then 
the simple root of the $\mathfrak{su}(2)$ in \eqref{decompE6} is 
$2L_0 - (L_1 + \cdots + L_6)$. The weights of the $\Lambda^3{\bf 6}$ 
representation are 
\begin{equation}
    L_0 - (L_i + L_j + L_k), \qquad - \left(L_0 - (L_l + L_m + L_m)\right) \,,
    \label{lambda6}
\end{equation}
where the two $\Lambda^3{\bf 6}$'s in \eqref{lambda6} transform as a doublet of 
the $\mathfrak{su}(2)$ when $\{i, j, k, l, m, n\}$ is a permutation of 
$\{1, 2, 3, 4, 5, 6\}$.


\begin{figure}
    \centering
    \includegraphics[width=14cm]{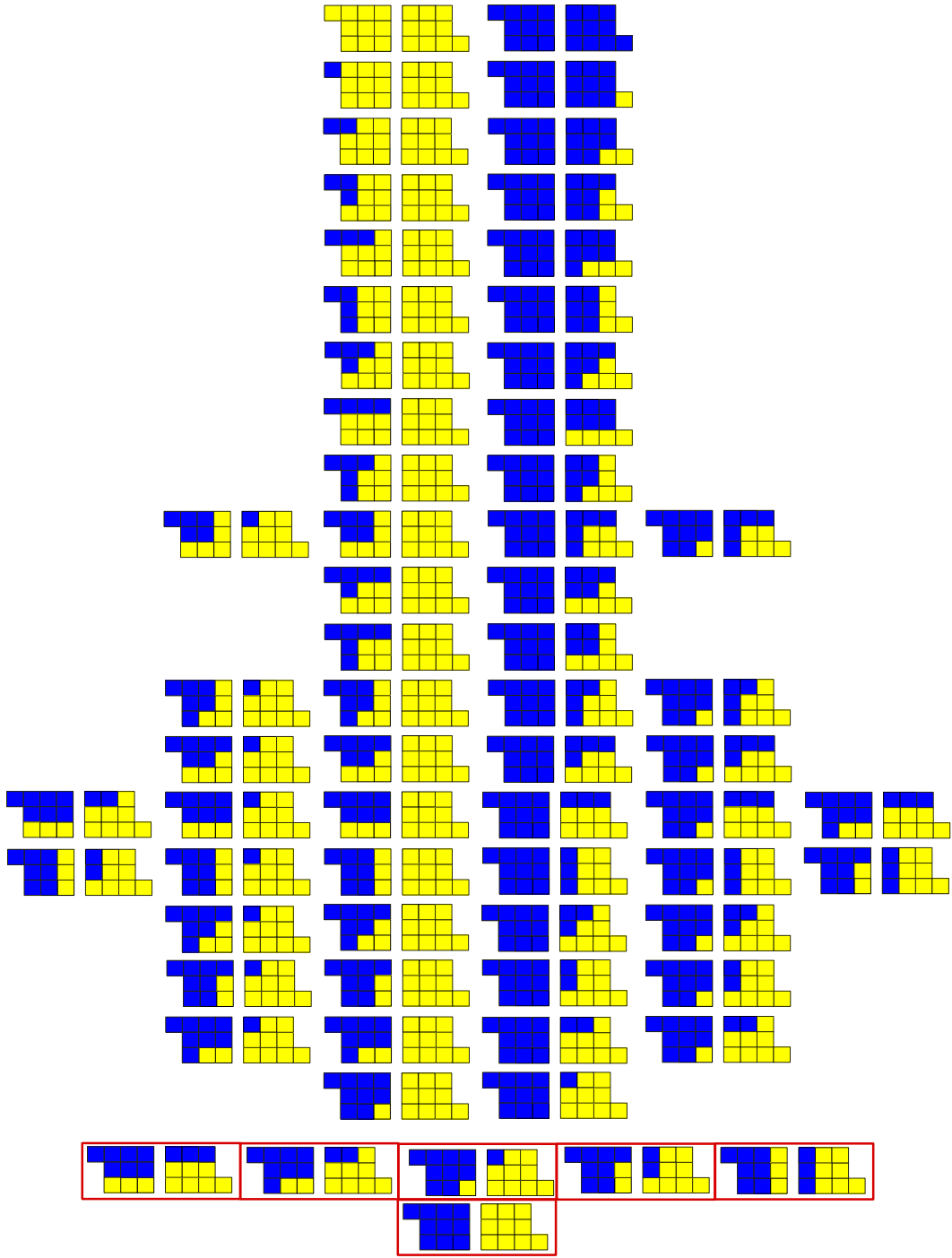}
    \caption{Phases of the $SU(6)\times SU(2)$ theory with  $\Lambda^3{\bf 6}$ of $SU(6)$ 
    and singlets $\pm \sum_{i=1}^6 L_i$.  The boxed phases are the $SU(6)$ 
    phases for which $\sum_{i=1}^6L_i$ is not fixed from the other weights, 
    and thus can be either positive or negative, i.e. these diagrams appear twice in the phases. There are 72 box graphs, of which 60 are not invariant under the $\mathbb{Z}_2$ and are mapped into each other. This correspond to reflection along the central axis in the above diagram. }
    \label{fig:U6Lambda3}
\end{figure}
 


\begin{figure}
    \centering
    \includegraphics[width=8cm]{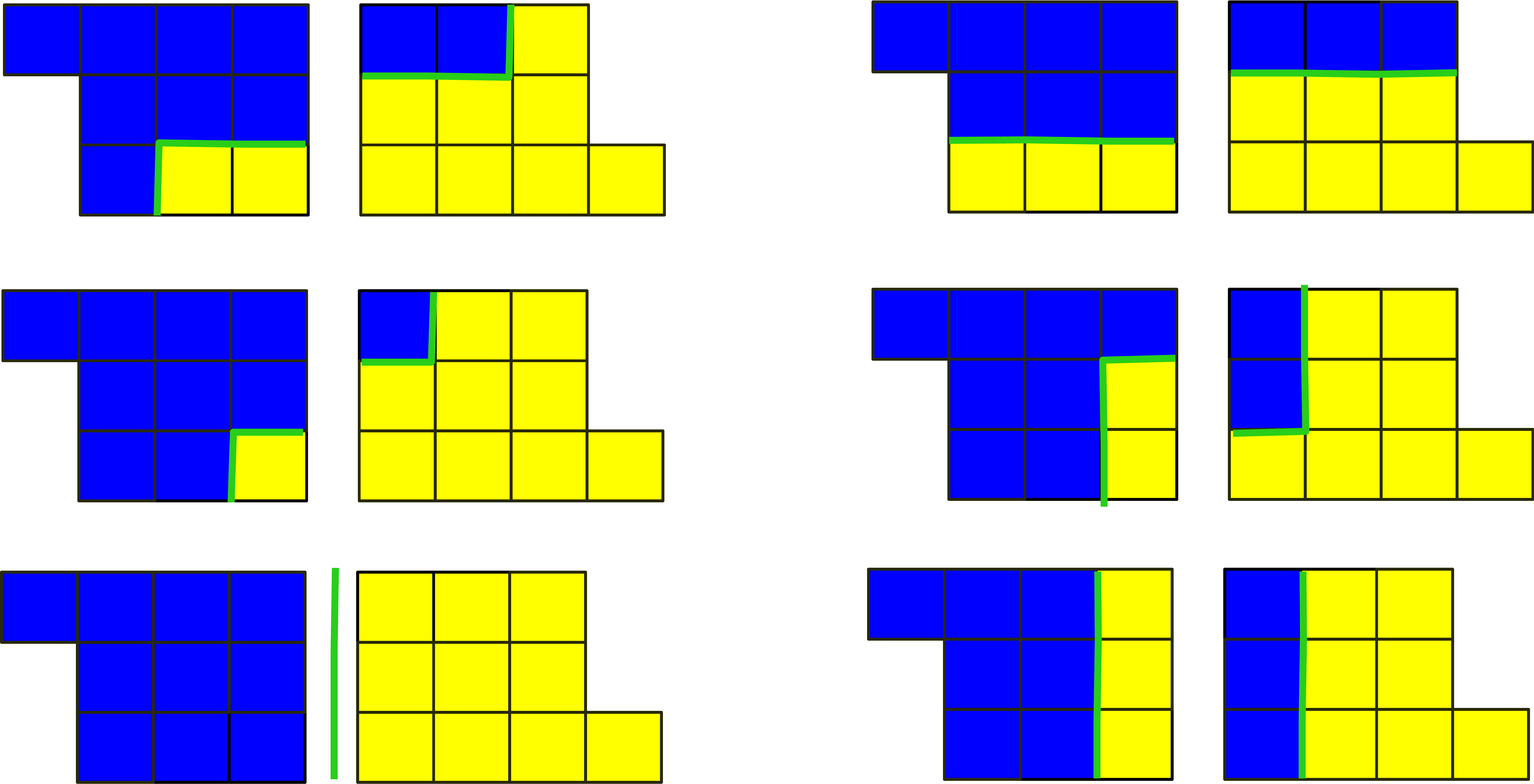}
    \caption{Phases of the $SU(6)$ theory with  $\Lambda^3{\bf 6}$ of 
    $SU(6)$, i.e., these satisfy in addition to the flow rules that the sum of 
    all $L_i$ vanishes. In particular these are also the diagrams which are invariant under the $\mathbb{Z}_2$ automorphism. The green lines show the corresponding anti-Dyck paths. }
    \label{fig:SU6Lambda3}
\end{figure}
 

Let us first discuss the phase of a gauge theory associated with the embedding \eqref{decompE6}. In order to define a standard phase, we consider an $SU(6) \times U(1)$ gauge symmetry.  Then, the matter content of the theory is ${\bf 20}_{3} + {\bf 20}_{-3}$ and ${\bf 1}_{6} + {\bf 1}_{-6}$ where the 
subscript denotes the charge of the overall $\mathfrak{u}(1)$. The number of  phases can be determined 
from the Weyl group quotient
\begin{equation}
    \frac{|W_{\mathfrak{e}_6}|}{|W_{\mathfrak{su}(6)}|} = 72. 
\end{equation}
The $72$ phases in terms of the box graphs are depicted in figure 
\ref{fig:U6Lambda3}. 
The right half of figure \ref{fig:U6Lambda3} corresponds 
to phases where $\sum_{i=1}^6 L_i > 0$ and the left half to 
$\sum_{i=1}^6 L_i < 0$. One can clearly see a symmetry associated with the 
Weyl reflection from figure \ref{fig:U6Lambda3}. The Weyl reflection 
$\mathbb{Z}_2$ of the $SU(2)$ changes the weight $L_i + L_j + L_k$ into 
$-(L_l + L_m+L_n)$ where $\{i, j, k, l, m, n\}$ is a permutation of 
$\{1, 2, 3, 4, 5, 6\}$. There are $30$ pairs of phases, which are related 
by this transformation. The remaining $6$ pairs, which are surrounded by a 
red square (and are shown only once as they are invariant), are related by the $\mathbb{Z}_2$ transformation, but it 
does not change the signs of any weight $L_i + L_j + L_k$.  

\begin{figure}
    \centering
    \includegraphics[width=2.6cm]{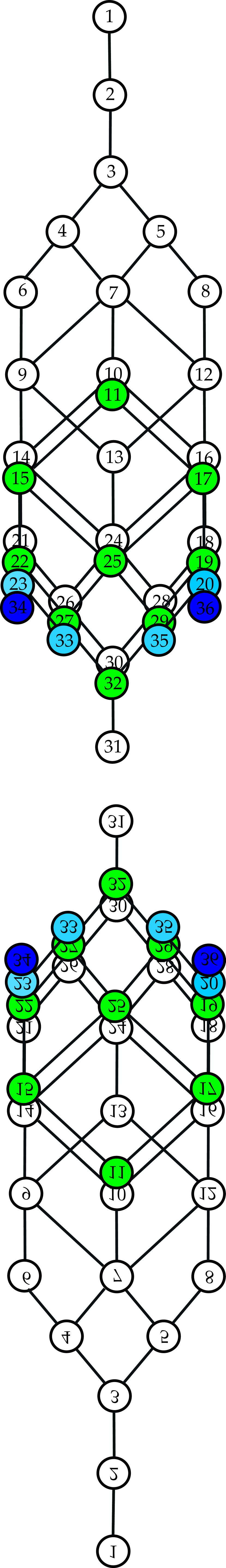}
    \caption{Flop diagram of the $SU(6) \times SU(2)$ theory with  $\Lambda^3{\bf 6}$ of 
    $SU(6)$. The top part depicts the flop diagram for the right half of figure \ref{fig:U6Lambda3} 
    as well as the 6 phases which are labeled by 31-36, which correspond to the invariant diagrams at 
    the bottom of figure \ref{fig:U6Lambda3}. The  mirrored numbers correspond to the left hand 36 box graphs. 
    The $\mathbb{Z}_2$ automorphism acts by reflection along the horizontal axis. Nodes which overlap are connected, and each color indicates one layer.  This flop graph is exactly the quasi-minuscule representation of $E_6$ with the zero weights removed.  }
    \label{fig:SU6E6Flops}
\end{figure}

The flop transitions among these are shown in figure 
\ref{fig:SU6E6Flops}. Note that the flop diagram is exactly the quasi-minuscule representation of $E_6$ except for 
the zero weights.

Let us move on to the case of the reduction to $\mathfrak{su}(6)$. This can be
achieved by considering the $\mathbb{Z}_2$ Weyl group action associated with
the root space of $\mathfrak{su}(2)$.  Then, the $\mathfrak{u}(1)$ Cartan of
$\mathfrak{su}(2)$ as well as the simple root map to minus of themselves.
Hence, they do not appear in the $\mathbb{Z}_2$ invariant theory. Furthermore,
the two $\Lambda^3 {\bf 6}$'s are identified. Putting it altogether, 
we consider phases of an $SU(6)$ theory with the $\Lambda^3{\bf 6}$ 
representation. To determine the phases for $SU(6)$ with $\Lambda^3 {\bf 6}$, 
in addition to the flow rules we need to impose consistency with the 
tracelessness condition $L_1 + L_2 + L_3 + L_4 + L_5 + L_6 = 0$. Note that 
the tracelessness condition implies that the $SU(6)$ phases should be the 
$\mathbb{Z}_2$ invariant phases. If a phase is not $\mathbb{Z}_2$ invariant, 
then it means that we have some weights $L_i + L_j + L_k > 0$ and 
$L_l + L_m + L_n > 0$ in the phase where $\{i, j, k, l, m, n\}$ is a 
permutation of $\{1, 2, 3, 4, 5, 6\}$, which contradicts the tracelessness 
condition. Therefore, the consistent sign assignments for the 
$\Lambda^3 {\bf 6}$ representations are those specified by the box graphs 
that are enclosed in a red rectangle in figure \ref{fig:U6Lambda3}. The number of the $SU(6)$ phases is then
\begin{equation}
    \frac{\text{$\#$ of $\mathbb{Z}_2$ invariant phases}}
        {|W_{\mathfrak{su}(2)}|} = 6.
\end{equation}
Those box graphs are depicted in figure \ref{fig:SU6Lambda3}, including the
anti-Dyck paths that equally define these phases, and which will be useful in
determining the extremal generators of the subwedges in the Weyl chamber.


\subsection{$Sp(n)$ with $V$ or $\Lambda^2V$}
\label{sec:Sp}

\begin{figure}
\centering
    \includegraphics[scale=0.6]{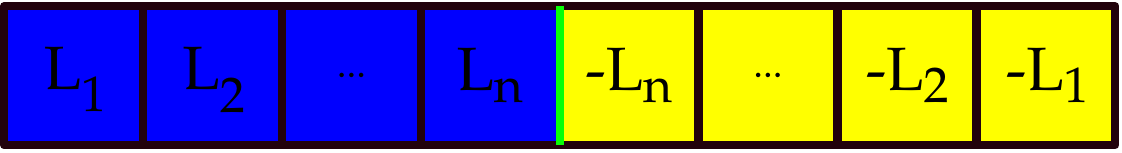}
    \caption{The fundamental representation of the $Sp(n)$ theory. 
    The only consistent phase of the $Sp(n)$ theory is given by this coloring. Blue indicates
    that the decoration of the weight is plus, and yellow negative.}
    \label{IMG:SpVPhase}
\end{figure}

The Lie algebra $\mathfrak{sp}(n)$ can be realized as the quotient by the outer
automorphism of $\mathfrak{su}(2n)$. This outer automorphism is the $\mathbb{Z}_2$
symmetry arising from the invariance of the Dynkin diagram under reflection;
concretely it is realized by the action of the map
\begin{equation}
    L_{i} \rightarrow -L_{n+1-i} \,.
\end{equation}
Generically, under the quotient by this map, the $\Lambda^iV$ representation
of $\mathfrak{su}(2n)$ becomes, where it exists, the $\Lambda^iV$ representation of 
$\mathfrak{sp}(n)$, however, in $\mathfrak{sp}(n)$ the $\Lambda^iV$ representations are not
irreducible; we shall use the notation of \cite{FultonHarris} and refer to the
relevant irreducible subrepresentation as $\Gamma_{1,0,\cdots}$,
$\Gamma_{0,1,0,\cdots}$, etc. The phases of the $Sp(n)$ theory are then those
phases of the $SU(2n)$ theory consistent under this quotient. Consider the
fundamental representation of $\mathfrak{sp}(n)$ arising from the quotient of the
$\mathfrak{su}(2n)$ fundamental representation, shown in figure \ref{IMG:SpVPhase}. It is
clear that the only decoration of $\mathfrak{su}(2n)$
\footnote{See section \ref{Sec:SUNF} for details of the phases of $\mathfrak{su}(n)$ 
with the fundamental representation.}
which consistently descends to the 
$\mathfrak{sp}(n)$ representation is the one marked in figure \ref{IMG:SpVPhase}. There
is thus exactly one phase of the $\mathfrak{sp}(n)$ theory with respect to the
fundamental representation.

\begin{figure}
    \centering
    \includegraphics[width= 4cm]{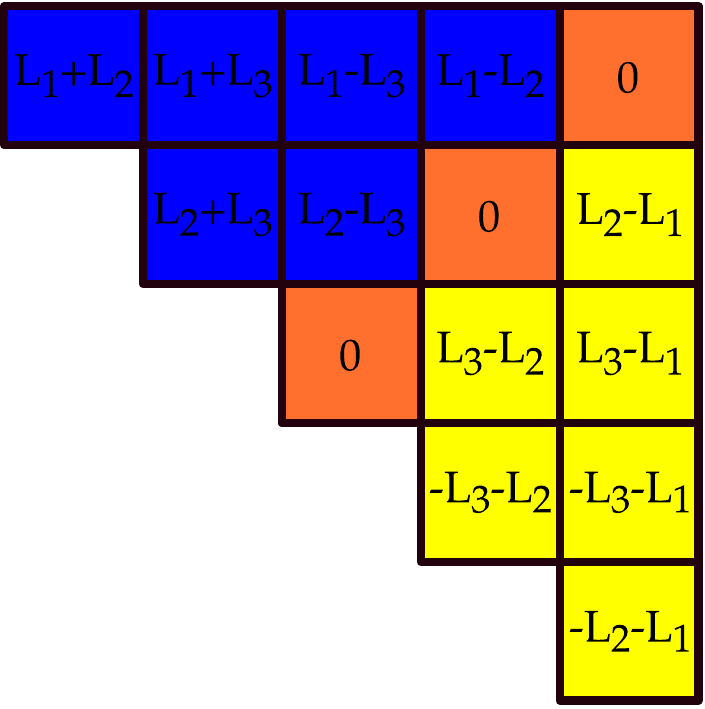}
    \caption{The box graph of the $Sp(3)$ theory with $\Lambda^2{\bf 6}$. Blue
    boxes indicate the sign of the corresponding weight is plus, yellow
    indicates negative, and orange that there is no way to associate a
    definite sign. }
    \label{IMG:Sp2VPhase}
\end{figure}

Equally we can consider $\mathfrak{sp}(n)$ with the $\Lambda^2V$ representation. This theory can be understood by embedding $\mathfrak{su}(2n)$ into $\mathfrak{so}(4n)$. As we learned from section \ref{subsec:AR}, the decomposition of the adjoint representation of $\mathfrak{so}(4n)$ under $\mathfrak{su}(2n)$ gives the $\Lambda^2 V \oplus \Lambda^2 \bar{V}$ representation of $\mathfrak{su}(2n)$. Hence, they further reduce to the $\Lambda^2 V$ representation of $\mathfrak{sp}(n)$ by the $\mathbb{Z}_2$ outer automorphism of $\mathfrak{su}(2n)$. The $\mathbb{Z}_2$ outer automorphism of $\mathfrak{su}(2n)$ can be also considered as an element of the Weyl group of $\mathfrak{so}(4n)$. In fact, $\Lambda^2 V$ is not an irreducible representation of $\mathfrak{sp}(n)$, but its subgroup $\Gamma_{0,1,0,\cdots,}$ is irreducible. 
As an example, the weights of $\Lambda^2 {\bf 6}$ of $\mathfrak{sp}(3)$ are depicted in figure \ref{IMG:Sp2VPhase}. It is clear from figure \ref{fig:ExamplePhase}
that weights, which are mapped to each other under reflection in the diagonal,
will be identified (with a minus sign) in the quotient. Note that there are
singlets in the $\Lambda^2 V$ representation as well as the
$\Gamma_{0,1,0,\cdots}$ representation of $\mathfrak{sp}(n)$, which are
depicted as the orange boxes in figure \ref{IMG:Sp2VPhase}. We cannot assign a
definite sign to the singlets, which means that
there are still massless chiral multiplets in the bulk of the Coulomb branch of the $Sp(n)$ gauge theory. 


Finally we can ask if there exists some consistent phase structure when
considering $\mathfrak{sp}(n)$ with both the $V$ and $\Lambda^2V$ representations. This can be seen by the embedding of $\mathfrak{su}(2n+1)$ into $\mathfrak{so}(4n+4)$ with an intermediate embedding by $\mathfrak{su}(2n+2)$ like
\be
\mathfrak{so}(4n+4) \supset \mathfrak{su}(2n+2) \supset \mathfrak{su}(2n+1).
\ee
The outer automorphism of $\mathfrak{su}(2n+2)$, which is again an element of
the Weyl group of $\mathfrak{so}(4n+4)$ reduces $\mathfrak{su}(2n+1)$ to
$\mathfrak{sp}(n)$. As we have the $\Lambda^2 V \oplus \Lambda^2\bar{V}$
representation from the embedding $\mathfrak{so}(4n+4) \supset
\mathfrak{su}(2n+2)$, and the $V \oplus \bar{V}$ representation from the
embedding $\mathfrak{su}(2n+2) \supset \mathfrak{su}(2n+1)$\footnote{The $V
    \oplus \bar{V}$ representation of $\mathfrak{su}(2n+1)$ can also arise
    from the decomposition of the $\Lambda ^2 V \oplus \Lambda^2 \bar{V}$
    representation of $\mathfrak{su}(2n+2)$.}, the resulting theory has both the $V$ and $\Lambda^2 V$ representations of $\mathfrak{sp}(n)$. As we have no phase for the $Sp(n)$ gauge theory with the $\Lambda^2 V$ representation, we also do not have a phase in this case.

The fact that there is no phase for the $Sp(n)$ gauge theories indicates (via the correspondence in section
\ref{Sec:PhasesGeom})
that there is generically no network of small resolutions resolving the $Sp(n)$ singularity with $\Lambda^2 V$ associated with a higher codimension enhanced singularity.


\subsection{$Sp(3)$ with $\Lambda^3{\bf 6}$}

\begin{figure}
    \centering
    \includegraphics[width=10cm]{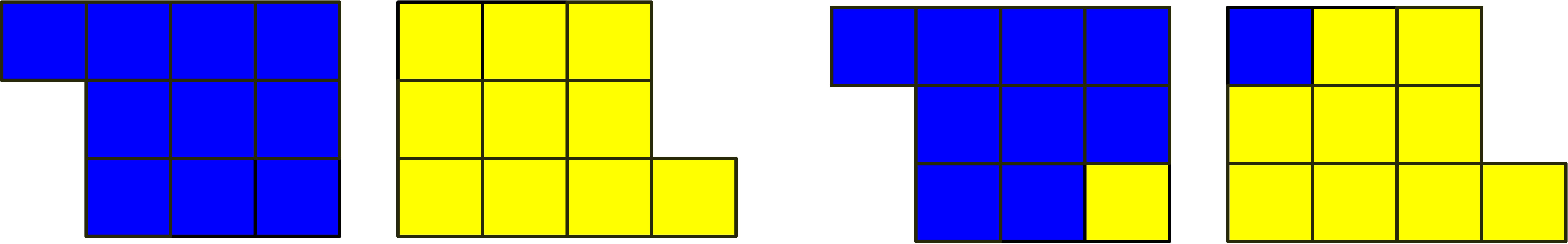}
    \caption{The phases of the $Sp(3)$ theory with $\Lambda^3{\bf 6}$. Blue indicates the
             box is decorated with a plus sign and yellow with a negative.}
    \label{IMG:Sp3VPhase}
\end{figure}

In section \ref{Sec:Su6L3} we considered the phase of the $SU(6)$ theory with
respect to the $\Lambda^3{\bf 6}$ representation, which we can exploit now to
study the $Sp(3)$ theory with the $\Gamma_{0,0,1}$ irreducible representation.
The $\Gamma_{0,0,1}$ representation is, up to multiplicity of the weights,
identical to the reducible $\Lambda^3{\bf 6}$ representation, which arises
under the quotient by the outer automorphism of the $\Lambda^3{\bf 6}$
representation of $\mathfrak{su}(6)$. It is here where we first observe a non-trivial
phase structure for the $Sp$ series. The two phases of the $SU(6)$ theory
(figure \ref{fig:SU6Lambda3}) which consistently descend to the $Sp(3)$ theory
are depicted in figure \ref{IMG:Sp3VPhase}.

\section{Box Graphs and  Elliptic Fibrations}
\label{Sec:PhasesGeom}

\subsection{The Lie group of an elliptic fibration}

So far we studied the Coulomb phases of  three-dimensional $\mathcal{N}=2$ gauge theories. We now move on to the corresponding geometric analysis. The basic setup closely follows \cite{fiveDgauge,LieF}.

The Lie group associated to an elliptic fibration $\pi:\overline{X}\to B$ 
is determined via the
compactification of F-theory to M-theory.  Let $X$
be a resolution of singularities of the total space of $\overline{X}$
with trivial canonical bundle.  In the M-theory model on $X$,
the gauge group is abelian and the coweight lattice is the lattice of
classes $[D]$ of divisors $D$ on $X$, which naturally lie in
$H^2(X,\mathbb{Z})$:
the corresponding gauge fields arise from the M-theory $3$-form reduced
on these cohomology classes.
On the other
hand, M2-branes wrapping the curves $C$ on $X$, 
whose classes $[C]$ 
belong to the lattice $H_2(X,\mathbb{Z})$, determine massive particles which
are charged under the gauge fields; the charges are naturally given
by the negative of the intersection pairing
\be \langle [D], [C] \rangle = -\#(D\cap C).\ee
(We are changing conventions from \cite{fiveDgauge,LieF}, and putting a minus sign here for better harmony between algebraic
geometry and Lie theory).  Thus, we identify the weight lattice in M-theory
with $H_2(X,\mathbb{Z})$.

We have to modify these lattices slightly for F-theory: in the F-theory
limit, the classes
in $\pi^*H^2(B)$ correspond naturally to $2$-form fields in the effective action (by reducing
the type IIB self-dual $4$-form field on the cohomology class).
For Calabi-Yau fourfolds, these $2$-forms in $d=4$ can be dualized to pseudo-scalars, and hence do not correspond to the vector fields that we are interested in (and they 
do not participate in the nonabelian gauge symmetry enhancement).
Thus, the only relevant classes which survive to the F-theory limit
are those with intersection number $0$ with the fiber $E$ of $\pi$.
In particular, the relevant coweight lattice in F-theory
is $\Lambda^\vee=\operatorname{Ann}([E])\subset
H^2(X,\mathbb{Z})/\pi^*H^2(B,\mathbb{Z})$, and the  relevant F-theory
weight lattice is 
$\Lambda=\operatorname{Ann}(\pi^*H^2(B,\mathbb{Z}))\subset H_2(X,\mathbb{Z})/\mathbb{Z}.[E]$.

The nonabelian data (i.e., the roots and coroots) are determined by considering
which curves $C$ move in families that sweep out divisors $D$.  
For such a curve,
by Witten's analysis of the quantization of wrapped branes \cite{WitMF} 
(see also \cite{4d-transitions}),
the spectrum contains a massive vector with the same gauge charges as the
curve.  In the limit where this curve has zero area, the vector becomes
massless 
and we get nonabelian gauge symmetry (unless lifted by a superpotential,
a possibility which we ignore for this discussion).  
Following \cite{fiveDgauge},
we associate the class of the curve $C$ to a root, and the 
class of the divisor $D$ swept out by $C$ to the corresponding coroot.
The pairing between the two satisfies
\be \label{DCProd}
\langle [D], [C] \rangle = -\#(D\cap C) = 2\,,
\ee
as expected from the group theory.
The geometric pairing between divisors and curves is generally asymmetric:
the way this corresponds to the group theory (and to the possibility of
gauge groups whose root systems are not simply-laced)
 is spelled out in detail for the classical
groups in \cite{fiveDgauge} (with some further explanation in
\cite{LieF}), and for the exceptional groups in 
\cite{Diaconescu:1998cn}.

\subsection{Representation associated to an elliptic fibration}

The representations given by other curves can be worked out as well.
One thing that is important to remember is that the total representation
is given by wrapping both holomorphic and anti-holomorphic curves,
obtaining a complex scalar for each \cite{WitMF,4d-transitions}.
For example, although in five-dimensional theories
one often speaks of ``matter in the fundamental
representation $n$ of ${SU}(n)$,'' the representation actually
being considered\footnote{This must be modified for a quaternionic
representation such as the fundamental 
representation of ${Sp}(r)$.  For such representations in five-dimensional theories, one
speaks of a ``half-hypermultiplet in the representation.'' Geometrically,
to build up such a representation requires wrapping both holomorphic and
anti-holomorphic curves.  In three-dimensional theories, there is a chiral multiplet
for each kind of wrapping.}
 is the sum of that representation and its complex conjugate,
${\bf n} \oplus \overline{\bf n}$.
We can either first analyze the geometry and calculate the matter 
representation, or we can start with a representation and learn what the
geometric properties must be which will lead to that representation.
Since we are working in M-theory with everything resolved (i.e. on
the Coulomb branch of the gauge theory) the dictionary between geometry
and gauge theory will depend on which phase of the Coulomb branch
we are in. 

To this end, we consider the possible K\"ahler classes 
$\phi\in H^2(X,\mathbb{R})$.  For any class  
$c\in H_2(X,\mathbb{Z})$, the
sign of $\langle \phi, c\rangle$ determines whether $c$ has a chance
of being an effective curve, since 
\be
\langle \phi, [C]\rangle=-\int_{C} \phi<0 
\ee
for an effective curve $C$ (i.e., the pairing gives the negative of
the area).  Conversely, deep results of Kleiman \cite{kleiman-ampleness}, 
Mori \cite{MR662120},
and others tell us that on Calabi--Yau varieties of low dimension, classes
whose area is positive will be effective classes (up to a rational multiple).

We will focus on curves $C$ which have nonzero intersection number
with one of the ``coroot''
divisors responsible for nonabelian gauge symmetry.  The corresponding
weight is then charged under the coroot, and so must form part of a 
representation of the nonabelian part of the gauge group.  Turning this
around, if we have a representation of the nonabelian part we can determine
the geometric properties of the curves which are involved in the
representation.


\subsection{Geometry for $SU(n)$ with fundamental representation}

Consider the case of a group whose nonabelian part is 
$SU(n)$.
The simple roots 
$\alpha_k= L_k- L_{k+1}$ 
which we have chosen 
are represented by effective curves.
Define the 
curves associated to the weights of the fundamental and anti-symmetric representations, associated to the positive or negative weights, as follows 
\be\ba
C_i^{\pm}: & \quad \pm L_i  \cr
C_{i,j}^{\pm}: & \quad \pm (L_i + L_j)  \,.
\ea
\ee
The Cartan divisors $D_{i}$  are ruled by effective curves associated to the simple roots
\be
F_i:\quad \alpha_i = L_i - L_{i+1} \,.
\ee
Their inner products must  satisfy
$\langle \phi, F_k\rangle<0$.  
Thus, we have
\be \langle \phi, C^+_k \rangle < \langle \phi, C^+_{k+1}\rangle\ee
for $k=1,\dots, n-1$.  This condition on $\phi$ places $\phi$ within
the (co-)Weyl chamber determined by our choice of positive roots.

Now suppose that in addition, there are curves in the fundamental 
representation of $\mathfrak{su}(n)$.  We label the classes of
these curves as
$ C^+_1$, \dots, $ C^+_n$ and ask which ones are effective.
Because the inner products increase as $k$ increases
and because negative inner products correspond
to effective curves, 
there must be some $\ell$ such that $C^+_k$ is effective
for $k\le \ell$, and $C^-_k$ is effective for $k>\ell$.  We can thus
write the effective curves as a sequence
\be\label{FundamentalCurveSplits}
\ba
& C^+_\ell +F_1+\cdots+F_{\ell-1}\cr
& C^+_\ell+ F_2+\cdots+F_{\ell-1} \cr
&\qquad \vdots \cr
& C^+_\ell+ F_{\ell-1}\cr
& C^+_\ell \cr 
& C^-_{\ell+1} \cr
& C^-_{\ell+1}+ F_{\ell+1} \cr
& C^-_{\ell+1}+ F_{\ell+1}+ F_{\ell+2}\cr
& \qquad \vdots \cr
& C^-_{\ell+1}+F_{\ell+1}+\cdots+F_{n-1},
\ea
\ee
and note that the entire representation ${\bf n}\oplus \overline{{\bf n}}$ is given
by wrapping these holomorphic curves and their anti-holomorphic counterparts.

\begin{figure}
    \centering
    \includegraphics[scale=0.2]{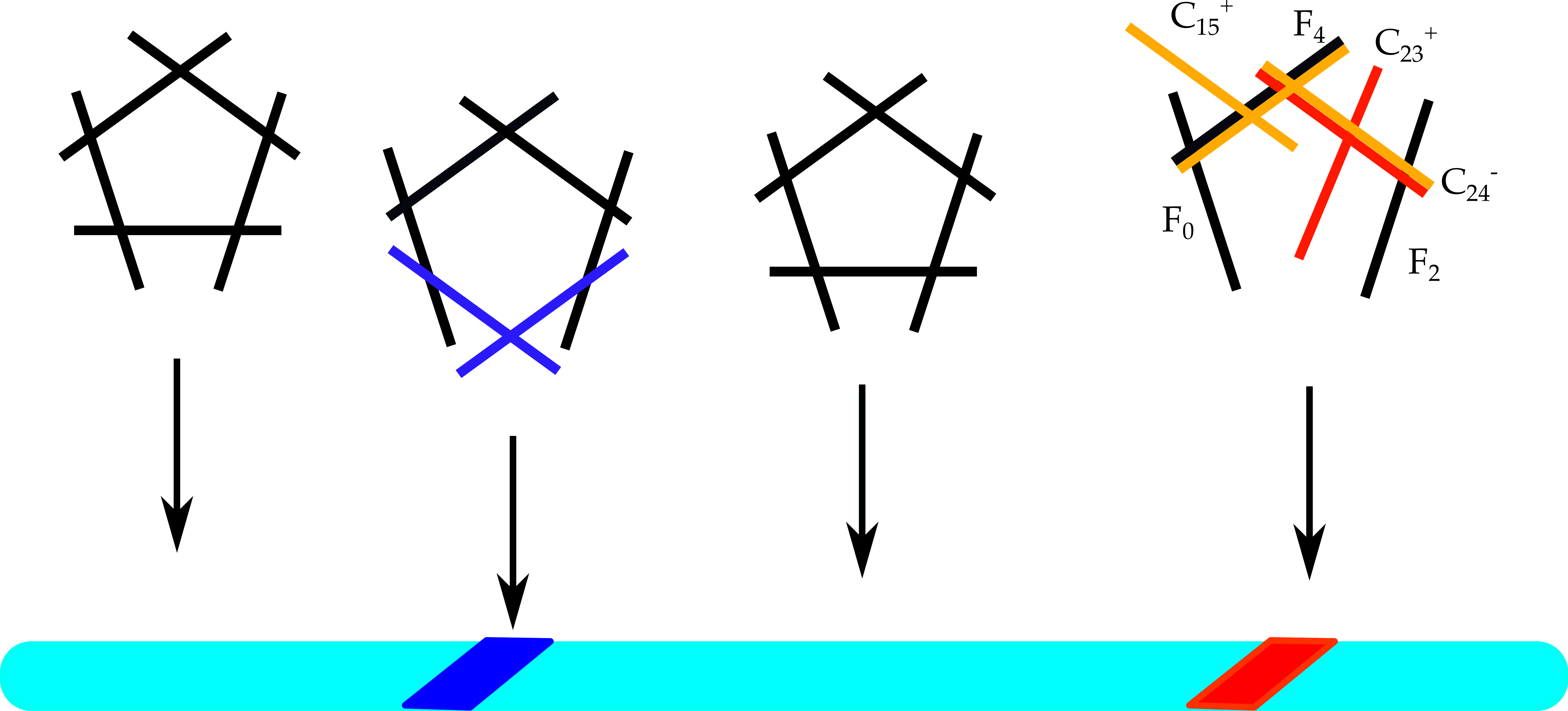}
    \caption{Surfaces for $SU(5)$. The fiber above the blue
    locus on the base expresses the irreducible effective curves when
    including classes of curves in the fundamental representation. The red locus
    indicates the irreducible effective curves when the antisymmetric
    representation is included, with effective curves specified by phase 9,
    figure \ref{fig:SU5AFiber}.
} \label{fig:sun}
\end{figure}

Since the $F_k$ are all classes of irreducible curves, we easily see
 that $C^+_\ell$ and $C^-_{\ell+1}$ must
 be classes of irreducible effective curves as well.  Since 
$ C^+_\ell+C^-_{\ell+1}=F_\ell$, we conclude that there must be a
particular fiber of the ruling on $D_\ell$ which splits into two curves.
This is illustrated in Figure~\ref{fig:sun}.

Note that there are $n$  curves in
this story  $C^+_1$, \dots, $C^+_n$, but only
$n-1$ divisors $D_1$, \dots, $D_{n-1}$.  There are thus two possibilities:
either (i) there is an additional divisor $D_0$ (a coweight which is linearly
independent of the coroots)
which enables the areas of $C^+_k$ to be linearly independent,
or (ii) there is no such divisor so there must be a relation among the areas,
which is determined by \eqref{eq:su-restrict} to be
\be \langle \phi, C^+_1\rangle + \cdots + \langle \phi, C^+_n\rangle=0.\ee
In the second case, it is not possible for all of the quantities
$\langle \phi, C^+_k\rangle$ to have the same sign, so the
index $\ell$ above is restricted to $1\le \ell\le n-1$.

On the other hand, in the first case when there is an additional divisor, $\ell=0$ and $\ell=n$ are both possible.
Moreover, if we add $D_0$ to our coweights and assume that
$\langle D_0,  C^+_k\rangle=1$ for some $k$, then
it is easy to see that we
now have the coweights of the larger algebra $\mathfrak{u}(n)$
with the weights $L_1$, \dots, $ L_n$ corresponding
to the fundamental representation of $\mathfrak{u}(n)$.
The geometry of these two extra cases is that when $\ell=0$,
the curve $C^+_1$ sticks out of $D_1$ without being on another surface,
while when $\ell=n$, the curve $C^-_m$ sticks out of $D_{n-1}$
without being on another surface.

Notice that if there is more than one fiber of a ruling which splits
into two components to generate a fundamental representation, then,
depending on the number of coweights, there may be a new, independent
homology class for each such split fiber, i.e., for each such 
$ L_\ell$.  It is even possible for the values of
$\ell$ determining the effective curves to be different:  the differences
in areas 
$\langle \phi,  L_k\rangle - \langle\phi,  L_{k+1}\rangle$ 
and 
$\langle \phi, \widetilde L_k\rangle - \langle\phi, \widetilde L_{k+1}\rangle$ 
must be the same, but there can be an overall additive shift of the areas 
$-\langle \phi, \widetilde L_k\rangle$ relative to
$-\langle \phi,  L_k\rangle$ which can lead to a different index
value at which the sign of the area changes.


\subsection{Geometry for $SU(n)$ from decorated box graphs}
\label{sec:GeofromBox}

The geometry described in the last subsection has a counterpart in the phase story and a very efficient description in terms of decorated box graphs. 
We first draw the connection with the box graphs for the fundamental representation, which reproduce the geometries that we discussed in the last subsection. The box graphs also allow us to construct more general geometries in an efficient way, for instance for the anti-symmetric representation, which we will also discuss. 

\begin{figure}
\centering
\includegraphics[width=7cm]{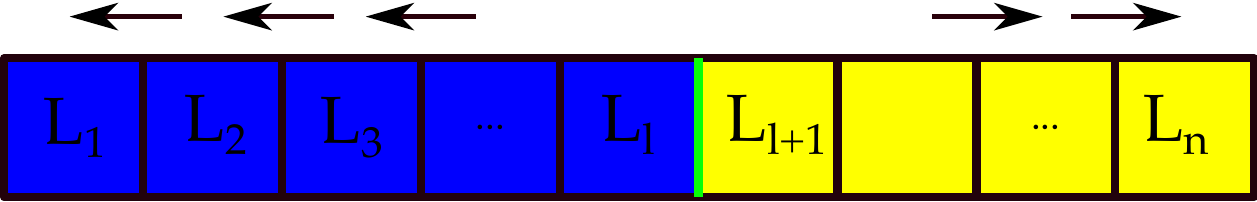}
\caption{Box graph for $U(n)$ with fundamental representation. The arrows indicate the direction of the action of the simple roots, as they generate the phase. The green line is the ``anti-Dyck path" in this case, which separates the blue/yellow, i.e., positive and negative, weights. This is the phase diagram corresponding to the effective curves in (\ref{FundamentalCurveSplits}). 
}\label{fig:sunBox}
\end{figure}

The phases of $U(n)$ with fundamental ${\bf n}$, are given by (\ref{UnF}), corresponding to decorated box graphs based on the representation graph in  figure \ref{fig:SU8F}. The box graph corresponding to the geometry in figure \ref{fig:sun} is shown in figure \ref{fig:sunBox}. The central fiber can be read off by considering the curves adjacent to the Dyck path, where the signs change between $+$ and $-$. The effective curves are
\be
\mathcal{K}_{U(n), {\bf n}, \text{phase }l} = \{F_1, \cdots, F_{l-1}, C^+_l, C^{-}_{l+1}, F_{l+1}, \cdots, F_{n-1} \} \,.
\ee
The splitting occurs at $F_l = C^{+}_{l}+ C^-_{l+1}$.
We can read off the Dynkin labels of the curves $C^\pm$ from the box graph. 
The intersection of a curve with a Cartan divisor $D_i$, which are fibered by $F_i$ associated to the simple root $\alpha_i$,  is obtained by determining whether  adding the simple root changes/maintains the sign of the box, in which case the
intersection is $-1/+1$, and if it does not give rise to a weight in a
neighboring box, then the intersection vanishes.
From this we obtain the intersections
\be\label{CDIntersect}
\ba
C^{+}_l \cdot D_{l-1} &= +1  \cr
C^{+}_l \cdot D_{l} &= -1  \cr
C^{-}_{l+1} \cdot D_{l}& = -1  \cr
C^{-}_{l+1} \cdot D_{l+1} &= +1  \,.
\ea
\ee
From this it follows that the intersections are precisely as in figure \ref{fig:sun}.


\begin{figure}
\centering
\includegraphics[width=4.5cm]{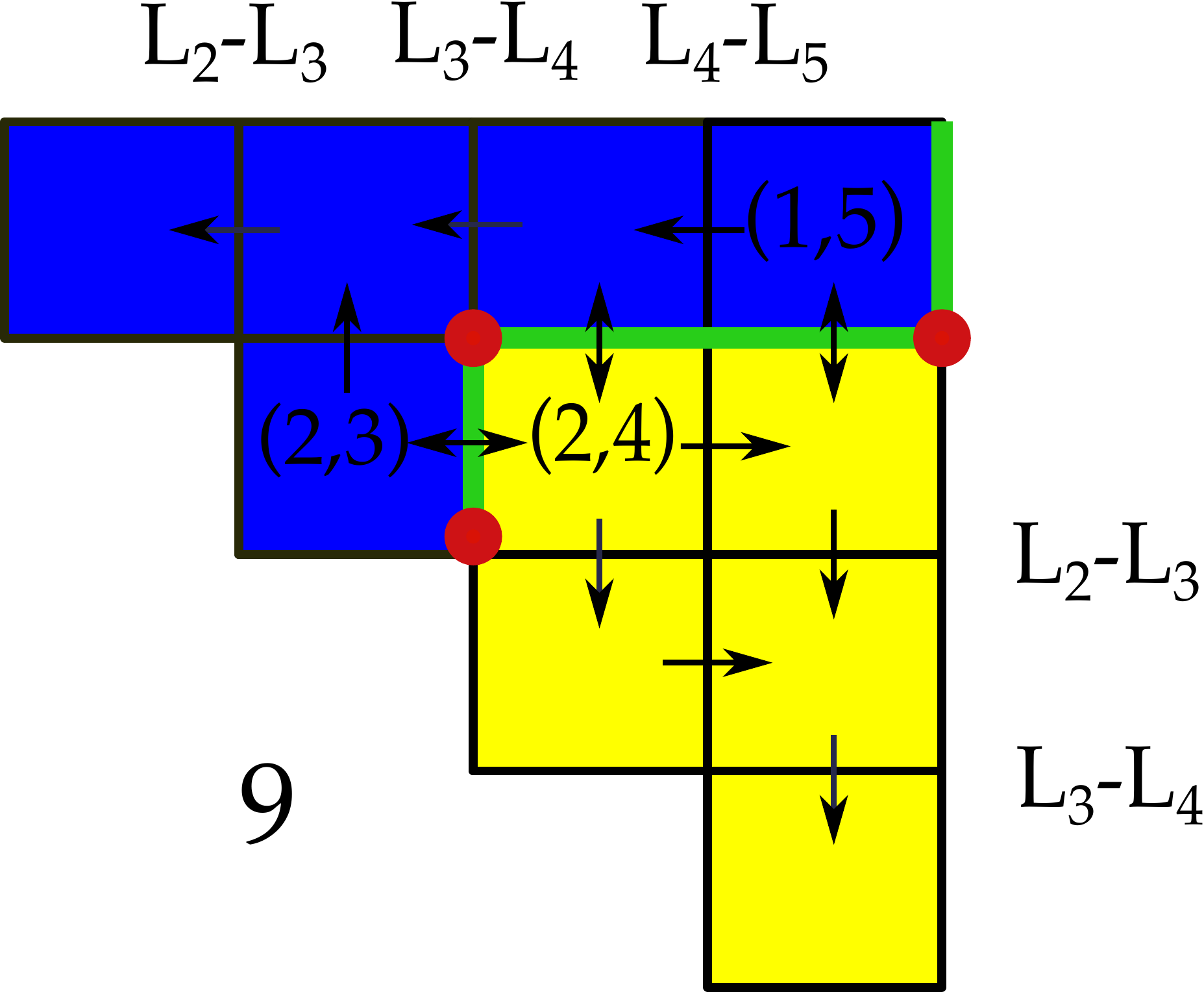}
\caption{Again, this is phase 9 of $SU(5)$, with entries $(i,j)$ in the boxes denoting the weights $L_i + L_j$. The double-headed arrows indicate where crossing from $+$ to $-$ (blue to yellow), the corresponding roots, namely $\alpha_1 = L_1-L_2$ and $\alpha_3= L_3-L_4$, split. The single arrows correspond to the action of the simple roots, that remain irreducible, as they generate the representation from the extremal weights, along the anti-Dyck path. 
}\label{fig:SU5AFiber}
\end{figure}


This process has a direct generalization for other representations, for instance consider the anti-symmetric representation for 
$SU(n)$. 
Each decorated box graph or, equivalently, anti-Dyck path defines a codimension-two fiber corresponding to a $D_n$ enhancement from an $A_{n-1}$ singularity in codimension one. The geometry  of the central fiber associated to such a diagram is read off as follows: consider for instance phase 9 for $SU(5)$ shown in  figure \ref{fig:SU5AFiber}. 
The irreducible curves are located along the anti-Dyck path (in fact they are one-to-one with the extremal points, shown in red), as well as the roots that do not correspond to crossing the path
\be
\mathcal{K}_{\includegraphics[width=.6cm]{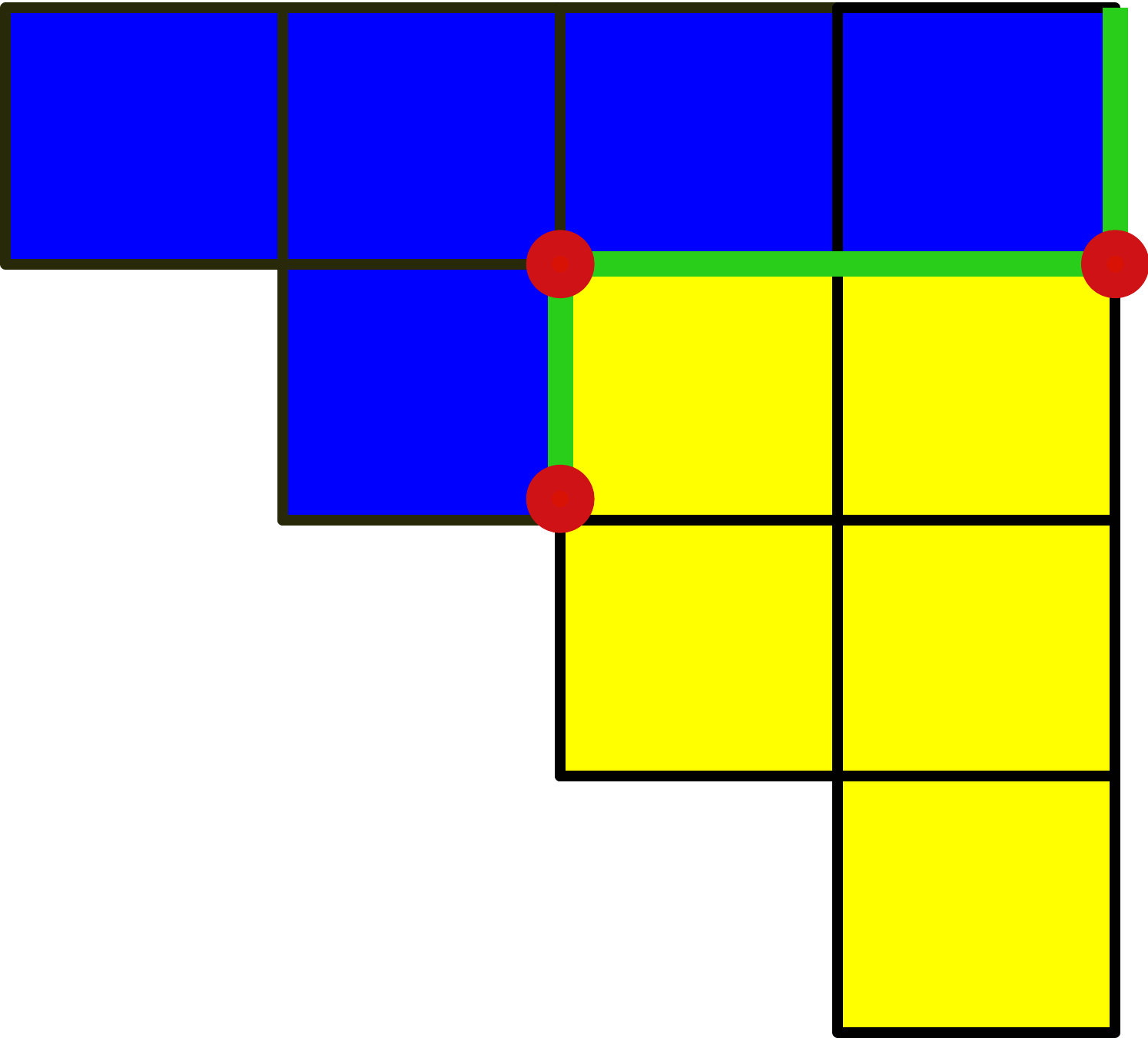}}= \{ C^+_{1,5}\,,\ C^+_{2,3}\,,\ C^-_{2,4} \,, \  F_{2}\,,\ F_{4} \} \,.
\ee
The remaining curves $F_{1}$ and $F_{3}$ are reducible, as they correspond to crossing the anti-Dyck path. 
In particular $C^+_{2,3}+C^-_{2,4} = F_3$, $C^+_{1,4}+C^-_{2,4}= F_1$ and $C^+_{1,5} + C_{2,5}^+ = F_1$. 
The splitting in terms of the irreducible components is then
\be
\ba
C^+_{1,4}+F_4+C^-_{2,4} &= F_1 \cr
C^+_{2,3}+C^-_{2,4}& = F_3\,.
\ea
\ee
We can read off the intersections from the box graphs using the same rules as above (\ref{CDIntersect})\footnote{Note that these are the intersection numbers, and thus the
    negative of the inner product defined in (\ref{DCProd}). } 
\be
\ba
C^+_{1,5} \cdot D_{1} &=-1 \cr
C^+_{1,5} \cdot D_{4} &=+1 \cr
C^+_{2,3} \cdot D_{1} &= +1 \cr
C^+_{2,3} \cdot D_{3} & = -1  
\ea
\qquad
\ba
C^-_{2,4} \cdot D_1 & = -1 \cr
C^-_{2,4} \cdot D_2 & = +1 \cr
C^-_{2,4} \cdot D_3 & = -1 \cr
C^-_{2,4} \cdot D_4 & = +1 \,.
 \ea
\ee
With these intersections the fiber associated to the phase 9 diagram is given
by figure \ref{fig:sun}, which also shows the multiplicity of the fiber components. Positive intersections correspond to the curve meeting the divisor transversally, negative intersections mean it is contained in the divisor.

Using this method we can also determine the extremal set of generators for the combined phases for $SU(5)$ 
with ${\bf 5}$ and ${\bf 10}$ representation, whose decorated box diagrams were determined in figure \ref{fig:SU5FlopsAF}. Given the symmetry of the problem, we only need to consider one half of the phases, e.g. in the first two columns of figure \ref{fig:SU5FlopsAF}. 
The extremal generators for the phases with only ${\bf 10}$ matter, with labels as in figure \ref{fig:SU5FlopsAF} are
\be
\ba
\mathcal{K}_{4} &= \{C_{2,5}^+\,,\ C_{3,4}^-\,,\ F_1\,,\ F_3\,,\ F_4 \} \cr
\mathcal{K}_{7} &= \{C_{2,4}^+\,,\ C_{2,5}^-\,,\ C_{3,4}^-\,,\ C_{1,5}^+\,,\ F_3 \} \cr 
\mathcal{K}_{9}  &= \{ C_{2,3}^+\,,\ C_{2,4}^-\,,\ C_{1,5}^+\,,\ F_2\,,\ F_4 \} \cr
\mathcal{K}_{11}  &= \{C_{2,3}^- \,,\ C_{1,5}^+\,,\ F_2\,,\ F_3\,,\ F_4 \}  \,.
\ea
\ee
Likewise the relevant extremal generators for the phases of the ${\bf 5}$ matter case with labels as in figure \ref{fig:SU5FPhases} are
\be
\ba
\mathcal{K}_{II} &= \{ C_{3}^+\,,\ C_4^-\,,\ F_1\,,\ F_2\,,\ F_4\}\cr
\mathcal{K}_{III} &= \{ C_{2}^+\,,\ C_3^-\,,\ F_1\,,\ F_3\,,\ F_4\}\cr
\mathcal{K}_{IV} &= \{ C_{1}^+\,,\ C_2^-\,,\ F_2\,,\ F_3\,,\ F_4\}\,.
\ea
\ee
On the other hand the phase with the combined representation has four generators in each case, given by
\be\label{K510}
\ba
\mathcal{K}_{9, III} &= \{C_3^- \,,\ C_{2,3}^+\,,\ C_{2,4}^-\,,\ F_4 \}\cr
\mathcal{K}_{9, II} &= \{C_{3}^+\,,\ C_{1,5}^+\,,\  F_2\,,\ F_4 \}\cr
\mathcal{K}_{11, IV} &= \{ C_2^-\,,\ F_2\,,\ F_3 \,,\ F_4 \}\cr
\mathcal{K}_{11, III} &= \{C_2^+ \,,\ C_{2,3}^- \,,\ F_3 \,,\ F_4 \}\cr
\mathcal{K}_{4, III} &= \{C_{2,5}^+\,,\ F_1\,,\ F_3\,,\ F_4 \}\cr
\mathcal{K}_{7, III} &= \{  C_{1,5}^+ \,,\ C_{2,4}^+ \,,\ C_{2,5}^- \,,\ F_3 \}\,.
\ea\ee
Note that these generators are only in the cone if, in the sense of section
\ref{subsec:reducibility}, all splittings are allowed in codimension 3, i.e. 
both those compatible with $E_6$ as well as $SO(12)$. 
We shall determine the fibers at the $E_6$ codimension 3 loci in section
\ref{subsec:E6Codim3}, in which case not all splittings are compatible with only $E_6$, and the cone has five generators.


\subsection{Flops}

The possible choices 
of $\ell$ in the previous example gives a decomposition of the
Weyl chamber into smaller chambers.  These are easy to interpret: they
correspond to different resolutions of $\overline{X}$, related by flops.

This is easily seen by considering the area of various curve classes.  
To illustrate this, consider the previous example where we have two irreducible
effective curves in the divisor $D_\ell$ represented by $ L_\ell$
and $- L_{\ell+1}$.  We can move
$\phi$ within $H^2(X,\mathbb{R})$ in order to decrease
$\langle \phi,  L_\ell\rangle$ to zero.  If we continue to move in the
same direction, we will make $\langle \phi,  L_\ell\rangle$ negative,
while keeping $\langle \phi,  L_{\ell-1}\rangle$ positive.  The geometric
interpretation is that the irreducible curve corresponding to $ L_\ell$
becomes smaller and smaller until it is blown down to a conifold point.
Continuing to move in the same direction causes a flop, and the conifold
point is blown up with its alternate small resolution.  The new blowup creates
a new curve of class $- L_\ell$ in the divisor $D_{\ell-1}$, and
the proper transform of the old fiber in $D_{\ell-1}$ becomes an irreducible
curve in the class $ L_{\ell-1}$.  (In figure~\ref{fig:sunBox},
this process flops the left-most reducible
curve towards the component to its left.)

Note that the geometric flop transition changes one of the signs of $\langle
\phi, C \rangle$. This is nothing but the flop defined in the decorated box
graph or the anti-Dyck path in section \ref{sec:flopgauge}. The weights
associated with the extremal points correspond to the flopped curves. As we
demonstrated the phases of the $SU(5)$ gauge theory with the anti-symmetric
representation and the fundamental representation in figure
\ref{fig:SU5AFPhaseDiag}, each box graph corresponds to a resolution, and a
single-box sign change among the box graphs or equivalently a flop of a corner
of the anti-Dyck path corresponds to a flop transition between distinct resolutions.


\section{Elliptic fibrations in codimensions two and three}
\label{sec:EllipticFibsCodim}

\subsection{Local models for fibers in codimension two}

\begin{figure}\centering
\includegraphics[width=10cm]{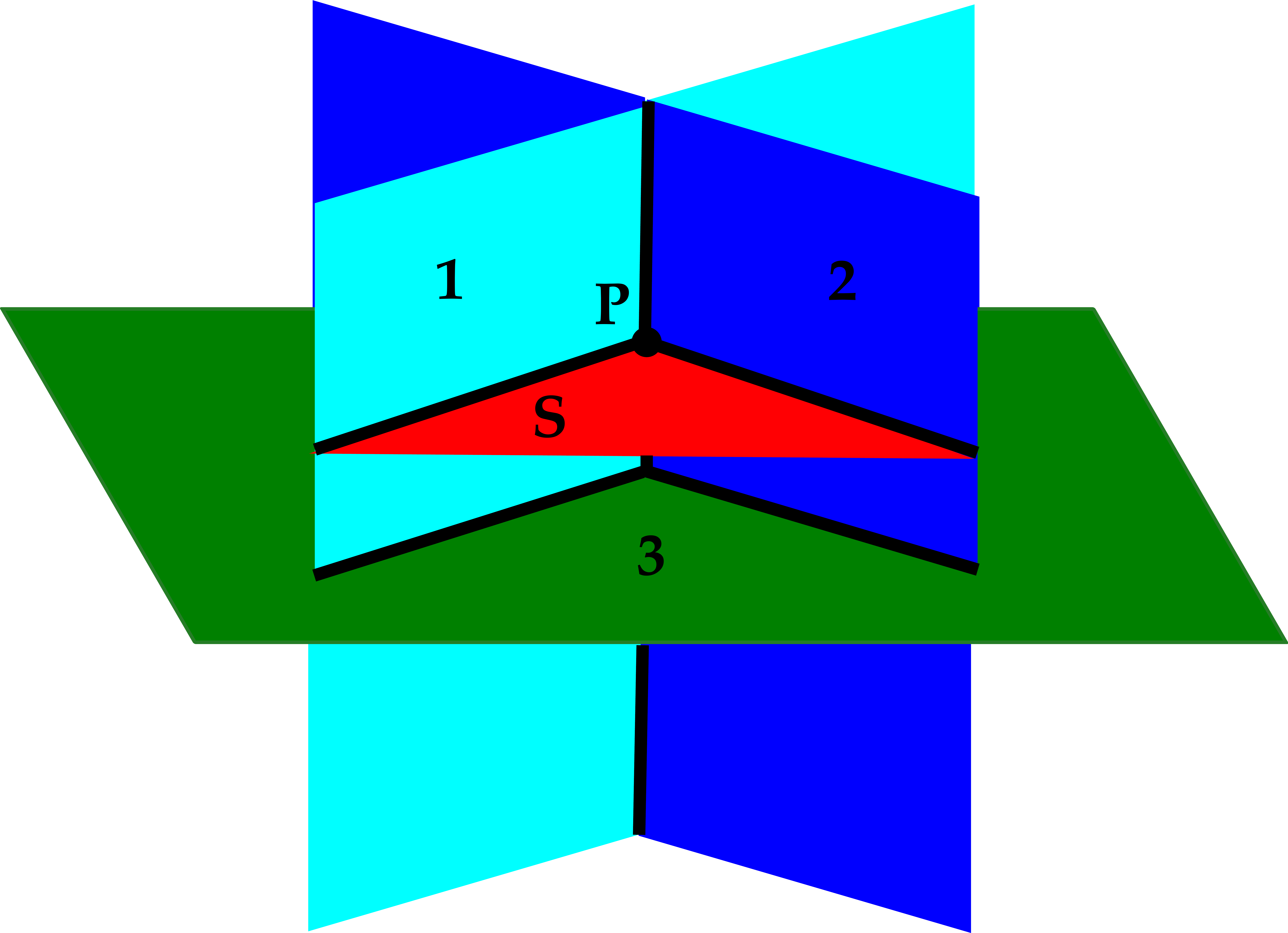}
\caption{Discriminant in higher codimension.} \label{fig:disc}
\end{figure}

Consider an elliptic fibration\footnote{Recently, a generalization to
genus-one fibrations has been discussed \cite{genus-one}, but we
have no need of that generalization here.}
 $\pi:\overline{X}\to B$ over a base $B$
in Weierstrass form $y^2=x^3+fx+g$.
In codimension one on the
base (along components of the discriminant locus $\Delta\subset B$,
as in figure~\ref{fig:disc}), 
we find various types of singular fibers as classified by Kodaira,
and we can determine the monodromy of these singular fibers by using
Tate's algorithm.  The upshot is a determination of the nonabelian part
of the (geometric\footnote{For F-theory models in four dimensions, part
of this ``geometric'' gauge algebra may be lifted by a superpotential,
resulting in the actual gauge algebra being smaller.}) gauge 
algebra of the corresponding F-theory model.

The singularities are enhanced in codimension two, which we discuss 
following \cite{gorenstein-weyl,Katz:1996xe} (see also \cite{Bershadsky:1996nh,MR2681776,Morrison:2011mb,anomalies}).  
Let $\Sigma_\alpha\subset B$ be
an irreducible subvariety of codimension two along which some
enhancement occurs: necessarily, $\Sigma_\alpha\subset\Delta$.
We make a local model for the singularity enhancement by choosing a general
point $P_\alpha$ of $\Sigma_\alpha$, a general surface $S_\alpha\subset B$
which is transverse to $\Sigma_\alpha$ at $P_\alpha$, and a general
function $\varphi_\alpha:S_\alpha\to \mathbb{C}$ such that 
$\varphi_\alpha(P_\alpha)=0$.  For each $u\in \mathbb{C}$ with 
$|u|<\epsilon$, we can restrict our
elliptic fibration to the curve $\varphi_\alpha^{-1}(u)$ and use the
Kodaira classification to ask what type of singular fiber appears in
the Weierstrass model.  This is illustrated in figure~\ref{fig:disc},
where a transverse surface $S$
at a point $P$ retains information about the discriminant locus components
which pass through the point.

If $f$ vanishes to order at least $4$ and $g$ vanishes to order at least $6$
at $P_\alpha$, then on any Calabi-Yau variety $\rho:X\to\overline{X}$ 
which resolves the singularities of $\overline{X}$, the fiber of
$\pi\circ \rho$ over $P_\alpha$ contains a surface.  When we
compactify M-theory on $X$ we find an infinite tower of massive Kaluza-Klein
states corresponding to curves in that surface, all of which become
massless in the F-theory limit, signaling a tensionless string in the
F-theory model
\cite{WitMF}. 

Thus, in order to avoid tensionless strings in our F-theory compactification,
we insist that the orders of vanishing of $f$ and $g$ at $P_\alpha$ do not
simultaneously exceed $4$ and $6$. It then follows that the elliptic fibration
over $\varphi_\alpha^{-1}(0)$ has a well-defined
 Kodaira type, which determines the
singularity in the Weierstrass model:  it is one of the ADE singularities,
which are also known as ``rational double points.''
We can regard the family of surfaces $\varphi_\alpha^{-1}(u)$ (for $u$ near
$0$ in $\mathbb{C}$) as a deformation of the singularity 
in $\varphi_\alpha^{-1}(0)$.  The total space of the family is
a threefold $\overline{X}=\pi^{-1}(S_\alpha)$ 
which is fibered by surfaces 
$\overline{X}_t=\pi^{-1}(\varphi_\alpha^{-1}(t))$, 
all of which have rational double points.

To get a good F-theory model, we assume that there is a Calabi--Yau variety
$X$ and a map $\rho:X\to \overline{X}$
which resolves the singularities of $\overline{X}$ in such a way that
the induced family $\pi\circ \rho: \overline{X}\to B$ is flat.\footnote{This
technical condition ensures the absence of tensionless strings
in the associated M-theory compactification by ensuring that all fibers
are one-dimensional;
it is well-understood for elliptic Calabi--Yau threefolds \cite{Grassi91}
but not for elliptic Calabi--Yau fourfolds \cite{codimthree}.  }
The Kodaira fiber over the general
point of any component of $\Delta$ gets resolved by this procedure.
Restricting
$\rho$ to $\rho^{-1}(\overline{X})$, we find a resolution $X$ of the threefold
 $\overline{X}$ which resolves
each of the surfaces $\overline{X}_t$ 
in the family when $t\ne0$, although it may fail
to resolve the limiting surface $\overline{X}_0$.  This phenomenon is known as
{\em partial simultaneous resolution of rational double points}
\cite[Section 8]{[P]} (see also \cite[Theorem 1.14]{pagoda} 
and \cite{gorenstein-weyl}), and we now explain it in detail.

We associate to the singularity on $\overline{X}_0$  
a simply-laced root system $\widetilde{\rootsystem}$ (the one with the same
name as the  ADE type 
of the singularity on $\overline{X}_0$),
and to the singularity on $\overline{X}_t$ for $t\ne0$ 
a sub-root system $\rootsystem\subset \widetilde{\rootsystem}$ (also simply-laced, of the ADE type of the singularity on $\overline{X}_t$).
For each subgroup $\Gamma$ of the Weyl group $W_{\widetilde{\rootsystem}}$ of $\widetilde{\rootsystem}$ which fixes
a Weyl chamber of $\rootsystem$, there is a ``universal'' family of surfaces
$\overline{\mathcal X} \to {\mathcal U}$ (depending on $\rootsystem$, $\widetilde{\rootsystem}$, and $\Gamma$)
with a partial resolution ${\mathcal{X}}\to\overline{\mathcal{X}}$
such that (possibly after shrinking $\epsilon$ and $S$)
our given family $\overline{X}$ and its resolution $X$
are  obtained from $\overline{\mathcal{X}}$ and $\mathcal{X}$
using some embedding
$\{|u|<\epsilon\}\to\mathcal{U}$.
Moreover, the classes of algebraic curves on $\overline{X}$ are generated
by the $\Gamma$-invariant sublattice of $\widetilde{\rootsystem}$.

For each curve in the central fiber, there is a divisor on $X$ meeting
this curve only once and not meeting any of the other curves in the central
fiber.  We can see this by describing $X$ as a union of small neighborhoods
of each of the curves in the central fiber.  Away from the intersections
with other curves, these neighborhoods are locally a product of the curve
with a small neighborhood of the origin in $\mathbb{C}^2$; choosing a point
on the curve, that ``small neighborhood of the origin'' gives a divisor
on $X$ meeting only this curve.

As a consequence, in a small neighborhood of $P_\alpha$, there are the
same number of linearly independent classes of curves as there are linearly
independent classes of divisors.  As we will see in the next section,
in a global F-theory model there can be additional relations among the
divisor classes, and hence fewer linearly independent divisors.
However, in a local model, the choice of resolution $X$ determines which
curve classes are effective, and thereby determines a phase as in 
section~\ref{Sec:PhasesGeom}.

We will first consider
the case that the singularities on $\overline{X}_0$ are completely
resolved in $X$, and later return to consider the case when the
singularities are only partially resolved.
In the completely resolved case,
the effective curve classes on the resolved surface $X_0$ 
correspond to the positive roots in
the simply-laced root system $\widetilde{\rootsystem}$.
Both effective curves and anti-effective curves
can be used for wrapping M2-branes, allowing us to identify
matter fields in the theory
corresponding to both positive and negative roots in $\widetilde{\rootsystem}$.
The possible gauge charges on these matter fields are naturally
identified with the
coweight lattice of the root system $\widetilde{\rootsystem}$.

The effective curves on the nearby fiber $X_t$ form the positive roots
in the sub-root system $\rootsystem\subset \widetilde{\rootsystem}$, and
again, M2-branes can be wrapped on the curves corresponding to both positive
and negative roots.  These curves move in a larger family, and the spectrum
of the wrapped M2-brane is correspondingly different, containing a
(massive) vector multiplet as well as hypermultiplets (the number of which
depends on the genus of the parameter space for the curve in question).

Since the M2-brane spectra
of curves corresponding to $\rootsystem$ include vector multiplets,
we identify $G_\rootsystem$ as the potential gauge group\footnote{We
refer to this as the ``potential'' gauge group because although it locally
reflects the correct gauge symmetry, there may be some changes in the
group due to global effects, as we shall explain below.} associated to
these singularities.
Because the vector multiplets are massive
when the singularities are resolved, this group is broken to its
Cartan subgroup $H_\rootsystem$ at the generic point of the Coulomb branch.
  However, at the origin of the Coulomb branch
all of these curve classes will have zero area, i.e., they will be blown
back down, and the nonabelian gauge symmetry is restored.

We hasten to point out one subtlety:  there may be global relations among
some of the curve classes in $\rootsystem$, leading to a different nonabelian
gauge group.  This is because there can be an outer automorphism of the Lie algebra
$\mathfrak{g}_\rootsystem$ under which certain roots are identified;
when such an automorphism acts geometrically,
the ``correct'' gauge group is a subgroup of $G_\rootsystem$,
which typically has a non-simply-laced algebra \cite{Bershadsky:1996nh,anomalies}.

The cohomology classes of the
curves in $X_0$ corresponding to roots  of the larger root system 
$\widetilde{\rootsystem}$ may span a larger space than those from 
the root system $\rootsystem$,
and the weight lattice can be enlarged to study these additional classes.
We thus consider the larger Lie group generated by $G_\rootsystem$
and the weight lattice $H_{\widetilde{\rootsystem}}$ of 
$\widetilde{\rootsystem}$.  This is a potential gauge group associated
to the curves on $X_0$ as well as those on $X_t$, although global effects may
cause the actual gauge group to be smaller.
The Lie algebra of this larger group takes the form
\be\label{U1Commutant}
\mathfrak{g}_\rootsystem \oplus \mathfrak{u}(1)^{\text{rk}(\widetilde{\rootsystem}) 
-\text{rk}(\rootsystem)} \subset \mathfrak{g}_{\widetilde{\rootsystem}}\,,
\ee
which is a subalgebra of $\mathfrak{g}_{\widetilde{\rootsystem}}$.

To see how this works in some examples, consider first the case in which
$\widetilde{\rootsystem}=A_n$ and $\rootsystem=A_{n-1}$ with $\Gamma$ trivial.  Then $G_{\rootsystem}=SU(n)$
and it is not hard to see that $H_{\widetilde{\rootsystem}}$ and $G_{\rootsystem}$ together generate\footnote{Note that the embedding of $U(n)$ into
$SU(n+1)$ sends a matrix $g$ to the matrix $\operatorname{diag}(g, \det(g)^{-1})$.}
$U(n)$.  
The simple roots of the root system $A_n$ are irreducible curves on $X$; however, only
the roots of $A_{n-1}$ are roots of the gauge group of $X$ since the others do not sweep
out divisors on $X$, i.e., do not have associated coroots.
Geometrically, we have a family of Kodaira fibers of type $I_n$
degenerating to a fiber of type $I_{n+1}$, which has various resolutions
of singularities as illustrated in figure~\ref{fig:sun}.  

The curve classes
corresponding to positive roots of $A_n$ are those which one can wrap
an M2-brane with positive orientation; the negative roots are those
which can be wrapped with negative orientation \cite{WitMF}.  These two sets
are the nonzero weights in the 
adjoint representation of $\mathfrak{g}_{A_n}=\mathfrak{su}(n+1)$.  To see how these classes
are related to the gauge group of our local model, we decompose
the adjoint representation of 
  $G_{\widetilde{\rootsystem}}=SU(n+1)$
under the subalgebra $\mathfrak{u}(n)$, and find
the adjoint representation of $\mathfrak{u}(n)$ together
with ${\bf n}\oplus \overline{{\bf n} }$.  In the Katz--Vafa approach
\cite{Katz:1996xe}, 
this corresponds to
``matter in the fundamental representation of $\mathfrak{u}(n)$.''

As another example, consider $\widetilde{\rootsystem}=D_n$ and $\rootsystem=A_{n-1}$.  We again have
that $H_{\widetilde{\rootsystem}}$ and $G_{\rootsystem}$ together generate $\mathfrak{u}(n)$.  This time, when we
restrict the adjoint representation of $G_{\widetilde{\rootsystem}}=SO(2n)$ to $U(n)$,
we get the adjoint representation of $U(n)$ together with
$\Lambda^2{\bf n} \oplus \overline{\Lambda^2{\bf n}}$, which corresponds to matter in the
antisymmetric representation.
As we have seen earlier, the decorated box graphs provide the phase structure in this case for both $\mathfrak{u}(n)$ and $\mathfrak{su}(n)$. 

The story is somewhat more complicated if $\Gamma$ is nontrivial. 
For non-trivial $\Gamma$ there are two instances: either $\Gamma$ acts on the root system $\rootsystem$ as a non-trivial outer automorphism, in which case the codimension-two fibers are affected. This is one way of generating non-simply-laced gauge groups. For instance consider $G_{\widetilde{\rootsystem}} = SU(2n)$ and $G_{\rootsystem} = Sp(n)$, where $\Gamma =\mathbb{Z}_2$ is the outer automorphism of the $A_{2n-1}$ root system. The fibers in codimension two are obtained as the $\Gamma$-invariant fibers, or said in terms of the phases, only the $\Gamma$-invariant phases descend to phases of the theory with gauge group $G_{\rootsystem}$. Other examples are $G_{\widetilde{\rootsystem}}= E_6$ and $G_{\rootsystem}= F_4$ as well as $G_{\widetilde{\rootsystem}}= SO(8)$ and $G_{\rootsystem}= G_2$, where the quotient is by the triality symmetry of the $SO(8)$ Dynkin diagram. We have discussed $Sp(n)$ in section \ref{sec:Sp}, where  a non-trivial network of phases remains for $G_{\rootsystem}$.

Non-trivial monodromy arises also when the commutant $G_{\perp}$ of $G_{\rootsystem}$ inside $G_{\widetilde{\rootsystem}}$ is non-abelian, i.e. instead of (\ref{U1Commutant}) consider more generally
\be\label{NACom}
\mathfrak{g}_{\rootsystem} \oplus \mathfrak{g}_\perp \subset 
\mathfrak{g}_{\widetilde{\rootsystem}} \,.
\ee
Then the Weyl group of $\mathfrak{g}_\perp$ can act non-trivially on the codimension-two fibers and thereby give monodromy-reduced fibers instead of standard Kodaira fibers in higher codimension. 
The interesting triplets $(\mathfrak{g}_{\widetilde{\rootsystem}}, \mathfrak{g}_{\rootsystem},  \mathfrak{g}_{\perp})$ involving the exceptional Lie algebras\footnote{There are more examples involving non-simply-laced Lie algebras, which we will not consider here, as well as higher rank ADE examples. See e.g. \cite{Slansky}. } are as follows
\be\label{Triplets}
\ba
\mathfrak{g}_{\widetilde{\rootsystem}} &\quad\rightarrow\quad \mathfrak{g}_{\rootsystem} \oplus \mathfrak{g}_{\perp} \cr
\mathfrak{e}_6 &\quad\rightarrow\quad  \mathfrak{su}(6) \oplus \mathfrak{su}(2) \cr
\mathfrak{e}_6 &\quad\rightarrow\quad  \mathfrak{su}(3) \oplus \left(\mathfrak{su}(3) \oplus \mathfrak{su}(3)\right) \cr
\mathfrak{e}_7  &\quad\rightarrow\quad \mathfrak{so}(12) \oplus \mathfrak{su}(2 ) \cr 
\mathfrak{e}_7 &\quad\rightarrow\quad \mathfrak{su}(6) \oplus \mathfrak{su}(3) \cr
\mathfrak{e}_8 &\quad\rightarrow\quad  \mathfrak{e}_7 \oplus \mathfrak{su}(2) \cr
\mathfrak{e}_8 &\quad\rightarrow\quad  \mathfrak{e}_6 \oplus \mathfrak{su}(3) \cr
\mathfrak{e}_8&\quad\rightarrow\quad  \mathfrak{su}(5)  \oplus \mathfrak{su}(5) \,.
\ea
\ee
Of course these decompositions can also be read in reverse, such as $\mathfrak{g}_{\rootsystem} = \mathfrak{su}(2)$ and $\mathfrak{g}_\perp= \mathfrak{su}(6)$, etc. 
We will see that unless there are extra rational sections in the elliptic fibration, the fiber in codimension two will always be monodromy-reduced.  In the following we exemplify this for $\mathfrak{e}_6 \rightarrow \mathfrak{su}(6) \oplus \mathfrak{su}(2)$. In this case, there is a non-trivial monodromy in $\Gamma= W_{\mathfrak{su}(2)} = \mathbb{Z}_2$. We will see that this affects the codimension-two fibers, which are not standard Kodaira $IV^*$ fibers, unless the fibration allows for additional sections. Global issues of this kind will be discussed in detail in section \ref{sec:FibsNoMono}. 
Likewise the decomposition of $\mathfrak{e}_8\rightarrow \mathfrak{su}(5)\oplus \mathfrak{su}(5)_\perp$ requires generically  4 extra sections in order to have a standard type $II^*$ fiber in codimension two.


\subsection{Fibers of $E_6$ type with monodromy}
\label{sec:FibE6Mono}
\begin{figure}
    \centering
    \includegraphics[width=16.5cm]{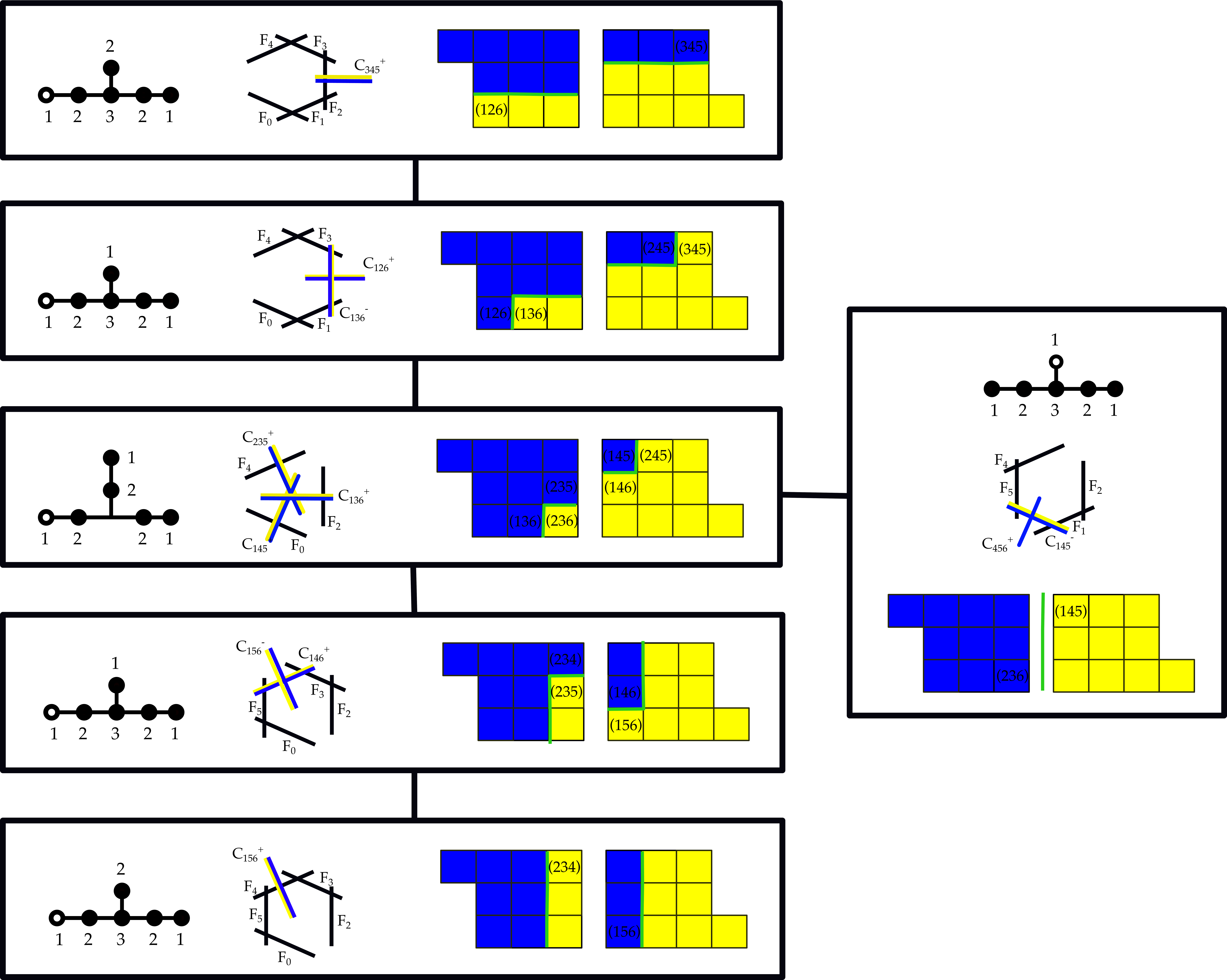}
    \caption{Phases and Fiber types of the $SU(6)$ theory with  $\Lambda^3{\bf 6}$ of 
    $SU(6)$. In each box the left most diagram is the intersection graph of the fiber with multiplicities and the white node denoting the zero-section. The middle diagram is the fiber graph with the extremal rays, and on the left we show the decorated box graph, where the weights $(ijk)$ that are explicitly labeled in the blue/yellow boxes correspond curves $C_{ijk}^\pm$ in the fibers, which together with the irreducible Cartans $F_i$ are the extremal rays. 
    Note that the $\mathbb{Z}_2$ quotient identifies these weights pairwise, which is shown also in the fiber diagrams in terms of the blue/yellow double lines.  
    Black lines connecting the various phases correspond to flops. The intersection graphs are precisely obtained by deleting one (non-affine) node of the $IV^*$ Kodaira fiber. }   \label{fig:SU6Lambda3Fibs}
\end{figure}

An example of phases with non-trivial monodromy was discussed in section \ref{Sec:Su6L3}, for $\mathfrak{e}_6 \rightarrow \mathfrak{su}(6)\oplus \mathfrak{su}(2)$ and non-trivial $\Gamma=\mathbb{Z}_2$. In this case, the fibers are not of affine $E_6$ type, i.e. Kodaira type $IV^*$,  however give rise to generalized fiber types. 
Using the same methods as in section  \ref{sec:GeofromBox}, the fiber types can be determined for all diagrams in figure \ref{fig:SU6Lambda3} and are shown together with the flop transitions between them in figure \ref{fig:SU6Lambda3Fibs}. 
Note, as we explained in section \ref{sec:MiniRep}, in this instance the monodromy-reduced phases have a flop diagram given by the non-affine $E_6$ Dynkin diagram. This is consistent with the fact that all the monodromy-reduced fibers are obtained by deleting one of the non-affine nodes of the $IV^*$ Kodaira fiber, as we will now show by explicit computation.


Consider the phase at the top of figure \ref{fig:SU6Lambda3Fibs}. The extremal rays are 
\be
\mathcal{K}_{\includegraphics[width=1cm]{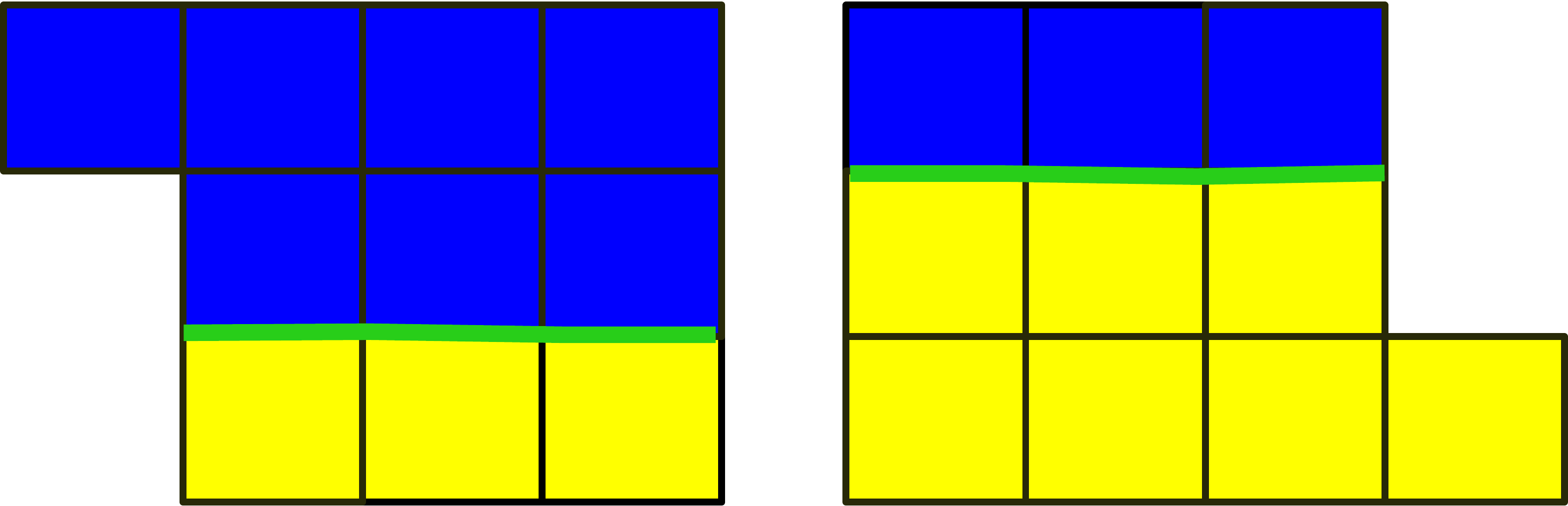}} = \left\{ F_1, F_2, F_3, F_4,  C_{126}^- \right\} \,.
\ee
The extended node is always obtained by the linear combination $F_0 = - \sum_i F_i$. 
It is clear that $F_5$ splits, as adding $\alpha_5$ crosses the anti-Dyck path (i.e. changes the sign of the weights)
\be
F_5 \quad \rightarrow \quad C_{126}^- + C_{345}^+ +F_1+ 2 F_2 + F_3 \,,
\ee
which implies multiplicities $2, 3, 2$ for $F_1, F_2, F_3$, respectively. 
Under the $\mathbb{Z}_2$ automorphism, the two generators $C_{126}^-$ and $C_{345}^+$ are identified, and it has multiplicity 2. 
To determine the intersections between the $C_{126}^-$ and the remaining generators, we apply the rules derived in section \ref{sec:GeofromBox}, i.e. if adding a root $\alpha_i$ to the weight $L_1 + L_2 + L_6$ changes/retains the sign of curve, the intersection with $F_i$ is $\mp1$, and thus the inner product $\langle C_{126}^-,  D_i\rangle =\pm1$. 
From the diagram we obtain $C_{126}^- \cdot D_2 = +1$, i.e. these intersect transversally, and has trivial intersections with the other $F_i$ except
\be
 C_{126}^-  \cdot D_5  = -1 \quad \Rightarrow \quad C_{126}^- \cdot  C_{345}^+ = +1  \,.
\ee
After the quotient, the fiber is as shown in figure \ref{fig:SU6Lambda3Fibs}. 

The second fiber type has extremal rays 
\be
\mathcal{K}_{\includegraphics[width=1cm]{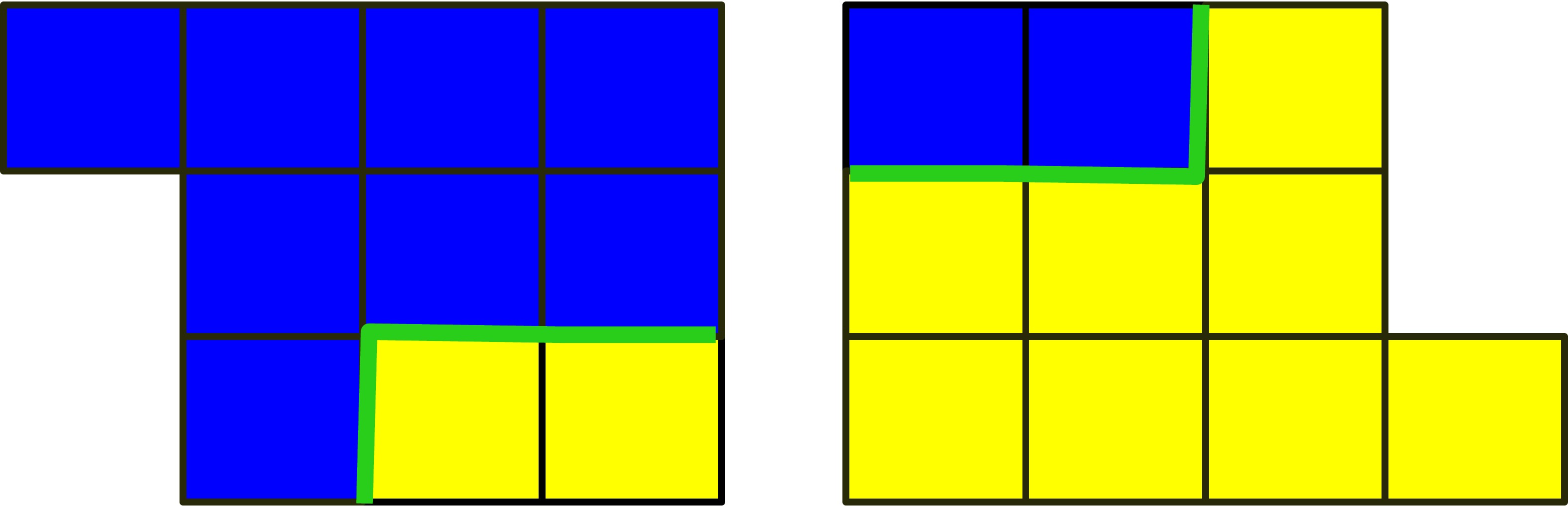}} = \left\{ F_1, F_3, F_4, C_{126}^+, C_{136}^-\right\} \,,
\ee
and the following curves become reducible
\be
\ba
F_{2} \quad &\rightarrow\quad C_{126}^+ + C_{136}^- \equiv C_{245}^+ + C_{345}^- \cr
F_{5}  \quad &\rightarrow\quad C_{136}^- + C_{245}^+ + F_1 + F_3 \,.
\ea
\ee
Under the $\mathbb{Z}_2$ quotient the curves $C_{126}^+$ and $C_{345}^-$ as well as $C_{136}^-$ and $C_{245}^+$ are identified. The multiplicities are then $2,2,3$ for each of $F_1, F_3, C_{136}^+$ and $1$ for the remaining generators. 
The intersections are obtained by noting that
\be
C_{126}^+ \cdot F_5 = +1 \,,\qquad C_{126}^+ \cdot F_2 = -1 \,,\qquad 
C_{136}^- \cdot F_2  = -1 \,,\qquad  C_{136}^- \cdot F_5 = -1 \,,\qquad C_{136}^- \cdot F_1 = +1 \,.
\ee
This again gives rise to a non-standard fiber, which is not an affine $E_6$ Dynkin diagram. 

The third fiber type has extremal rays 
\be
\mathcal{K}_{\includegraphics[width=1cm]{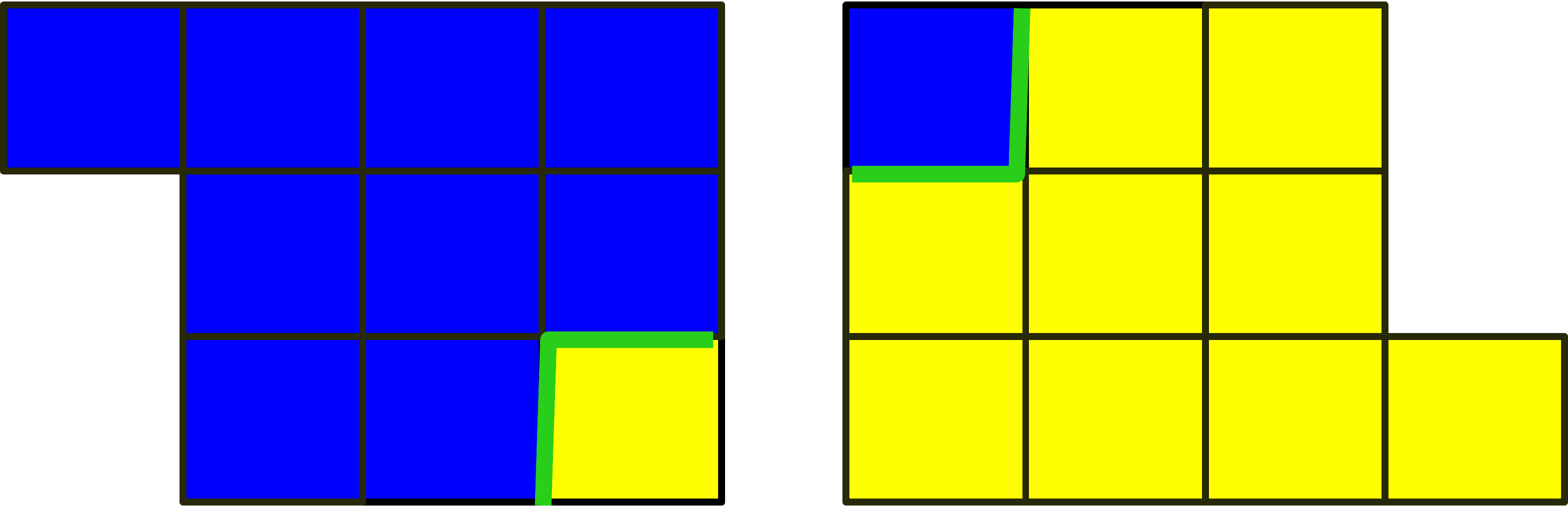}} = \left\{  F_2, F_4, C_{145}^+, C_{136}^+, C_{235}^+  \right\}
\ee 
where again we quotiented out the $\mathbb{Z}_2$ action, to identify $C_{145}^+$ with $C_{236}^-$ etc. The following curves are reducible
\be
\ba
F_{1} \quad &\rightarrow \quad C_{145}^+ + C_{245}^- \equiv C_{136}^+ + C_{236}^- \cr
F_{3} \quad &\rightarrow \quad C_{136}^+ + C_{146}^- \equiv C_{235}^+ + C_{245}^- \cr
F_{5} \quad &\rightarrow \quad C_{145}^+ + C_{146}^- \equiv C_{235}^+ + C_{236}^- \,.
\ea
\ee 
Thus each of the $C_{ijk}$ appears with multiplicity $2$ in the fiber, and the irreducible $F_i$ with multiplicity $1$. 
Intersections are again obtained from the diagram as usual and are 
\be
\ba
&C_{236}^- \cdot F_1 = -1 \,,\quad C_{236}^- \cdot F_3 = +1 \,,\quad C_{236}^- \cdot F_5 = -1 \cr
&C_{136}^+ \cdot F_1 = -1 \,,\quad C_{136}^+ \cdot F_3 = -1 \,,\quad C_{136}^+ \cdot F_5 = +1 \,,\quad C_{136}^+\cdot F_2= +1\cr
& C_{235}^+ \cdot F_1= +1 \,,\quad  C_{235}^+ \cdot F_3= -1 \,,\quad C_{235}^+ \cdot F_5= -1 \,,\quad C_{235}^+ \cdot F_4= +1 \,.
\ea
\ee
The resulting fiber does not correspond to a  standard Dynkin diagram, but could be described as an affine $E_6$ without the middle node. 

The remaining fiber types on the left column in figure \ref{fig:SU6Lambda3Fibs} are simple extensions of the analysis done so far. The remaining case is the one on the right hand side. In this case the extremal rays are
\be
\mathcal{K}_{\includegraphics[width=1cm]{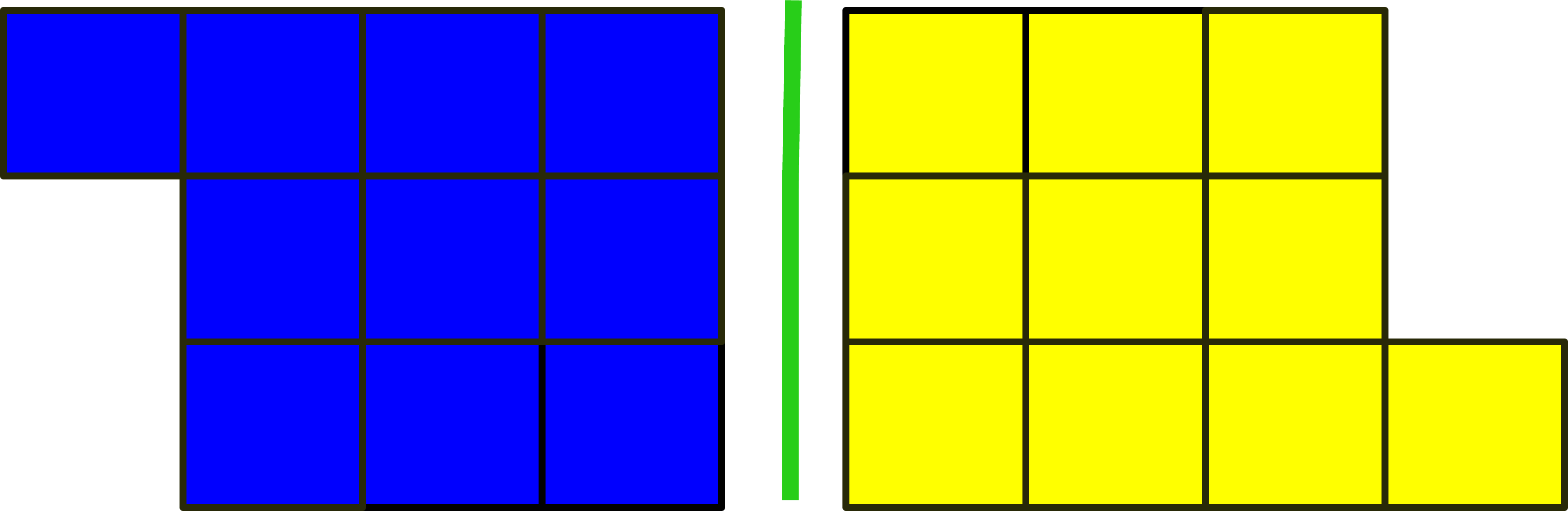}} = \left\{F_1, F_2, F_4, F_5, C_{145}^- \right\}  \,,
\ee
where $C_{145}^-$ and $C_{236}^+$ are identified under the $\mathbb{Z}_2$ quotient. 
In this case the splits are 
\be
F_{3} \quad \rightarrow \quad C_{236}^+ + C_{145}^- + F_1 + F_5 \,,
\ee
so that $C_{145}^+$ has multiplicity $3$, and $F_1$ and $F_5$ each multiplicity $2$. 
The intersections are
\be
C_{145}^- \cdot F_1 = + 1 \,,\quad 
C_{145}^- \cdot F_3= -1 \,,\quad 
C_{145}^- \cdot F_5 = +1 \,.
\ee
From these it follows that 
\be
F_0 \cdot C_{145}^- = -1  \,,
\ee
which means that $F_0$ (which corresponds to $L_6-L_1$) does not intersect this extremal ray transversally, but splits
\be
F_0 \quad \rightarrow \quad C_{145}^- + C_{456}^+ \,.
\ee
The resulting additional curve, $C_{456}^+$ is the component of the fiber that remains large in the singular limit, and thus corresponds to the zero section. 

In summary the new fiber types can be characterized by deleting one (non-extended) node in the affine Dynnkin diagram of the $IV^*$ Kodaira fiber.

\begin{figure}
    \centering
    \includegraphics[width=15cm]{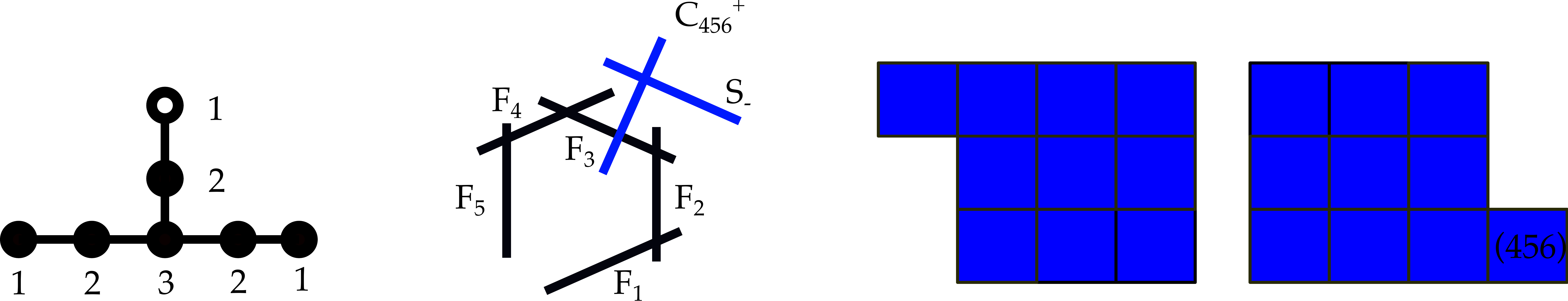}
    \caption{Example phase and fiber type of the $U(6)$ theory with  $\Lambda^3{\bf 6}$, corresponding to trivial monodromy action, with standard Kodaira $IV^*$ fiber. Description of the various diagrams is as in figure \ref{fig:SU6Lambda3Fibs}.}   \label{fig:U6Lambda3Fib}
\end{figure}

Finally, recall that for trivial $\Gamma=\mathbb{Z}_2$ the fibers will be of Kodaira $IV^*$ type. These correspond to phases of the $U(6)$ theory, which include a singlet given by $S_-= -\sum_{i=1}^6 L_i$.  It is a bit subtle to see this so that we will exemplify it with the all $+$ (blue) box graph in figure \ref{fig:U6Lambda3}. 
The extremal rays are
\be
\mathcal{K}_{\includegraphics[width=1cm]{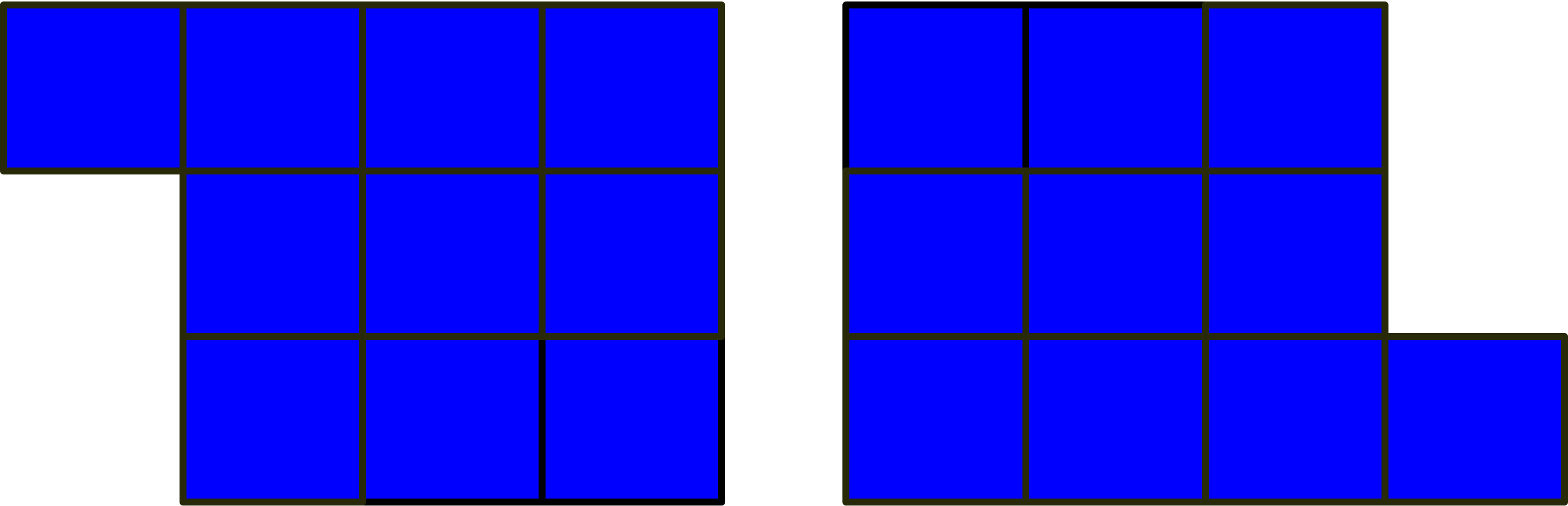}} = \left\{ F_1, F_2, F_3, F_4, F_5, C_{456}^+ \right\} \,.
\ee
Intersections are $C_{456}^+ \cdot F_3 = +1$ and 0 for all other Cartans $F_i$, so that $C_{456}^+ \cdot F_{0} = -1$, which implies that $F_0$ splits in this case
\be
F_0 \quad \rightarrow \quad  C_{456}^+ + S_-   + C_{236}^+ \,.
\ee
 Furthermore $C_{236}^+$ can be written in terms of the extremal generators as $C_{236}^+ = F_4 + 2 F_3 + F_2 + C_{456}^+$, so that overall
\be
F_0 \quad \rightarrow \quad 2 C_{456}^+ + F_4 + 2 F_3 + F_2 + S_- \,,
\ee
which precisely results in the correct multiplicities for a type $IV^*$ fiber, and the intersections comply with this as well: $S_-$ is the extended node, and intersects $C_{456}^+$, which has multiplicity 2, and intersects $F_3$ (multiplicity 3), as shown in figure \ref{fig:U6Lambda3Fib}.


\subsection{Local models for fibers in codimension three}
\label{subsec:FibCodim3}

The singularity enhancement in codimension three follows a similar pattern.
Along certain codimension-three loci $\Psi_{\mu}$ in the base, contained
in the union of all the codimension-two loci $\Sigma_{\alpha}$, there
are further increases in singularity type.  For such $\Psi_{\mu}$,
this increase  can be measured
by choosing a general point $P_\mu$ of $\Psi_\mu$,
and taking a general threefold $T_\mu\subset B$ which is transverse to
$\Psi_\mu$ at $P_\mu$.  We choose a general map $\varphi_\mu:T_\mu\to \mathbb{C}^2$ such that $\varphi_\mu(P_\mu)=0$.

If the Weierstrass coefficients $f$ and $g$
have multiplicities at $P_\mu$ which exceed $4$ for $f$ and $6$ for $g$,
then there is no desingularization of the Weierstrass model which 
is flat \cite{codimthree}, which means that there would be tensionless
strings in the low-energy theory.  Thus, for a well-behaved F-theory
model, those multiplicities will not be exceeded, and the Weierstrass
model over the curve $\varphi_\mu^{-1}(0)$ will have a well-defined
Kodaira type at $P_\mu$.

There is again a partial simultaneous resolution of singularities for
the two-parameter family of surfaces $\varphi_\mu^{-1}(u)$, $u\in\mathbb{C}^2$,
$\|u\|<\epsilon$.  So, although the Kodaira classification tells us about
the singularity type of the singular fiber, it does not predict the resolution
(as explicitly shown in \cite{Esole:2011sm, MS, Lawrie:2012gg}).  

The central fiber will be associated to a root system $\widehat{\mathfrak{g}}$,
each codimension-two locus which passes through our codimension-three
locus will be associated to a root system
$\widetilde{\mathfrak{g}}_j\subset \widehat{\mathfrak{g}}$,
and the gauge algebra $\mathfrak{g}$ of the local model will be 
contained in the intersection of all of the $\widetilde{\mathfrak{g}}_j$s.
The analysis of the phases proceeds as we explain in section~\ref{subsec:reducibility}.


\subsection{Fibers of $E_6$ type in codimension three}
\label{subsec:E6Codim3}


\begin{figure}
    \centering
    \includegraphics[width=14cm]{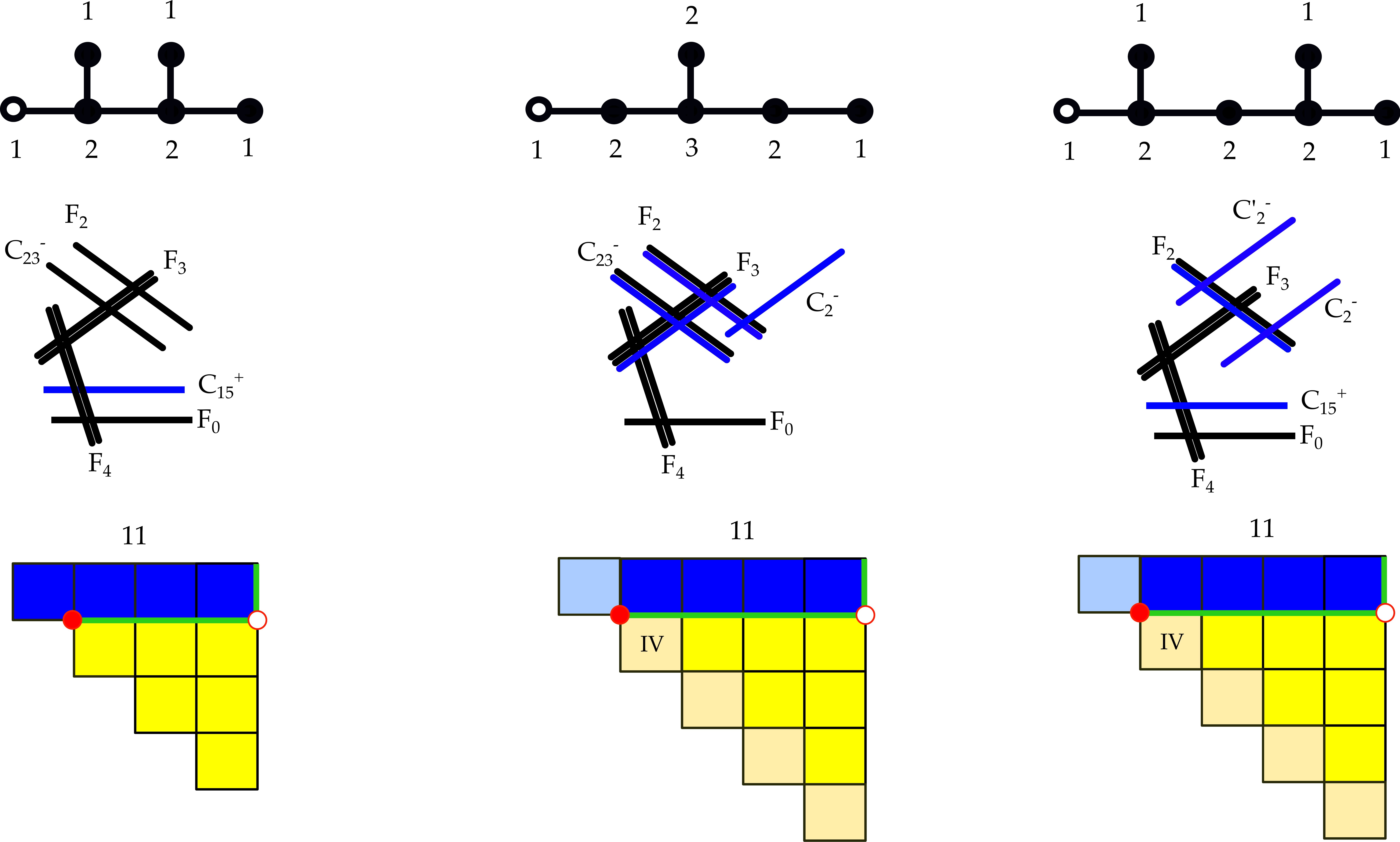}
    \caption{On the left hand side the fiber type and box graph of the $SU(5)$ with {\bf 10} matter in phase 11 is shown, which is a standard $I_1^{*s}$ fiber corresponding to a $D_5$ enhancement. Along the $E_6$ locus, obtained by combining this phase with the  {\bf 5} matter in phase IV as explained in section \ref{subsec:reducibility}, the curve $C_{1,5}^+$ becomes reducible, and  the fiber is of the type shown in the middle: note that this fiber type has multiplicity $2$ along the single-node leg. The fiber type is shown in the central line including the components that arise from the splitting of $C_{1,5}^+$. On the right hand side we show the analogous splitting along the $SO(12)$ locus for the same phase diagram. In this case $C_{2,3}^{-}$  becomes reducible, yielding a Kodaira type $I_2^{*s}$ fiber. }   \label{fig:Phase11IVFib}
\end{figure}
 

We shall now given an explicit example for the codimension-three phenomenon explained in the last section. 
The box graphs for $SU(5)$  with ${\bf 5}$ and ${\bf 10}$ matter that are relevant for determining the fibers in codimension three were discussed in section \ref{subsec:reducibility}. In particular, the combined box graphs for fundamental and anti-symmetric representation are shown in figure \ref{fig:SU5FlopsAF}. The codimension-three fibers that arise at the $E_6$ enhancement loci are known to not be Kodaira type $IV^*$ fibers \cite{Esole:2011sm, MS, Lawrie:2012gg}. 
As was observed in \cite{Hayashi:2013lra}, there are 12 distinct small resolutions, of which only six correspond to the ones in \cite{Esole:2011sm}. It is therefore interesting to determine all possible codimension-three fibers of $E_6$ type using the description in terms of decorated box graphs. 

The analysis in section \ref{subsec:reducibility} shows how to determine which extremal generators of the relative K\"ahler cone become non-extremal along codimension-three loci, i.e. which of these curves become reducible and split into a combination of effective curves. There are two group theoretic ways that this happens, either the splitting is compatible with $E_6$ or with $SO(12)$, which are characterized by either having two curves carrying weights of the {\bf 10} matter\footnote{Note that generically they will then be identified under monodromy, however this case is distinct from two {\bf 5} and one ${\bf 10}$ curve intersecting in that the relevant Yukawa coupling is consistent with the $E_6$ algebra. } or two of the ${\bf 5}$ involved in the splitting, respectively. To actually determine the fibers above the $E_6$ codimension-three loci, we need to consider only the former type of splitting. We consider the intersection of two codimension loci giving rise to the  ${\bf 10} \times {\bf 10}\times {\bf 5}$ interaction.
 
\begin{figure}
    \centering
    \includegraphics[width=16cm]{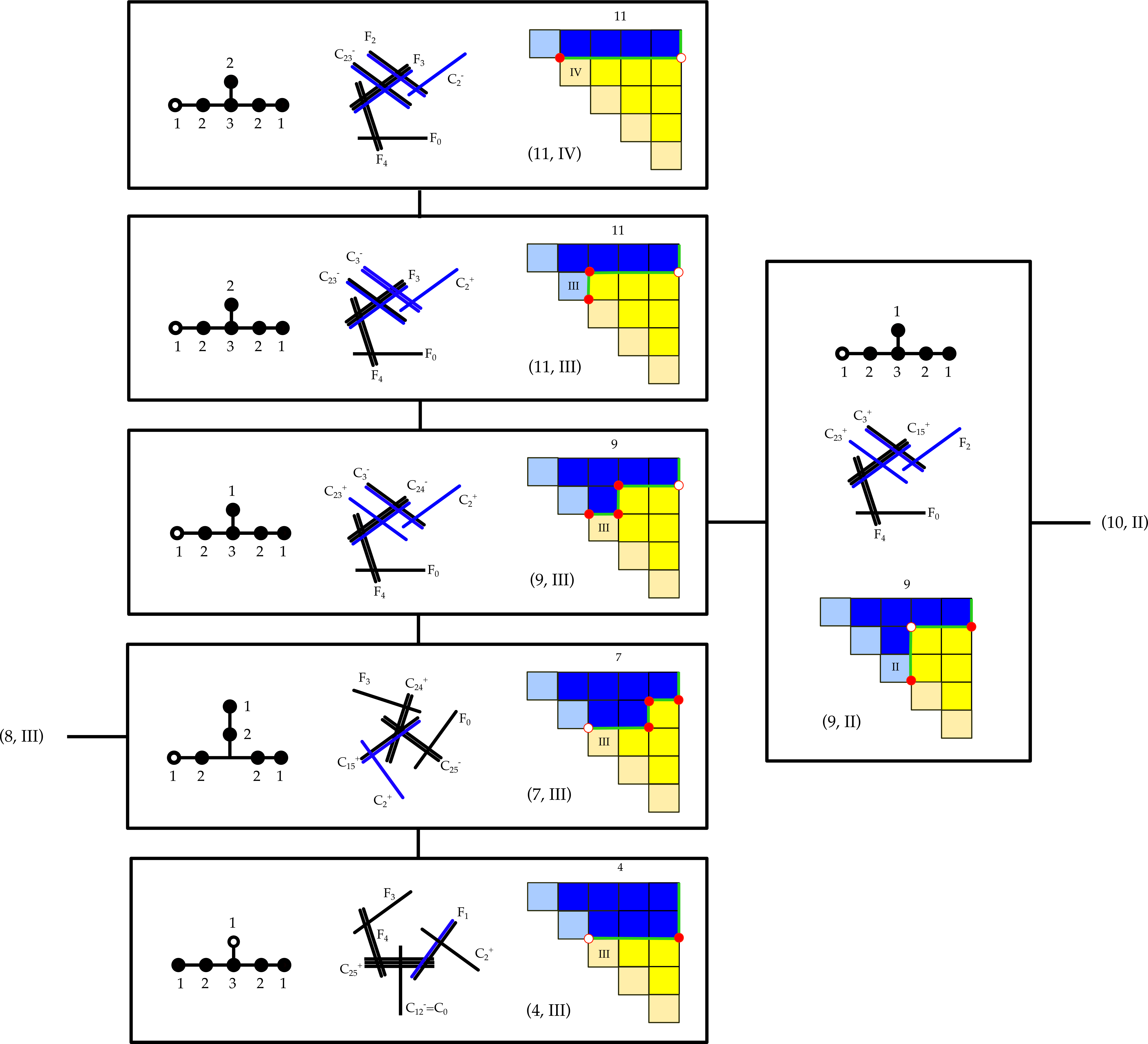}
    \caption{Codimension-three fiber degenerations along $E_6$ locus of a codimension-one $SU(5)$ fiber, obtained from the combined decorated box graphs $(i, I)$, where $i$ indicates the phase with matter in the antisymmetric representation, and $I$ the one for the fundamental. As in figure 
    \ref{fig:SU6Lambda3Fibs} from left to right the intersection graph, fiber (including multiplicities) and corresponding box graphs are shown. Black boxes are connected along  flop transitions. Note that this is half of the phases appearing in the flop diagram figure \ref{fig:SU5AFPhaseDiag}, the other half is simply obtained by relabeling the simple roots of $SU(5)$ in the reverse order. Again the flop transitions are shown as lines between the black boxes, including the flops into the other half of the flop diagram in figure \ref{fig:SU5AFPhaseDiag}. } 
     \label{fig:SU5AFCodim3Fibs}
\end{figure}

The fiber types for each of the resolutions corresponding to the phases in figures  \ref{fig:SU5AFPhaseDiag} and \ref{fig:SU5FlopsAF} are shown in figure \ref{fig:SU5AFCodim3Fibs}, including the flops which are shown as connecting the boxes. We only present half of the fibers appearing in  figures  \ref{fig:SU5AFPhaseDiag} and \ref{fig:SU5FlopsAF}, as the other half is obtained by a simple relabeling of the roots of $SU(5)$:  ($1\leftrightarrow 4$, $2\leftrightarrow 3$). The ``hexagon" in figure \ref{fig:SU5AFPhaseDiag} is obtained by combining two of these $E_6$-Dynkin diagram shaped flop diagrams in figure \ref{fig:Phase11IVFib}.  Note that the fibers for the phases (9, II), (9, III) and (7, III) correspond precisely to the ones obtained by Esole and Yau in \cite{Esole:2011sm}, however the fibers appearing in (4, III), (11, III) and  (11, IV) are previously unknown. The fiber types agree with the ones obtained from $SU(6)$ with $\Lambda^3{6}$ in codimension two with monodromy in figure \ref{fig:SU6Lambda3Fibs}, however the flop transitions connect them differently, which is not surprising from the point of view of the Weyl group action. Note that the fibers along the $SO(12)$  codimension-three loci are standard Kodaira fibers.

From the analysis, which is again very similar to the one in section \ref{sec:FibE6Mono} we can determine the fiber types in codimension three. 

First consider the phase  (11, IV) in figure \ref{fig:SU5FlopsAF}. Phase  11 of $SU(5)$ with only matter in the ${\bf 10}$ representation has extremal generators $\mathcal{K}_{11}=\{C_{2,3}^-, C_{1,5}^+, F_2, F_3, F_4 \}$ as we determined in (\ref{K510})  and the splitting in codimension two is 
\be
F_{1} \quad \rightarrow \quad C_{2,3}^- + C_{1,5}^+ + F_3 + F_4 \,.
\ee
The fiber type is shown in figure \ref{fig:Phase11IVFib}.
Including ${\bf 5}$ matter corresponds to considering the combined box graphs that are consistent with the flow rules, as explained in section \ref{subsec:reducibility}. For phase $11$, there are two choices of ${\bf 5}$ matter phases: IV and III, as shown in figure \ref{fig:SU5FlopsAF}. Consider the case (11, IV).  
 The extremal generators above the codimension-three $E_6$ locus are\footnote{Note that for the analysis of the fibers above the $E_6$ locus, we do not allow splittings that correspond to $D_6$ points, e.g. a {\bf 5} curve splitting into another ${\bf 5}$ and a ${\bf 10}$ matter curve. The extremal points shown in figure \ref{fig:SU5FlopsAF} take both splittings into account. For instances, if we take $C_3^+$ in this case as extremal generator, the splittings would realize the $D_6$, not $E_6$, point. Also, these are then different from the generators that we discussed in (\ref{K510}). } 
 \be
 \mathcal{K}^{E_6}_{ \includegraphics[width=0.7cm]{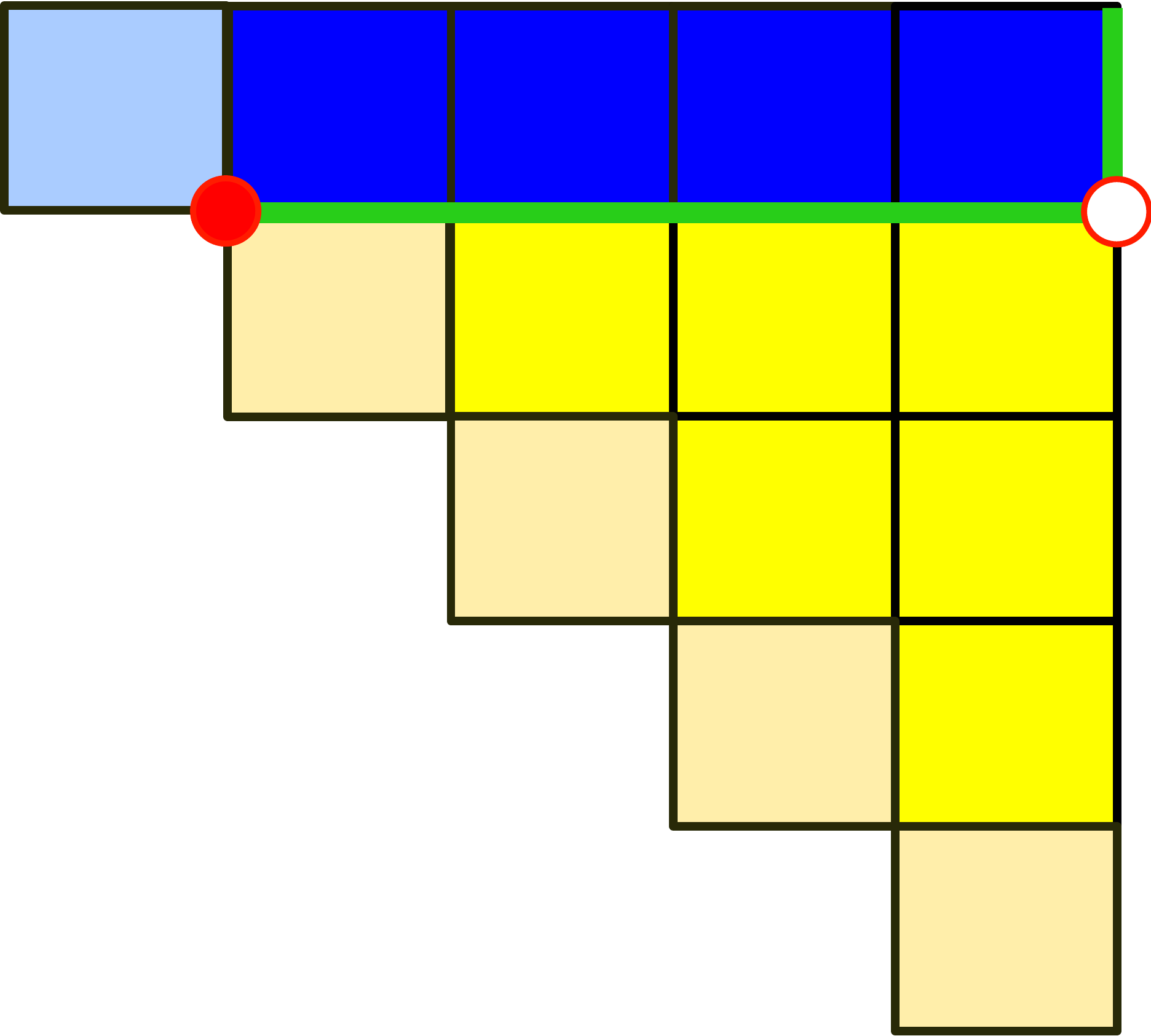}}= \{C_{2,3}^-, C_{2}^-, F_2, F_3, F_4\} \,.
 \ee
 and $C_{15}^+$ splits as follows
 \be
 C_{15}^+ \quad \rightarrow \quad  C_{23}^- + F_2 + F_3 + C_2^- \,.
 \ee
Intersections are determined as explained e.g. in section \ref{sec:GeofromBox}.

The case of phase (11, III), the new extremal set above the $E_6$ codimension 3 locus is
\be
 \mathcal{K}^{E_6}_{\includegraphics[width=0.7cm]{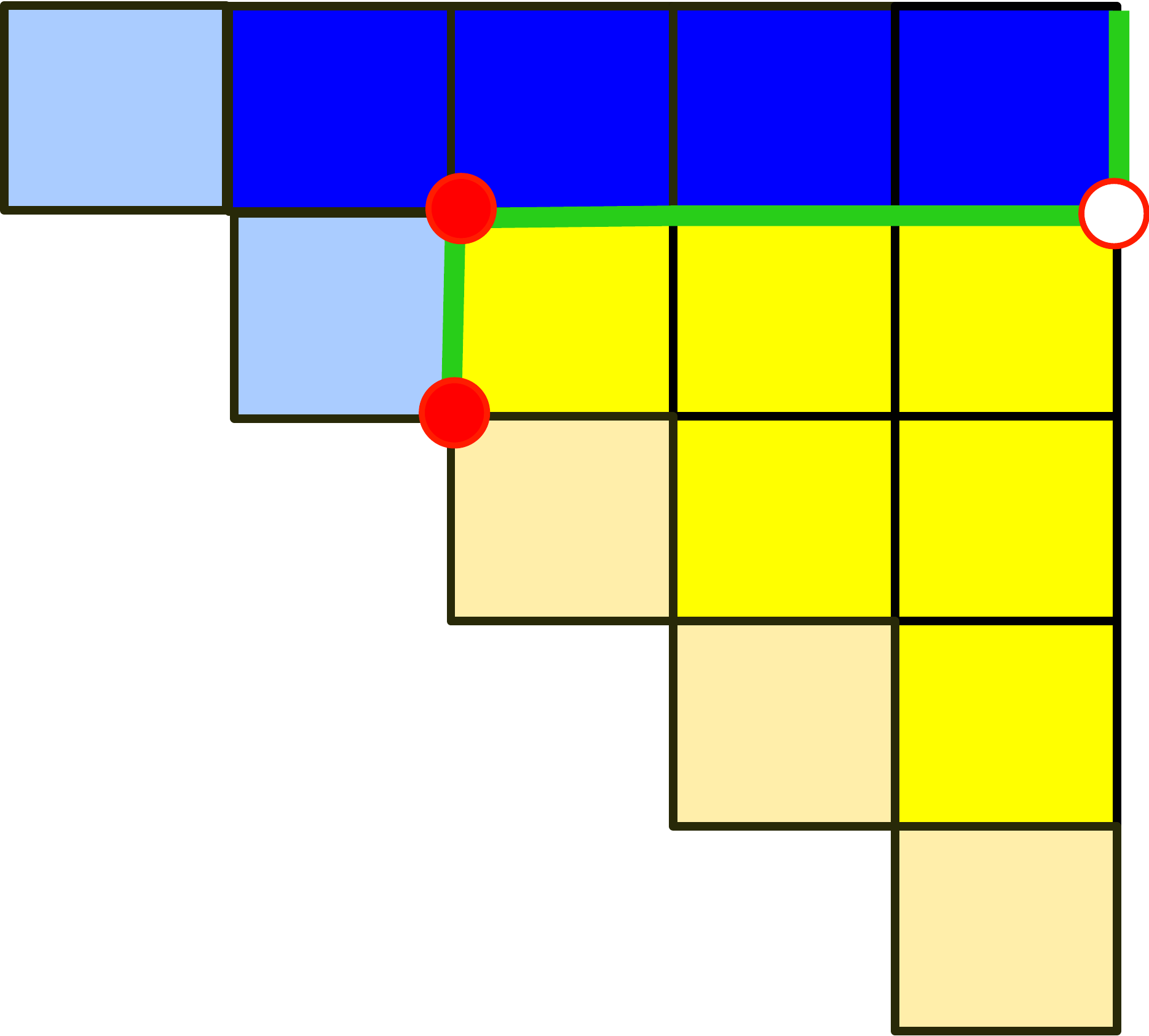}}=  \{C_{23}^-, C_2^+, C_3^-, F_3, F_4 \}
 \ee
and the splitting is
\be
\ba
C_{15}^+ &\quad \rightarrow \quad C_{23}^-+ F_3 + C_3^-  \cr
F_2 &\quad \rightarrow \quad C_2^+ + C_3^- \,.
\ea
\ee
The intersections are determined to be
\be
C_{3}^- \cdot F_3 = +1  \,,\qquad 
C_2^+ \cdot (C_2^+ + C_3^- )= C_{2}^+ \cdot F_2  = -1 \quad \Rightarrow \quad  C_2^+ \cdot C_3^- = 1
\ee
resulting in the intersection graph shown in figure \ref{fig:SU5AFCodim3Fibs}.


Next consider the combinations of phase (9, III). Note that the extremal generators are 
$\mathcal{K}_9 = \{C_{2,3}^+, C_{2,4}^-, C_{1,5}^+, F_2, F_4\}$. Along the $E_6$ locus in the phase 
$(9, III)$ the extremal generators of the cone of curves is
\be
\mathcal{K}^{E_6}_{\includegraphics[width=0.7cm]{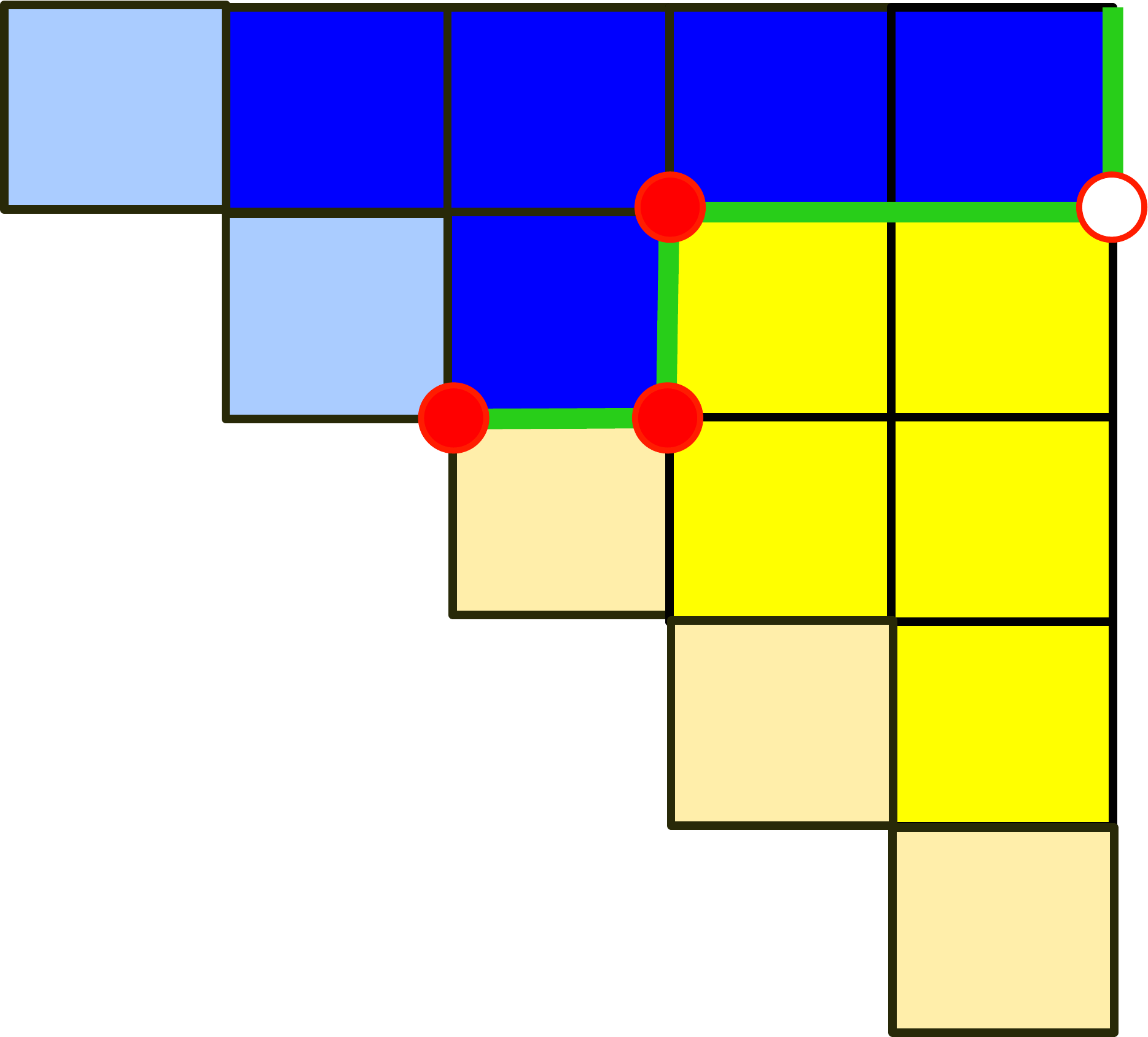}}
=  \{C_{2}^+ , C_{3}^- , C_{2,3}^+ ,  C_{2,4}^-,   F_4 \}
\ee
with splitting
\be\ba
C_{1,5}^+ &\quad\rightarrow\quad C_{2,4}^- + C_3^- \cr
F_2 & \quad\rightarrow\quad  C_{2}^+ + C_3^- \,.
\ea\ee
The intersections are
\be
C_{3}^- \cdot C_2^+ = 1 \,,\qquad  C_3^- \cdot C_{2,4}^- = 1\,,
\ee
with the resulting fiber type as in figure  \ref{fig:SU5AFCodim3Fibs}. 
Starting with phase 9, we can also construct the combined phase (9, II), which at the $E_6$ codimenion three locus  implies that the extremal generators 
\be
\mathcal{K}^{E_6}_{\includegraphics[width=0.7cm]{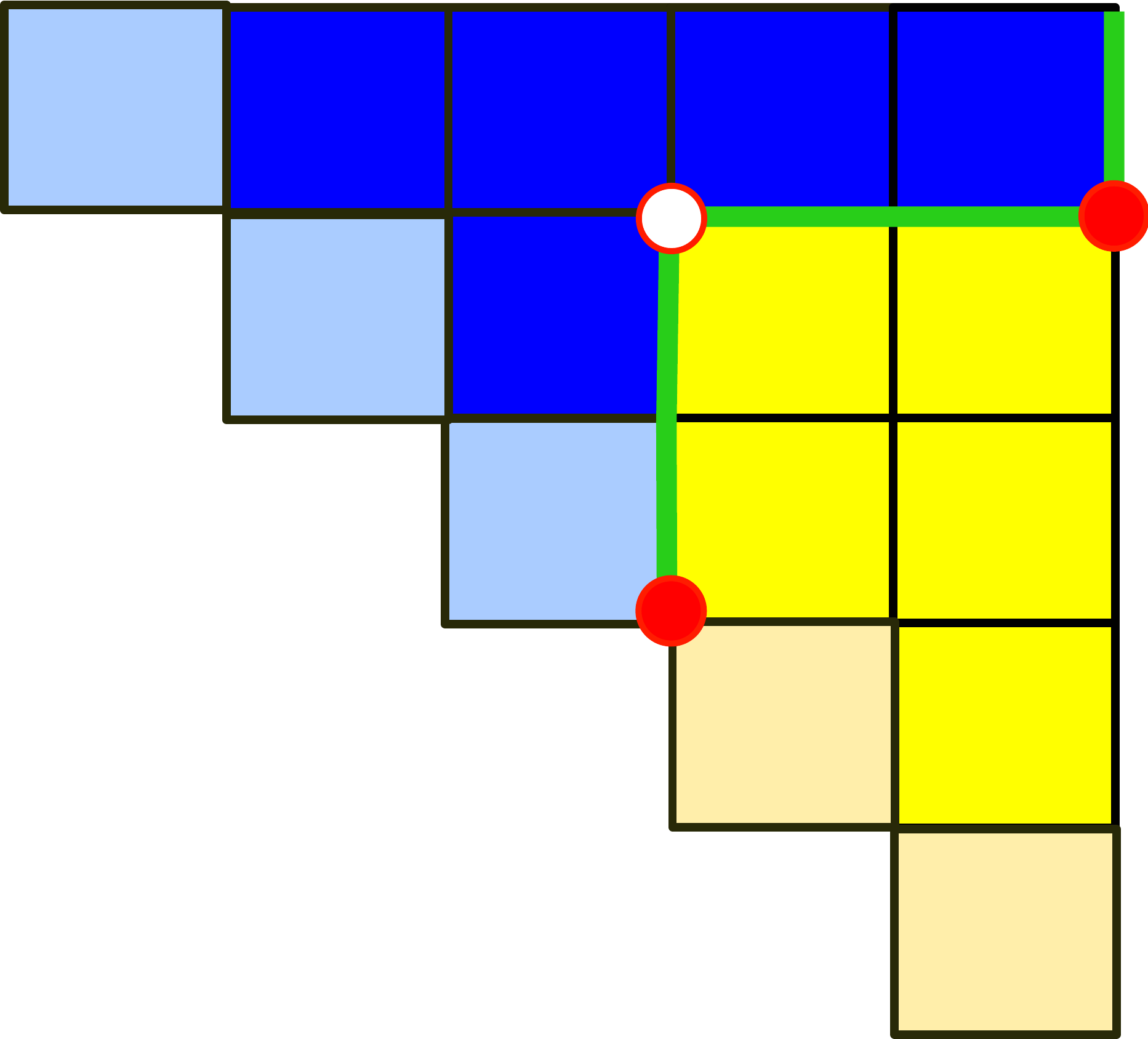}}
=  \{C_{3}^+, C_{2,3}^+ , C_{1, 5}^+ , F_2, F_4 \}
\ee
with the splitting of $E_6$ type given by
\be
C_{2,4}^- \quad \rightarrow \quad C_{3}^+ + C_{1, 5}^+ \,.
\ee
The relevant intersections are read off from the box graph as
\be
C_3^+ \cdot F_2 = 1 \,,\qquad C_{3}^+ \cdot C_{1,5}^+ = 1 \,,
\ee
resulting in the intersection and fiber type shown in figure \ref{fig:SU5AFCodim3Fibs}. 

For phase 7 with the cone generated by $\mathcal{K}_{7} = \{C_{2,4}^+, C_{2,5}^-, C_{3,4}^-, C_{1,5}^+, F_3 \}$ the only consistent combined phase is (7, III), which has extremal generators along the $E_6$ locus 
\be
\mathcal{K}^{E_6}_{\includegraphics[width=0.7cm]{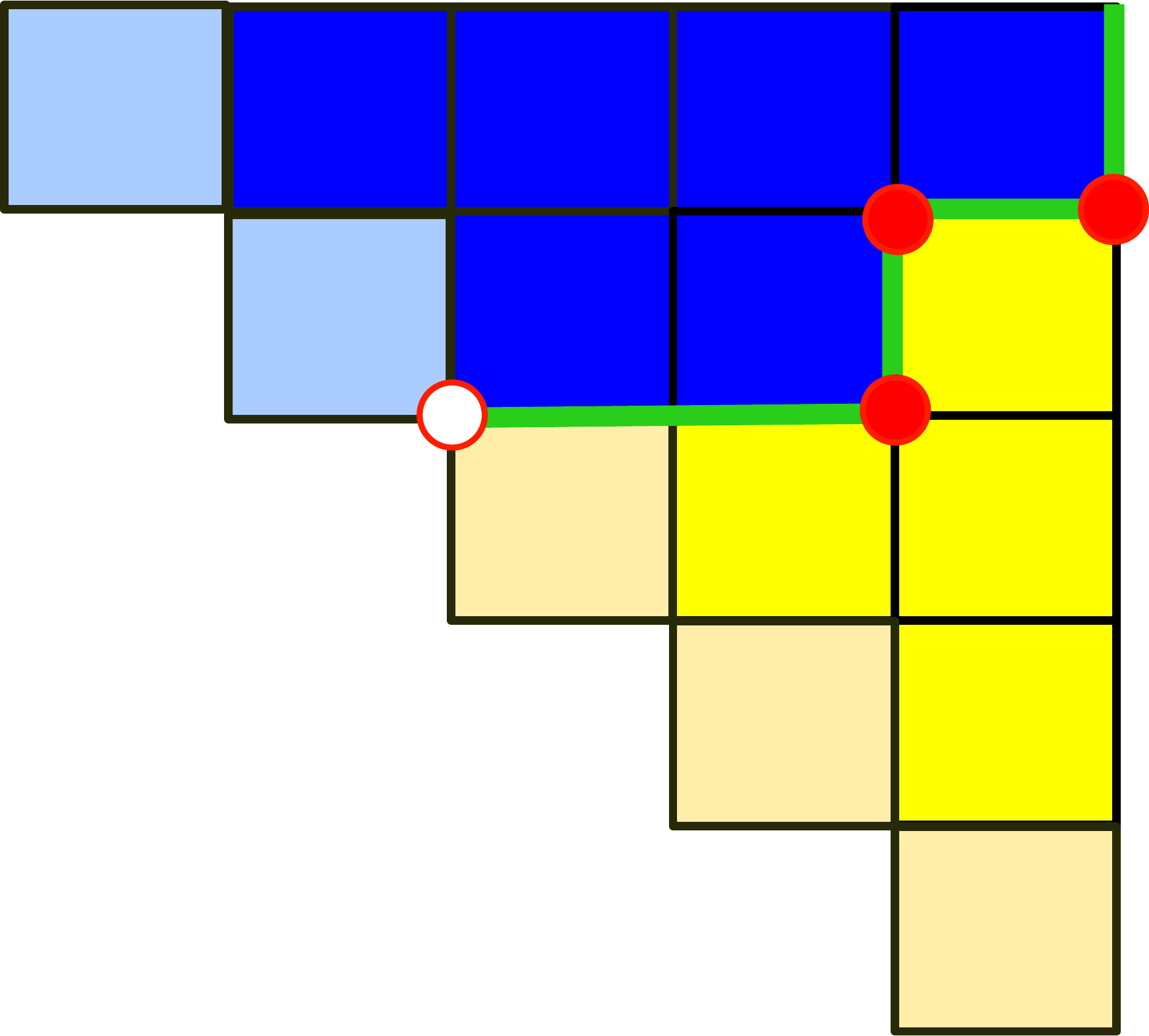}}
=  \{C_{2}^+, C_{2,5}^- , C_{1, 5}^+ , C_{2,4}^+ , F_3 \} \,.
\ee
The curves that split starting from the fiber of the $SO(10)$ locus are
\be
C_{3,4}^- \quad\rightarrow\quad C_2^+ +  C_{1,5}^+ \,,
\ee
which results in the  intersections $C_2^+ \cdot C_{1,5}^+ = 1$ resulting in the fiber shown in figure \ref{fig:SU5AFCodim3Fibs}.

Finally consider (4, III) with cone of phase 4 generated by $\mathcal{K}_4=  \{C_{2,5}^+, C_{3,4}^-, F_1, F_3, F_4 \}$, then  the extremal generators along the $E_6$ locus are
\be
\mathcal{K}^{E_6}_{\includegraphics[width=0.7cm]{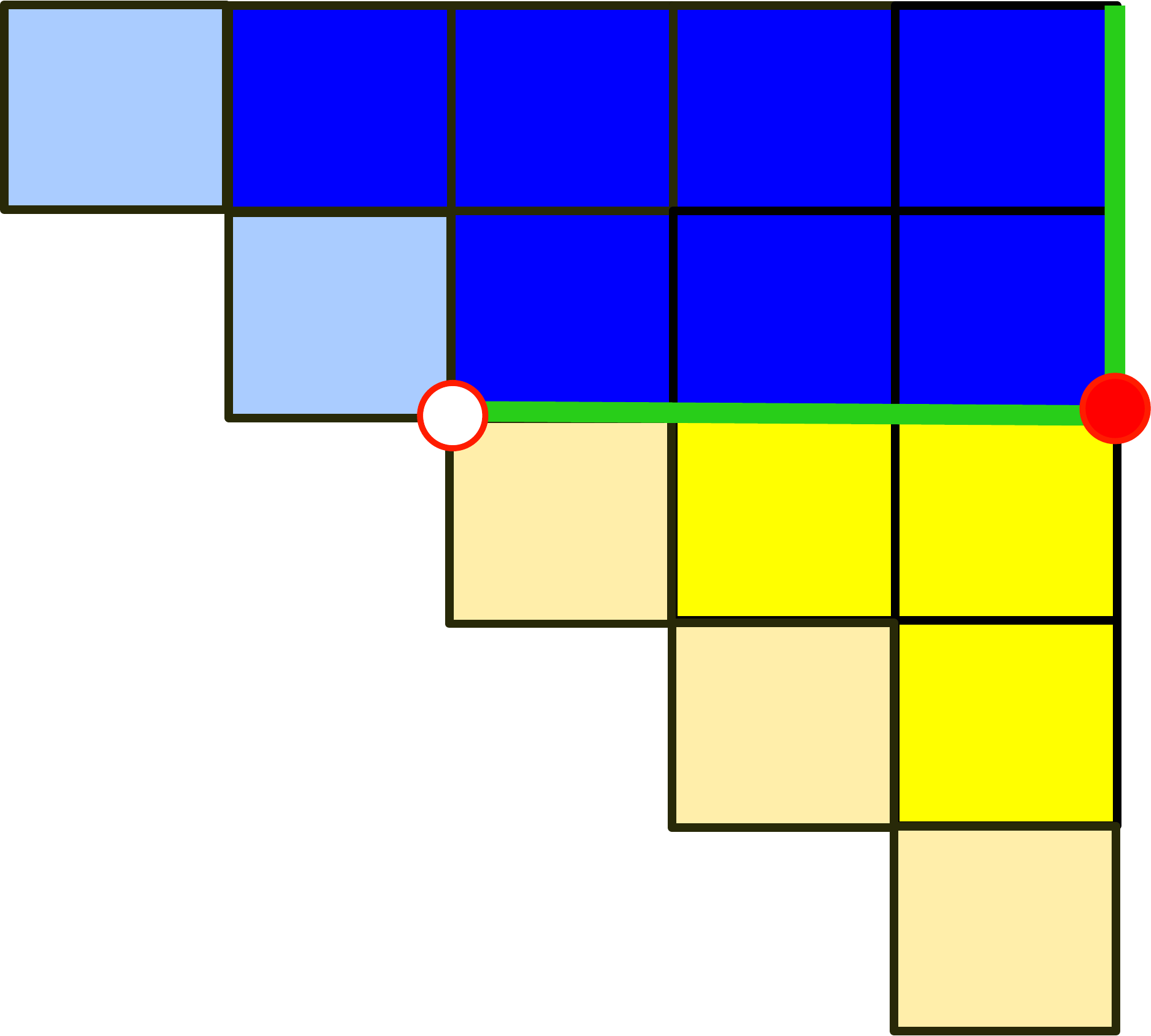}}=  
\{C_2^+ , C_{2,5}^+ , F_1, F_3, F_4 \}  \,.
\ee
In this case $F_0$ splits as well as $C_{3,4}^-$
\be
\ba
C_{3,4}^- & \quad \rightarrow\quad C_{2,5}^++ F_1 + C_2^+ \cr
F_0 & \quad \rightarrow\quad C_{2,5}^+ + C_0 \,,
\ea
\ee
where $C_0 = C_{1,2}^-$ realizes the new zero-section. The relevant intersections are $C_{0} \cdot C_{1,5}^+ = 1$ and $C_2^+ \cdot F_1 = 1$, resulting in the codimension three fibers in figure  \ref{fig:SU5AFCodim3Fibs}.

In all cases, note, that we could also consider the enhancement to $SO(12)$, and it is not too difficult to see that all of the fibers are of standard $I_2^{* s}$ type. For example for phase $(11, IV)$ the extremal generators of the cone of curves along the $SO(12)$ point are 
\be
\mathcal{K}^{SO(12)}_{\includegraphics[width=0.7cm]{SU5APhases11IV.pdf}}
=  \{ C_2^-, {C'}_2^-, C_{1,5}^+, F_2,  F_3, F_4 \}
\ee
and the additional splitting starting from the phase 11 of $SO(10)$ is given by
\be
C_{2,3}^-  \quad \rightarrow \quad C_2^+ + {C'}_2^+ + F_2 \,,
\ee
where the two curves with weight $L_2$ correspond to two distinct codimension two cones for an $SU(6)$ enhancement. The intersection is then as shown in figure \ref{fig:Phase11IVFib}.


\subsection{Fibers of $E_7$ type with monodromy}
\label{subsec:E7Mono}

\begin{figure}
    \centering
    \includegraphics[width=10cm]{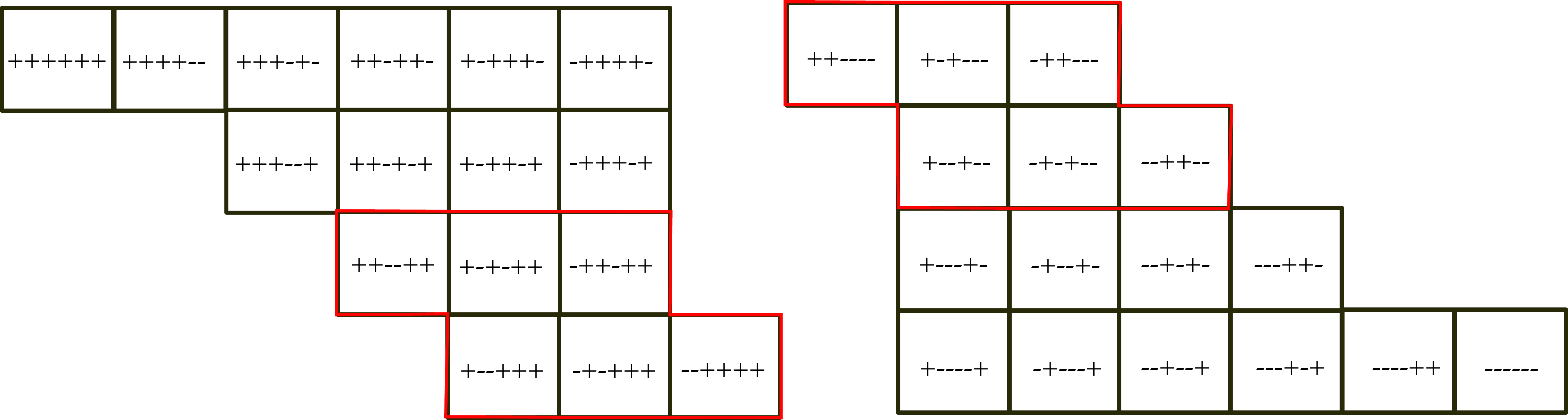}
    \caption{Representation graph for the ${\bf 32}$ spin representation of $SO(12)$. The red boxes are layered on top of each other. The $i$th entry in boxes correspond to the signs of ${1\over 2}L_i$.}\label{fig:SO1232}
\end{figure}

\begin{figure}
    \centering
    \includegraphics[width=13cm]{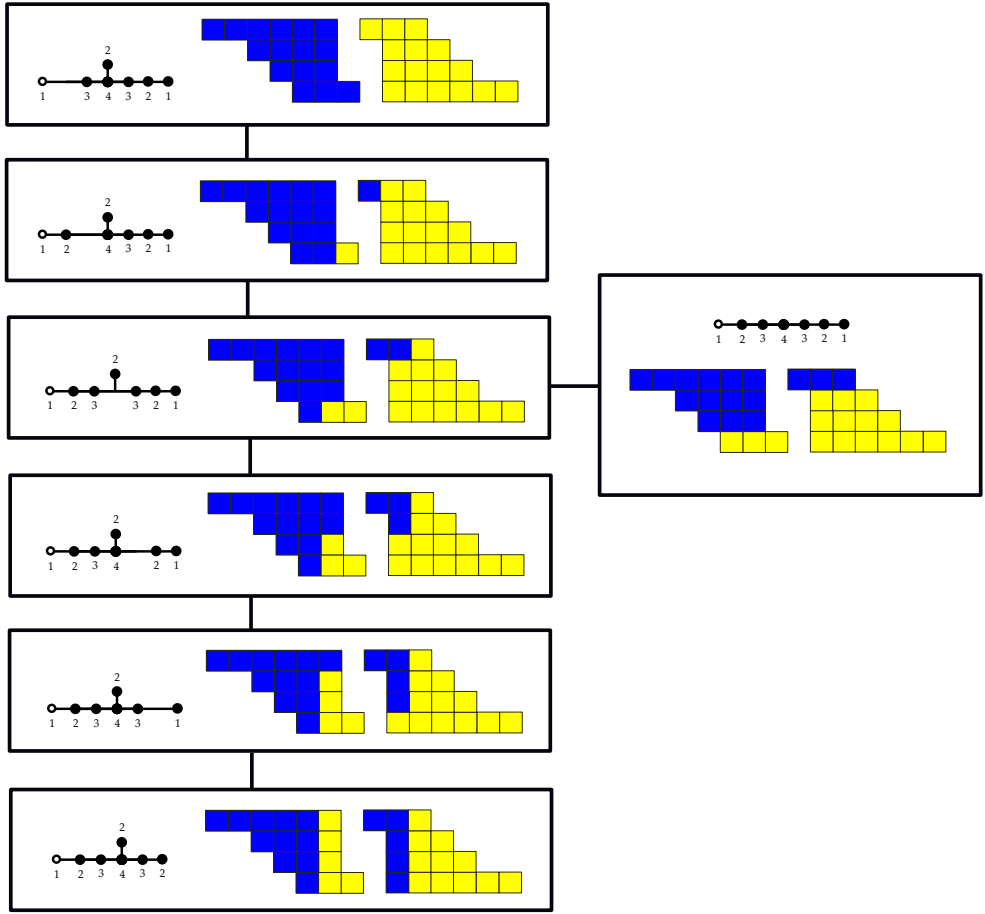}
    \caption{Box graphs and fiber types for $\mathbb{Z}_2$ monodromy-reduced $E_7$, obtained from  $SO(12)$ with ${\bf 32}$ representation. The fibers are obtained by deleting a single node from the Kodaira type $III^*$ fiber, and are connected by the flops indicated by the lines connecting the different phases. The phase diagram is the Dynkin diagram of $E_7$.} 
     \label{fig:SO12E7}
\end{figure}

Similarly, we can consider the case of $\mathfrak{e}_7$ with monodromy from the decomposition 
\be
\ba
\mathfrak{e}_7  &\quad \rightarrow \quad \mathfrak{so}(12) \oplus \mathfrak{su}(2)\cr
{\bf 133} &\quad \mapsto \quad ( {\bf 66}, {\bf 1}) \oplus  (  {\bf 1}, {\bf 3}) \oplus {({\bf 32}, {\bf 2})  }
\ea
\ee
The representation graph is shown in figure \ref{fig:SO1232}. The Weyl group of $\mathfrak{su}(2)$ acts as a sign change in the weights. Again, we can determine all the possible $\mathfrak{so}(12)\oplus \mathfrak{su}(2)$ fibers from the box graphs, and the number of phases is
\be
\left|{W_{\mathfrak{e}_7 }\over W_{\mathfrak{so}(12)}}\right|= 133 - 7 = \hbox{dim} (\mathfrak{e}_7) - \hbox{rank} (\mathfrak{e}_7) \,,
\ee
in agreement with our general arguments in section \ref{sec:MiniRep}: the phases form the quasi-minuscule representation (minus the zero weights). In addition we can consider the phases of the $\mathfrak{so}(12)$ theory with ${\bf 32}$ matter, which corresponds to imposing tracelessness. The resulting fibers are shown in figure \ref{fig:SO12E7}, and correspond to the monodromy-reduced $E_7$ fibers, which are characterized by deleting a single node in the standard $III^*$ Kodaira fiber. Note that the flop diagram in this case is given by the Dynkin diagram of $\mathfrak{e}_7$, as excepted from section \ref{sec:MiniRep}. 

\section{Comparing local and global models}
\label{sec:GlobalFib}
\subsection{Mordell-Weil group and $U(n)$ Phases }
\label{subsec:U1s}

Consider now a global Weierstrass model of an elliptic fibration.  For each codimension-two 
enhancement of singularities, we can carry out the local analysis
described in the previous section and find a pair of root systems
$\rootsystem\subset \widetilde{\rootsystem}$ associated to the codimension-two locus.  There may
be more than one such $\widetilde{\rootsystem}$ for a given $\rootsystem$.  For example, if $\rootsystem=A_{n-1}$
and there are $k$ fundamentals in the spectrum, then our
local models associate $k$ different enhancements 
$\widetilde{\rootsystem}_1$, \dots, $\widetilde{\rootsystem}_k$,
each isomorphic to $A_n$.  The point is that the total local weight
and coweight lattices associated to these enhancements must be generated
by all of the Cartan subgroups $H_j\subset G_{\widetilde{\rootsystem}_j}$, which of course
all intersect in $H$, the Cartan subgroup of $G_\rootsystem$.  Each enhancement
determines a curve in a particular fiber, and the classes of these
curves may or may not be independent of each other (or of the roots in
$\rootsystem$).

However, for a global F-theory model (on a good resolution of
a Weierstrass model, i.e., a smooth elliptic fibration
$X$ which is flat over its base $B$ and has a section),
we have a global description of the
lattice of divisors on $X$:  the divisors are generated by the section,
the pullbacks of divisors on the base, the divisors formed by
components of reducible fibers,
and the Mordell-Weil group of the elliptic fibration.  Since the F-theory
coweight lattice does not include the section or the pullbacks of divisors
on the base, the coweight lattice is precisely the divisors formed by
components of reducible fibers together with the Mordell-Weil group.
Now the divisors formed by components of reducible fibers are precisely
the coroots of the F-theory model (since each such component is a ruled
variety swept out by some rational curve whose class gives the corresponding
root).
Thus, any divisors on the coweight lattice beyond the coroots must arise
from elements of the Mordell-Weil group, i.e., additional rational sections of
$X\to B$ \cite{pioneG}.

In particular, we can determine from global properties of the model
whether or not there is a $\mathfrak{u}(1)$ factor in the gauge group, and if
not, there are relations among the local divisors near the codimension-two locus.
To illustrate this we consider $SU(5)$ models with an  extra rational section in section \ref{sec:SU5U1Ex}, and show that phases of the $U(5)$ theory that are not $SU(5)$  phases are realized in these geometries. 


\subsection{Higher-codimension fibers with trivial monodromy}
\label{sec:FibsNoMono}

Another global question arises in the context of the codimension-two and codimension-three fibers with possibility of monodromy, as we have discussed in terms of phases in section \ref{sec:OtherPhases} and geometry in section \ref{sec:EllipticFibsCodim}. In codimension two this occurs when the commutant $\mathfrak{g}_\perp$ of the gauge algebra in the higher rank group $\widetilde{\mathfrak{g}}$ is non-abelian as in (\ref{NACom}), and the Weyl group of the commutant can introduce monodromy in the fibers. 
If the elliptic fibration does not have extra rational sections, the fibers in codimension two  are monodromy-reduced. However, if there are extra sections, these can result in reducing the monodromy, realizing standard Kodaira type in codimension two. 
The number of additional sections that are required to construct the standard Kodaira fibers is given by the rank of the commutant $\mathfrak{g}_\perp$. 

For example, consider the  codimension-two locus with $E_6$ enhancement for an $SU(6)$ fibration in codimension one. As we have seen in section \ref{sec:FibE6Mono}, there is monodromy from the Weyl group action of the commutant, which yields non-Kodaira fibers of $E_6$ type, which are not standard type $IV^*$ fibers.  
Similarly issues can arise in codimension three, for instance for $SU(5)$ with an $E_6$ codimension-three locus.  The resolution by \cite{Esole:2011sm, MS} at the codimension-three $E_6$ singularity point does not yield an affine $E_6$ Dynkin diagram. 

To realize a Kodaira type $IV^*$ fiber in either codimension two or three, i.e. a phase of the theory with trivial monodromy, the complex structure needs to be tuned. 
The local Katz--Vafa field theory interpretation of the non-affine $E_6$ Dynkin diagram was given in \cite{Braun:2013cb}. Namely, we need to tune the complex structure of the Calabi-Yau fourfold such that the monodromy associated with the Weyl group of $SU(2)$, which is in  the orthogonal complement to $SU(5)$ or $SU(6)$ inside $E_6$, becomes trivial. 

Globally our analysis shows that a standard Kodaira type $IV^*$ fiber  can be obtained in codimension 2 or 3, if the elliptic fibration has  an extra rational section associated with the Cartan $U(1)$ of the $SU(2)$, which in particular trivializes the Weyl group of $SU(2)$. 
In practice, starting with a singular fibration with only $SU(5)$ gauge symmetry, in fact, the two requirements are that the {\bf 10} matter locus factors, i.e. in the standard Tate form $b_1 \rightarrow b_{1a} b_{1b}$, and furthermore that the model has an extra section, i.e. a $U(1)$ symmetry, under which the two {\bf 10} curves are charged differently\footnote{Note that in  \cite{Braun:2013cb} the tuning which resulted in just a factored ${\bf 10}$ curve was not enough to result in a type $IV^*$ fiber, and they had to further tune the complex structure. This additional tuning exactly corresponds to realizing the additional section in our discussion. }. 
We show this explicitly in section \ref{Sec:AffE6Example} by constructing a codimension-three fiber of type $IV^*$ for a model with extra section and factorized {\bf 10} curve.

Similarly for the $SU(6)$ enhancement to $E_6$ in codimension-two along the $\Lambda^3{\bf 6}$ matter locus, in order to realize one of the phases with trivial monodromy, e.g. the one in figure \ref{fig:U6Lambda3Fib}, the matter locus has to split and the model needs to have an extra section.


\subsection{$SU(5) \times U(1)$ models: extra flops and  $IV^*$ fibers}
\label{sec:SU5U1Ex}

In elliptic fibrations with extra rational sections there are additional flop transitions as we have discussed in section \ref{subsec:U1s}. Furthermore, extra rational sections are instrumental for the realization of standard Kodaira type fibers in higher codimension, in particular in cases with monodromy. 
We now give the geometric setup and an example for realizing both of these aspects. 
Elliptic Calabi--Yau varieties with multiple sections were studied recently in \cite{MW} \footnote{For explicit construction of example fiber types for Calabi-Yau fourfolds with extra sections see \cite{Mayrhofer:2012zy, Braun:2013yti, Borchmann:2013jwa, Cvetic:2013nia, Grimm:2013oga, Braun:2013nqa, Cvetic:2013uta, Borchmann:2013hta, Cvetic:2013qsa, KLS}.
} 
We shall restrict ourselves to $SU(5)$ models with one extra rational section. 
In \cite{KLS} the Tate forms for $SU(5)$ were obtained for $SU(5)\times U(1)$ models realized in $\mathbb{P}^{112}$, or more precisely in the blowup $\mathrm{Bl}_{[0,1,0]}\mathbb{P}_{112}[4]$. 
The singularities along $z=0$ in the base can be characterized in terms the equation
\be
\ba
    \mathcal{Q}(i_1,i_2,i_3,i_4,i_5,i_6,i_7): \qquad & 
    s y^2 + \mathfrak{b}_{0,i_5} z^{i_5} y x^2 + \mathfrak{b}_{1,i_6} z^{i_6}  s y w x + \mathfrak{b}_{2,i_7} z^{i_7} s^2 y w^2 \cr
    &\qquad = \mathfrak{c}_{0,i_1} z^{i_1} s^3 w^4 + \mathfrak{c}_{1,i_2} z^{i_2} s^2 w^3 x + \mathfrak{c}_{2,i_3} z^{i_3} s w^2 x^2 + \mathfrak{c}_{3,i_4} z^{i_4} w x^3 \,,
\ea
\ee 
where $i_j$ indicates the   vanishing order in $z$ of the respective terms. Unlike for  the standard Tate models in $\mathbb{P}^{123}$  there are several models for each non-abelian gauge group, which differ by the location of the two sections (zero section and the additional section, which is given by $s=0$ in $\mathcal{Q}$) as well as the codimension two fiber structure. 
The possible fiber types of $I_5$ models are obtained from the following vanishing orders \cite{KLS}
\be\label{eqn:canSU5}
\ba
I_5^{(01)}:&\qquad  \mathcal{Q} (5,3,1,0,0,0,2)\cr
I_5^{(0|1)}: &\qquad  \mathcal{Q} (4,2,1,1,0,0,2)\cr
I_5^{(0|1)}: &\qquad  \mathcal{Q} (4,3,2,1,0,0,1)\cr
I_5^{(0||1)}: &\qquad  \mathcal{Q} (3,2,2,2,0,0,1)\,,
\ea
\ee
where  $I_5^{(01)}$, $I_5^{(0 | 1)}$ and  $I_5^{(0 || 1)}$ indicates that the two sections are located on the same, next or next-to-nearest divisors, respectively. 


\subsubsection{New Flops from Extra Section}

So to see a phase of ${U}(n)$ corresponding to the flop which takes a
curve outside the ``end'' divisor on a chain (i.e., a phase of ${U}(n)$
which is not visible in ${SU}(n)$), we will need an additional section
of the fibration.  

For the standard Tate form the flops were studied in \cite{Hayashi:2013lra} for $SU(5)$ with ${\bf 5}$ and ${\bf 10}$ matter. Restricting this to the case of fundamental matter only, there were exactly four inequivalent resolutions, which are connected by flops, and are reproduced in table \ref{tab:SU5FunTab} in appendix \ref{app:U5}. To see the additional two phases, which come from flops at the ``end" divisor, we need to consider models with additional sections, which can be realized in $\mathbb{P}^{112}$. 

Consider $\mathcal{Q} (4,2,1,1,0,0,2)$, which has an enhancement to $SU(6)$ along 
\be
P_0 = b_0 c_2  - b_1 c_3\,.
\ee
This can be resolved by 
\be
\ba
( x,y , z; \zeta_1) &\cr
(x, y, \zeta_1 ;\zeta_2) &\cr
(y, z;\delta_0) &\cr
(y, \zeta_1;\delta_1) &\,,
\ea
\ee
where the simple roots are associated to the divisors as follows \footnote{Note that the weights/roots assigned to curves are associated via the inner product (\ref{DCProd}), which is the negative of the actual intersection number, which is usually assigned to the curves e.g. in \cite{Lawrie:2012gg}. }
\be
(\alpha_0, \alpha_1, \alpha_2, \alpha_3, \alpha_4) \leftrightarrow (z, \zeta_1, \zeta_2, \delta_1, \delta_0) \,.
\ee
The notation is as in \cite{Lawrie:2012gg}, i.e., $(x_1, x_2, x_3;\zeta)$ stands for the resolution $x_i\rightarrow x_i \zeta $ and $[x_1, x_2, x_3]$ are projective coordinates of the blowup $\mathbb{P}^2$. 
This resolution realizes the phase $I$ in table \ref{tab:SU5FunTab} in appendix \ref{app:U5}\footnote{Note that in $\mathbb{P}^{123}$ this phase was obtained by flop in patches from algebraic resolutions in \cite{Hayashi:2013lra}.}.
The fundamental matter is located at $P_0=0$, along which the divisor associated to the root $\alpha_4$  splits as 
\be
{\alpha_4}  = (-L_5)+L_4  \quad \longrightarrow  \quad -{\bf w}_5  +{\bf w}_4 \,,
\ee
where in Cartan-Weyl basis
\be
 -{\bf w}_5 = (0,0,0,1) \,,\qquad  {\bf w}_4 =  (0, 0,-1,1)  \,.
\ee
To reach the resolution $0$, which corresponds to $U(5)$, we need to flop the curve ${\bf w}_5$. 
We follow the same procedure as in \cite{Hayashi:2013lra}, and consider a patch in which this curve is realized
$w=x=z=\zeta_1=\zeta_2=\delta_1=1$. 
The equation for the resolved model  $\mathcal{Q}(4,2,1,1,0,0,2)$ in this patch is
\be
{P_0  + b_0 c_2 \over b_1} 
+ b_0 y  +b_1 s y + b_2 \delta_0^2 s^2 y - c_0 \delta_0^3 s^3 - s_1 \delta_0 s^2 - c_2 s + \delta_0 s y^2 =0 \,.
\ee
Introducing the coordinates
\be
 u_1 = y \,,\quad u_3 = \delta_0 y \,,\qquad u_2 = s \,,\qquad u_4 = \delta_0 s  \,,
\ee
the equation can be rewritten as 
\be
{P_0  + b_0 c_2 \over b_1} 
+ b_0 u_1  +b_1 u_3 + b_2 u_4^2 u_1 - c_0 u_4^3 - s_1 u_2 u_4 - c_2 u_2 + u_1^2 u_2 =0 
\ee
under the condition
\be\label{ConiFold}
u_1 u_4 = u_2 u_3 \,,
\ee
which is precisely a conifold equation. We can now blow down the curve corresponding to ${\bf w}_5$, which is given by $c_3= u_1=u_2= u_3= u_4=0$. The flopped geometry is obtained by resolving this in terms of $[\beta_1, \beta_2]$, which in the patch $\beta_1 \not=0$ can be rewritten in terms of  $\beta= \beta_2/\beta_1$ 
\be
u_1 = \beta u_2 \,,\qquad u_3 = \beta u_4 \,.
\ee
The flopped curve which carries the weight ${\bf w}^5$ is given by $z=0$, which means 
\be
{\bf w}^5:\qquad u_2 =u_4=0 
\ee
Note that in this flop the extra section $s$ was instrumental, as it allowed the rewriting in terms of the conifold equation in (\ref{ConiFold}).

\subsubsection{Example for $U(5)$ phases}

We can also directly  realize the $U(5)$ phases, that are not $SU(5)$ phases, i.e. phases 0 and V in table \ref{tab:SU5FunTab} in  appendix \ref{app:U5} from the following model with extra section $\mathcal{Q} (5,3,1,0,0,0,2)$. This can be resolved by 
\be
\ba
(w, x, z; \zeta_1) &\cr
(x, y, \zeta_1 ;\zeta_2) &\cr
(y, \zeta_1;\delta_1) &\cr
(y, \zeta_2;\delta_2) &\,.
\ea
\ee
The fundamental matter is located at $c_3=0$, along which the divisor associated to the affine root $\alpha_0$  splits as 
\be
-{\alpha_0}   \quad \longrightarrow  \quad  - {\bf w}_5 + {\bf w}_1 \,,
\ee
where 
\be
- {\bf w}_5 =  -L_5 \,,\qquad  {\bf w}_1 = L_1 \,.
\ee
The equations are
\be
c_3=z=0:\qquad
\ba
-{\bf w}_5:\qquad 0&= y  \cr
{\bf w}_1:\qquad 0&=\delta_1 \delta_2 (b_0 \delta_2 x^2+s y)+b_1 s x \,.
\ea
\ee
Depending on which section we choose to remain large in the singular limit, we now either shrink $-{\bf w}_5$, and get phase with all the weights ${\bf w}_{i}>0$ (this is when we keep the section $s=0$ large), or we keep the standard zero section $w=0$ large, which results in the phase with all weights being negative ${\bf w}_i<0$. These are exactly the phases 0 and V in table \ref{tab:SU5FunTab}, which are $U(5)$ phases, that are not $SU(5)$.

\subsubsection{Example for codimension-three affine $E_6$ fiber}
\label{Sec:AffE6Example}

We argued in section \ref{subsec:U1s} that the type $IV^*$ fibers whose intersection graph is an affine $E_6$ Dynkin diagram, which corresponds to absence of monodromy in an  $E_6$ enhancement, can be realized in codimension three starting with an $SU(5)$ model only if the Mordell Weil group has rank at least one and the locus of the ${\bf 10}$ matter is factored. This can be exemplified with the Tate forms for $SU(5)\times U(1)$ models obtained in  \cite{KLS}.
The purpose of this section is the elucidation of such an example.

In equation (\ref{eqn:canSU5}), 4 models for $I_5$ fibers are given which come
from the application of Tate's algorithm, however there are branches of
Tate's algorithm where the resulting model cannot be globally shifted so as
to just have as its data a set of vanishing orders $\mathcal{Q}(\cdots)$.
To give an example one can have an $I_4$ model for which the discriminant
enhances to $\mathcal{O}(z^5)$ at some polynomial locus, and there does not
exist a coordinate shift which absorbs this polynomial into the vanishing
orders. The $I_5$ models which arise in this way, as some lower rank model
and a set of polynomial constraints, are called {\it non-canonical models}, and
are explicated in \cite{KLS}. It is within these models that we find the
occurrence of full affine $E_6$ fibers above a codimension-three locus.

\begin{figure}
    \centering
    \includegraphics[scale=0.3]{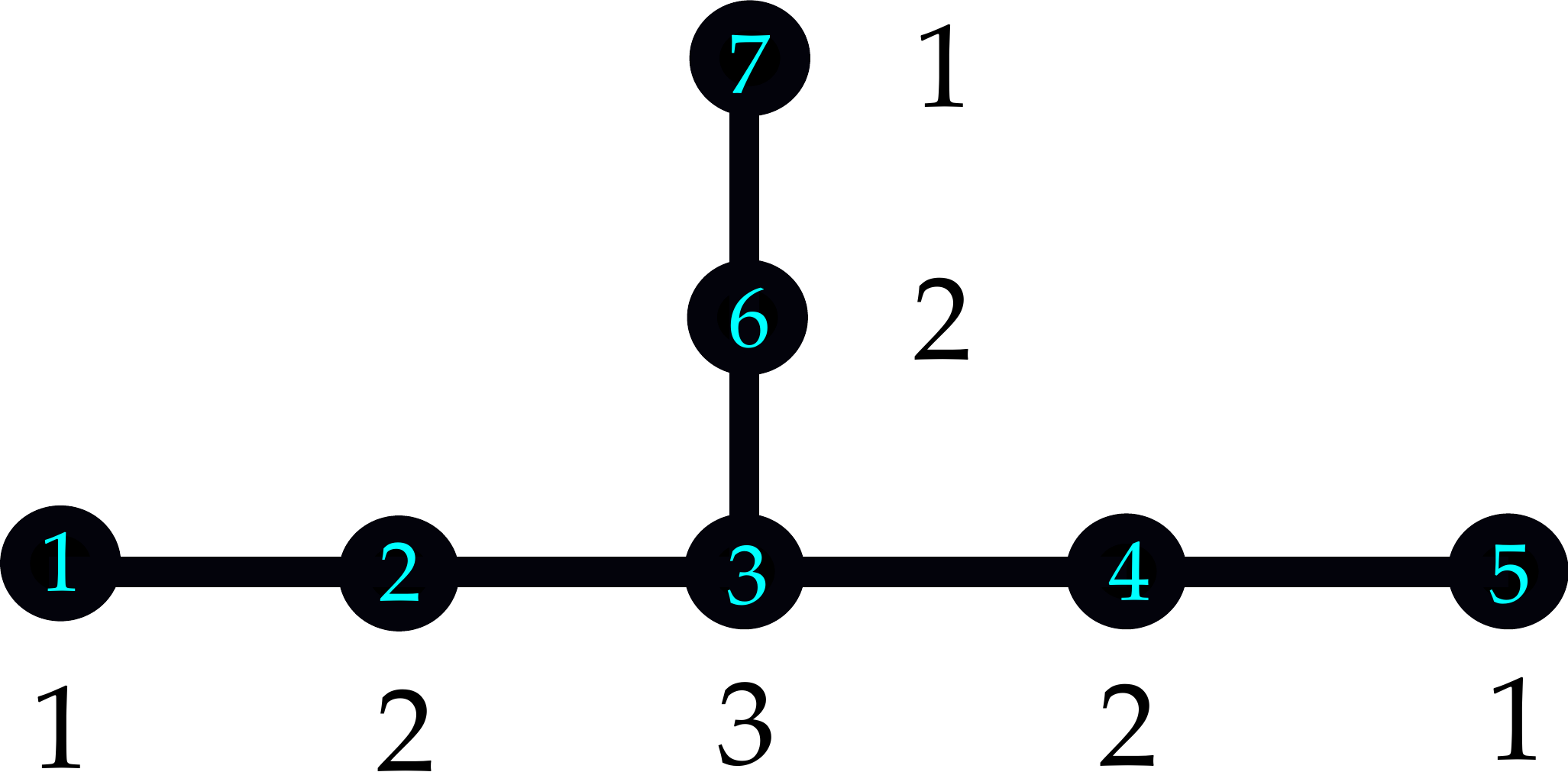}
    \caption{The intersection diagram of the fiber above a codimension-three
        locus for the model described in section \ref{Sec:AffE6Example}. The
        blue numerals indicate fiber components from (\ref{eqn:E6examplecpts}), the black numerals the multiplicities, which are those of a type $IV^*$ Kodaira fiber.}
    \label{fig:E6Dynk}
\end{figure}

Consider the non-canonical $SU(5)$ model, described in \cite{KLS}, which is obtained from an $I_4$ singularity that is described in terms of vanishing orders $\mathcal{Q}(3,2,1,1,0,0,1)$ and the additional condition $b_0 c_2 - b_1 c_3 = 0$, which enhances this to an $I_5$ singularity
\begin{equation}
        \mathcal{Q}(3,2,1,1,0,0,1)|_{     b_0 c_2 - b_1 c_3}=0  \,.
\end{equation}
The extra condition is solved generally, as in \cite{KLS}, by 
\begin{equation}
    \begin{aligned}
        &b_0 \rightarrow \sigma_1 \sigma_2, \quad &c_2 \rightarrow \sigma_3
        \sigma_4, \cr
        &b_1 \rightarrow \sigma_1 \sigma_3, \quad &c_3 \rightarrow \sigma_2
        \sigma_4 \,.\cr
    \end{aligned}
\end{equation}
Note that there is no shift that brings this model back into a canonical form. Also, it is clear that the codimension-two locus that enhances to $SO(10)$, which is given by $b_1 = \sigma_1 \sigma_3=0$,  factors, as required for obtaining the codimension-three $IV^*$ fiber.  
This example can be resolved by the following series of resolutions
\begin{equation}
    \begin{aligned}
        (x, y, z; \zeta_1) &\cr
        (y, z; \delta_0) &\cr
        (y, \zeta_1; \delta_1) &\cr
        (\delta_0, A = \sigma_2 \delta_1 \zeta_1 x + \sigma_3 s w; \delta_2) &\,,
    \end{aligned}
\end{equation}
using again the notation as in \cite{Lawrie:2012gg}. 
In this model there is a codimension-three locus where the  vanishing
order of the
discriminant increases $\mathcal{O}(5) \rightarrow \mathcal{O}(8)$
indicating that this is the locus containing the $E_6$ enhancement. We
consider the locus $\sigma_1 = \sigma_3 = 0$. We are interested in the
structure of the fiber above this locus, so we study to what irreducible
components the Cartan divisors degenerate.
\begin{equation}
    \label{eqn:E6examplecpts}
    \begin{aligned}
        &1: &\zeta_0 = s = \delta_2A - \sigma_2 = 0 \cr
        &2: &\zeta_0 = \delta_0 = \delta_2 - \sigma_2\zeta_1 = 0 \cr
        &3: &\zeta_1 = \delta_0 = \delta_2 = 0 \cr
        &4: &\zeta_1 = \delta_1 + b_2\zeta_0 = \delta_2 = 0 \cr
        &5: &\zeta_1 = \delta_1 + b_2\zeta_0 = A = 0 \cr
        &6: &\delta_1 = \delta_2 = b_2y\delta_0 - \zeta_1(\delta_0(c_1 +
            c_0\delta_0) + \sigma_4A) = 0 \cr
        &7: &\delta_1 = A = b_2y - \zeta_1(c_1x + c_0) = 0 \cr
    \end{aligned}
\end{equation}
These can be seen to intersect as in figure \ref{fig:E6Dynk}, realizing in the fiber
the full dual graph to the affine $E_6$ Dynkin diagram, with the correct
multiplicities.

\section*{Acknowledgments}

We thank  Antonella Grassi, Moritz K\"untzler, Joe Marsano, Konstanze Rietsch and in particular Martin Weidner for interesting and helpful discussions. HH would like to 
thank King's College London and Korea Institute for Advanced Study for hospitality and support 
during the work. The work of HH is supported by the REA grant agreement 
PCIG10-GA-2011-304023 from the People Programme of FP7 (Marie Curie Action), the grant 
FPA2012-32828 from the MINECO, the ERC Advanced Grant SPLE under contract 
ERC-2012-ADG-20120216-320421 and the grant SEV-2012-0249 of the ``Centro de Excelencia 
Severo Ochoa" Programme. 
The work of CL and SSN is supported in part by the STFC grant ST/J002798/1. We thank the European COST action ``The String Theory Universe" for a short-term visiting grant. 
The work of DRM is supported by National Science Foundation grant
PHY-1307513.


\appendix


\section{Group Theoretic Setup}
\label{app:Group}

\subsection{Root systems}

The combinatorics of a compact Lie group and its representations
are captured in large part by the notion of a  root system,
which we present following \cite{MR0240238}.
A {\em reduced root system} \/ in a real vector space $V$ is a finite subset 
$\rootsystem\subset V$ 
such that for each $\alpha\in \rootsystem$ there exists
 $\alpha^\vee \in V^*$ satisfying
\begin{enumerate}
\item $\langle \alpha^\vee, \alpha
\rangle=2$, and $\langle \alpha^\vee, \beta \rangle \in \mathbb{Z}$ for
any $\beta\in \rootsystem$, 
\item the map from $V$ to $V$ defined by 
\[ s_\alpha : x \mapsto x - \langle \alpha^\vee, x\rangle \, \alpha\]
(called the ``reflection in $\alpha$'') maps $\rootsystem$ to $\rootsystem$, and
\item
if $\alpha\in \rootsystem$ then $2\alpha\not\in \rootsystem$.
\end{enumerate}
The real vector space  spanned by $\rootsystem$
is called the  {\em root space}, and its
 dimension is called the {\em rank}\/
of the root system.  (We are making a small departure from 
\cite{MR0240238} by allowing $V$ to be larger than the root space.)
The elements $\alpha\in \rootsystem$ are called {\em roots},
and the associated elements $\alpha^\vee \in V^*$ are called {\em coroots}.
Note that the reflection $s_\alpha$ fixes the hyperplane
$\hyperplane_\alpha = \operatorname{Ker}(\alpha^\vee)\subset V$.

The group of automorphisms of the root space which leave the root system invariant
is denoted by $\operatorname{Aut}(\rootsystem)$.  It contains a subgroup $W(\rootsystem)$,
the {\em Weyl group of the root system}, generated by the reflections
$s_\alpha$.  (More generally, we can regard $W(\rootsystem)$ as a group of automorphisms
of the larger space $V$.)  The only reflections which appear
in $W(\rootsystem)$ are the reflections 

The {\em dual root system} of $\rootsystem$ is the subset $\rootsystem^\vee = \{ \alpha^\vee \ |\
\alpha\in \rootsystem\}$ of $V^*$.  The map $u\mapsto {}^tu^{-1}$
gives an isomorphism between $W(\rootsystem)$ and $W(\rootsystem^\vee)$, and we can use it to
identify 
the two groups; in this way, $W(\rootsystem)$ acts on $V^*$ as well as on $V$.

The connected components of the
set $\{v\in V\ |\ v \not\in \hyperplane_\alpha \text{ for any } \alpha\}$ are called
the {\em Weyl chambers}\/ of the root system, and are acted upon simply
transitively by the Weyl group.  Picking one such chamber $\mathcal{C}$
determines a set of {\em positive roots $\rootsystem^+$:} the ones for which the
coroot $\alpha^\vee$ takes positive values on $\mathcal{C}$. (Since none of the
coroots can take the value $0$ on $\mathcal{C}$, every root is either
positive or negative.)  There is
also a set of {\em simple roots}\/ determined by $\mathcal{C}$:  these
are the positive roots $\alpha$ whose coroot
$\alpha^\vee$ is zero along a codimension-one
face of the closure $\overline{\mathcal{C}}$.

Equally important for us will be the Weyl chambers  $\mathcal{C}^*$ of the dual root system,
which are subsets of $V^*$ and permuted by the Weyl group in exactly the
same way.  Given a Weyl chamber in $V$ and the corresponding set of simple
roots, the associated coroots are a set of simple roots in the dual root
system, and determine a dual Weyl chamber in the coroot space.

\subsection{Compact Lie groups and their representations}

Let $G$ be a compact  Lie group.
It is known
that finite-dimensional complex representations of such a group are always
the direct sum of irreducible representations 
(see \cite{MR0252560}, for example). For simplicity,
we assume that $G$ is connected.  

The complex representations  of $G$ can be analyzed by means of a
Cartan subgroup $H\subset G$, which is a maximal torus contained in $G$;
$H$ is itself a compact (abelian) Lie group.
Irreducible representations of $H$ are all one-dimensional, and
correspond to elements of the {\em weight lattice}\/
$\Lambda_H:=\operatorname{Hom}_{\mathbb{Z}}(H,U(1))$, which is a finite abelian group.
Given a representation of $H$ on a complex
vector space $V$ and a weight $\alpha\in \Lambda_H$, the subspace
\be
V_\alpha:=\{ v\in V\ |\ h\cdot v=\alpha(h) v \ \forall h\in H\}
\ee
(where we have used a dot to denote the action of $H$ on $V$)
is called the {\em weight space of $V$ with weight $\alpha$}.  The
representation can be recovered from its weight spaces:
$V=\bigoplus_{\alpha\in \Lambda_H} V_\alpha$.

From the definition, it may appear that we should denote the group
operation on $\Lambda_H$ multiplicatively, but if we let $\mathfrak{h}$ be the
Lie algebra of $H$, the action on the tangent space at the identity
element determines a natural inclusion
\be \Lambda_H \subset \operatorname{Hom}_{\mathbb{R}}(\mathfrak{h},\mathbb{R})
=\mathfrak{h}^*,\ee
and additive notation becomes appropriate.  It is common to use the
description in terms of Lie algebras when weights must be added 
together.\footnote{In fact, our discussion of roots and weights can be
formulated equally well for (complex) reductive Lie algebras and
we have used that formulation in the body of the paper.}

Closely related is the {\em coweight lattice}\/
\be\Lambda^\vee_H:=\operatorname{Hom}_{\mathbb{Z}}(U(1),H)\subset
\operatorname{Hom}_{\mathbb{R}}(\mathbb{R},\mathfrak{h})=\mathfrak{h},\ee
which can be naturally identified with the fundamental group of $H$
since $U(1)$ is topologically a circle.
This identification enables us to 
recover $H=\mathfrak{h}/\pi_1(H)=\mathfrak{h}/\Lambda^\vee_H$; 
we can also recover $H$ from $\Lambda_H$ via Pontryagin duality, as
$H=\operatorname{Hom}_{\mathbb{Z}}(\Lambda_H,U(1))$.
Note that there is a natural pairing between the coweight and weight lattices
\be \Lambda^\vee_H\times \Lambda_H\to \mathbb{Z}\ee
given by composition, 
since $\operatorname{Hom}_{\mathbb{Z}}(U(1),U(1))\cong\mathbb{Z}$.

Sometimes one speaks of the ``weight lattice of $G$'' $\Lambda_G$ and the
``coweight lattice of $G$'' $\Lambda^\vee_G$, although strictly speaking
one must choose $H$ before these are defined.  When there is no danger
of confusion, we omit the subscript $G$ (or $H$).

For any complex representation $G\to {GL}(V)$ 
we can restrict to $H$ and
decompose $V$ into weight spaces; the corresponding weights are called
the {\em weights of the representation}.  As a particular case of this,
we can consider the {\em adjoint representation}\/ of $G$ on its
Lie algebra $\mathfrak{g}$.  This is a real representation, but we
can complexify to get a representation 
$G\to {GL}(\mathfrak{g_{\mathbb{C}}})$.
The weight space of $\mathfrak{g}_{\mathbb{C}}$ with weight $0\in \Lambda_G$
is the complex Lie algebra $\mathfrak{h}_{\mathbb{C}}$,
and the nonzero weights of $\mathfrak{g}_{\mathbb{C}}$
are called the
{\em roots of $G$}; these  exist only when $G$ is nonabelian.
It turns out that the weight space for each root is one-dimensional.
Each root $\alpha\in \Lambda$  has an associated
{\em coroot} $\alpha^\vee \in \Lambda^\vee$, and
the set of roots of $G$ (and associated coroots)
satisfies the conditions for a {\em reduced root
system $\rootsystem_G$}\/ as described in the previous subsection (with $V=\mathfrak{h}^*$).  
In particular, there is a Weyl group $W(\rootsystem_G)$ which is generated by
reflections in the roots.  The Weyl group has another interpretation as
well: it is 
isomorphic to $N(H)/H$ where $N(H)$ is the normalizer of $H$ in $G$.

An irreducible representation always has a {\em highest}, or {\em dominant}\/
weight $\varpi$ once a set of positive roots has been chosen.  It has the
property that under the induced action of the Lie algebra $\mathfrak{g}$,
the action of the root space $\mathfrak{g}_{\alpha}$
on $V_{\varpi}$ is trivial for every positive root $\alpha$.

\subsection{Root systems for the classical Lie groups}

We will set up some notation for the representation theory of
the classical Lie groups ${SO}(m)$, ${Spin}(m)$,
${U}(n)$, $SU(n)$, and
${Sp}(r)$; 
this could also be formulated
in terms of the corresponding Lie algebras
$\mathfrak{so}(m)$, $\mathfrak{u}(n)$, $\mathfrak{su}(n)$,
and $\mathfrak{sp}(r)$.
(These groups act on different kinds of spaces:
$\mathbb{R}^m$, $\mathbb{C}^n$, and $\mathbb{H}^r$, which
is why different letters are being used for the dimensions, as in
\cite{anomalies}.
In this notation, we have 
${Sp}(r)\subset SU(2r)$ and
${U}(n) \subset {SO}(2n)$.)

We begin with $U(n)$.
 Let $z_1$, \dots, $z_n$
be a basis for a complex
vector space of dimension $n$ on which 
$U(n)$ acts
by matrix multiplication, giving the so-called
{\em fundamental representation}\/ of
complex dimension $n$.  Using the diagonal unitary matrices
as a Cartan subgroup $H\subset U(n)$,
the weight spaces are the one-dimensional subspaces spanned by the
individual basis vectors $v_k$; we let
$ \weight_k:\mathfrak{h}\to \mathbb{R}$ be the corresponding weight.
Then $ \weight_1$, \dots , $ \weight_n$ forms a basis for the
weight lattice of $\mathfrak{u}(n)$.  There is a natural dual basis
$\coweight_1$, \dots, $\coweight_n$ of the coweight lattice.

A Cartan subgroup $H_0$ of $SU(n)$
is given by the diagonal unitary matrices of determinant $1$; its
Lie algebra $\mathfrak{h}_0$ consists of diagonal Hermitian matrices
of trace $0$.  Weights of $U(n)$ can be restricted
to $SU(n)$, where they satisfy
\begin{equation}
\label{eq:su-restrict}
 ( \weight_1+\cdots+ \weight_n)|_{\mathfrak{h}_0}=0 
\end{equation}
(here we indicate the Lie algebra since we are using additive notation).
We will suppress the explicit restriction to $\mathfrak{h}_0$ and
continue to use $ \weight_k$ to denote a weight of 
$SU(n)$.
In fact, the weights of the fundamental
representation of $SU(n)$ are
precisely $ \weight_1$, \dots, $ \weight_n$.

The Weyl group of $SU(n)$
is the permutation group
$\mathfrak{S}_n$ acting on $\{ \weight_1, \dots,  \weight_n\}$
(and preserving \eqref{eq:su-restrict}).
The roots are $ \weight_k- \weight_\ell$, $k\ne \ell$, and the corresponding
coroots are $\coweight_k-\coweight_\ell$.
One choice
of simple roots for $SU(n)$ is given by 
\be\{\alpha_k:= \weight_k- \weight_{k+1},\ 1\le k\le n-1\}.\ee
The corresponding simple coroots are
\be \alpha^\vee_k:=\coweight_k-\coweight_{k+1},\ee
and these satisfy
\be \langle \alpha^\vee_\ell, \alpha_k \rangle = \begin{cases}
2 & \text{if } k=\ell \\
-1 & \text{if } k=\ell\pm1 \\
0 & \text{otherwise}
\end{cases}.
\ee
The root system is type $A_{n-1}$.  When $n\ge3$, it has an automorphism
of order $2$
not contained in the Weyl group, given by $\alpha_k\mapsto \alpha_{n-k}$.

From the fundamental representation $V$ we can construct other irreducible
representations $\Lambda^jV$ as exterior powers.  The highest weight of
$\Lambda^jV$ is $\varpi_j:=\weight_1+\cdots+\weight_j$ for $j=1,\dots, n{-}1$.

We next consider ${SO}(m)$.  Let $n=[m/2]$, and let
$x_1,\dots,x_{[(m+1)/2]}, y_1, \dots, y_{[m/2]}$ be a basis for
$\mathbb{R}^m$ (which is called the {\em vector representation}\/
of ${SO}(m)$).  We use $H={SO}(2)^n\subset {SO}(m)$ as a Cartan subgroup,
where the $k^{\text{th}}$ copy of ${SO}(2)$ acts on the space spanned
by $x_k, y_k$ by rotations.  The weight spaces in the complexification
$\mathbb{C}^m$
of $\mathbb{R}^m$ are then spanned by
$x_k+y_k\sqrt{-1}$ and $x_k-y_k\sqrt{-1}$ 
(as well as $x_{n+1}$ if $m$ is
odd); we call the corresponding weights $ \weight_k$ and 
$- \weight_k$ (and note that when $m$ is odd, the weight for $x_{n+1}$
is $0$).
The weight lattice of ${SO}(m)$ 
is spanned by $ \weight_1$, \dots, $ \weight_n$, and there is 
a natural dual basis $\coweight_1$, \dots, $\coweight_n$ for the coweight lattice.

The adjoint representation of ${SO}(m)$ is
the second anti-symmetric power of the vector representation.
We can thus describe the roots as sums of
distinct weights from the vector representation, whenever the sum
is nonzero.  Note that 
$ \weight_k+(- \weight_k)=0$ so we get $n$ zeros among
the weights of the adjoint representation, which agrees with the dimension
of the Cartan subgroup (as expected).  Nonzero roots are given by
$\pm  \weight_k \pm  \weight_\ell$, $k\ne\ell$; if $m$ is odd,
we also get $\pm  \weight_k$ (by adding  the zero weight in the
vector representation to the other weights).  
The corresponding coroots are $\pm \coweight_k\pm \coweight_\ell$, and if $m$ is odd,
$\pm2\coweight_k$.
Thus, if $m=2n$ 
the dimension of the group is 
$n + 2n(n-1) = \frac12m(m-1)$, while if $m=2n+1$ 
the dimension of the group is 
$n+2n(n-1)+2n =\frac12m(m-1)$.

The Weyl group of ${SO}(m)$ is 
$\mathfrak{S}_{[m/2]}\rtimes (\mathbb{Z}/2\mathbb{Z})^{[(m-1)/2]}$.
The group permutes the $ \weight_k$'s and multiplies them by signs;
when $m$ is even, the number of minus signs must be even.  
One choice of simple roots is given by $\alpha_k=\weight_k-\weight_{k+1}$
for $1\le k\le n-1$, together with
\be \alpha_n:=\begin{cases} \weight_n & \text{if } m=2n+1 \\
\weight_{n-1}+\weight_n & \text{if } m=2n
\end{cases}.
\ee
The root system is type $B_n$ is $m=2n+1$, and type $D_n$ if $m=2n$.
There are no automorphisms other than the Weyl group for $m$ odd,
but for ${SO}(8)$ there is an automorphism group $\mathfrak{S}_3$ which
permutes $\{\alpha_1, \alpha_3, \alpha_4\}$, while for 
${SO}(2n)$, $n\ge5$, there is an automorphism of order
two exchanging $\alpha_{n-1}$ and $\alpha_n$ while leaving the other
simple roots fixed.

From the vector representation $V$ we can construct other irreducible
representations $\Lambda^jV$ as exterior powers.  The highest weight of
$\Lambda^jV$ is $\varpi_j:=\weight_1+\cdots+\weight_j$ for $j=1,\dots, n{-}2$.

The group ${SO}(m)$ has a double cover ${Spin}(m)$, and the Cartan subgroup
of ${Spin}(m)$ is also a double cover of the Cartan subgroup of ${SO}(m)$.
This implies that the weight and coweight lattice are different (although
the roots and coroots do not change).  The weight lattice is enlarged
to include the weights of the spinor representation(s), which are
\be \frac12(\pm \weight_1\pm \cdots\pm  \weight_{[m/2]}).\ee
(If $m$ is even, each of these weights occurs in exactly one of the two spinor
representations, depending on the parity of the number of minus signs.)
The weight lattice is therefore
\be \Lambda_{{Spin}(m)} = \{ \sum a_k  \weight_k \ |\ 
a_k \in \frac12\mathbb{Z},\ a_k - a_\ell \in \mathbb{Z}\}.\ee
It follows that the coweight lattice is
\be \Lambda^\vee_{{Spin}(m)} = \{ \sum b_\ell \coweight_\ell \ |\
b_\ell\in\mathbb{Z},\ \sum b_\ell\in 2\mathbb{Z}\}.\ee
When $m$ is odd, the highest weight of the spinor representation is
\be\varpi_{(m-1)/2}= \frac12\left(L_1+\cdots+L_{(m-1)/2}\right),\ee 
while when $m=2n$
is even, the highest weights of the two spinor representations are
\be\varpi_{n-1} = \frac12\left(L_1+\cdots+L_{n-1}-L_n\right),\ee
and
\be\varpi_{n} = \frac12\left(L_1+\cdots+L_{n-1}+L_n\right).\ee

Finally we consider ${Sp}(r)$.  Using the standard quaternions
$i$, $j$, and $k=ij$, we can write 
\begin{equation}\label{eq:quaternions}
\mathbb{H}=\mathbb{C}\oplus \mathbb{C}\cdot j.
\end{equation}
We choose a basis of $\mathbb{H}^r$ of the form
\be z_1+w_1\cdot j,\ z_2+w_2\cdot j,\ \dots,\ z_r+w_r\cdot j.\ee
We choose ${Sp}(1)\subset {Sp}(r)$ by letting the $\ell^{\text{th}}$ copy
of ${Sp}(1)$ act on $z_\ell+w_\ell\cdot j$.  Since ${Sp}(1)=SU(2)$, we can choose
a Cartan subgroup $U(1)\subset SU(2)$ compatible with the decomposition
(\ref{eq:quaternions}).  Then $U(1)^r\subset {Sp}(r)$ is a maximal torus.  The roots are
$ \weight_\ell$ with weight space spanned by $z_\ell$, and
$- \weight_\ell$ with weight space spanned by
$w_\ell$.  The complex dimension of the fundamental representation
is $2r$.

The adjoint representation of ${Sp}(r)$ is the second symmetric power 
of the fundamental representation, so the roots are given by the nonzero
sums of
pairs of weights of the fundamental representation, not necessarily distinct.
We obtain zero as an adjoint weight
 via $ \weight_\ell+(- \weight_\ell)=0$,
$\ell =1 ,\dots, r$ (which implies that the dimension of the Cartan subgroup
is $r$, as expected)
but all other sums are nonzero.  The roots take the form
$\pm 2 \weight_\ell$ and $\pm  \weight_\ell \pm  \weight_m$ for
$\ell\ne m$.  The corresponding coroots are $\pm \coweight_\ell$ and 
$\pm \coweight_\ell \pm \coweight_m$.

The Weyl group is $\mathfrak{S}_r\rtimes (\mathbb{Z}/2\mathbb{Z})^r$,
which acts by permuting the $\weight_k$'s and multiplying them
by signs.  One choice of simple roots is given by
$\alpha_\ell=\weight_\ell-\weight_{\ell+1}$ for $1\le \ell\le r-1$,
$\alpha_r=2\weight_r$.
The root system is type $C_r$.  There are no automorphisms other than
the Weyl group.



\subsection{Root systems for the 
simple Lie groups of type $E_n$}
\label{app:ERoots}

For reference, we will set up some similar notation for the representation
theory of the exceptional compact simple Lie groups $E_6$, $E_7$, 
and $E_8$.
We use the simply-connected form of each of these.

For $E_n$, $n=6,7,8$,
we follow the presentation
of \cite{demazure-del-pezzo}.  We begin with the vector space spanned
by $n+1$ vectors $\weight_0, \weight_1, \dots, \weight_n$ as well as
the dual space spanned by $\coweight_0, \coweight_1, \dots, \coweight_n$.
The root space for $E_n$ will be
$\operatorname{Ker}(3\coweight_0 - \sum_{j=1}^n\coweight_j)$.
In particular, just as in the case of $SU(n)$, we can regard $\weight_j$
as a root by restricting it to this space.

\begin{figure}
 \centering
    \includegraphics[width=7cm]{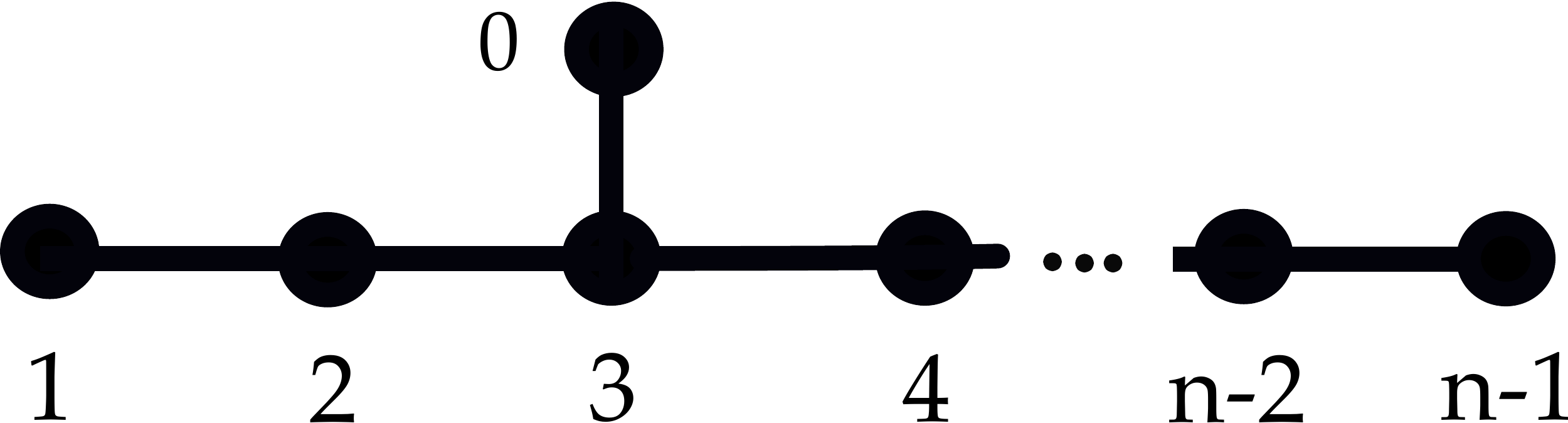}

\caption{Dynkin diagram for $E_n$, establishing our notation, with numbers $i$ labeling the simple roots $\alpha_i$.}\label{fig:En}
\end{figure}

There are four kinds of positive roots: $\weight_j-\weight_k$ ($0<j<k$);
$\weight_0-\sum_{i=1}^3\weight_{j_i}$; 
$2\weight_0-\sum_{i=1}^6\weight_{j_i}$; and
$3\weight_0-\sum_{i=1}^7\weight_{j_i}-2\weight_k$,
where in the last three cases,
the $j_i$ are all distinct and differ from $0$ and $k$.
A set of simple roots is given by $\alpha_0=\weight_0-\weight_1-\weight_2
-\weight_3$ and $\alpha_j=\weight_j-\weight_{j+1}$ for $0<j<n$,
with dual expressions giving the corresponding coroots.
The Dynkin diagram, which summarizes the intersection properties between
simple roots and simple coroots,  is shown in Figure~\ref{fig:En}.


The number of positive roots of each type depends on $n$:
\begin{center}
\begin{tabular}{c|ccc} 
& $n=6$ & $n=7$ & $n=8$ \\ \hline
$\weight_j-\weight_k$ & $15$ & $21$ & $28$ \\
$\weight_0-\sum_{i=1}^3\weight_{j_i}$ & $20$ & $35$ & $56$ \\
$2\weight_0-\sum_{i=1}^6\weight_{j_i}$ & $1$ & $7$ & $28$ \\
$3\weight_0-\sum_{i=1}^7\weight_{j_i}-2\weight_k$ & $0$ & $0$&$8$ \\ \hline
Total & $36$&$63$ &$120$  \\ \hline
$\dim G$ & $78$ & $133$ & $248$ \\
$|W|$& $51840$ & $2903040$ & $696729600$ \\
\end{tabular}
\end{center}
The dimension of the group is twice the number of positive roots, plus $n$ 
(the rank).  We have also listed the order $|W|$ of each Weyl group.

The minimum-dimension representations for these groups can be described
in terms of another set of weights, for $n=6, 7$.  There are four types
of weights which occur:  $\weight_j$ ($j>0$), $\weight_0-\weight_j-\weight_k$,
$2\weight_0-\sum_{i=1}^5 \weight_{j_i}$, 
$3\weight_0-\sum_{i=1}^6\weight_{j_i}-2\weight_k$.  (This analysis can actually be extended to $n=8$ as well,
but three additional types of weights occur.)  The number of weights of
each type is:
\begin{center}
\begin{tabular}{c|cc}
&$n=6$&$n=7$\\ \hline
$\weight_j$&$6$&$7$\\
$\weight_0-\weight_j-\weight_k$&$15$&$21$\\
$2\weight_0-\sum_{i=1}^5 \weight_{j_i}$&$6$&$21$\\
$3\weight_0-\sum_{i=1}^6\weight_{j_i}-2\weight_k$&$0$&$7$\\ \hline
Total & $27$ & $56$ \\
\end{tabular}
\end{center}

The highest weight vectors for these representations can also be identified.
For $E_6$, one of the minimum-dimension representations has highest weight
\be \varpi_5:=\alpha_0+\frac23\alpha_1+\frac43\alpha_2+2\alpha_3+\frac53\alpha_4+\frac43\alpha_5= -L_6,\ee
while the other has highest weight
\be \varpi_1:=\alpha_0+\frac43\alpha_1+\frac53\alpha_2+2\alpha_3+\frac43\alpha_4+\frac23\alpha_5=2L_0-\sum_{j=2}^6L_j.\ee
For $E_7$, the minimum-dimension representation has highest weight
\be \varpi_6:=\frac32\alpha_0+\alpha_1+2\alpha_2+3\alpha_3+\frac52\alpha_4+2\alpha_5+\frac32\alpha_6 = -L_7.\ee
In all of these definitions, the weights $L_j$ must be restricted to the
subspace $\operatorname{Ker}(3\coweight_0 - \sum_{j=1}^n\coweight_j)$.


\section{Phases of $U(5)$ from the Weyl group quotient}
\label{app:U5}

In this appendix we give the explicit example of $U(5)$, with the ${\bf 5}$ and with ${\bf 10}$ 
representation, respectively,  for phases determined by the Weyl group quotient. 
Consider first $U(5)$ with the ${\bf 5}$ representation. 
The phases are determined from the embedding of  the simple roots of $SU(5)$ into 
the simple roots of $SU(6)$, by identifying the first four simple roots with the $SU(5)$ ones, 
$\alpha_i$, $i=1, \cdots 4$.  Furthermore, the simple roots of $SU(6)$ are denoted by $\hat\alpha$. The initial embedding is shown in table \ref{tab:SU5FunTab} in the box labeled $0$. 

 We act with the Weyl group of $SU(6)$, keeping the constraints that the image projected back to $SU(5)$ gives rise to positive roots of $SU(5)$ only. 
Projecting back to $SU(5)$ results in each step in a set of positive roots and  weights of the ${\bf 5}$. 
Denote by $\alpha_{i_1\cdots i_n}= \sum_j \alpha_{i_j}$, and define the corresponding Weyl reflection with respect to $\sum_{j}\hat\alpha_{i_j}$ as 
 $\sigma_{\hat\alpha_{i_1\cdots i_n}}$, i.e. $\sigma_{\alpha}(\beta) = \beta- \langle \alpha, \beta \rangle \alpha$. The block in the middle denotes the generators of the cone (weights/roots of $SU(5)$),  the last column completes these to $SU(6)$ roots. 

\begin{table}
{\footnotesize
\begin{equation}
\begin{array}{rl}
{\begin{array}{|c|l|cccc|c|}
	\hline
	+\alpha_1 &+\hat\alpha_1 &2 & -1 & 0 & 0 &0\cr
	+\alpha_2 &+\hat\alpha_2 &-1 & 2 & -1 & 0& 0\cr
	+\alpha_3 &+\hat\alpha_3 &0 & -1 & 2 & -1 &0\cr
	+\alpha_4 &+\hat\alpha_4 &0 & 0 & -1 & 2 &-1\cr
	+w_{\bf 5}^5 &+\hat\alpha_5 &0 & 0 & 0 & -1 & {2}\cr
	\hline
\end{array}
{\boxed{0\phantom{0}}}}
 \stackrel{\sigma_{\hat\alpha_5} }{ \xrightarrow{\hspace*{1cm}}}
&
\textcolor{blue}{
{\boxed{I\phantom{0}}}
\begin{array}{|c|l|cccc|c|}
\hline
+\alpha_1 &+\hat\alpha_1 &2 & -1 & 0 & 0 &0\cr
+\alpha_2 &+\hat\alpha_2 &-1 & 2 & -1 & 0& 0\cr
+\alpha_3 &+\hat\alpha_3 &0 & -1 & 2 & -1 &0\cr
+w_{\bf 5}^4 &+\hat\alpha_{45} &0 & 0 & -1 & 1& {1}\cr
-w_{\bf 5}^5 &- \hat\alpha_5 & 0 & 0 & 0 & 1& {-2}\cr
\hline
\end{array}
}
\cr\cr
&\qquad\qquad\qquad \qquad  \downarrow  \sigma_{\hat\alpha_{45}}
\cr\cr
\textcolor{blue}{
\begin{array}{|c|l|cccc|c|}
\hline
+\alpha_1 &+\hat\alpha_1 &2 & -1 & 0 & 0 &0\cr
+w_{\bf 5}^2 &+\hat\alpha_{2345}  &-1 & 1 & 0 & 0& {1}\cr
-w_{\bf 5}^3 &-\hat\alpha_{345} & 0 & 1 & -1 & 0& {-1}\cr
+\alpha_3 &+\hat\alpha_3 &0 & -1 & 2 & -1& 0\cr
+\alpha_4 &+\hat\alpha_ 4&0 & 0 & -1 & 2 &-1\cr
\hline
\end{array}{\boxed{III\phantom{0}}}
}
\stackrel{ \sigma_{\hat\alpha{345}}}{\xleftarrow{\hspace*{1cm}}}
&
\textcolor{blue}{
{\boxed{II\phantom{0}}}
\begin{array}{|c|l|cccc|c|}
\hline
+\alpha_1 &+\hat\alpha_1 &2 & -1 & 0 & 0 &0\cr
+\alpha_2 &+\hat\alpha_2 &-1 & 2 & -1 & 0& 0\cr
+w_{\bf 5}^3 &+\hat\alpha_{345}&0 & -1 & 1 & 0&{1} \cr
-w_{\bf 5}^4 &-\hat\alpha_{45} &0 & 0 & 1 & -1&-1\cr
+\alpha_4 &+\hat\alpha_4 &0 & 0 & -1 & 2&-1\cr
\hline
\end{array}
}
\cr\cr
 \downarrow  \sigma_{\hat\alpha_{2345}} \qquad\qquad\qquad \qquad\qquad\qquad &
\cr\cr
\textcolor{blue}{
\begin{array}{|c|l|cccc|c|}
\hline
+w_{\bf 5}^1 &+\hat\alpha_{12345} &1 & 0 & 0 & 0& {1}\cr
-w_{\bf 5}^2 &-\hat\alpha_{2345}& 1 & -1 & 0 & 0& {-1}\cr
+\alpha_2 &+\hat\alpha_2 &-1& 2 & -1 & 0 &0\cr
+\alpha_3 &+\hat\alpha_3 &0 & -1 & 2 & -1& 0\cr
+\alpha_3 &+\hat\alpha_3 &0 & 0 & -1 & 2 &-1\cr
\hline
\end{array}{\boxed{IV\phantom{0}}}
}
\stackrel{ \sigma_{\hat\alpha_{12345}}}{\xrightarrow{\hspace*{1cm}}}
&
{\boxed{V\phantom{0}}\begin{array}{|c|l|cccc|c|}
\hline
-w_{\bf 5}^1 &-\hat\alpha_{12345} &-1 & 0 & 0 & 0&-1\cr
+\alpha_1 &+\hat\alpha_1 &2 & -1 & 0 & 0 &0\cr
+\alpha_2 &+\hat\alpha_2 &-1 & 2 & -1 & 0& 0\cr
+\alpha_3 &+\hat\alpha_3 &0 & -1 & 2 & -1&0\cr
+\alpha_4 &+\hat\alpha_4 &0 & 0 & -1 & 2&-1 \cr
\hline
\end{array}}
\cr
\cr
\end{array}\nonumber
\end{equation}}
\caption{Phases/resolutions for $SU(5)$ with fundamental representation ${\bf 5}$ shown in blue. Phases 0 and V  are the two phases for $U(5)$, which are not $SU(5)$ phases. }
\label{tab:SU5FunTab}
\end{table}

To explain the process, begin with the first phase, which is given in terms of the standard embedding of $\alpha_i = \hat\alpha_i$. The projection of $\hat\alpha_5$ results in the $SU(5)$ weight $w_{\bf 5}^{5}$. In order not to introduce negative roots by acting with Weyl reflections, the only option is to act with a Weyl reflection $\sigma_{\hat\alpha_5}$. In the next step, the $SU(6)$ roots, projected to $SU(5)$ give the two weights $-w_{\bf 5}^5$ and $w_{\bf 5}^4$. We cannot perform a Weyl reflection along $\hat\alpha_1, \hat\alpha_2, \hat\alpha_3, \hat\alpha_4$, as these would yield negative roots, and thus would not generate consistent phases. The only Weyl reflection which does not generate such roots is $\sigma_{\hat{\alpha}_{45}}$. This picture generalizes directly for any fundamental representation. 
Likewise we can consider the  ${\bf 10}$ representation, for which we now embed the simple roots of $SU(5)$ into those of $SO(10)$, as shown in table \ref{SU5ATable}.  

\begin{table}
{\footnotesize
$$
\begin{array}{rl}
\begin{array}{|c|l|cccc|c|}
	\hline
	+\alpha_1 &+\hat\alpha_1 & 2 &-1 &0 &0 &0  \cr
	+\alpha_2 &+\hat\alpha_2 & -1 &2 &-1 &0 & 0 \cr
	+\alpha_3 &+\hat\alpha_3 & 0 &-1 &2 &-1 &-1  \cr
	+\alpha_4 &+\hat\alpha_4 & 0 &0 &-1 &2 & 0 \cr
	+w_{\bf 10}^{10} &+\hat\alpha_5 & 0 &0 &-1 &0 &2  \cr
	\hline
\end{array}
{\boxed{1\phantom{0}}}
&
{\boxed{16}}
\begin{array}{|c|l|cccc|c|}
	\hline
	+\alpha_1 &+\hat\alpha_1 & 2 & -1& 0&0 &0  \cr
	+\alpha_2 &+\hat\alpha_2 & -1 &2 &-1 &0 &0  \cr
	+\alpha_3 &+\hat\alpha_3 &  0&-1 &2 &-1 & -1  \cr
	+\alpha_4 &+\hat\alpha_4 & 0 &0 &-1 &2 & 0 \cr
	-w_{\bf 10}^{1} &-\hat\alpha_{1223345} &  0& -1& 0& 0& 0 \cr
	\hline
\end{array}
\cr
\cr
\textcolor{blue}{
\begin{array}{|c|l|cccc|c|}
	\hline
	+\alpha_2 &+\hat\alpha_2 & -1 &2 &-1 & 0&0  \cr
	+\alpha_3 &+\hat\alpha_3 &  0&-1 & 2& -1& -1 \cr
	+\alpha_4 &+\hat\alpha_4 &  0&0 &-1 &2 &0  \cr
	+w_{\bf 10}^4 &+\hat\alpha_{1235} &  1& 0&0 & -1& 1 \cr
	\textcolor{blue}{-w_{\bf 10}^{5}} &-\hat\alpha_{23345} & 1 &0 &-1 &0 & 0 \cr
	\hline
\end{array}
{\boxed{11}}
}
&
{\boxed{2\phantom{0}}}
\begin{array}{|c|l|cccc|c|}
	\hline
	+\alpha_1 &+\hat\alpha_1 & 2 &-1 &0 & 0& 0 \cr
	+\alpha_2 &+\hat\alpha_2 &  -1& 2& -1&0 & 0 \cr
	+\alpha_4 &+\hat\alpha_4 & 0 &0 &-1 &2 &0  \cr
	+w_{\bf 10}^{9} &+\hat\alpha_{35} & 0 &-1 &1 &-1 &+1  \cr
	-w_{\bf 10}^{10} &-\hat\alpha_5 & 0 &0 &1 &0 &-2  \cr
	\hline
\end{array}
\cr\cr
\textcolor{blue}{
\begin{array}{|c|l|cccc|c|}
	\hline
	+\alpha_1 &+\hat\alpha_1 & 2 &-1 &0 & 0& 0 \cr
	+\alpha_3 &+\hat\alpha_3 &  0& -1& 2& -1& 0 \cr
	+\alpha_4 &+\hat\alpha_4 &  0& 0& -1& 2& 0 \cr
	{+w_{\bf 10}^{7}} &+\hat\alpha_{235} &  -1& 1& 0& -1& 1 \cr
	-w_{\bf 10}^{8} &-\hat\alpha_{345} & 0 &1 &0 &-1 &-1  \cr
	\hline
\end{array}
{\boxed{4\phantom{0}}}
}
&
\textcolor{blue}{
{\boxed{6\phantom{0}}}
\begin{array}{|c|l|cccc|c|}
	\hline
	+\alpha_1 &+\hat\alpha_1 & 2 &-1 &0 &0 &0  \cr
	+\alpha_2 &+\hat\alpha_2 & -1 &2 &-1 &0 & 0 \cr
	+\alpha_3 &+\hat\alpha_3 &  0& -1&2 &-1 &-1  \cr
	-w_{\bf 10}^{4} &-\hat\alpha_{1235} & -1 &0 &0 &1 &-1  \cr
	\textcolor{blue}{+w_{\bf 10}^{8}} &+\hat\alpha_{345} &  0& -1& 0&1 &+1  \cr
	\hline
\end{array}
}
\cr\cr
\textcolor{blue}{
\begin{array}{|c|l|cccc|c|}
	\hline
	+\alpha_1 &+\hat\alpha_1 &  2& -1& 0& 0& 0 \cr
	+\alpha_2 &+\hat\alpha_2 &  -1& 2& -1& 0& 0 \cr
	+\alpha_4 &+\hat\alpha_4 & 0 &0 &-1 &2 &-1  \cr
	\textcolor{blue}{-w_{\bf 10}^{3}} &-\hat\alpha_{12345} &  -1& 0& 1& -1& -1 \cr
	+w_{\bf 10}^{5} &+\hat\alpha_{23345} &  -1& 0& 1& 0& 0 \cr
	\hline
\end{array}
{\boxed{13}}
}
&
{\boxed{15}}
\begin{array}{|c|l|cccc|c|}
	\hline
	+\alpha_1 &+\hat\alpha_1 &  2& -1& 0& 0&0  \cr
	+\alpha_3 &+\hat\alpha_3 &  0& -1& 2& -1& -1 \cr
	+\alpha_4 &+\hat\alpha_4 &  0& 0& -1& 2& 0 \cr
	-w_{\bf 10}^{2} &-\hat\alpha_{123345} & -1 &1 &-1 &0 &0  \cr
	+w_{\bf 10}^{1} &+\hat\alpha_{1223345} &  0&1 &0 &0 &0  \cr
	\hline
\end{array}
\cr\cr
\begin{array}{|c|l|cccc|c|}
	\hline
	+\alpha_1 &+\hat\alpha_1 &  2& -1& 0& 0&0  \cr
	+\alpha_3 &+\hat\alpha_3 & 0 &-1 &2 &-1 &-1  \cr
	+w_{\bf 10}^{7} &+\hat\alpha_{235} & -1 &1 &0 &-1 &+1  \cr
	-w_{\bf 10}^{9} &-\hat\alpha_{35} & 0 &1 &-1 &1 &-1  \cr
	+w_{\bf 10}^{8} &+\hat\alpha_{345} &  0& -1& 0& 1& +1 \cr
	\hline
\end{array}
{\boxed{3\phantom{0}}}
&
{\boxed{5\phantom{0}}}
\begin{array}{|c|l|cccc|c|}
	\hline
	+\alpha_2 &+\hat\alpha_2 & -1 &2 &-1 &0 &0  \cr
	+\alpha_3 &+\hat\alpha_3 & 0 &-1 &2 &-1 &0  \cr
	+w_{\bf 10}^{4}&+\hat\alpha_{1235} &  1& 0& 0&-1 &+1  \cr
	-w_{\bf 10}^{7} &-\hat\alpha_{235} &  1& -1& 0& 1& -1 \cr
	+w_{\bf 10}^{8} &+\hat\alpha_{345} &  0& -1&0 &1 &+1  \cr
	\hline
\end{array}
\cr\cr
\textcolor{blue}{
\begin{array}{|c|l|cccc|c|}
	\hline
	+\alpha_2 &+\hat\alpha_2 & -1 &2 &-1 &0 &0  \cr
	+\alpha_4 &+\hat\alpha_4 & 0 &0 & -1& 2& 0 \cr
	{+w_{\bf 10}^{4}} &+\hat\alpha_{1235} & 1 &0 &0 &-1 &+1  \cr
	{-w_{\bf 10}^{6}} &-\hat\alpha_{2345} &  1& -1& 1& -1& -1 \cr
	{+w_{\bf 10}^{5}} &+\hat\alpha_{23345} &  -1& 0& 1& 0& 0 \cr
	\hline
\end{array}
{\boxed{9\phantom{0}}}
}
&
\textcolor{blue}{
{\boxed{8\phantom{0}}}
\begin{array}{|c|l|cccc|c|}
	\hline
	+\alpha_1 &+\hat\alpha_1 & 2 &-1 &0 &0 &0  \cr
	+\alpha_3 &+\hat\alpha_3 & 0 & -1& 2& -1& -1 \cr
	{-w_{\bf 10}^{4}} &-\hat\alpha_{1235} &  -1& 0& 0& 1& -1 \cr
	{+w_{\bf 10}^{6}} &+\hat\alpha_{2345} &  -1& 1& -1& 1& +1 \cr
	{-w_{\bf 10}^{8}} &-\hat\alpha_{345} & 0 &1 &0 &-1 &-1  \cr
	\hline
\end{array}
}
\cr\cr
\begin{array}{|c|l|cccc|c|}
	\hline
	+\alpha_2 &+\hat\alpha_2 &  -1& 2& -1& 0& 0 \cr
	+\alpha_3 &+\hat\alpha_3 & 0 &-1 &2 &-1 &-1  \cr
	-w_{\bf 10}^{4} &-\hat\alpha_{1235} &  -1& 0& 0& 1& -1 \cr
	+w_{\bf 10}^{3} &+\hat\alpha_{12345} & 1 &0 &-1 &1 &+1  \cr
	-w_{\bf 10}^{5} &-\hat\alpha_{23345} &  1& 0& -1& 0& 0 \cr
	\hline
\end{array}
{\boxed{12}}
&
{\boxed{14}}
\begin{array}{|c|l|cccc|c|}
	\hline
	+\alpha_2 &+\hat\alpha_2 & -1 &2 &-1 &0 &0  \cr
	+\alpha_4 &+\hat\alpha_4 & 0 &0 &-1 &2 &0  \cr
	-w_{\bf 10}^{3} &-\hat\alpha_{12345} & -1 &0 &1 &-1 &-1  \cr
	+w_{\bf 10}^{2} &+\hat\alpha_{123345} & 1 &-1 &1 &0 &0  \cr
	-w_{\bf 10}^{5} &-\hat\alpha_{23345} &1  &0 &-1 &0 &0  \cr
	\hline
\end{array}
\cr\cr
\textcolor{blue}{
\begin{array}{|c|l|cccc|c|}
	\hline
	+\alpha_3 &+\hat\alpha_3 &  0& -1& 2& -1& 0 \cr
	{+w_{\bf 10}^{4}} &+\hat\alpha_{1235} &1  &0 &0 &-1 &+1  \cr
	{-w_{\bf 10}^{7}} &-\hat\alpha_{235} &  1& -1& 0& 1& -1 \cr
	{+w_{\bf 10}^{6}} &+\hat\alpha_{2345} & -1 &1 &-1 &1 &+1  \cr
	-w_{\bf 10}^{8} &-\hat\alpha_{345} &0  &1 &0 &-1 &-1  \cr
	\hline
\end{array}
{\boxed{7\phantom{0}}}
}
&
\textcolor{blue}{
{\boxed{10}}
\begin{array}{|c|l|cccc|c|}
	\hline
	+\alpha_2 &+\hat\alpha_2 &  -1& 2& -1& 0& 0 \cr
	{-w_{\bf 10}^{4}} &-\hat\alpha_{1235} &  -1& 0& 0& 1& -1 \cr
	{+w_{\bf 10}^{3}} &+\hat\alpha_{12345} & 1 &0 &-1 &1 &+1  \cr
	{-w_{\bf 10}^{6}} &-\hat\alpha_{2345} & 1 &-1 &1 &-1 &-1  \cr
	+w_{\bf 10}^{5} &+\hat\alpha_{23345} & -1 &0 &1 &0 &0  \cr
	\hline
\end{array}
}
\end{array}
$$}
\caption{Phases/resolutions for $SU(5)$ with anti-symmetric representation ${\bf 10}$. The blue boxes indicate the $SU(5)$, the remaining ones are $U(5)$ phases, that are not $SU(5)$ phases. The labels are as in figure \ref{fig:U5A}, which shows the flops between the phases, and the $SU(5)$ phases in figure \ref{fig:SU5Flops}. }
\label{SU5ATable}
\end{table}

\section{Phases of $SO(2n)$ with Fundamental Matter}
\label{app:SO}


\begin{figure}
    \centering
    \includegraphics[width=10cm]{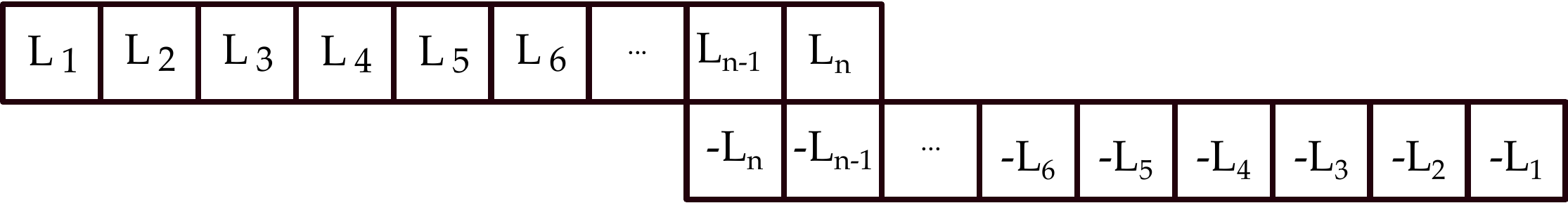}
    \caption{The standard representation of $SO(2n)$.}
    \label{IMG:SOVrep}
\end{figure}
\begin{figure}
    \centering
    \includegraphics[width=10cm]{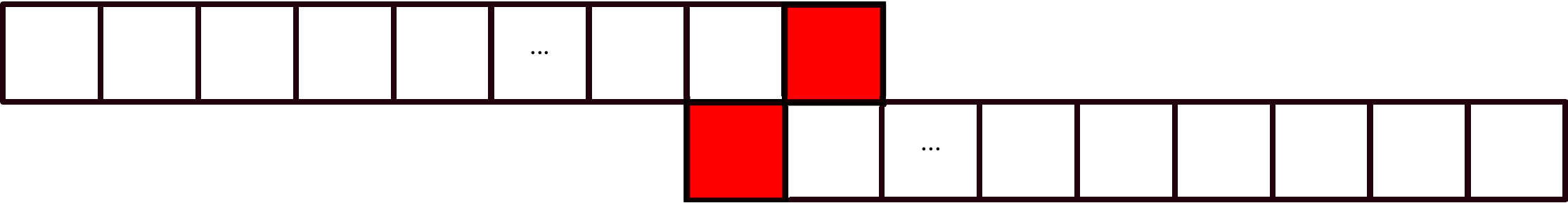}
    \caption{The sign condition for the $SO(2n)$ theory.}
    \label{IMG:SOSignC}
\end{figure}

In this section we consider the phases of the $SO(2n)$ theory with respect to
the vector representation, $V$. We consider the decomposition 
\begin{equation}
    \mathfrak{so}(2n+2) \rightarrow \mathfrak{so}(2n) \oplus \mathfrak{u}(1)
    \,,
\end{equation}
such that
\begin{equation}
    \text{adj}(\mathfrak{so}(2n+2)) \rightarrow
    \text{adj}(\mathfrak{so}(2n)) \oplus \text{adj}(\mathfrak{u}(1)) \oplus V
    \oplus V \,.
\end{equation}

From section \ref{Sec:WGQ} the phases of the $SO(2n)\times U(1)$ theory 
are determined by the quotiented Weyl group
\begin{equation}
    \frac{|W_{\mathfrak{so}(2n+2)}|}{|W_{\mathfrak{so}(2n)}|} = 2(n+1) \,.
\end{equation}
As for $\mathfrak{su}(n)$ we introduce the following representation of the positive roots
for $SO(2n)$.


\begin{figure}
    \centering
    \includegraphics[width=10cm]{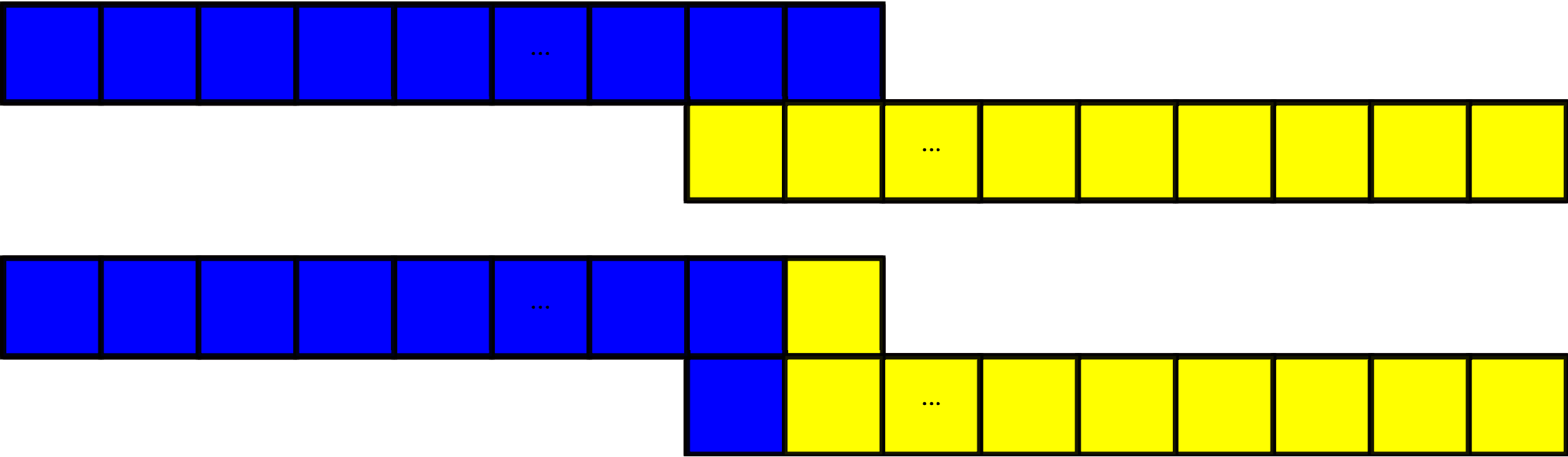}
    \caption{The phases of the $SO(2n)$ theory. Blue indicates the box is 
        decorated with a plus sign and yellow with a negative.}
    \label{IMG:SOPhases}
\end{figure}

\begin{equation}
    \rootsystem^+ = \{L_i - L_j\,|\,i=1,\cdots,n\,,i<j\} \cup
    \{L_i + L_j\,|\,i=1,\cdots,n\,,i<j\} \,.
\end{equation}
The weights of the vector representation are then
\begin{equation}
    V = \{\pm L_i \,|\, i=1,\cdots,n\} \,.
\end{equation}
In fact the weights of the representation of $SO(2n)\times U(1)$ are 
$w^i$, $i=1, \cdots, 2n$, where the $\mathfrak{u}(1)$ generator is given by 
identifying $w^i$ with $-w^{n+i}$
\begin{equation}
    U(1): \qquad \sum_{i=1}^{2n} w^{i} - w^{n+i} \,.
\end{equation}
One can present the representation $V$ as in figure \ref{IMG:SOVrep}.
Similarly to section \ref{Sec:FlowRules} there exist flow rules which
determine whether the decorated box graph gives a consistent phase. For
$SO(2n)\times U(1)$ these are
\begin{equation}\label{FlowsRulesV1}
        \begin{array}{cc}
            \boxed{+} \quad \leftarrow& \boxed{+} \cr
            & \cr
                 & \uparrow \cr
                     &        \cr
                         & \boxed{+} 
        \end{array} \qquad
        \begin{array}{cc}
            \boxed{-} \quad \rightarrow& \boxed{-} \cr
            \cr 
                 & \downarrow \cr
                     &            \cr
                         & \boxed{-}
        \end{array}
\end{equation}
One can do some combinatorics with the flow rules and so count the number of
$SO(2n)\times U(1)$ phases, which agrees with the order of the Weyl
group quotient.

The sign condition for the $SO(2n)$ phases is that the sign decorating the
boxes corresponding to the $L_n$ and $-L_n$ must be different, as depicted in
figure \ref{IMG:SOSignC}. The only phases consistent with both the flow rules
and the sign condition are shown in figure \ref{IMG:SOPhases}, therefore, 
\begin{equation}
    SO(2n), \text{ with } {\bf 2n}: \quad \#\text{Phases} = 2\,.
\end{equation}
The two phases can be characterized by the sign of $L_n$, however there is a
$\mathbb{Z}_2$ outer automorphism of $SO(2n)$ which swaps $\alpha_{n-1}
\leftrightarrow \alpha_n$. This swaps $L_n \leftrightarrow -L_n$. From the
geometry we would thus not expect to distinguish the two phases.


\section{Phases of the $E$-type groups}
\label{app:EPhases}

In this appendix we consider the phases of the exceptional $E_n$ type 
theories with matter, deriving the flow rules for the representation graphs,
the decorated box graphs, as well as the flop diagrams. The examples we consider are 
\begin{equation}\label{E678}
    \begin{aligned}
        \mathfrak{e}_7 &\quad \rightarrow \quad \mathfrak{e}_6 \oplus \mathfrak{u}(1) \cr
        {\bf 133}&\quad \rightarrow \quad {\bf 78}_0 \oplus {\bf 1}_0 \oplus {\bf 27}_2 \oplus {\overline{\bf 27}}_{-2 }\cr
        \mathfrak{e}_8 & \quad \rightarrow \quad \mathfrak{e}_7 \oplus \mathfrak{su}(2) \cr
        {\bf 248}&\quad \rightarrow \quad ({\bf 133}, {\bf 1} ) \oplus ({\bf 3}, {\bf 1})  \oplus ({\bf 56}, {\bf 2}) 
    \end{aligned}
\end{equation}


\subsection{Phases of $E_6$ with ${\bf 27}$ matter}

\begin{figure}
    \centering
    \includegraphics[width=7cm]{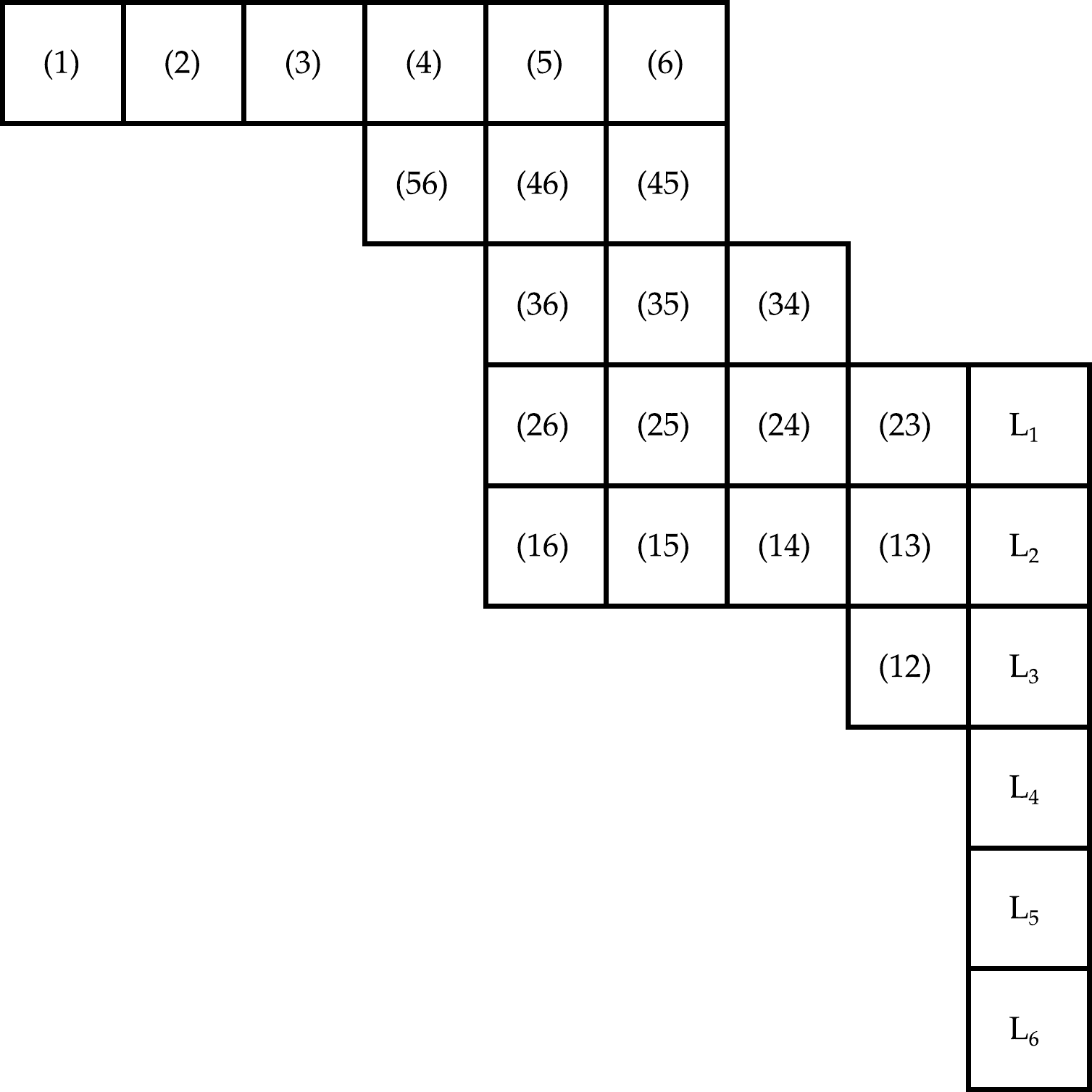}
    \caption{The undecorated box graph of $E_6$ with the ${\bf 27}$
        representation. $(ij)$ denotes the weight $L_0 - L_i - L_j$, and 
        $(i)$ the weight $2L_0 - \sum_{k=1}^5L_{j_k}$ where $j_k \neq i$ are
        distinct.
    }
    \label{fig:E6RepGraph27}
\end{figure}

\begin{figure}
    \centering
    \includegraphics[width=7cm]{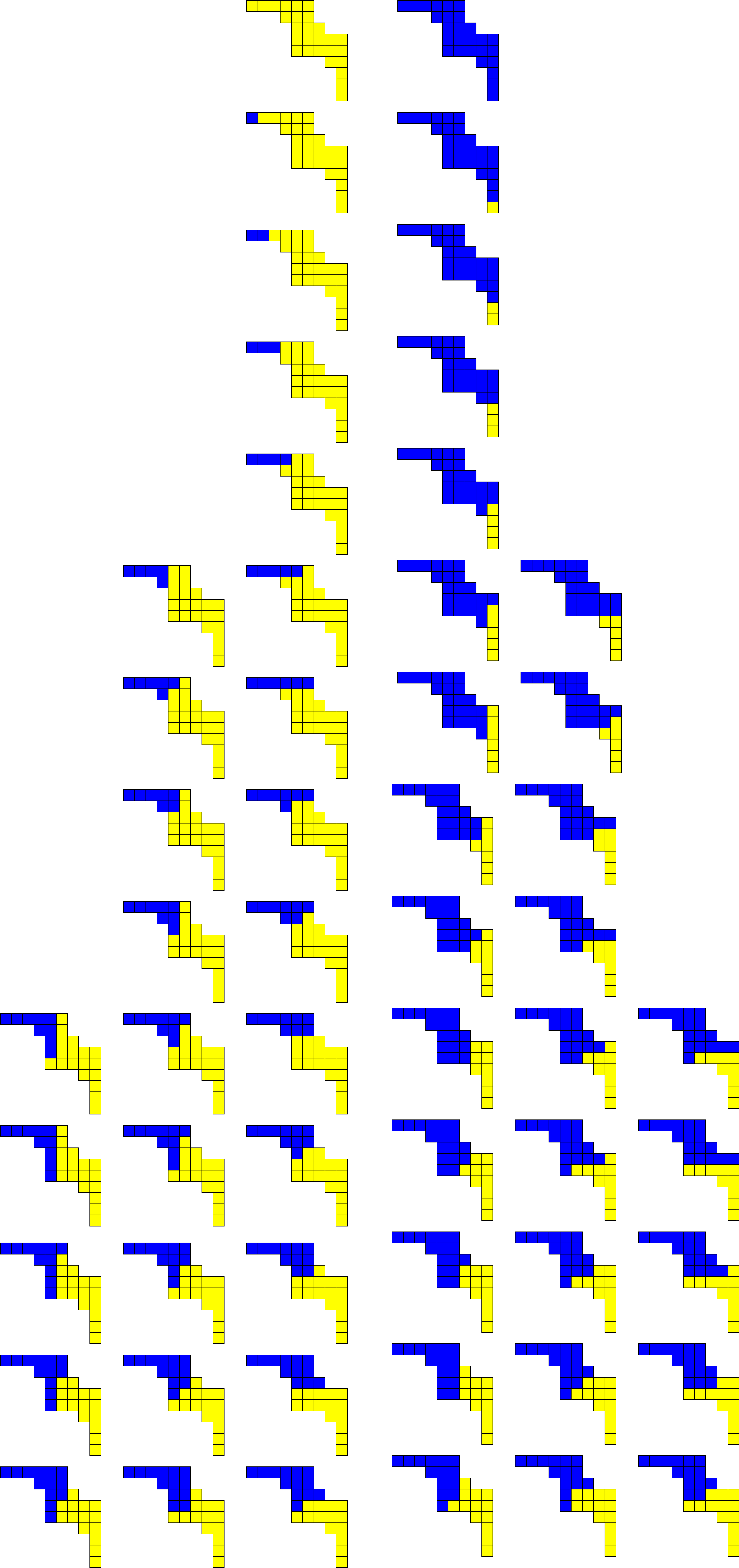}
    \caption{Phases in terms of decorated box graphs of the
        $\mathfrak{e_6}\oplus\mathfrak{u}(1)$ theory with matter in the ${\bf 27}$. Blue are +, yellow are -.   }   \label{fig:E6Phases27}
\end{figure}
\begin{figure}
    \centering
    \includegraphics[width=4.2cm]{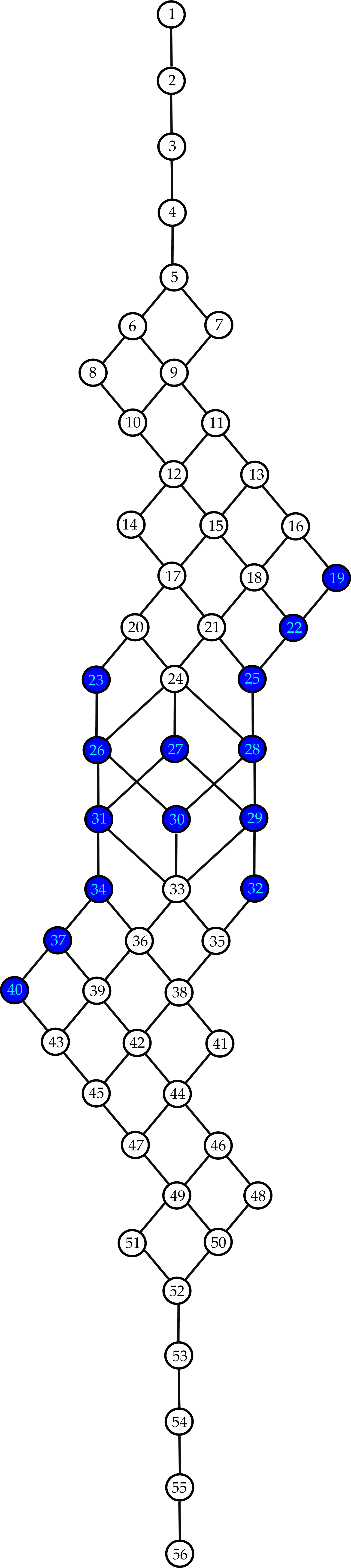}
    \caption{Phase diagram of the  $\mathfrak{e}_6\oplus \mathfrak{u}(1)$
    theory with matter in the ${\bf 27}$. The flop diagram agrees with the
    representation graph of the {\bf 56} of $E_7$. The nodes colored in blue
    are the phases of the $\mathfrak{e}_6$ theory. The numbering corresponds to the decorated box graphs in figure \ref{fig:E6Phases27}, by considering each column in that diagram, read from left to right, top to bottom. }   
   \label{fig:E6Flops27}
\end{figure}

The phases of the $E_6$ theory with ${\bf 27}$ has the standard flow rules
given in (\ref{Flows1}) acting on the representation graph in figure 
\ref{fig:E6RepGraph27}. The weights of the representation shown there are in the standard representation in terms of $L_i$ is given in appendix \ref{app:ERoots}. In addition in order to be phases of the $E_6$, not the $E_6 \times U(1)$ theory, we require that 
\begin{equation}\label{E6Trace}
    \mathcal{E}_{E_6} = 3L_0 - (L_1 +L_2 + L_3 + L_4 + L_5 + L_6 )=0 \,. 
\end{equation}
The phases written in terms of decorated box diagrams satisfying the flow 
rules, without necessarily satisfying (\ref{E6Trace}), are shown in figure 
\ref{fig:E6Phases27}. From the Weyl group quotient we indeed expect there to be
\begin{equation}
\left| {W_{\mathfrak{e}_7} \over W_{\mathfrak{e}_6}}\right| = 56 \,.
\end{equation}
The flop transitions between phases are obtained from these diagrams by 
considering single sign box changes, that are compatible with the flow rules.
The resulting flop diagram is shown in figure \ref{fig:E6Flops27}. 
Again, the flop diagram for the theory with gauge group $E_6\times U(1)$ is 
the representation graph of $E_7$ with ${\bf 56}$. 

The phases, which satisfy (\ref{E6Trace}) are shown in blue and are the 
actual phase diagrams of $E_6$ (not $E_6 \times U(1)$).


\subsection{Phases of $E_7$ with {\bf 56} matter}
\label{sec:E756}

\begin{figure}
    \centering
    \includegraphics[width=16cm]{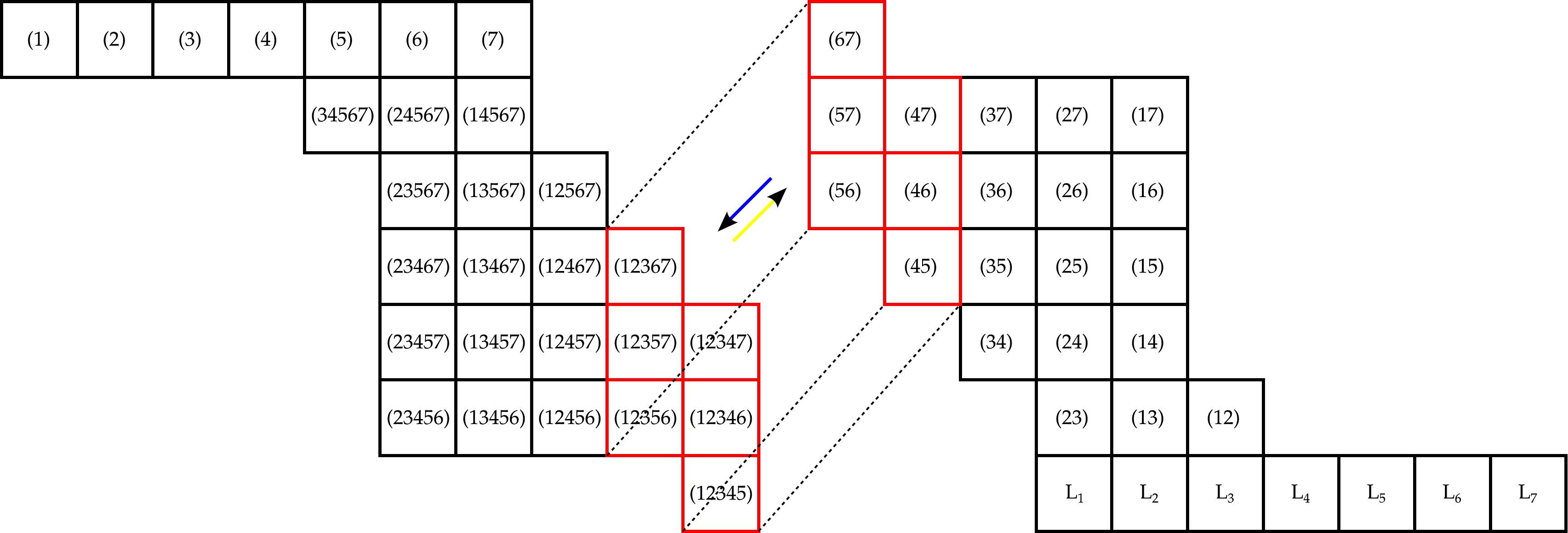}
    \caption{The undecorated box graph for $E_7$ with ${\bf 56}$. $(ij)$
denotes weight $L_0 - L_i - L_j$, $(ijklm)$ denotes weight $L_0 - L_i - L_j -
L_k - L_l - L_m$, and $(i)$ denotes weight $3L_0 - \sum_{k=1}^5L_{j_k} -
2L_i$, where the $j_k \neq i$ are distinct. The flow rules are as in
(\ref{Flows1}) and between layers $+$, resp. $-$, flows in the direction
of the blue, resp. yellow, arrows.}
    \label{fig:E7RepGraph56}
\end{figure}

\begin{landscape}
\begin{figure}
    \centering    \includegraphics[width=19cm]{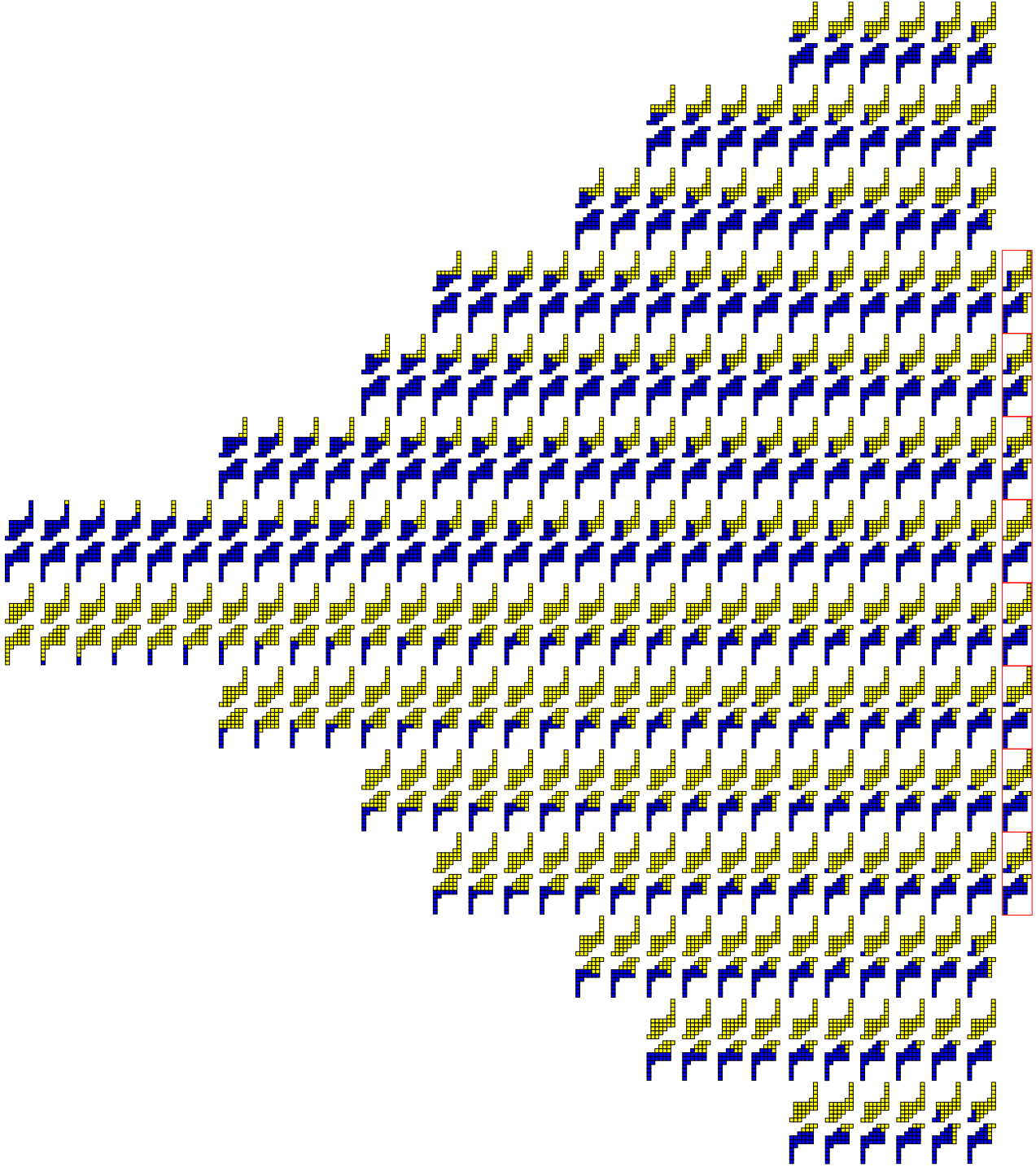}
    \caption{Phases in terms of decorated box graphs of the
        $\mathfrak{e}_7\oplus\mathfrak{su}(2)$ theory with matter in the ${\bf 56}$. Blue are +, yellow are -.   }   \label{fig:E7Phases56}
\end{figure}
\end{landscape}

For ${E_7}$ with ${\bf 56}$ matter the representation graph is shown in 
figure \ref{fig:E7RepGraph56}. An additional constraint differentiating 
the $E_7$ phases from the $E_7 \times SU(2)$ phases is
\begin{equation}\label{E7Trace}
    \mathcal{E}_{E_7} = 3L_0 - (L_1 +L_2 + L_3 + L_4 + L_5 + L_6 + L_7)=0 \,. 
\end{equation}
In figure \ref{fig:E7Phases56} we show the phases of the $E_7\times SU(2)$ 
theory with ${\bf 56}$ matter, as well as, boxed in red, the phases that 
satisfy \ref{E7Trace}.


\begin{figure}
    \centering    \includegraphics[width=12cm]{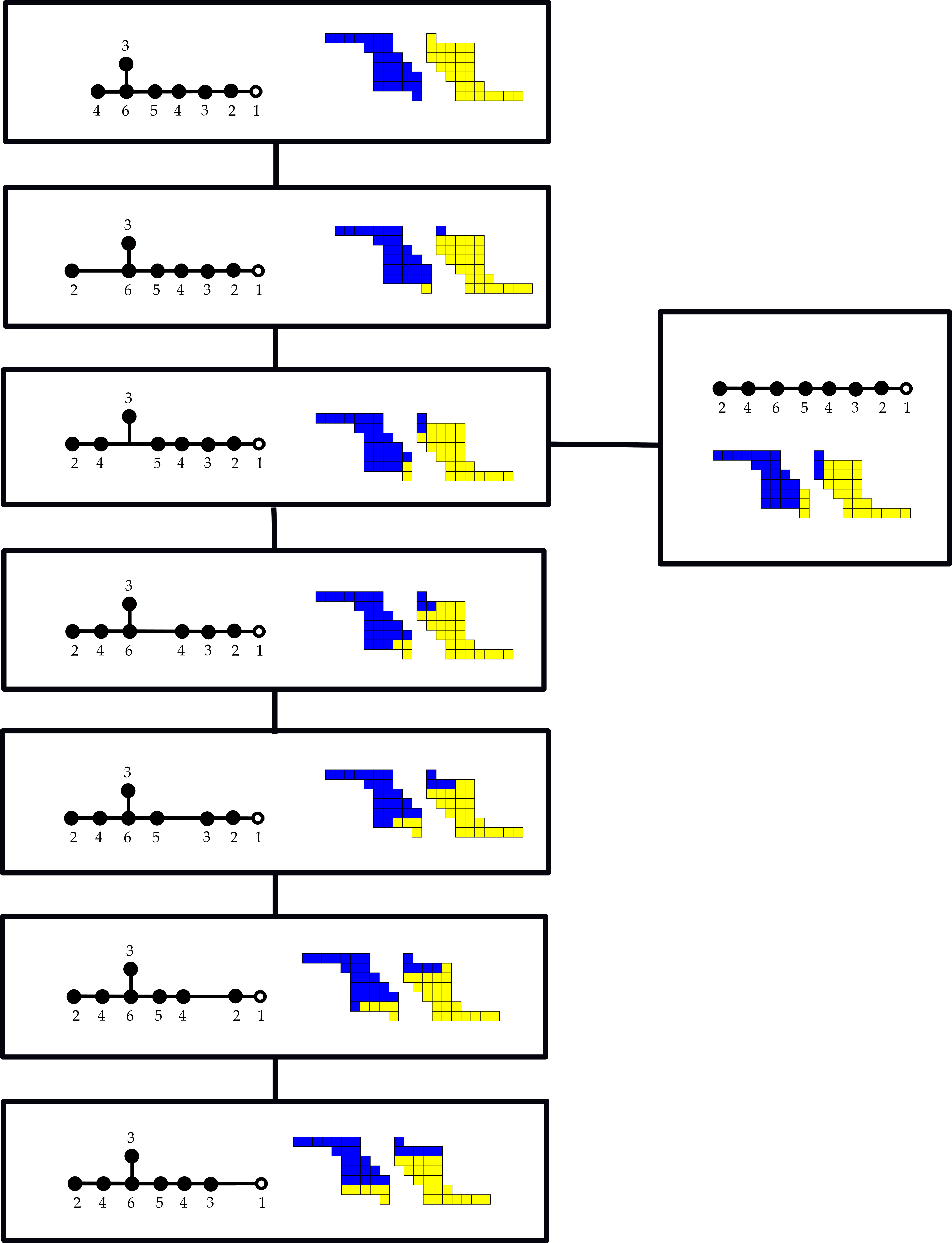}
    \caption{Monodromy-reduced fibers for $E_8$, we shown the intersection graph including multiplicities for each of the decorated box graphs. Note the the fibers correspond to type $II^*$ Kodaira fibers, with one non-affine node deleted. The lines connecting black boxes  correspond to flop transitions, and form a (non-affine) $E_8$ Dynkin diagram. }   \label{fig:E8MonoFibs}
\end{figure}

We shall now determine the fiber structure for the eight $E_8$ phases, those which
satisfy (\ref{E7Trace}). These will give the $E_8$ rank one monodromy fiber
types. The results of this section are summarized in figure
\ref{fig:E8MonoFibs}, and we notice again that the flop diagram is exactly the
$E_8$ Dynkin diagram; each monodromy-reduced fiber corresponds to an affine
$E_8$ fiber structure with exactly one (non-affine) node removed.

We begin by specifying the convention for the $E_7$ data, following appendix
\ref{app:ERoots}; we list here the intersections among the curves corresponding
to roots
\begin{equation}\label{eqn:E7Intersections}
    \begin{aligned}
        F_1 \cdot F_2 = +1, \quad F_2 \cdot F_3 = +1, \quad F_3 \cdot F_4 =
        +1, \cr
        F_3 \cdot F_7 = +1, \quad F_4 \cdot F_5 = +1, \quad F_5 \cdot F_6 =
        +1. \cr
    \end{aligned}
\end{equation}
The multiplicity of each curve is $\#F_i = (2, 3, 4, 3, 2, 1, 2)$.
Additionally there is the affine root
\begin{equation}
    F_0 = - \sum_{i=1}^7 n_i F_i \,,
\end{equation}
where the $n_i$ are the multiplicities of the roots. This curve has
multiplicity $1$, and only intersects the other roots through $F_0 \cdot F_1 =
+1$.

Let us consider first the phase at the top of figure
\ref{fig:E8MonoFibs}. The splitting of the roots of the $E_7$ can be read off
from the box graph as being just
\begin{equation}
    \begin{aligned}
        F_7 \rightarrow C^+_{12345} + C^-_{67} + F_4 + 2F_5 + F_6 \,.
    \end{aligned}
\end{equation}
Under the $\mathbb{Z}_2$ action coming from the Weyl group of the
$\mathfrak{su}(2)$ the curves $C^+_{12345}$ and $C^-_{67}$ are identified. The
extremal curves in this phase are then
\begin{equation}
    \mathcal{K}_{\includegraphics[width=1.5cm]{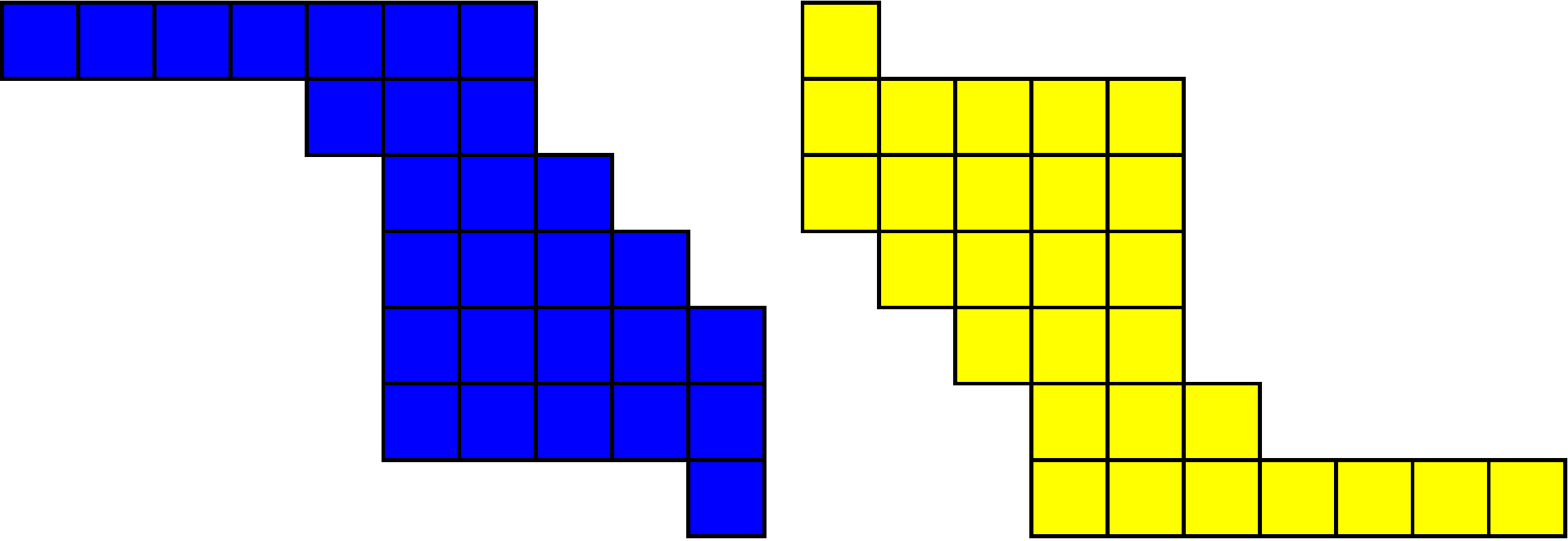}} = \{F_1, F_2, F_3,
        F_4, F_5, F_6, C^+_{12345}\} \,,
\end{equation}
with corresponding multiplicities $\{2, 3, 4, 5, 6, 3, 4\}$. The box graph
tells us the following intersections of the curve $C^+_{12345}$ with the roots
\begin{equation}
    \begin{aligned}
        &C^+_{12345} \cdot F_1 = 0, \quad &C^+_{12345} \cdot F_2 = 0, \quad
        &C^+_{12345} \cdot F_3 = 0, \cr
        &C^+_{12345} \cdot F_4 = 0, \quad &C^+_{12345} \cdot F_5 = +1, \quad
        &C^+_{12345} \cdot F_6 = 0. \cr
    \end{aligned}
\end{equation}
The fiber type is then seen to be that of the topmost phase in figure
\ref{fig:E8MonoFibs}.

Let us now consider the second top phase in figure \ref{fig:E8MonoFibs}, where
the curves that become reducible are
\begin{equation}
    \begin{aligned}
        &F_5 \rightarrow C^+_{12346} + C^-_{12345} \equiv C^+_{67} + C^-_{57}
        \cr
        &F_7 \rightarrow C^+_{12346} + C^-_{57} + F_4 + F_6 \,.
    \end{aligned}
\end{equation}
The $\mathbb{Z}_2$ symmetry identifies the curves $C^+_{12346} \leftrightarrow
C^-_{57}$ and $C^-_{12345} \leftrightarrow C^+_{67}$. The extremal curves are
then
\begin{equation}
    \mathcal{K}_{\includegraphics[width=1.5cm]{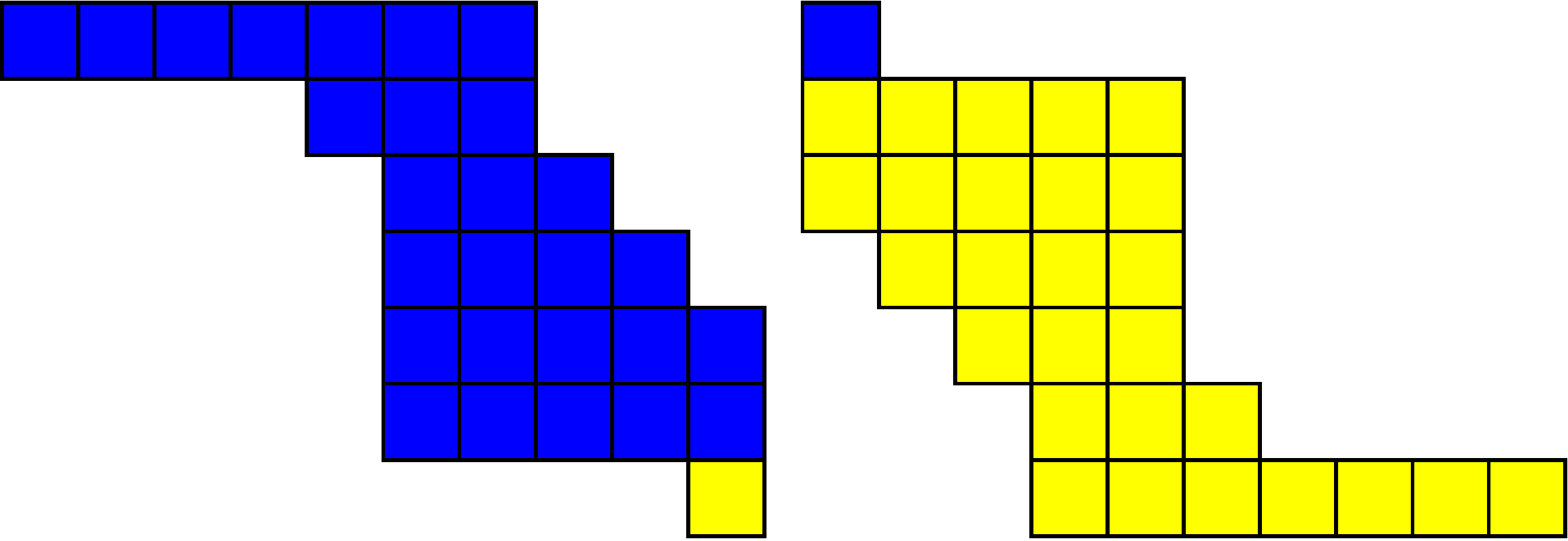}} = \{F_1, F_2,
        F_3, F_4, F_6, C^+_{12346}, C^+_{67}\} \,,
\end{equation}
where the multiplicities are $\{2, 3, 4, 5, 3, 6, 2\}$. We read off of the box
graph the non-trivial intersections involving the new extremal curves 
\begin{equation}
    \begin{aligned}
        C^+_{12346} \cdot F_4 = +1, \quad C^+_{12346} \cdot F_6 = +1, \quad
        C^+_{67} \cdot C^+_{12346} = +1. \cr
    \end{aligned}
\end{equation}
Using these intersections in addition to those given for the $E_7$ roots in
\ref{eqn:E7Intersections} we produce the intersection graph depicted in the
second top box of figure \ref{fig:E8MonoFibs}.

Moving down figure \ref{fig:E8MonoFibs} from the previously considered phase
we reach the phase for which the splitting into irreducible curves takes the
form
\begin{equation}
    \begin{aligned}
        &F_4 \rightarrow C^+_{12356} + C^-_{12346} \equiv C^+_{57} + C^-_{47} \cr
        &F_6 \rightarrow C^+_{12347} + C^-_{12346} \equiv C^+_{57} + C^-_{56} \cr
        &F_7 \rightarrow C^+_{12347} + C^-_{47} \equiv C^+_{12356} + C^-_{56} \,.
    \end{aligned}
\end{equation}
The $\mathbb{Z}_2$ symmetry identifies the curves
\begin{equation}
    \begin{aligned}
        &C^+_{12356} \leftrightarrow C^-_{47} \cr
        &C^+_{57} \leftrightarrow C^-_{12346} \cr
        &C^+_{12347} \leftrightarrow C^-_{56} \,,
    \end{aligned}
\end{equation}
leaving the set of extremal rays
\begin{equation}
    \mathcal{K}_{\includegraphics[width=1.5cm]{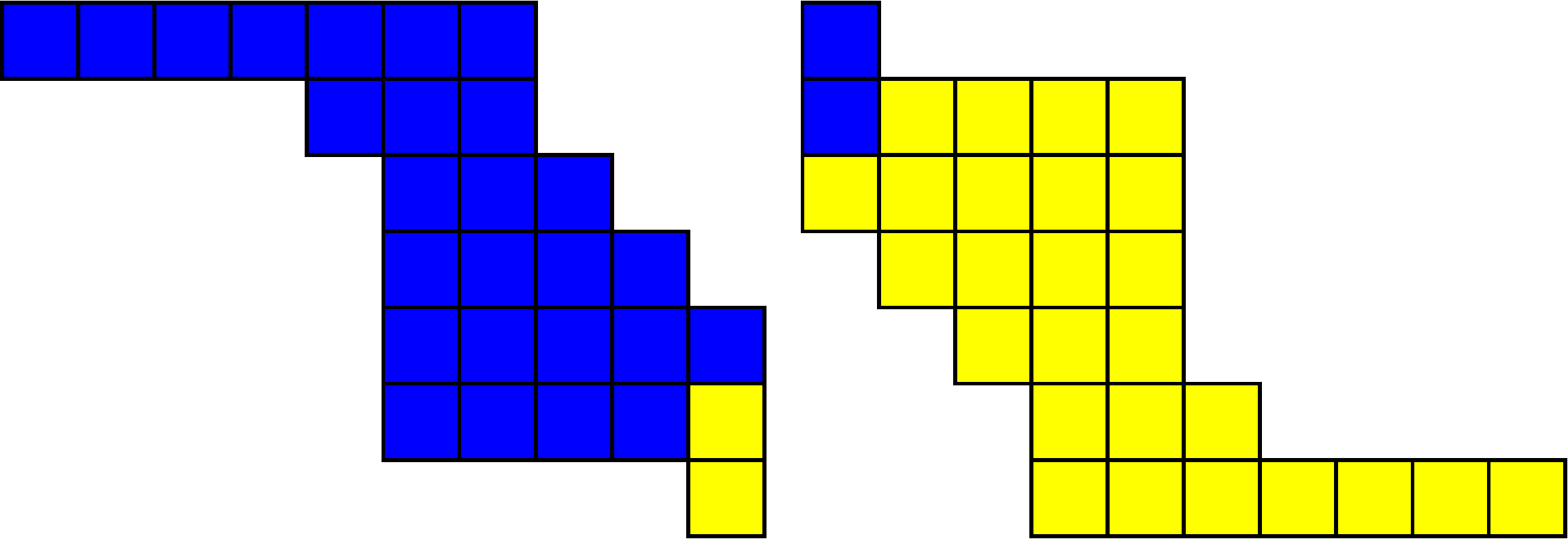}} = \{F_1, F_2, F_3,
        F_5, C^+_{12347}, C^+_{12356}, C^+_{57}\} \,,
\end{equation}
with the associated multiplicities $\{2, 3, 4, 2, 3, 5, 4\}$. The box graph
gives the following intersection numbers involving the new irreducible curves
\begin{equation}
    \begin{aligned}
        &C^+_{12356} \cdot F_3 = +1, \quad C^+_{57} \cdot F_5 = +1, \cr
        C^+_{12356} \cdot C^+_{57} = &+1, \quad C^+_{12356} \cdot C^+_{12347} =
        +1, \quad C^+_{12347} \cdot C^+_{57} = +1 \,.
    \end{aligned}
\end{equation}
It is now straightforward to read off that these irreducible curves intersect
in the diagram associated to this phase in figure \ref{fig:E8MonoFibs}; it
takes the form of an affine $E_8$ fiber type with the central node excised.

    
The majority of the other phases depicted in figure \ref{fig:E8MonoFibs} are
calculated by application of the methods described for the previous three
examples. There is, however, one phase, corresponding to the bottom-most phase
in figure \ref{fig:E8MonoFibs}, which is slightly more subtle, which, for this
purpose, we explain here. In this case we observe that the affine root $F_0$
is one of the curves that become reducible
\begin{equation}
    \begin{aligned}
        &F_6 \rightarrow C^+_{17} + C^-_{23456} + 2F_1 + 2F_2 + 2F_3 + F_4 \cr
        &F_0 \rightarrow C^+_{17} + C^-_{(7)} \,.
    \end{aligned}
\end{equation}
The curve denoted by $C^-_{(7)}$ is associated to the weight represented by
$(7)$ in figure \ref{fig:E7RepGraph56}, and, as it remains large in the singular
limit, takes its place as the new affine curve. The $\mathbb{Z}_2$ identifies
the curves $C^+_{17} \leftrightarrow C^-_{23456}$, and the extremal curves are
thus
\begin{equation}
    \mathcal{K}_{\includegraphics[width=1.5cm]{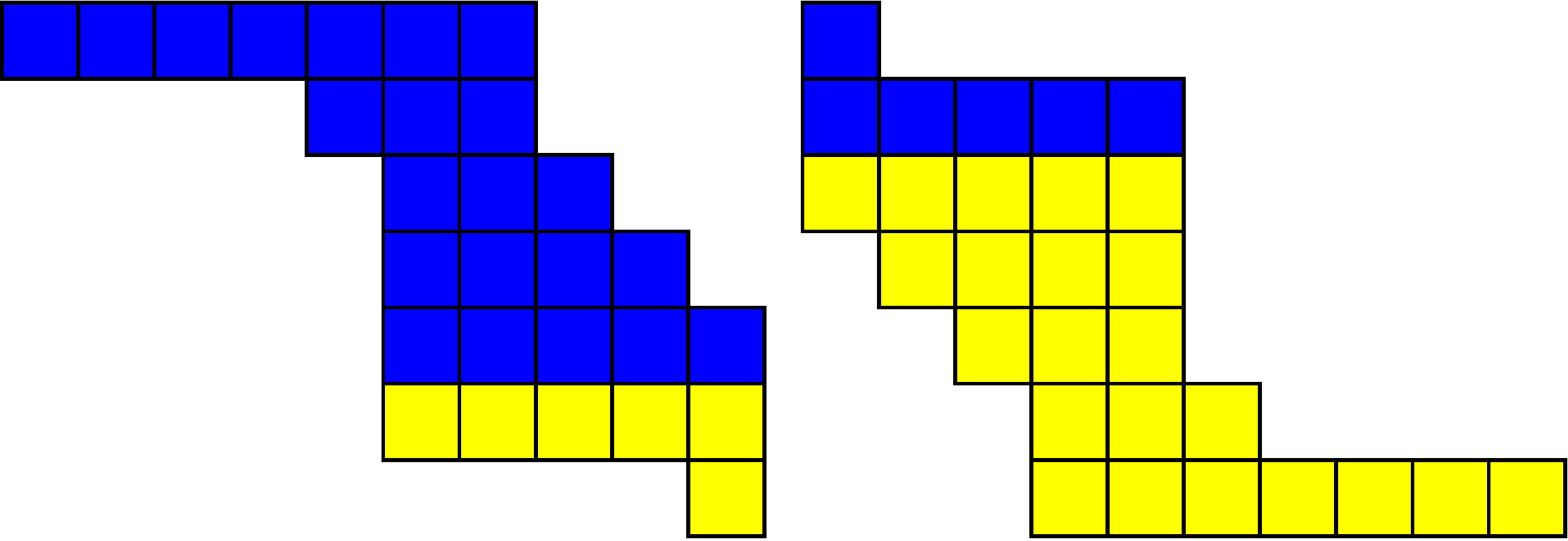}} = \{F_1, F_2,
        F_3, F_4, F_5, F_7, C^+_{17}\} \,,
\end{equation}
with multiplicities $\{4, 5, 6, 4, 2, 3, 3\}$ and non-trivial intersections
$C^+_{17} \cdot F_1 = +1$. Putting this information together results in the
fiber type shown in the bottom-most box of figure \ref{fig:E8MonoFibs}.

For the purposes of completeness we shall now briefly detail the pertinent
data of the remaining possible fiber structures. We begin with the phase 
\includegraphics[width=1.5cm]{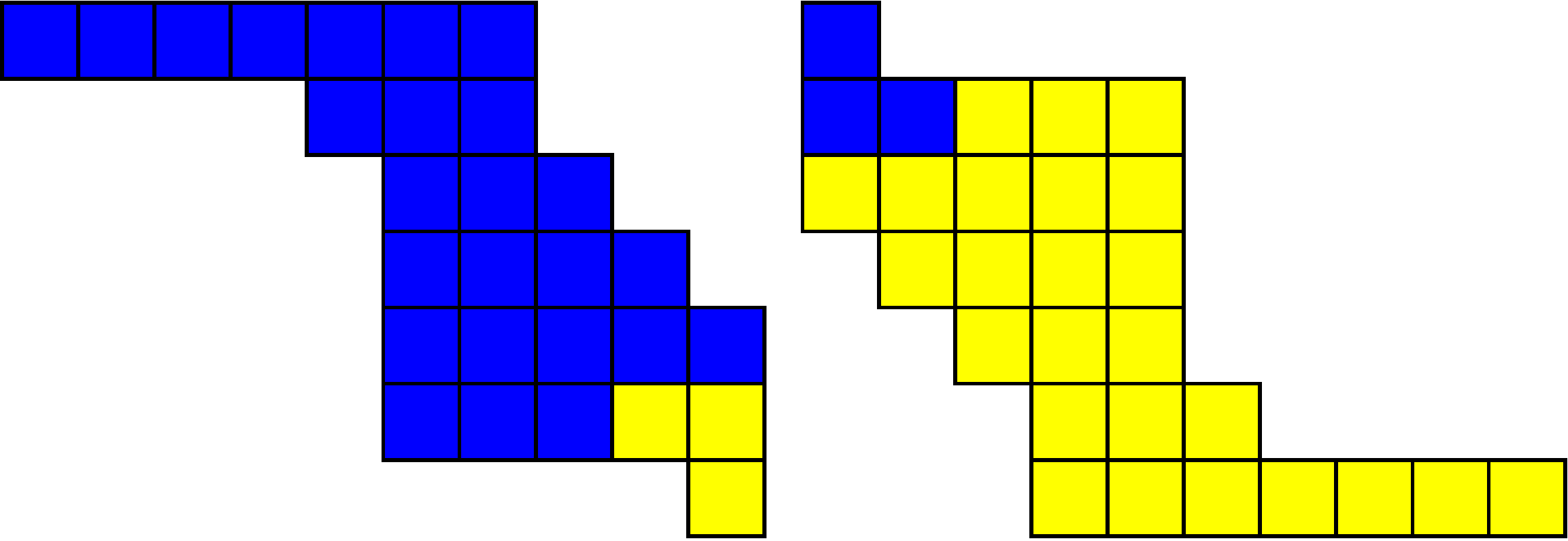}, which has the splitting
\begin{equation}
    \begin{aligned}
        &F_3 \rightarrow C^+_{12456} + C^-_{12356} \equiv C^+_{47} + C^-_{37}
        \cr
        &F_6 \rightarrow C^+_{47} + C^-_{12356} + F_4 + F_7 \,.
    \end{aligned}
\end{equation}
The $\mathbb{Z}_2$ symmetry identifies the curves $C^-_{12356}
\leftrightarrow C^+_{47}$ and $C^+_{12456} \leftrightarrow C^-_{37}$, which
makes the extremal generators
\begin{equation}
    \mathcal{K}_{\includegraphics[width=1.5cm]{E7Phase1.pdf}} = \{F_1, F_2,
        F_4, F_5, F_7, C^+_{12456}, C^+_{47}\} \,,
\end{equation}
with multiplicities $\{2, 3, 4, 2, 3, 4, 6\}$. The non-trivial intersections
involving the new irreducible curves are
\begin{equation}
    \begin{aligned}
        &C^+_{12456} \cdot F_2 = +1, \quad C^+_{47} \cdot F_4 = +1, \cr
        &C^+_{47} \cdot F_7 = +1, \quad C^+_{12456} \cdot C^+_{47} = +1 \,.
    \end{aligned}
\end{equation}
The phase \includegraphics[width=1.5cm]{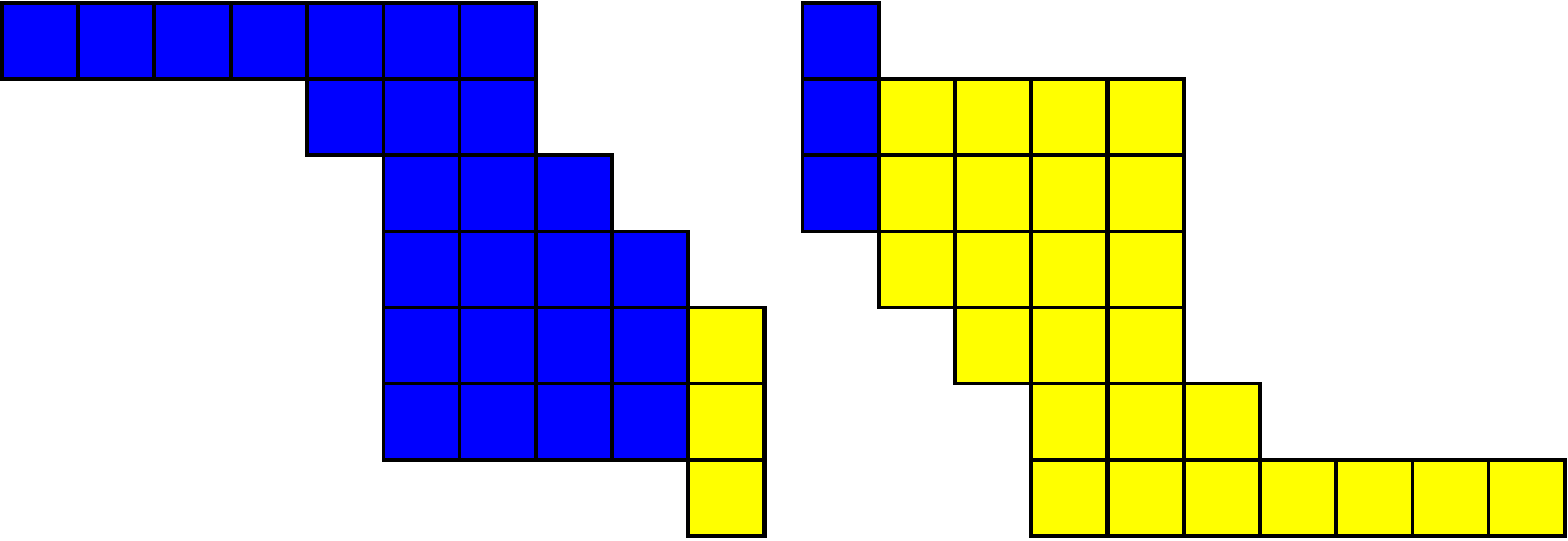} splits as
\begin{equation}
    \begin{aligned}
        F_4 \rightarrow C^+_{56} + C^-_{12347} + F_6 + F_7 \,.
    \end{aligned}
\end{equation}
The $\mathbb{Z}_2$ identifies the curves $C^+_{56} \leftrightarrow
C^-_{12347}$, and the extremal rays are
\begin{equation}
    \mathcal{K}_{\includegraphics[width=1.5cm]{E7Phase3.pdf}} = \{F_1, F_2,
        F_3, F_5, F_6, F_7, C^+_{56}\} \,,
\end{equation}
with multiplicities $\{2, 3, 4, 2, 4, 5, 6\}$, and intersections
\begin{equation}
    C^+_{56} \cdot F_6 = +1, \quad C^+_{56} \cdot F_7 = +1 \,.
\end{equation}
Let us now consider the phase \includegraphics[width=1.5cm]{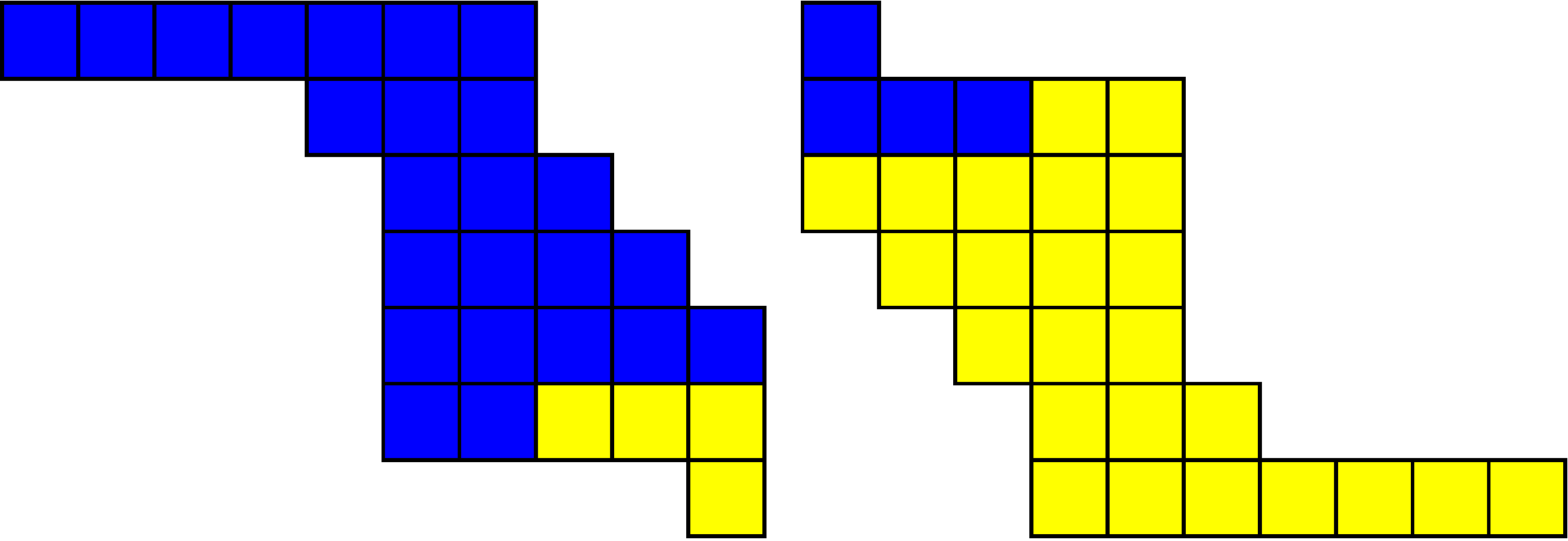}.
The curves split via
\begin{equation}
    \begin{aligned}
        &F_2 \rightarrow C^+_{13456} + C^-_{12456} \equiv C^+_{37} + C^-_{27}
        \cr
        &F_6 \rightarrow C^+_{37} + C^-_{12456} + 2F_3 + F_4 + F_7 \,,
    \end{aligned}
\end{equation}
where the curves $C^+_{37} \leftrightarrow C^-_{12456}$ are identified. The
extremal curves are
\begin{equation}
    \mathcal{K}_{\includegraphics[width=1.5cm]{E7Phase6.pdf}} = \{F_1, F_3,
        F_4, F_5, F_7, C^+_{37}, C^+_{13456}\} \,,
\end{equation}
and the multiplicities are $\{2, 6, 4, 2, 3, 5, 3\}$. The non-trivial
intersections are
\begin{equation}
    C^+_{37} \cdot F_3 = +1, \quad C^+{13456} \cdot F_1 = +1, \quad C^+_{37}
    \cdot C^+_{13456} = +1 \,.
\end{equation}
The final $E_7$ phase corresponds to the box graph
\includegraphics[width=1.5cm]{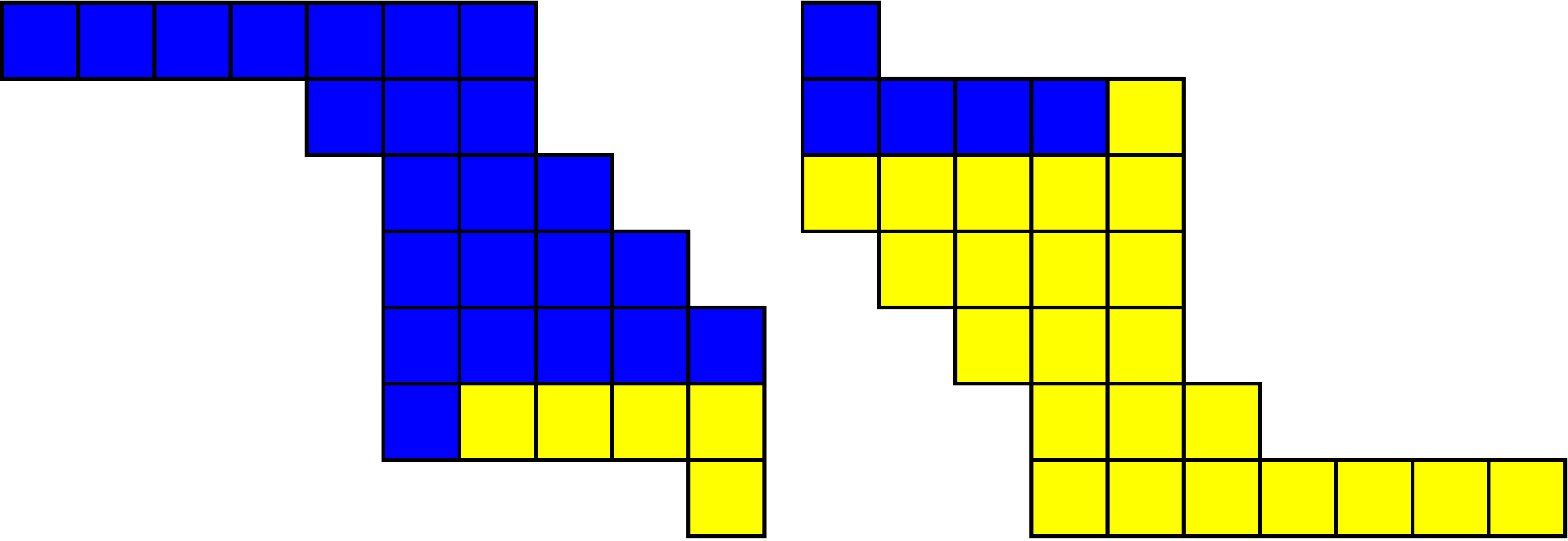}. One reads off from the box
graph that the curves split into the following components
\begin{equation}
    \begin{aligned}
        &F_1 \rightarrow C^+_{23456} + C^-_{13456} \equiv C^+_{27} + C^-_{17}
        \cr
        &F_6 \rightarrow C^+_{27} + C^-_{13456} + 2F_2 + 2F_3 + F_4 + F_7
        \,.
    \end{aligned}
\end{equation}
The $\mathbb{Z}_2$ symmetry from the Weyl group of the $\mathfrak{su}(2)$
identifies the curves
\begin{equation}
    \begin{aligned}
        &C^+_{23456} \leftrightarrow C^-_{17} \cr
        &C^-_{13456} \leftrightarrow C^+_{27} \,.
    \end{aligned}
\end{equation}
The extremal rays of this phase are
\begin{equation}
    \mathcal{K}_{\includegraphics[width=1.5cm]{E7Phase7.pdf}} = \{F_2, F_3,
        F_4, F_5, F_7, C^+_{27}, C^+_{23456}\} \,,
\end{equation}
which have respective multiplicities $\{5, 6, 4, 2, 3, 4, 2\}$. The
non-trivial intersections of the extremal curves involving the new
irreducible components are
\begin{equation}
    C^+_{27} \cdot F_2 = +1, \quad C^+_{27} \cdot C^+_{23456} = +1 \,.
\end{equation}
Using the intersections given above for each of these phases, and the
general intersection information given in (\ref{eqn:E7Intersections}),
allows the reproduction of the remaining fibers in figure
\ref{fig:E8MonoFibs}.


\newpage



\begin{thebibliography}{10}

\bibitem{Kodaira}
K.~Kodaira, {\it On compact complex analytic surfaces. {I and II}},  {\em Ann.
  of Math. (2)} {\bf 71} (1960) 111--152.

\bibitem{Neron}
A.~N{\'e}ron, {\it Mod\`eles minimaux des vari\'et\'es ab\'eliennes sur les
  corps locaux et globaux},  {\em Inst. Hautes \'Etudes Sci. Publ.Math. No.}
  {\bf 21} (1964) 128.

\bibitem{Cadavid:1995bk}
A.~Cadavid, A.~Ceresole, R.~D'Auria, and S.~Ferrara, {\it {Eleven-dimensional
  supergravity compactified on Calabi-Yau threefolds}},  {\em Phys.Lett.} {\bf
  B357} (1995) 76--80, [\href{http://xxx.lanl.gov/abs/hep-th/9506144}{{\tt
  hep-th/9506144}}].

\bibitem{Ferrara:1996hh}
S.~Ferrara, R.~R. Khuri, and R.~Minasian, {\it {M theory on a Calabi-Yau
  manifold}},  {\em Phys.Lett.} {\bf B375} (1996) 81--88,
  [\href{http://xxx.lanl.gov/abs/hep-th/9602102}{{\tt hep-th/9602102}}].

\bibitem{WitMF}
E.~Witten, {\it Phase transitions in {$M$}-theory and {$F$}-theory},  {\em
  Nuclear Phys. B} {\bf 471} (1996) 195--216,
  [\href{http://xxx.lanl.gov/abs/arXiv:hep-th/9603150}{{\tt
  arXiv:hep-th/9603150}}].

\bibitem{Ferrara:1996wv}
S.~Ferrara, R.~Minasian, and A.~Sagnotti, {\it {Low-energy analysis of M and F
  theories on Calabi-Yau threefolds}},  {\em Nucl.Phys.} {\bf B474} (1996)
  323--342, [\href{http://xxx.lanl.gov/abs/hep-th/9604097}{{\tt
  hep-th/9604097}}].

\bibitem{Becker:1996gj}
K.~Becker and M.~Becker, {\it M-theory on eight-manifolds},  {\em Nucl. Phys.
  B} {\bf 477} (1996) 155--167,
  [\href{http://xxx.lanl.gov/abs/arXiv:hep-th/9605053}{{\tt
  arXiv:hep-th/9605053}}].

\bibitem{SVW}
S.~Sethi, C.~Vafa, and E.~Witten, {\it Constraints on low-dimensional string
  compactifications},  {\em Nucl. Phys. B.} {\bf 480} (1996) 213--224,
  [\href{http://xxx.lanl.gov/abs/arXiv:hep-th/9606122}{{\tt
  arXiv:hep-th/9606122}}].

\bibitem{Morrison:1996xf}
D.~R. Morrison and N.~Seiberg, {\it {Extremal transitions and five-dimensional
  supersymmetric field theories}},  {\em Nucl.Phys.} {\bf B483} (1997)
  229--247, [\href{http://xxx.lanl.gov/abs/hep-th/9609070}{{\tt
  hep-th/9609070}}].

\bibitem{Witten:1996md}
E.~Witten, {\it {On flux quantization in M-theory and the effective action}},
  {\em J. Geom. Phys.} {\bf 22} (1997) 1--13,
  [\href{http://xxx.lanl.gov/abs/hep-th/9609122}{{\tt hep-th/9609122}}].

\bibitem{fiveDgauge}
K.~Intriligator, D.~R. Morrison, and N.~Seiberg, {\it Five-dimensional
  supersymmetric gauge theories and degenerations of {C}alabi--{Y}au spaces},
  {\em Nuclear Phys. B} {\bf 497} (1997) 56--100,
  [\href{http://xxx.lanl.gov/abs/arXiv:hep-th/9702198}{{\tt
  arXiv:hep-th/9702198}}].

\bibitem{Diaconescu:1998ua}
D.-E. Diaconescu and S.~Gukov, {\it {Three-dimensional N=2 gauge theories and
  degenerations of Calabi-Yau four folds}},  {\em Nucl.Phys.} {\bf B535} (1998)
  171--196, [\href{http://xxx.lanl.gov/abs/hep-th/9804059}{{\tt
  hep-th/9804059}}].

\bibitem{Gukov:1999ya}
S.~Gukov, C.~Vafa, and E.~Witten, {\it {CFT}'s from {C}alabi--{Y}au
  four-folds},  {\em Nucl. Phys. B} {\bf 584} (2000) 69--108,
  [\href{http://xxx.lanl.gov/abs/arXiv:hep-th/9906070}{{\tt
  arXiv:hep-th/9906070}}].

\bibitem{Vafa:1996xn}
C.~Vafa, {\it {Evidence for F-Theory}},  {\em Nucl. Phys.} {\bf B469} (1996)
  403--418, [\href{http://xxx.lanl.gov/abs/hep-th/9602022}{{\tt
  hep-th/9602022}}].

\bibitem{Morrison:1996na}
D.~R. Morrison and C.~Vafa, {\it {Compactifications of F-Theory on Calabi--Yau
  Threefolds -- I}},  {\em Nucl. Phys.} {\bf B473} (1996) 74--92,
  [\href{http://xxx.lanl.gov/abs/hep-th/9602114}{{\tt hep-th/9602114}}].

\bibitem{Morrison:1996pp}
D.~R. Morrison and C.~Vafa, {\it {Compactifications of F-Theory on Calabi--Yau
  Threefolds -- II}},  {\em Nucl. Phys.} {\bf B476} (1996) 437--469,
  [\href{http://xxx.lanl.gov/abs/hep-th/9603161}{{\tt hep-th/9603161}}].

\bibitem{Bershadsky:1997zs}
M.~Bershadsky, A.~Johansen, T.~Pantev, and V.~Sadov, {\it {On four-dimensional
  compactifications of F theory}},  {\em Nucl.Phys.} {\bf B505} (1997)
  165--201, [\href{http://xxx.lanl.gov/abs/hep-th/9701165}{{\tt
  hep-th/9701165}}].

\bibitem{deBoer:1997kr}
J.~de~Boer, K.~Hori, and Y.~Oz, {\it {Dynamics of N=2 supersymmetric gauge
  theories in three-dimensions}},  {\em Nucl.Phys.} {\bf B500} (1997) 163--191,
  [\href{http://xxx.lanl.gov/abs/hep-th/9703100}{{\tt hep-th/9703100}}].

\bibitem{Aharony:1997bx}
O.~Aharony, A.~Hanany, K.~A. Intriligator, N.~Seiberg, and M.~Strassler, {\it
  {Aspects of N=2 supersymmetric gauge theories in three-dimensions}},  {\em
  Nucl.Phys.} {\bf B499} (1997) 67--99,
  [\href{http://xxx.lanl.gov/abs/hep-th/9703110}{{\tt hep-th/9703110}}].

\bibitem{Grimm:2011fx}
T.~W. Grimm and H.~Hayashi, {\it {F-theory fluxes, Chirality and Chern-Simons
  theories}},  {\em JHEP} {\bf 1203} (2012) 027,
  [\href{http://xxx.lanl.gov/abs/1111.1232}{{\tt 1111.1232}}]. 53 pages, 5
  figures/ v2: typos corrected, minor improvements.

\bibitem{Cvetic:2012xn}
M.~Cvetic, T.~W. Grimm, and D.~Klevers, {\it {Anomaly Cancellation And Abelian
  Gauge Symmetries In F-theory}},  {\em JHEP} {\bf 1302} (2013) 101,
  [\href{http://xxx.lanl.gov/abs/1210.6034}{{\tt 1210.6034}}].

\bibitem{Hayashi:2013lra}
H.~Hayashi, C.~Lawrie, and S.~Schafer-Nameki, {\it {Phases, Flops and F-theory:
  SU(5) Gauge Theories}},  {\em JHEP} {\bf 1310} (2013) 046,
  [\href{http://xxx.lanl.gov/abs/1304.1678}{{\tt 1304.1678}}].

\bibitem{gorenstein-weyl}
S.~Katz and D.~R. Morrison, {\it Gorenstein threefold singularities with small
  resolutions via invariant theory for {W}eyl groups},  {\em J. Algebraic
  Geom.} {\bf 1} (1992) 449--530,
  [\href{http://xxx.lanl.gov/abs/arXiv:alg-geom/9202002}{{\tt
  arXiv:alg-geom/9202002}}].

\bibitem{MS}
J.~Marsano and S.~Schafer-Nameki, {\it {Yukawas, G-flux, and Spectral Covers
  from Resolved Calabi-Yau's}},  {\em JHEP} {\bf 1111} (2011) 098,
  [\href{http://xxx.lanl.gov/abs/1108.1794}{{\tt 1108.1794}}].

\bibitem{Katz:1996xe}
S.~Katz and C.~Vafa, {\it Matter from geometry},  {\em Nucl. Phys. B} {\bf 497}
  (1997) 146--154, [\href{http://xxx.lanl.gov/abs/arXiv:hep-th/9606086}{{\tt
  arXiv:hep-th/9606086}}].

\bibitem{Morrison:2011mb}
D.~R. Morrison and W.~Taylor, {\it {Matter and singularities}},  {\em JHEP}
  {\bf 1201} (2012) 022, [\href{http://xxx.lanl.gov/abs/1106.3563}{{\tt
  1106.3563}}].

\bibitem{Esole:2011sm}
M.~Esole and S.-T. Yau, {\it {Small resolutions of SU(5)-models in F-theory}},
  \href{http://xxx.lanl.gov/abs/1107.0733}{{\tt 1107.0733}}.

\bibitem{Lawrie:2012gg}
C.~Lawrie and S.~Schafer-Nameki, {\it {The Tate Form on Steroids: Resolution
  and Higher Codimension Fibers}},  {\em JHEP} {\bf 1304} (2013) 061,
  [\href{http://xxx.lanl.gov/abs/1212.2949}{{\tt 1212.2949}}].

\bibitem{birgeoRDP}
D.~R. Morrison, {\it The birational geometry of surfaces with rational double
  points},  {\em Math. Ann.} {\bf 271} (1985) 415--438.

\bibitem{Blumenhagen:2009yv}
R.~Blumenhagen, T.~W. Grimm, B.~Jurke, and T.~Weigand, {\it {Global F-theory
  GUTs}},  {\em Nucl.Phys.} {\bf B829} (2010) 325--369,
  [\href{http://xxx.lanl.gov/abs/0908.1784}{{\tt 0908.1784}}].

\bibitem{Grimm:2009yu}
T.~W. Grimm, S.~Krause, and T.~Weigand, {\it {F-Theory GUT Vacua on Compact
  Calabi-Yau Fourfolds}},  {\em JHEP} {\bf 1007} (2010) 037,
  [\href{http://xxx.lanl.gov/abs/0912.3524}{{\tt 0912.3524}}].

\bibitem{Grimm:2010ez}
T.~W. Grimm and T.~Weigand, {\it {On Abelian Gauge Symmetries and Proton Decay
  in Global F-theory GUTs}},  {\em Phys.Rev.} {\bf D82} (2010) 086009,
  [\href{http://xxx.lanl.gov/abs/1006.0226}{{\tt 1006.0226}}].

\bibitem{Krause:2011xj}
S.~Krause, C.~Mayrhofer, and T.~Weigand, {\it {$G_4$ flux, chiral matter and
  singularity resolution in F-theory compactifications}},  {\em Nucl.Phys.}
  {\bf B858} (2012) 1--47, [\href{http://xxx.lanl.gov/abs/1109.3454}{{\tt
  1109.3454}}]. 53 pages, 2 figures.

\bibitem{enumerative-combinatorics-2}
R.~P. Stanley, {\em Enumerative combinatorics. {V}ol. 2}, vol.~62 of {\em
  Cambridge Studies in Advanced Mathematics}.
\newblock Cambridge University Press, Cambridge, 1999.

\bibitem{MatsukiWeyl}
K.~Matsuki, {\it Weyl groups and birational transformations among minimal
  models},  {\em Mem. Amer. Math. Soc.} {\bf 116} (1995), no.~557 vi+133.

\bibitem{Esole:2014bka} 
  M.~Esole, S.~-H.~Shao and S.~-T.~Yau,
  {\it Singularities and Gauge Theory Phases},
    [\href{http://xxx.lanl.gov/abs/1402.6331}{{\tt 1402.6331}}].

\bibitem{CombinatoricsOfCoxeter}
A.~Bj{\"o}rner and F.~Brenti, {\em Combinatorics of {C}oxeter groups}, vol.~231
  of {\em Graduate Texts in Mathematics}.
\newblock Springer, New York, 2005.

\bibitem{gal-78}
N.~Bourbaki, {\em {G}roupes et alg\`ebres de {L}ie, {C}hap. {VII}, {VIII}}.
\newblock Hermann, Paris, 1975.

\bibitem{QuasiMinirepRef}
V.~Lakshmibai and K.~N. Raghavan, {\em Standard monomial theory}, vol.~137 of
  {\em Encyclopaedia of Mathematical Sciences}.
\newblock Springer-Verlag, Berlin, 2008.
\newblock Invariant theoretic approach, Invariant Theory and Algebraic
  Transformation Groups, 8.

\bibitem{FultonHarris}
W.~Fulton and J.~Harris, {\em Representation Theory}.
\newblock Springer-Verlag, 1991.

\bibitem{fibonacci-lattice}
R.~P. Stanley, {\it The {F}ibonacci lattice},  {\em Fibonacci Quart.} {\bf 13}
  (1975) 215--232,
  [\href{http://www.fq.math.ca/13-3.html}{{\tt
  http://www.fq.math.ca/13-3.html}}].

\bibitem{LieF}
P.~S. Aspinwall, S.~Katz, and D.~R. Morrison, {\it Lie groups, {C}alabi--{Y}au
  threefolds, and {F}-theory},  {\em Adv. Theor. Math. Phys.} {\bf 4} (2000)
  95--126, [\href{http://xxx.lanl.gov/abs/arXiv:hep-th/0002012}{{\tt
  arXiv:hep-th/0002012}}].

\bibitem{4d-transitions}
K.~Intriligator, H.~Jockers, P.~Mayr, D.~R. Morrison, and M.~R. Plesser, {\it
  {Conifold Transitions in M-theory on Calabi-Yau Fourfolds with Background
  Fluxes}},  \href{http://xxx.lanl.gov/abs/1203.6662}{{\tt 1203.6662}}.

\bibitem{Diaconescu:1998cn}
D.-E. Diaconescu and R.~Entin, {\it {C}alabi--{Y}au spaces and five-dimensional
  field theories with exceptional gauge symmetry},  {\em Nucl. Phys. B} {\bf
  538} (1999) 451--484,
  [\href{http://xxx.lanl.gov/abs/arXiv:hep-th/9807170}{{\tt
  arXiv:hep-th/9807170}}].

\bibitem{kleiman-ampleness}
S.~Kleiman, {\it Toward a numerical theory of ampleness},  {\em Ann. of Math.
  (2)} {\bf 84} (1966) 293--344.

\bibitem{MR662120}
S.~Mori, {\it Threefolds whose canonical bundles are not numerically
  effective},  {\em Ann. of Math. (2)} {\bf 116} (1982), no.~1 133--176.

\bibitem{genus-one}
V.~Braun and D.~R. Morrison, {\it {F-theory on Genus-One Fibrations}},
  \href{http://xxx.lanl.gov/abs/1401.7844}{{\tt 1401.7844}}.

\bibitem{Bershadsky:1996nh}
M.~Bershadsky, K.~A. Intriligator, S.~Kachru, D.~R. Morrison, V.~Sadov, and
  C.~Vafa, {\it {Geometric singularities and enhanced gauge symmetries}},  {\em
  Nucl.Phys.} {\bf B481} (1996) 215--252,
  [\href{http://xxx.lanl.gov/abs/hep-th/9605200}{{\tt hep-th/9605200}}].

\bibitem{MR2681776}
A.~Degeratu and K.~Wendland, {\it Friendly giant meets pointlike instantons?
  {O}n a new conjecture by {J}ohn {M}c{K}ay},  in {\em Moonshine: the first
  quarter century and beyond}, vol.~372 of {\em London Math. Soc. Lecture Note
  Ser.}, pp.~55--127.
\newblock Cambridge Univ. Press, Cambridge, 2010.

\bibitem{anomalies}
A.~Grassi and D.~R. Morrison, {\it Anomalies and the {E}uler characteristic of
  elliptic {C}alabi--{Y}au threefolds},  {\em Commun. Number Theory Phys.} {\bf
  6} (2012), no.~1 51--127, [\href{http://xxx.lanl.gov/abs/arXiv:1109.0042
  [hep-th]}{{\tt arXiv:1109.0042 [hep-th]}}].

\bibitem{Grassi91}
A.~Grassi, {\it On minimal models of elliptic threefolds},  {\em Math. Ann.}
  {\bf 290} (1991), no.~2 287--301.

\bibitem{codimthree}
P.~Candelas, D.-E. Diaconescu, B.~Florea, D.~R. Morrison, and G.~Rajesh, {\it
  Codimension-three bundle singularities in {F}-theory},  {\em J. High Energy
  Phys.} {\bf 06} (2002) 014,
  [\href{http://xxx.lanl.gov/abs/arXiv:hep-th/0009228}{{\tt
  arXiv:hep-th/0009228}}].

\bibitem{[P]}
H.~Pinkham, {\it Factorization of birational maps in dimension 3},  in {\em
  Singularities} (P.~Orlik, ed.), vol.~40, part 2 of {\em Proc. Symp. Pure
  Math.}, pp.~343--371, American Mathematical Society, 1983.

\bibitem{pagoda}
M.~Reid, {\it Minimal models of canonical 3-folds},  in {\em Algebraic
  Varieties and Analytic Varieties} (S.~Iitaka, ed.), vol.~1 of {\em Adv. Stud.
  Pure Math.}, pp.~131--180, Kinokuniya, 1983.

\bibitem{Slansky}
R.~Slansky, {\it {Group Theory for Unified Model Building}},  {\em Phys.Rept.}
  {\bf 79} (1981) 1--128.

\bibitem{pioneG}
P.~S. Aspinwall and D.~R. Morrison, {\it Non-simply-connected gauge groups and
  rational points on elliptic curves},  {\em J. High Energy Phys.} {\bf 07}
  (1998) 012, [\href{http://xxx.lanl.gov/abs/arXiv:hep-th/9805206}{{\tt
  arXiv:hep-th/9805206}}].

\bibitem{Braun:2013cb}
A.~P. Braun and T.~Watari, {\it {On Singular Fibres in F-Theory}},  {\em JHEP}
  {\bf 1307} (2013) 031, [\href{http://xxx.lanl.gov/abs/1301.5814}{{\tt
  1301.5814}}].

\bibitem{MW}
D.~R. Morrison and D.~S. Park, {\it F-theory and the {M}ordell--{W}eil group of
  elliptically-fibered {C}alabi--{Y}au threefolds},  {\em J. High Energy Phys.}
  {\bf 10} (2012) 128, [\href{http://xxx.lanl.gov/abs/arXiv:1208.2695
  [hep-th]}{{\tt arXiv:1208.2695 [hep-th]}}].

\bibitem{Mayrhofer:2012zy}
C.~Mayrhofer, E.~Palti, and T.~Weigand, {\it {U(1) symmetries in F-theory GUTs
  with multiple sections}},  \href{http://xxx.lanl.gov/abs/1211.6742}{{\tt
  1211.6742}}.

\bibitem{Braun:2013yti}
V.~Braun, T.~W. Grimm, and J.~Keitel, {\it {New Global F-theory GUTs with U(1)
  symmetries}},  {\em JHEP} {\bf 1309} (2013) 154,
  [\href{http://xxx.lanl.gov/abs/1302.1854}{{\tt 1302.1854}}].

\bibitem{Borchmann:2013jwa}
J.~Borchmann, C.~Mayrhofer, E.~Palti, and T.~Weigand, {\it {Elliptic fibrations
  for SU(5) x U(1) x U(1) F-theory vacua}},  {\em Phys.Rev.} {\bf D88} (2013)
  046005, [\href{http://xxx.lanl.gov/abs/1303.5054}{{\tt 1303.5054}}].

\bibitem{Cvetic:2013nia}
M.~Cvetic, D.~Klevers, and H.~Piragua, {\it {F-Theory Compactifications with
  Multiple U(1)-Factors: Constructing Elliptic Fibrations with Rational
  Sections}},  {\em JHEP} {\bf 1306} (2013) 067,
  [\href{http://xxx.lanl.gov/abs/1303.6970}{{\tt 1303.6970}}].

\bibitem{Grimm:2013oga}
T.~W. Grimm, A.~Kapfer, and J.~Keitel, {\it {Effective action of 6D F-Theory
  with U(1) factors: Rational sections make Chern-Simons terms jump}},  {\em
  JHEP} {\bf 1307} (2013) 115, [\href{http://xxx.lanl.gov/abs/1305.1929}{{\tt
  1305.1929}}].

\bibitem{Braun:2013nqa}
V.~Braun, T.~W. Grimm, and J.~Keitel, {\it {Geometric Engineering in Toric
  F-Theory and GUTs with U(1) Gauge Factors}},  {\em JHEP} {\bf 1312} (2013)
  069, [\href{http://xxx.lanl.gov/abs/1306.0577}{{\tt 1306.0577}}].

\bibitem{Cvetic:2013uta}
M.~Cvetic, A.~Grassi, D.~Klevers, and H.~Piragua, {\it {Chiral Four-Dimensional
  F-Theory Compactifications With SU(5) and Multiple U(1)-Factors}},
  \href{http://xxx.lanl.gov/abs/1306.3987}{{\tt 1306.3987}}.

\bibitem{Borchmann:2013hta}
J.~Borchmann, C.~Mayrhofer, E.~Palti, and T.~Weigand, {\it {SU(5) Tops with
  Multiple U(1)s in F-theory}},  \href{http://xxx.lanl.gov/abs/1307.2902}{{\tt
  1307.2902}}.

\bibitem{Cvetic:2013qsa}
M.~Cvetic, D.~Klevers, H.~Piragua, and P.~Song, {\it {Elliptic Fibrations with
  Rank Three Mordell-Weil Group: F-theory with U(1) x U(1) x U(1) Gauge
  Symmetry}},  \href{http://xxx.lanl.gov/abs/1310.0463}{{\tt 1310.0463}}.

\bibitem{KLS}
M.~Kuentzler, C.~Lawrie, and S.~Schafer-Nameki, {\it {To appear}}, .

\bibitem{MR0240238}
N.~Bourbaki, {\em {G}roupes et alg\`ebres de {L}ie, {C}hap. {IV}, {V}, {VI}}.
\newblock Hermann, Paris, 1968.

\bibitem{MR0252560}
J.~F. Adams, {\em Lectures on {L}ie groups}.
\newblock W. A. Benjamin, Inc., New York-Amsterdam, 1969.

\bibitem{demazure-del-pezzo}
M.~Demazure, {\it Surfaces de {D}el {P}ezzo, {II, III, IV, V}},  in {\em
  S\'eminaire sur les {S}ingularit\'es des {S}urfaces}, vol.~777 of {\em
  Lecture Notes in Math.}, pp.~21--69.
\newblock Springer-Verlag, 1980.

\end{thebibliography}

\providecommand{\href}[2]{#2}\begingroup\raggedright\endgroup

\end{document}